\documentclass[epj]{svjour}
\usepackage{lipsum}
\usepackage{mathtools}
\usepackage{cuted}
\usepackage{amsmath}
\usepackage{tabularx}
\usepackage{bbm}
\usepackage{hyperref}
\usepackage{graphicx}
\newcommand{\beq}{\begin{equation}}
\newcommand{\eeq}{\end{equation}}
\newcommand{\beqa}{\begin{eqnarray}}
\newcommand{\eeqa}{\end{eqnarray}}
\newcommand{\nn}{\nonumber \\ }
{}\newcommand{\fet}[1]{\mbox{\boldmath $#1$}}

\newcommand{\II}[4]{I^{\left(#1\right)}_{\left(#2,#3,#4 \right)}}
\newcommand{\IIR}[4]{I_{\rm red \ \left(#2,#3,#4 \right)}^{\left( #1 \right)}}

\newcommand{\vek}[4]{\left(\vec{#1}_{#3}\cdot\vec{#2}_{#4} \right)}
\newcommand{\xp}[4]{\left[\vec{#1}_{#3}\times \vec{#2}_{#4} \right]^3}
\newcommand{\cp}[4]{\left[\vec{#1}_{#3}\times \vec{#2}_{#4} \right]}
\newcommand{\spr}[6]{\left(\vec{#1}_{#4} \cdot\left[ \vec{#2}_{#5}\times\vec{#3}_{#6}\right] \right)}
\newcommand{\one}{\mathbbm{1}}
\usepackage{graphics}
\begin{document}
\title{Nuclear Currents in Chiral Effective Field Theory}
\author{Hermann Krebs%\inst{1} %\and Second author\inst{2}% etc
%\thanks is optional - remove next line if not needed
}                     % Do not remove
\offprints{}          % Insert a name or remove this line
\institute{Institute  for Theoretical  Physics  II,\\
Faculty of Physics and Astronomy,\\
Ruhr-Universit\"at  Bochum,  D-44780  Bochum,  Germany}
\date{Received: date / Revised version: date}
\abstract{
In this article, we review the status of the calculation of nuclear
currents within chiral effective field theory. After formal discussion
of the unitary transformation technique and its application 
to nuclear currents we will give all available expressions for vector,
axial-vector currents. Vector and axial-vector
currents
 will be discussed up to order $Q$ with leading-order contribution starting
at order $Q^{-3}$. Pseudoscalar and scalar currents  will be discussed up to
order $Q^0$ with leading-order contribution starting at order
$Q^{-4}$. This is a complete set of expressions in
next-to-next-to-next-to-leading-order (N$^3$LO) analysis for
nuclear scalar, pseudoscalar, vector and axial-vector
current operators.  Differences between vector and axial-vector currents
calculated via transfer-matrix inversion and unitary transformation
techniques are discussed. The importance of consistent regularization is
an additional point which is emphasized: lack of consistent
regularization of axial-vector current
operators is shown to lead to a violation of the chiral symmetry in the chiral
limit at order $Q$. For this reason a hybrid approach at order $Q$,
discussed in various publications, is non-applicable. To
respect the chiral symmetry the same regularization procedure needs to be
used in the construction of nuclear forces and current operators. Although
full expressions of consistently regularized current operators are not
yet available an isoscalar part of the electromagnetic charge operator
up to order $Q$ has a very simple form and can be easily 
regularized in a consistent way. 
As an application, we review our recent high accuracy calculation of the
deuteron charge form factor with a quantified error estimate.  
 % end of PACS codes
} %end of abstract
\maketitle
\section{Introduction}
\label{intro}
The internal structure of nucleons and nuclei can be studied by probing
them with electromagnetic, weak, or even
scalar probes. Scalar probes play an important role in beyond the standard
model search of dark matter. The interactions of hadrons with the external
probes are well approximated by one photon, $W^\pm$, $Z^0$. In this case, the full scattering amplitude
factorizes in a leptonic and a hadronic part. In the case of
electroweak interaction, the amplitude can be written
as a multiplication of leptonic and hadronic four-current operators. Leptonic four-current can be well approximated
by perturbative calculations within the standard model. Hadronic four-current is less known. Electroweak nuclear
currents have been extensively studied in the last century within
boson-exchange (pions and heavier mesons) and soliton models,
see~\cite{Riska:2016cud,Kubodera:2010qx} for recent and~\cite{Carlson:1997qn,Riska:1989bh} for
earlier reviews on this topic. Electromagnetic nuclear currents have
been reviewed in~\cite{Phillips:2016mov,Bacca:2014tla,Golak:2005iy,Arenhovel:1995ie}. One of the simplest approximations of
the nuclear current operator is
Impulse approximation (IA) where only one nucleon in a nucleus is probed
by an external source and other nucleons act as spectators. IA can be
expected to work well at higher energies. However, this
approximation is not satisfactory in the low-energy
sector. Riska and Brown showed in their seminal paper~\cite{Riska:1972zz} on radiative
capture of a thermal neutron on a proton, $n+p\to \gamma+d$, that $10\%$ discrepancy between the IA prediction and experiment
can be explained by taking into account the leading pion exchange
electromagnetic current between two nucleons which was calculated by
Villars~\cite{Villars:1947hpa} and took additionally $\Delta(1232)$
resonance and $\omega\to\pi + \gamma$ channel in to
account~\cite{Chemtob:1971pu}. 
%The key point in this calculation was
%that $^3D_1$ state of the deuteron could not be neglected and gave
%comparable contribution as $^1S_0$ channel. 
This was a start for the development of more sophisticated
meson exchange currents where heavier mesons and nucleon resonances
have been taken into account.  
The currents have been studied both in 
relativistic and non-relativistic formalisms. Relativistic approach
is 
more complicated than a non-relativistic one and is reviewed
e.g. in~\cite{Marcucci:2015rca,Garcon:2001sz}, see also~\cite{Polyzou:2010kx} for
relativistic Hamiltonian approach. In a non-relativistic
formalism one usually performs a Foldy-Wouthousen unitary transformation~\cite{Foldy:1949wa} and
eliminates in this way antinucleon contributions. In practical
calculations, relativistic
corrections are then treated in terms of one-over-nucleon-mass
expansion. Based on the studies of Poincare algebra~\cite{Krajcik:1974nv,Friar:1975zza} one can give a
systematic one-over-nucleon-mass expansion of wave functions and
currents~\cite{Close:1971we,Coester:1975hj,Friar:1977xh}. One can even
block-diagonalize the full Poincare algebra simultaneously reducing in
this way, quantum field theoretical problem to quantum mechanical
one~\cite{Gloeckle:1981js,Krebs:2016rqz}\footnote{This
  statement was proven by Gl\"ockle and M\"uller
  in~\cite{Gloeckle:1981js} only for a restrictive model. The proof 
  for a general field theory was given only recently~\cite{Krebs:2016rqz}. }. In this way one can either keep everything relativistic or
perform a large nucleon mass expansion of block-diagonalized
operators. Through the phenomenological studies of the nuclear currents of the
last century, one could gain very important insights into a general
construction of nuclear currents. The interrelation between nuclear forces
and currents was clearly emphasized to keep gauge symmetry
exact~\cite{Siegert:1937yt}. 
Gauging technique of nuclear forces were developed to
derive consistent nuclear currents out of nuclear forces which respect
explicitly the 
gauge symmetry~\cite{Sachs:1948zz,Nyman:1967tbh}. 
Off-shell and energy-dependence of the
nuclear forces and currents had been extensively studied. Block-diagonalization techniques were developed to construct
energy-independent nuclear
forces~\cite{Okubo:1954zz,Suzuki:PTP1983}. Extension of these
techniques, in particular unitary transformation technique, to a
construction of nuclear currents had
been presented in~\cite{Gari:1976kj}. The advantage of the procedure
presented in~\cite{Gari:1976kj} is a systematic 
construction of the nuclear currents if perturbation theory would
work. Within this procedure, a vector current has been studied up to
one-loop level in a meson exchange model~\cite{Meissner:1983wn} which
is of comparable complexity as the state of the art calculations of nuclear
currents in chiral effective field theory.
%\footnote{Note that this
%  tedious calculation was
%  performed without powerful tools like Mathematica or FORM which are
%  available nowadays.} 

Already in the early studies of the
nuclear current operators, the prominent role of the
chiral symmetry (symmetry of QCD if the quark masses are set to zero) in the nuclear forces and currents was well
appreciated~\cite{Rho:1981zi}. Basically in all realistic models the  longest range interactions
are governed by one-pion-exchange. For this reason, the chiral
symmetry was respected in lowest order
approximation in the low energy-momentum expansion. How to further
systematically improve phenomenological models and in
particular their connection to QCD was rather unclear. A groundbreaking
idea that made systematically improvable calculations of nuclear
forces and currents possible came with the birth
of the chiral perturbation
theory~\cite{Weinberg:1978kz,Gasser:1983yg}. Gasser and
Leutwyler showed in~\cite{Gasser:1983yg} that perturbative expansion
in small momenta and masses of pions divided by the chiral symmetry
breaking scale $\Lambda_\chi$ can be systematically performed beyond a
tree-level approximation~\cite{Weinberg:1978kz}. To organize the infinite number of
possible interactions they used naive dimensional analysis (power
counting scheme) which was proposed by
Weinberg~\cite{Weinberg:1978kz}. The
price which one has to pay is the appearance of more and more complicated Lagrangians with
unknown coefficients, so-called low energy constants (LEC), if higher
precision is required. The procedure in~\cite{Gasser:1983yg} allows
one to
approximate Green functions of
QCD in the pionic sector by chiral perturbation theory  in a
systematically improvable way~\cite{Leutwyler:1993iq}. Degrees of freedom in chiral
perturbation theory are pointlike pions which gain their structure at
higher orders in the chiral expansion (loop effects). Only a few years
later chiral perturbation theory was formulated in the presence of
matter field allowing to extend the formalism to nucleon degrees of
freedom~\cite{Gasser:1987rb}. Nucleon states appeared in~\cite{Gasser:1987rb} as initial and
final states which are on-shell. Strictly speaking, the formalism does
not allow to make any statement about off-shell dynamics of the
nucleons with a clear connection to QCD. However, within QCD calculated matrix elements with on-shell
nucleons in the initial and final states can be approximated in a
systematically improvable way by chiral perturbation theory. One
technical difficulty which arises with the description of nucleons within
chiral perturbation theory is the appearance of the nucleon mass which
is a hard scale. As a consequence nucleon mass divided by chiral
symmetry breaking scale is not small but of the order one. Naive
application of dimensional regularization in loop diagrams would
generate also terms proportional to positive powers of nucleon mass
and would destroy in this way a power counting. There are two
solutions to this problem: the first one is to perform a field
redefinition and eliminate nucleon mass from the nucleon propagator on
the path integral level reducing the theory to a non-relativistic
approach. Poincar\'e invariance is restored order by order in the form of a systematic large
nucleon mass expansion. The method is called heavy-baryon
approach~\cite{Jenkins:1990jv,Bernard:1992qa} and was successfully
applied to various scattering observables in the single-nucleon
sector~\cite{Bernard:1995dp}. Another method, called infrared
regularization, respects the Lorentz-invariance of the theory resuming the whole
large nucleon mass expansion without violation of power
counting~\cite{Becher:1999he}. In this formulation, one introduces
nonphysical cuts far away from the applicability region of the
theory. Nevertheless, in practical calculations, these cuts might have
long tails such that it is advantageous not to have them. Another
formulation of the relativistic theory without violation of the power
counting scheme can be realized by modification of the subtraction scheme.
In this modified scheme all power counting violating
terms which are caused by hard nucleon mass scale are absorbed into
available LECs. The method is called
extended-on-mass-renormalization-scheme~\cite{Gegelia:1999gf,Fuchs:2003qc}. Applications
of relativistic and non-relativistic chiral perturbation theory methods in the single-nucleon
sector are reviewed in~\cite{Bernard:2007zu}.

Extension of chiral perturbation theory to two- and more-nucleon sector was
pioneered by Weinberg~\cite{Weinberg:1990rz,Weinberg:1991um,Weinberg:1992yk}. The
difficulty in the two- and more-nucleon sectors is the existence of bound
states which makes the perturbative approach impossible. As a way out of
this Weinberg suggested using chiral perturbation theory for the calculation
of an effective potential, which is called
nuclear force. Observables like nuclear spectra can be extracted out of the
non-perturbative numerical solution of the Schr\"odinger equation
with chiral nuclear forces as input. The effective potential was originally defined as a set of
time-ordered diagrams without two-nucleon or more-nucleon intermediate
states. The
absence of these states makes a perturbative approach applicable. This
idea was followed by several groups. Already one year after original
publication~\cite{Weinberg:1991um} 
nuclear forces have been studied up to next-to-leading-order (NLO)
in chiral expansion by~\cite{Ordonez:1992xp}. Soon after this
publication next-to-next-to-leading-order (NNLO) corrections have been calculated
in~\cite{Ordonez:1993tn,vanKolck:1994yi,Epelbaum:2002vt}. At this
order, one has to take two-pion-exchange corrections into account. For
two-nucleon operators, they appear as one loop corrections and for
three-nucleon forces as tree-level
diagrams. Time-ordered perturbation theory (TOPT) gives a nice graphical
interpretation of the forces but introduced a drawback of
energy-dependence in nuclear forces. This makes it difficult to apply
them in a few- and many-body simulations. This drawback, however, was
cured with the application of unitary transformation technique for
construction of nuclear forces~\cite{Epelbaum:1998ka,Epelbaum:1999dj} and lead to properly
normalized energy-independent nuclear forces. Next-to-next-to-next-to-leading-order (N$^3$LO) corrections
to two-nucleon forces
have been calculated more than a decade
ago~\cite{Kaiser:2001at,Kaiser:2001pc,Kaiser:1999jg,Kaiser:1999ff}. Numerical
studies of these contributions, including fits of various short-range
LECs which appear at this order, have been performed by
Bonn-Bochum \cite{Epelbaum:2004fk} and Idaho
group~\cite{Entem:2003ft}. We call these forces as first-generation
nuclear forces in further discussion. At the same order, there are corrections to leading
three-nucleon forces which have been calculated
in~\cite{Bernard:2007sp,Bernard:2011zr}. At N$^3$LO also four-nucleon
forces start to contribute. Their analytical expressions can be found
in~\cite{Epelbaum:2007us,Epelbaum:2006eu}. Density-dependent 
interactions which are needed for applications in nuclear matter
studies have
been derived from the N$^3$LO three-nucleon forces
in~\cite{Kaiser:2019yrc,Kaiser:2018ige} and from the N$^3$LO  four-nucleon forces
in~\cite{Kaiser:2015lsa}, see~\cite{Holt:2013fwa} for a review on this
direction. A first numerical estimate of $^4$He
expectation values of
four-nucleon forces has been performed in~\cite{Rozpedzik:2006yi}.  Numerical implementations of
N$^3$LO three- and four-nucleon forces in a few-nucleon sector are
non-trivial and still under investigation. Only exploratory
studies 
%with first-generation two-nucleon forces 
have been presented
in~\cite{Huther:2019ont,Golak:2014ksa,Skibinski:2011vi} and various perturbative applications
in many-body sector have been considered
in~\cite{Drischler:2017wtt,Drischler:2016cpy,Drischler:2016djf,Hebeler:2015hla}. In these studies, however, one did not pay any attention to consistency
issues of regularization between the two-, three- and
four-nucleon forces. Nowadays, we know that a mismatch of dimensional
and cut-off regularizations leads to a violation of the chiral symmetry at
a one-loop level in three-nucleon forces which is N$^3$LO, the same is
true for the axial vector currents~\cite{Krebs:CD18Proceeding}. So
a more careful investigation is needed which is work in
progress. Construction and application of nuclear forces in chiral EFT
are reviewed in several comprehensive review articles, see
e.g.~\cite{Epelbaum:2019kcf,Machleidt:2011zz,Epelbaum:2012vx,Epelbaum:2008ga,Epelbaum:2005pn}. Three-nucleon forces within chiral
EFT have been reviewed
in~\cite{Hammer:2012id,KalantarNayestanaki:2011wz}. By now two-nucleon
forces
have been calculated up to next-to-next-to-next-to-next-to-leading-order (N$^4$LO) for the two-nucleon forces~\cite{Entem:2014msa,Entem:2017gor,Epelbaum:2014sza,Epelbaum:2014efa,Reinert:2017usi}. Even partial N$^5$LO
contributions have been considered~\cite{Entem:2015xwa}. First
applications of these second-generation chiral two-nucleon forces can
be found in~\cite{Volkotrub:2020lsr,Witala:2019ffj,Epelbaum:2018ogq,Binder:2018pgl,Binder:2015mbz}. N$^4$LO
corrections to three-nucleon forces have been considered only
partly~\cite{Krebs:2013kha,Krebs:2012yv}. Longest and
intermediate-range contributions have been
calculated. Various short-range interactions, however, are still under
construction. Their numerical implementations are under construction like in the case of
N$^3$LO three-nucleon forces. 

In parallel to chiral EFT activities where numerical calculations are
performed within a finite cut-off range, there was activity
on non-perturbative renormalization of the theory for arbitrary values
of cut-offs. A pioneering work towards this direction was published by
Kaplan, Savage and Wise
(KSW)~\cite{Kaplan:1998tg,Kaplan:1998we}. Based on unnaturally large
nucleon-nucleon scattering length the authors suggested using a different
power counting and to reorganize a resummation of the
effective potential. In their power counting pion physics and higher-order short-range interactions are treated perturbatively. Only the
leading-order short-range interactions are resumed. Although this approach leads
to a non-perturbatively renormalizable theory it showed a poor
convergence in
description of $^3S_1-$$^3D_1$ channel in nucleon-nucleon
scattering \cite{Fleming:1999ee}, see also~\cite{Kaplan:2019znu} for
recent discussion. In the same framework,
electromagnetic form factors of the deuteron \cite{Kaplan:1998sz} and
radiative capture $n+p\to d+\gamma$ were analyzed up to next-to-leading-order. KSW power counting is also used in a pionless EFT
where pions are treated as heavy degrees of freedom and are integrated
out, see~\cite{vanKolck:1998bw} and references therein. The expansion
is performed around the unitary
limit where two-nucleon scattering length diverges. We are not going to discuss in this review all important
developments in the
pionless EFT. A comprehensive review on this topic can be found in~\cite{Hammer:2019poc}.

Soon after Weinberg's seminal papers on nuclear forces
~\cite{Weinberg:1990rz,Weinberg:1991um} Park et al. presented the
first study of nuclear electroweak currents based on chiral
EFT~\cite{Park:1993jf,Park:1995pn} up to N$^3$LO~\footnote{Note that
  there is no contribution at the order $Q^{-2}$ for vector- and
  axial-vector current operators. For this reason, order $Q^{-1}$
  contributions to vector- and axial-vector currents are denoted here as 
NLO. This convention is similar to nuclear
forces where order $Q^0$ and $Q^2$ contributions are denoted as
leading-order (LO) and
NLO, respectively. One advantage of this convention is that
axial-vector current at NNLO depends on the same LECs as the
three-nucleon force at NNLO.} in chiral
expansion. However, these first calculations were incomplete: only irreducible
one-loop diagrams were considered, 
%~\footnote{It was correct to
%  take only irreducible diagrams into account. Irreducible diagrams
%  were, however, originally defined to be time-ordered diagrams without purely
%  nucleonic states. So in every intermediate state, there should be at
%  least one pion. This definition leads to an energy-dependent nuclear
%  force which is difficult to deal with in three- and more-nucleon
%  sector. For this reason in later publications the irreducibility of
%  a diagram was redefined~\cite{Epelbaum:2004fk} which lead to hermitian energy-independent
%  potentials. This redefinition makes axial vector current
%  calculation~\cite{Park:1993jf} incomplete.}
fourth-order
pion Lagrangian contributions were not taken into account, in the case
of vector current considerations of two-pion-exchange diagrams were
restricted to magnetic moment operator. The calculated vector currents
lead to an excellent description of the total cross section in radiative
neutron-proton capture at thermal energy in the hybrid calculation with
Argonne $v_{18}$ nuclear
force~\cite{Park:1994sr,Park:1995pn}. They
also successfully calculated proton-proton fusion rate~\cite{Park:1998wq} showing
that at N$^3$LO meson-exchange currents make a $4\%$-effect on
proton-proton fusion rate compared to the leading single-particle
Gamow-Teller matrix element. Polarized neutron-proton capture
within N$^3$LO currents was presented later in~\cite{Park:1999sz} where the
authors included also a short-range current contribution which was
ignored in~\cite{Park:1994sr,Park:1995pn}. Deuteron electromagnetic form factors have been
studied in~\cite{Walzl:2001vb,Phillips:2003jz,Phillips:2006im}. Magnetic moment and
radiative capture of thermal neutrons for three-nucleon observables
had been studied in~\cite{Song:2008zf}. After fitting short-range
current to the magnetic moment of the deuteron the cut-off dependence of
the results was significantly reduced. Application of the 
%(somewhat
%modified~\footnote{In these currents not all expressions from N$^3$LO
%  contributions were included. Also, some contributions from N$^4$LO
%  were taken into account. This was a merge between the standard nuclear
%  physics approach and chiral EFT. The authors called this more
%  effective EFT.}) 
currents to the solar hep process followed where
the authors calculated S-factor with an accuracy smaller than
$20\%$~\cite{Park:2001jn,Park:2002yp}. Application to muon capture on deuteron can
be found in~\cite{Ando:2001es}. Although the absorption of the muon by the
deuteron leads to the energetically higher region, the part of the capture
rate where two neutrons carry higher energy is known to be small such
that the dominant contribution comes from the energy region where two
outgoing neutrons carry low energy such that the formalism is still
applicable. Also contributions of meson-exchange
currents to triton $\beta$-decay were studied in~\cite{Gazit:2008ma}
where the authors tried to extract the low energy constant $D$ which
governs chiral three-nucleon force at NNLO.
Development of the second-generation of chiral EFT currents started with
the work~\cite{Walzl:2001vb,Pastore:2008ui,Pastore:2009is}, where also reducible-like diagrams had been taken
into account. %These are the diagrams which look reducible but the
%purely nucleonic cuts are shifted to one- two- or more-pion-nucleon
%cuts that are already available in the diagrams leading to the higher
%power of pion-energy propagators. 
These diagrams show up when one defines an
effective potential as a transition amplitude with subtracted iterated
parts. In this way, one gets an energy-independent nuclear force which
is much easier to deal with in calculations of three- and more-nucleon
observables. In~\cite{Pastore:2009is} (TOPT currents) the authors also considered chiral
nuclear forces at NLO level in order to derive consistent
chiral forces and currents using only chiral EFT and leaving in this
way a hybrid approach. In parallel to these activities chiral nuclear
vector currents have been derived by using unitary transformation
technique~\cite{Kolling:2009iq,Kolling:2011mt} (UT currents)where the same
off-shell scheme had been used as in~\cite{Epelbaum:2004fk}. Various
applications of the second-generation currents followed: Deuteron
electromagnetic form factors have been studied with UT currents
in~\cite{Kolling:2012cs}. Application to $^2$H and $^3$He
photodisintegration with UT currents has been studied
in~\cite{Rozpedzik:2011cx}. TOPT currents have been applied to thermal
neutron captures on deuteron and $^3$He in~\cite{Girlanda:2010vm}. To
solve the three- and four-body problem the authors used
hyperspherical-harmonics technique, see e.g.~\cite{Kievsky:2008es} for
a review. Electromagnetic form factors of deuteron and $^3$H and
$^3$He and deuteron photo(electro)-disintegration
have been studied in~\cite{Piarulli:2012bn,Schiavilla:2018udt}. Electromagnetic moments
and transitions have been studied for nuclei with $A\leq 9$ by using
Quantum Monte Carlo (QMC) formalism
in~\cite{Pastore:2012rp,Pastore:2014oda}. The second-generation axial-vector current has been presented in~\cite{Baroni:2015uza} within TOPT and
in~\cite{Krebs:2016rqz} within UT methods. Application of the TOPT current
to tritium $\beta$-decay has been discussed in~\cite{Baroni:2016xll}
and~\cite{Baroni:2018fdn}. Inclusive neutrino scattering off the deuteron has been
analyzed in~\cite{Baroni:2017gtk} where the authors find that the predicted
cross-sections are consistently larger by a couple of percents than
those given in
phenomenological analysis of Nakamura et
al.~\cite{Nakamura:2002jg,Shen:2012xz}. They also found a very tiny cut-off
dependence of the cross sections. QMC calculation of
weak transitions for $A=6-10$ have been presented
in~\cite{Pastore:2017uwc} where the authors calculated $\beta$-decays
of $^6$He and $^{10}$C and electron capture in $^7$Be. They found an
excellent agreement with experimental data for the electron captures
in $^7$Be and an overestimate of the $^6$He and $^{10}$C data by
$\sim 2\%$ and $\sim 10\%$, respectively. In the latter case, a phenomenological
AV18+IL7 wave function has been used. Preliminary results presented
in~\cite{Pastore:2017uwc}, however, indicate that once chiral EFT wave
functions~\cite{Piarulli:2017dwd,Piarulli:2014bda,Piarulli:2016vel} are used the discrepancy for $^{10}$C decreases from $10\%$
to $4\%$. In  more recent QMC studies of weak transitions in $A\leq
10$ nuclei~\cite{King:2020wmp} the authors
find in most cases an
agreement with experimental data. As input they used
N$^3$LO axial-vector currents and chiral EFT wave
functions~\cite{Piarulli:2017dwd,Piarulli:2014bda,Piarulli:2016vel}. Two-body currents contribute at the
$2-3\%$ level with exception of $^8$Li, $^8$B, and $^8$He $\beta$-decays. In the latter cases, the contribution of the impulse
approximation of the Gamow-Teller transition operator  (LO approximation) is
suppressed. Two-body currents provide a $20-30\%$ correction which is,
however, insufficient to achieve the agreement with experimental
data. Extensive $\beta$-decay studies from  light-, medium-mass nuclei to
$^{100}$Sn have been presented in~\cite{Gysbers:2019uyb}.
The authors used interactions and currents from chiral EFT~\cite{Hebeler:2010xb,Ekstrom:2015rta} in combination
with no-core shell model, valence-space in-medium similarity renormalization group,
and coupled-cluster approaches to cover the whole light- and
medium-mass nuclei sectors. They found an overall good description of experimental
data for light nuclei. Similar to QMC studies, they found for $A\leq 7$ nuclei that two-body current
contributions to the Gamow-Teller operator are relatively small. They also
found a substantial enhancement of $^8$He Gamow-Teller matrix elements
due to two-body currents.  For medium-mass nuclei the authors found a remarkably good agreement
of Gamow-Teller matrix elements. The inclusion
of the two-body currents and three-nucleon forces was essential for
the 
description of the data in the medium-mass nuclei sector.

The purpose of this work is to review the construction of nuclear
currents within chiral EFT. Electroweak, as well as
pseudoscalar and scalar two-nucleon current operators, will be discussed up to one-loop (two-pion-exchange)
approximation. In the main part of this manuscript, we will
concentrate on the unitary transformation technique. Gauge and chiral
symmetries as well as four-vector relations will be 
discussed. Another purpose of this work is to quantify the differences
between the unitary transformation technique used in the derivation of all
currents by our group and the currents derived by time-ordered
perturbation theory in combination with subtraction technique by
JLab-Pisa group. In the last part of this review, we will concentrate on
the symmetry preserving
regularization. We will show that keeping the chiral symmetry at one-loop level
will require a consistent regularization of nuclear forces and
currents which is still work in progress. In all available
calculations, sofar dimensional regularization has been used in the
construction of the current. In the practical calculations of
observable, the current operators are usually multiplied with a cut-off
regulator. We will show that this mismatch of the regularizations leads
to the chiral symmetry violation at one-loop order. To cure this it is
necessary to calculate both nuclear forces and currents with the same
regulator which respects gauge and the chiral symmetry by construction. 

We start in Sec.~\ref{sec:blockdiagonalization} with the presentation
of the unitary transformation technique for  nuclear forces. Extension of
this technique to nuclear currents will be discussed in
Sec.~\ref{sec:nuclearcurrent}. In Sec.~\ref{sec:consistencycheck} we will discuss consistency
checks like a four-vector relation or continuity equations. Those have
to be satisfied with any effective current operators. Sec.~\ref{nuclear:chiralEFT} is
devoted to current operators within chiral EFT where we list all
expressions for vector, axial-vector, pseudoscalar, and scalar current
operators which are obtained within a unitary transformation technique
up to N$^3$LO. In Sec.~\ref{compare:with:pisa} we compare our results with those
obtained by JLab-Pisa group via using time-ordered perturbation theory
in a combination with a transfer-matrix inversion technique. In
Sec.~\ref{RegularizationPath} we discuss a path towards construction of
consistently regularized nuclear forces and currents. We will
demonstrate that a naive use of the dimensional regularization for currents
in combination with a cutoff regularization of all operators in the
Schr\"odinger or Lippmann-Schwinger equations leads to a violation of
the chiral symmetry at N$^3$LO. For this reason at this level of precision, we have to use
a consistent regularization with the same symmetry-preserving regulator in nuclear
forces and currents. In Sec.~\ref{first:application:deuteron:charge} we will discuss a deuteron charge
operator which is calculated within a consistent and symmetry-preserving
higher derivative regulator. A  high precision determination of the
deuteron charge form factor allows for very precise extraction of the
neutron radius. Lengthy expressions for two-pion exchange vector and
scalar currents as well as technical details about folded-diagram technique, transfer-matrix with time-dependent interactions, and
a derivation of the continuity equations are given in the Appendices.
\section{Block diagonalization of Hamilton operator}
\label{sec:blockdiagonalization}
The CHPT Hamiltonian up to a
given order in chiral expansion has a rather simple form but operates
on the full Fock-space which includes all possible pion-nucleon states. Nonperturbative
calculations of the amplitude with its input requires the quantum field
theoretical methods which
are very complicated. In order to reduce the complexity of the
calculation it is advantageous to decompose the full Fock space ${\cal
  F}$ into a
model space and a rest space. In our case the model space ${\cal M}$ will be
generated by states which include only nucleons. All other states like
the state with one or more pions or delta resonance belong to the rest
space ${\cal R}$
\beqa
{\cal F}&=&{\cal M} \oplus {\cal R}.
\eeqa
Let us denote by $\eta$ and $\lambda$ projector operators which
project ${\cal F}$ to ${\cal M}$ and ${\cal R}$, respectively. In the
absence of external sources there is no energy-momentum flow into the
system. Translation invariance guarantees energy conservation. For this reason, after
switching off pseudoscalar, vector and axial vector sources, the
Hamilton operator becomes time independent and we can start with the
stationary Schr\"odinger equation
\beqa
H\,|\psi\rangle&=&E\,|\psi\rangle.
\eeqa
In the first step we look for a unitary transformation which brings
the Hamilton operator $H$ into a block diagonal form such that the
stationary Schr\"odinger equation is restricted to model space
\beqa
\eta U^\dagger H U \eta |\phi\rangle&=& E \eta |\phi\rangle,
\eeqa
where 
\beq
|\phi\rangle=U^\dagger|\psi\rangle,
\eeq
and due to block - diagonalization we have
\beq
\eta U^\dagger H U\lambda =\lambda U^\dagger H U\eta = 0.\label{decopling_general}
\eeq
A unitary transformation which satisfies
Eq.~(\ref{decopling_general}) can be constructed via an ansatz of Okubo~\cite{Okubo:1954zz}
\beqa
\eta U \eta &=&\big(\eta+A^\dagger A\big)^{-1/2}, \quad \eta U \lambda
\,=\, - A^\dagger\big(1+ A A^\dagger\big)^{-1/2},\nn
\lambda U \eta &=& A\big(1+A^\dagger A)^{-1/2},\quad \lambda U\lambda \,=\,\big(\lambda+A A^\dagger\big)^{-1/2},\label{U_expressed_in_A}
\eeqa
where the operator $A$ satisfies
\beqa
A&=&\lambda A \eta,
\eeqa
and a nonlinear decoupling equation
\beqa
\lambda\big(H - \big[A, H\big] - A H A\big)\eta &=&0.\label{decoupling_eq_for_A_op}
\eeqa
Eq.~(\ref{decoupling_eq_for_A_op}) can be solved within chiral
perturbation theory~\cite{Epelbaum:1998ka,Epelbaum:1999dj}. The effective Hamiltonian is given by 
\beqa
H_{\rm eff}&=&  U^\dagger H U\label{UT_effective_potential_okubo}
\eeqa
 It
is important to note that
the unitary transformation $U$ of Eq.~(\ref{U_expressed_in_A}) is not
unique. Any additional transformation of the $\eta$-space for example will
not affect decoupling conditions of
Eq.~(\ref{decopling_general}). This degree of freedom can be used in
order to achieve  renormalizability of the effective potential.  Renormalizability of
$H_{\rm eff}$ means that it becomes finite after performing dimensional
regularization with beta functions taken from the pion and one-nucleon
sector which are specified in~\cite{Gasser:2002am,Fettes:1998ud,Ecker:1995rk}. Explicit construction of
the operator $U$ is reviewed in~\cite{Epelbaum:2005pn}.
Recent calculations of nuclear forces are performed
up to N$^4$LO in the chiral expansion which corresponds to the full two-loop
calculation for NN- and full one-loop calculation for 3N-operators. 
\section{Nuclear current operator}
\label{sec:nuclearcurrent}
In order to construct a nuclear current we start with the chiral
perturbation theory Hamiltonian in the presence of external sources,
see Appendix~\ref{ext_sources_app}, 
and define the effective Hamiltonian in a similar way via
\beqa
H_{\rm eff}[s,p,a,v]&=&U^\dagger H[s,p,a,v] U,\label{HeffWithsources}
\eeqa
where $s,p,a$ and $v$ denote  scalar, pseudoscalar, axial-vector
and vector sources, respectively. Here we use the same unitary
transformation $U$ which leads to a block-diagonal effective Hamiltonian
in the absence of external pseudoscalar, axial-vector and vector 
sources. Scalar source is set to the light quark mass matrix. Note
that the effective Hamiltonian of Eq.~(\ref{HeffWithsources}) is not
block-diagonal such that in general
\beq
\eta H_{\rm eff}[s,p,a,v] \lambda \neq 0 \neq \lambda H_{\rm eff}[s,p,a,v] \eta.
\eeq
Only the strong part of the Hamiltonian is block-diagonal:
\beqa
\eta H_{\rm eff}[m_q,0,0,0] \lambda&=& \lambda H_{\rm eff}[m_q,0,0,0] \eta \,=\, 0.
\eeqa
There is, however, no reason for block-diagonalization of $H_{\rm
  eff}[s,p,a,v]$ since we only want to consider expectation values of
the current operator and are not interested in its non-perturbative
iterations. We can derive a current operator out of the effective Hamiltonian in the presence of the
external sources by
\beqa
J_X&=&\frac{\delta}{\delta X}H_{\rm eff}[s,p,a,v]\Big|_{s=m_q,p=a=v=0},
\eeqa
where $X$ stays for $s,p,a$ or $v$ depending on which kind of nuclear
current we are interested in. With $H_{\rm eff}$ from
Eq.~(\ref{HeffWithsources}), however, we will get a singular current
which is non-renormalizable. In order to work with renormalizable
current we need to apply further unitary transformation on the effective
Hamiltonian which depends explicitly on external sources. Due to its explicit
dependence on external sources this additional unitary transformation
becomes time dependent. In order to understand how $H_{\rm eff}$
changes under time-dependent unitary transformation $U(t)$ 
consider a state in the Schr\"odinger picture which satisfies a time-dependent Schr\"odinger equation
\beq
i \frac{\partial}{\partial t}|\phi(t)\rangle = H_{\rm eff}[s,p,a,v]|\phi(t)\rangle.
\eeq
The state $|\phi(t)\rangle$ contains all information of the quantum
system in the presence of external sources. We can rewrite this
equation by multiplying left hand side and right hand side 
by $U(t)^\dagger$ and inserting a unity operator we get
\beq
 i \frac{\partial}{\partial t}|\phi^\prime(t)\rangle = H_{\rm eff}^\prime[s,p,a,v]|\phi^\prime(t)\rangle,
\eeq
where 
\beq
|\phi^\prime(t)\rangle = U(t)^\dagger |\phi(t)\rangle
\eeq
and
\beqa
H_{\rm eff}^\prime[s,\dot{s},p,\dot{p},a,\dot{a},v,\dot{v}]&=&U(t)^\dagger H_{\rm eff}[s,p,a,v] U(t)
\nn
&+&
\bigg(i\frac{\partial}{\partial t} U^\dagger(t)\bigg)U(t).\label{unitaryTrTimeDep}
\eeqa
The renormalizable current operator can be generated out of $H_{\rm
  eff}^\prime$. The momentum space currents are defined
\beqa
\tilde J_X(k)&=&\frac{\delta}{\delta \tilde X(k)}H_{\rm eff}^\prime, \label{JXcurrent}
\eeqa
where $H_{\rm eff}^\prime$ is taken at $x_0=0$ and $s=m_q,\dot
s=p=\dot p=a=\dot a=v=\dot v=0$. $X$ stays for $s,p,a$
or $v$ in dependence which current we are considering and
\beqa
\tilde X(k)=\int\frac{d^4 x}{(2\pi)^4} X(x) e^{i\, k\cdot x}.
\eeqa
Note, that due to time derivative term in
Eq.~(\ref{unitaryTrTimeDep}), the current $\tilde J_X(k)$ becomes energy-transfer
dependent.  The explicit form of the unitary transformations $U$ and
$U(t)$ can be found in~\cite{Krebs:2016rqz}.

The energy-transfer dependent current of Eq.~(\ref{JXcurrent}) has a specific general
structure. To see this, let us parametrize the unitary transformations
from Eq.~(\ref{unitaryTrTimeDep}) by
\beqa
U(t)&=&\exp\left(i\,{\cal U}(t)\right),
\eeqa
where the hermitian operator ${\cal U}$ has a form
\beqa
{\cal U}(t)&=&\int d^3 x \big[v_\mu^c(x) {\cal V}^{\mu,
      c}(\vec{x})+a_\mu^c(x) {\cal A}^{\mu, c}(\vec{x})\nn
&+&p_c(x)
    {\cal P}^c(\vec{x})+s_c(x) {\cal S}^c(\vec{x})\big],
\eeqa
where we use Einstein-convention for the space-time and
isospin-indices.The momentum-space form of this
operator is given by
\beqa
{\cal U}(t)&=&\int d^4p \int \frac{d^3 x}{(2\pi)^3} \,e^{-i\,p\cdot x} \big[{\tilde v}_\mu^c(p) {\cal V}^{\mu,
      c}(\vec{x})\nn
&+&{\tilde a}_\mu^c(p) {\cal A}^{\mu, c}(\vec{x})
+{\tilde p}_c(p)
    {\cal P}^c(\vec{x})+\tilde{s}_c(p) {\cal S}^c(\vec{x})\big]\nn
&=&\int d^4p  \,e^{-i\,p_0 t} \big[{\tilde v}_\mu^c(p) {\tilde{\cal V}}^{\mu,
      c}(-\vec{p})
+{\tilde a}_\mu^c(p) {\tilde{\cal A}}^{\mu, c}(-\vec{p})\nn
&+&{\tilde p}_c(p)
    {\tilde{ \cal P}}^c(-\vec{p})+\tilde{s}_c(p) {\tilde { \cal S}}^c(-\vec{p})\big].
\eeqa
For the current operator of Eq.~(\ref{JXcurrent}) we get
\beqa
\tilde J_X(k)&=&\frac{\delta}{\delta \tilde X(k)} H_{\rm eff}[s,p,a,v]\nn
&+&i\,\big[H_{\rm eff}[m_q,0,0,0],\frac{\delta}{\delta \tilde X(k)}
{\cal U}(0)\big]-\frac{\partial}{\partial t}{\cal U}(t)\bigg|_{t=0}\nn
&=&\frac{\delta}{\delta \tilde X(k)} H_{\rm eff}[s,p,a,v]\nn
&+&i\,\big[H_{\rm eff}[m_q,0,0,0], {\tilde {\cal X}}(-k)\big]-i k_0\,{\tilde {\cal X}}(-k),\label{k0DependenceJX}
\eeqa
where $ {\tilde {\cal X}}$ stays for ${\tilde { \cal S}}, {\tilde{ \cal
    P}}, {\tilde{\cal A}}$ or ${\tilde{\cal V}}$ dependent on which
current we consider, and all sources are set to zero after the
functional derivative is taken. Eq.~(\ref{k0DependenceJX}) shows a general
form of the energy-transfer dependent current. In the first line of
Eq.~(\ref{k0DependenceJX}) we see a current operator 
\beq
\frac{\delta}{\delta \tilde X(k)} H_{\rm eff}[s,p,a,v]
\eeq
which denotes the current with all phases of the time-dependent
transformations put to zero. The part proportional to the phases of the additional
time-dependent transformations is in the second line of
Eq.~(\ref{k0DependenceJX}). We see that the energy-transfer dependent part
of the current is always accompanied with the commutator of the same
structure with the nuclear force. An expectation  value of the
the second line of Eq.~(\ref{k0DependenceJX}) vanishes on-shell
when $k_0=E_f-E_i$ where $E_f$ and $E_i$ are final and initial
eigenenergies of the nuclear force $H_{\rm eff}[m_q,0,0,0]$,
respectively. This result is certainly expected since unitary transformations
can not affect observables.
\section{Consistency checks}
\label{sec:consistencycheck}
There are various consistency checks which the nuclear vector and axial-vector
current have to satisfy, in general. These relations are rooted in various symmetries of
the currents.
\subsection{Four-vector relation}
\label{sec:4vectorRel}
Vector and axial-vector currents are four-vectors and thus satisfy
\beqa
e^{-i \vec{e}\cdot\vec{K}\theta}{ J}_\mu^H(x)e^{i
\vec{e}\cdot\vec{K}\theta}&=&\Lambda(\theta)_\mu^{\,\,\,\,\,\nu}
{ J}_\nu^{H}(\Lambda(\theta)^{-1} x),
\eeqa
where ${ J}_\mu^H$ is a (axial) vector current in Heisenberg picture,  $\vec{K}$ is a boost generator, $\vec{e}$ is a boost
direction, $\theta$ is a boost angle and $\Lambda(\theta)$ is a
$4\times 4$ boost matrix. After a block-diagonalizing unitary
transformation this equation turns to
\beqa
e^{-i \vec{e}\cdot\vec{K}_{{\rm eff}}\theta}{ J}_\mu^{H_{{\rm eff}}}(x)e^{i
\vec{e}\cdot\vec{K}_{{\rm eff}}\theta}&=&\Lambda(\theta)_\mu^{\,\,\,\,\,\nu}
{ J}_\nu^{H_{{\rm eff}}}(\Lambda(\theta)^{-1} x),\quad\quad\label{boostGeneral}
\eeqa
where
\beqa
H_{{\rm eff}}&=&U^\dagger H U,\nn
{\vec K}_{{\rm eff}}&=&U^\dagger {\vec K} U,
\eeqa
and
\beqa
J_\mu^{H_{{\rm eff}}}(x)&=& e^{i\,H_{{\rm eff}} x_0} U^\dagger
  J_\mu(\vec{x}) U e^{-i\,H_{{\rm eff}} x_0}.\label{effectiveCurrentFormal}
\eeqa
In order to keep the notation short we denote from now on the effective
current $U^\dagger J_\mu(\vec{x}) U$ in the Schr{\"o}dinger picture by
$J_\mu(\vec{x})$ such that Eq.~(\ref{effectiveCurrentFormal}) turns in
this notation to
\beqa
J_\mu^{H_{{\rm eff}}}(x)&=& e^{i\,H_{{\rm eff}} x_0} 
  J_\mu(\vec{x}) e^{-i\,H_{{\rm eff}} x_0}.\label{effectiveCurrentFormalModified}
\eeqa
Note that the effective boost operator $\vec{K}_{\rm eff}$ has a block-diagonal form like
the effective Hamiltonian $H_{\rm eff}$. The reason is that the whole
Poincar\'e algebra get's a block-diagonal form after the application of
unitary transformation $U$. A perturbative proof of this statement to all orders can
be found in~\cite{Gloeckle:1981js} for a special model and in
\cite{Krebs:2016rqz} for an arbitrary local field theory.

Expanding Eq.~(\ref{boostGeneral}) in $\theta$ and comparing the
coefficients we get 
\beqa
&&-i\,\Big[\vec{e}\cdot\vec{K}_{\rm eff}, { J}_\mu^{H_{{\rm eff}}}(x)\Big]=
{ J}_\mu^{H_{{\rm eff}},\perp}(x) - x_\alpha^\perp\partial_x^\alpha{
  J}_\mu^{H_{{\rm eff}}}(x),\nn
&&\label{KeffJeffKommRelation}
\eeqa
where we used
\beqa
\Lambda(\theta) x &=& x + \theta x^\perp + {\cal O}(\theta^2 ),
\eeqa
and
\beqa
x^\perp&=&(\vec{e}\cdot\vec{x},\vec{e} \,x_0),
\eeqa
where ${\bf e}$ is a unit vector which is a boost direction.
In the next step, we use the Poincar\'e algebra relation to get
\beqa
e^{-i\,H_{{\rm eff}} x_0}\vec{K}_{{\rm eff}} e^{i\,H_{{\rm eff}}
  x_0}&=&\vec{K}_{{\rm eff}} - x_0 \vec{P},\label{KeffCommWithHeffRelation}
\eeqa
where $\vec{P}$ denotes the momentum operator. Using
Eq.~(\ref{KeffCommWithHeffRelation}) we can rewrite
Eq.~(\ref{KeffJeffKommRelation}) into a well known relation~\cite{Coester:1975hj} 
\beqa
-i\,\Big[\vec{e}\cdot\vec{K}_{\rm
  eff}, { J}_\mu(\vec{x})\Big]&=&
%-i\,x_0\Big[\vec{e}\cdot\vec{P}, {
%  j}_\mu(\vec{x})\Big] + 
{
  J}_\mu^\perp(\vec{x}) 
-
i\vec{e}\cdot\vec{x}\big[H_{{\rm eff}},J_\mu(\vec{x})\big] 
%-x_0\,\vec{e}\cdot\vec{\nabla}_x j_\mu(\vec{x})
,\quad\quad\label{poincareCoordinateSpaceGeneral}
\eeqa
where we used the relation~\footnote{For $\vec{x}=0$ e.g. this
  relation can be found in Eqs. (93a) and (93b) of the
seminal work of Friar~\cite{Friar:1977xh}, see also~\cite{Close:1971we} for more general case.}
\beqa
\Big[i\vec{e}\cdot\vec{P}, {
  J}_\mu(\vec{x})\Big]&=&-\vec{e}\cdot\vec{\nabla}_x J_\mu(\vec{x}).
\eeqa
Now we transform Eq.~(\ref{poincareCoordinateSpaceGeneral}) into
momentum space and get
\beqa
-i\,\Big[\vec{e}\cdot\vec{K}_{\rm
  eff}, \tilde{ J}_\mu(\vec{k})\Big]&=&
%-i\,x_0\Big[\vec{e}\cdot\vec{P}, {
%  j}_\mu(\vec{x})\Big] + 
{
  J}_\mu^\perp(\vec{k}) 
-
\vec{e}\cdot\vec{\nabla}_k\big[H_{{\rm eff}},\tilde{J}_\mu(\vec{k})\big] 
%-x_0\,\vec{e}\cdot\vec{\nabla}_x j_\mu(\vec{x})
,\quad\quad\label{poincareCoordinateSpaceGeneralMomentumSpace}
\eeqa
where
\beqa
\tilde{J}_\mu(\vec{k})&=&\int d^3 x \,e^{i\,\vec{k}\cdot\vec{x}} J_\mu(\vec{x}).
\eeqa
At this stage the effective current operator $\tilde J_\mu(\vec{k})$
does not depend on energy-transfer $k_0$ since sofar we did not apply any
time-dependent unitary transformation. As we are interested in a more
general current operator we apply now these transformations and get
\beqa
\tilde{J}_\mu(\vec{k})&\to&\tilde{J}_\mu(\vec{k}) + i\,k_0
\tilde{Y}_\mu(\vec{k}) -i\, \big[H_{{\rm eff}}, \tilde{Y}_\mu(\vec{k})\big],\quad\quad
\eeqa
where $\tilde Y_\mu(\vec{k})$ is some local hermitian operator. Our goal is
to derive a consistency relation for the operator $\tilde{J}_\mu(k)$
which should be a generalization of
Eq.~(\ref{poincareCoordinateSpaceGeneralMomentumSpace}). The only information about
an operator $\tilde Y_\mu(\vec{k})$ which we will use is its locality property
\beqa
\big[\vec{P},\tilde Y_\mu(\vec{k})\big]&=&-\vec{k}\, \tilde Y_\mu(\vec{k}).
\eeqa
In particular, we do \underline{not} require the operator $\tilde Y_\mu(\vec{k})$ to
be a four-vector. In order to derive the relation we rewrite the
commutator
\beqa
\big[\vec{K}_{{\rm eff}}, \big[H_{{\rm
    eff}},\tilde{Y}_\mu(\vec{k})\big]\big]&=&\big[H_{{\rm eff}},
\big[\vec{K}_{{\rm eff}},\tilde{Y}_\mu(\vec{k})\big]\big] \nn
&+& 
\big[\big[\vec{K}_{{\rm eff}}, H_{{\rm
    eff}}\big],\tilde{Y}_\mu(\vec{k})\big]\big].
\eeqa
Using Poincar\'e algebra relation
\beqa
\big[\vec{K}_{{\rm eff}}, H_{{\rm eff}}\big]&=&-i\,\vec{P},
\eeqa
we get
\beqa
\big[\vec{K}_{{\rm eff}}, \big[H_{{\rm
    eff}},\tilde{Y}_\mu(\vec{k})\big]\big]&=&\big[H_{{\rm eff}},
\big[\vec{K}_{{\rm eff}},\tilde{Y}_\mu(\vec{k})\big]\big] \nn
&+& i\,\vec{k} \tilde{Y}_\mu(\vec{k}).
\eeqa
Using this result in combination with Eq.~(\ref{poincareCoordinateSpaceGeneralMomentumSpace}) we get
\beqa
&-&i\,\Big[\vec{e}\cdot\vec{K}_{\rm eff},\tilde { J}_\mu(k)\Big]\,=\,\tilde
{ J}_\mu^\perp(k) - \vec{e}\cdot\vec{\nabla}_k\Big[H_{\rm eff},\tilde
{ J}_\mu(k)\Big]\nn
&-&\vec{e}\cdot\vec{k}\frac{\partial}{\partial k_0}\tilde { J}_\mu(k)
+ i\Big[H_{\rm eff},\tilde{ X}_\mu(\vec{k})\Big] - i k_0 \tilde{ X}_\mu(\vec{k}),\label{PoincareGeneralConsistencyRelation}
\eeqa
where a hermitian operator ${ X}_\mu$ is defined by
\beqa
\tilde{X}_\mu(\vec{k})&=&\tilde{Y}_\mu^\perp(\vec{k})
+i\,\big[\vec{e}\cdot\big(\vec{K}_{{\rm eff}} + i\, H_{{\rm
    eff}} \vec{\nabla}_k \big),\tilde{Y}_\mu(\vec{k})\big]. \quad\quad\label{XmuDefinition}
\eeqa
Note, $\tilde{J}_\mu(k)$ is linear in
$k_0$, so we have
\beqa
\tilde{Y}_\mu(\vec{k})&=&-i\frac{\partial}{\partial k_0}\tilde{J}_\mu(k).
\eeqa
For this reason, we can rewrite the $\tilde{X}_\mu(\vec{k})$ operator in
terms of $\tilde{J}_\mu(k)$. 

Eq.~(\ref{PoincareGeneralConsistencyRelation}) is a final consistency relation
which builds on a four-vector property of the (axial) vector
current. A somewhat different derivation of this result can be found
in~\cite{Krebs:2016rqz} where, however, we did not specify the form of
the operator $\tilde{X}_\mu(\vec{k})$. In this respect, 
Eq.~(\ref{PoincareGeneralConsistencyRelation}) with the additional Eq.~(\ref{XmuDefinition}) includes more information
than Eq. (2.78) of~\cite{Krebs:2016rqz}.

Eq.~(\ref{PoincareGeneralConsistencyRelation}) relates the charge and
current operators with each other. In particular it allows to extract
a charge operator out of the current operator. To see this we multiply
the Eq.~(\ref{PoincareGeneralConsistencyRelation}) by $(0,-\vec{e})$
and get
\beqa
&&\tilde{J}_0(k)+\bigg[H_{\rm eff},\frac{\partial}{\partial k_0}\tilde
J_0(k)\bigg]-k_0\frac{\partial}{\partial k_0}{\tilde J}_0(k)
=\nn
&&-i\,\bigg[Z,\bigg(1-k_0\frac{\partial}{\partial k_0}\bigg)\vec{e}\cdot\vec{\tilde
  J}(k)\bigg]+\vec{e}\cdot\vec{k}\frac{\partial}{\partial
  k_0}\vec{e}\cdot\vec{\tilde J}(k)\nn
&&-i\,\bigg[H_{\rm eff}, \big[Z,\frac{\partial}{\partial k_0}\vec{e}\cdot\vec{\tilde
  J}(k)\big]\bigg], \label{j0::OffShellBoost}
\eeqa
where
\beq
Z=\vec{e}\cdot\big(\vec{K}_{{\rm eff}} + i\, H_{{\rm
    eff}} \vec{\nabla}_k \big). 
\eeq
Since the sum of the second and third
term on the left hand side of  Eq.~(\ref{j0::OffShellBoost}) is
unobservable,
\beqa
&&\langle\alpha|\bigg(\bigg[H_{\rm eff},\frac{\partial}{\partial
  k_0}\tilde J_0(k)\bigg]-k_0 \frac{\partial}{\partial k_0}\tilde
J_0(k)\bigg)|\beta\rangle=\nn
&&(E_\alpha-E_\beta-k_0)\langle\alpha|\frac{\partial}{\partial k_0}\tilde
J_0(k)|\beta\rangle = 0,
\eeqa
Eq.~(\ref{j0::OffShellBoost}) determines the charge operator (modulo
unobservable off-shell effects) once the
current operator is known. So for practical calculations one can
always use the right hand side of Eq.~(\ref{j0::OffShellBoost}) as an
energy transfer independent charge operator. This operator, however,
can not be used to test the continuity equation since in that case the
off-shell information of the charge operator is essential, see next
paragraph for explanation. It is interesting that the boost
transformation constrains the charge operator in such a way that one
can express it either by longitudinal current, for the choice
$\vec{e}=\vec{k}/|\vec{k}|$, or by transverse current for the choice
$\vec{e}=\vec{k}^\perp$, or by linear combination of them for other
choices of boost direction. 
\subsection{Continuity equation}
\label{sec:continuityeq:general}
Chiral effective field theory is by construction invariant under chiral ${\rm
  SU}(2)_L\times{\rm SU}(2)_R$ as well as $U(1)_V$
transformations. As a consequence this leads to various Ward -
identities for amplitudes and continuity equations for current
operators. Continuity equation for the currents follows directly from
the requirement that the Hamilton operator in the presence of external
sources is unitary equivalent to the Hamiltonian in the presence of
transformed external sources. This means that there exists a
(time-dependent) unitary transformation ${\cal U}(t)$ such that
\beqa
 &&H_{\rm eff}^\prime[s^\prime,\dot{s}^\prime, p^\prime,\dot{p}^\prime,
 a^\prime,\dot{a}^\prime,v^\prime,\dot{v}^\prime]=\bigg(i\frac{\partial}{\partial
   t}{\cal U}(t)^\dagger\bigg){\cal U}(t)\nn
&+&{\cal U}(t)^\dagger H_{\rm eff}^\prime[s,\dot{s}, p,\dot{p},
 a,\dot{a},v,\dot{v}]{\cal U}(t).\label{HamiltonChiralTransfGeneral}
\eeqa
Here, the primed external sources denote the transformed
sources. Considering infinitesimal chiral or $U(1)_V$ transformations, 
expanding both sides of Eq.~(\ref{HamiltonChiralTransfGeneral}) up
to the first order in transformation angles, and comparing the
coefficients in front of transformation angles we get  the continuity
equation. For the vector current, we get
\beqa
{\cal C}_{V,A}(\vec{k},0) + \big[H_{\rm eff},\frac{\partial}{\partial
  k_0}{\cal C}_{V,A}(\vec{k},k_0)\big]&=&0, \label{continuity_va}
\eeqa 
where for the electromagnetic vector current, the quantity ${\cal
  C}_V$ is defined via 
\beqa
{\cal C}_V(k)&=&\big[H_{\rm eff},\tilde{ V}_0(k)\big] -
\vec{k}\cdot\vec{\tilde V}(k),
\eeqa
and for the axial vector current, ${\cal C}_A$ is 
\beqa
{\cal C}_A(k)&=&\big[H_{\rm eff},\tilde{ A}_0^a(k)\big] -
\vec{k}\cdot\vec{\tilde A}^a(k) + i m_q \tilde {P}^a(k).
\eeqa
Here we denote vector, axial vector and pseudoscalar currents in
momentum space by $\tilde {V}_\mu(k), \tilde {A}_\mu^a(k)$ and
$\tilde {P}^a(k)$, respectively. In the derivation of the continuity equation
(\ref{continuity_va}), we used the fact that the energy-transfer
dependence of the current is at most linear. For more general
energy-transfer dependence the continuity equation gets more
complicated form with increasing number of nested commutators if the
power of energy-transfer dependence increases. There is, however, a way
to give a general continuity equation for currents without
specification of their energy-transfer dependence. In Appendix~\ref{app:continuity_equation} we prove
the following general continuity equations: for the vector current one gets
\beqa
&&\exp\left(H_{\rm eff}\frac{\overrightarrow{\partial}}{\partial k_0}\right)k^\mu \tilde{\fet
    V}_\mu(k)\exp\left(-H_{\rm eff}\frac{\overleftarrow{\partial}}{\partial k_0}\right)\Bigg|_{k_0=0}=0,\nn
&&\label{continuity_vector_main}
\eeqa
and for the
axial-vector current
\beqa
&&\exp\left(H_{\rm eff}\frac{\overrightarrow{\partial}}{\partial k_0}\right)\Big[k^\mu \tilde{\fet
    A}_\mu(k)+i\,m_q \tilde{\fet P}(k)\Big]\nn
&&\times\exp\left(-H_{\rm eff}\frac{\overleftarrow{\partial}}{\partial
    k_0}\right)\Bigg|_{k_0=0}=0.
\label{continuity_axial_main}
\eeqa 
Between exponential operators in Eqs.~(\ref{continuity_vector_main})
and (\ref{continuity_axial_main}), we find structures which should vanish in
the classical limit as a consequence of the continuity equation. $k_0$-derivatives in
the exponentials generate an increasing number of nested commutators
with the effective Hamiltonian. If we sandwich continuity equations
~(\ref{continuity_vector_main}) and (\ref{continuity_axial_main})
between initial and final eigenstates of the full
Hamiltonian 
\beq
H_{\rm eff} |i\rangle\,=\, E_i \,|i\rangle, \quad H_{\rm eff}
|f\rangle\,=\, E_f \,|i\rangle, 
\eeq 
we get the classical continuity equations
\beqa
k^\mu \tilde{\fet
    V}_\mu(k)|_{k_0=E_f-E_i}&=&0,\nn
k^\mu \tilde{\fet
    A}_\mu(k)+i\,m_q \tilde{\fet P}(k) |_{k_0=E_f-E_i}&=&0.\label{continuity_eq_field_th}
\eeqa
For derivation of Eq.~(\ref{continuity_eq_field_th}) we used the
relation for energy-shift operator
\beqa
\exp\left(E_{f}\frac{\overrightarrow{\partial}}{\partial
    k_0}\right) f(k_0)&=&f(k_0 + E_f),\nn
f(k_0) \exp\left(-E_{i}\frac{\overleftarrow{\partial}}{\partial
    k_0}\right)&=&f(k_0 - E_f),
\eeqa
which are valid for any infinitely differentiable function $f$. So one
can interpret the continuity equations~(\ref{continuity_vector_main})
and (\ref{continuity_axial_main}) as an energy-independent form
of classical continuity
equation~(\ref{continuity_eq_field_th}). The energies are replaced by
corresponding effective Hamiltonians by using energy-shift operator.
\section{Nuclear currents in chiral EFT}
\label{nuclear:chiralEFT}
In this section, we summarize all expressions of the nuclear currents
up to order $Q$ in chiral expansion. $Q$ denotes momenta
and masses which are much smaller than the chiral symmetry breaking
scale. We skip here the discussion of
their construction. As an example,
we discuss here Feynman diagrams which contribute to the two-nucleon
vector current at leading order $Q^{-1}$. All other details about
Feynman diagrams and
specification of unitary phases 
can be found
in~\cite{Krebs:2016rqz,Kolling:2009iq,Kolling:2011mt,Hoferichter:2015ipa,Krebs:2020plh}. 
\subsection{Power counting}
\label{subs:powercounting}
To organize chiral EFT calculations of nuclear current operators we
follow Weinberg's 
analysis~\cite{Weinberg:1990rz,Weinberg:1991um}. A Feynman diagram contributing to the current operator counts as
$Q^\nu$. To derive the expression for the chiral dimension $\nu$, we
consider a generic Feynman diagram which is
proportional to the integral written symbolically as~\footnote{This expression is valid for irreducible
  Feynman diagrams. Nuclear forces and currents are derived within the
  Hamiltonian approach where one has to deal with time-ordered
  structures. Propagators are given in form of energy
  denominators, one has to take into account phase space factors, and
  loop integrals are three-dimensional. However, dimensional
  counting in the Hamiltonian approach
  leads to
  the same expression for $\nu$ as if one would deal with an irreducible Feynman
  diagram in a four-dimensional formalism~\cite{Weinberg:1990rz,Weinberg:1991um}.}
\beqa
%\delta^{(4)}(p)J&\sim& 
&&\delta^{(4)}(p)^{C}\int
(d^4q)^L\frac{1}{(q^2)^{I_p}}\frac{1}{q_0^{I_n}}\prod_i(q^{d_i})^{V_i},\quad\label{power_counting_structure}
\eeqa
where $L$ is the number of loops, $I_p$ and $I_n$ are number of
internal pion and nucleon lines, respectively. $d_i$ is the number of
derivatives or pion mass insertions in the vertex $"i"$, $V_i$ denotes how many
times the vertex $"i"$ appears in a given diagram, and $C$ denotes the
number of connected pieces in the diagram. From
Eq.~(\ref{power_counting_structure}) we read off the index $\nu$:
\beqa
\nu&=&4-4 C+4 L-2 I_p - I_n+\sum_{i} V_i d_i.\label{powercounting:general}
\eeqa
The couplings of external sources are not taken into account in
Eq.~(\ref{powercounting:general}). We treat them as small but count them separately. They are also not
taken into account in $d_i$. For example, $d_i=0$ for the leading-order photon-nucleon coupling. Using the
identities
\beqa
\label{viniEq}
\sum_{i} V_i n_i &=& 2 I_n + E_n\\
\sum_{i} V_i p_i &=& 2 I_p + E_p,\label{vipiEq}
\eeqa
where $n_i$ and $p_i$ denotes the number of nucleon and pion fields in
the vertex $"i"$, respectively. $E_n$ and $E_p$ denote the number of
external nucleon and pion lines, respectively. A well known
topological identity which connects the number of loops with the
number of internal lines is given by
\beqa
L&=& C + I_p + I_n - \sum_i V_i.\label{topological_L_eq}
\eeqa
Using Eqs.~(\ref{viniEq}), (\ref{vipiEq}) and (\ref{topological_L_eq})
we get
\beqa
\nu&=&4 - \frac{3}{2} E_n - E_p + \sum_i V_i \tilde\kappa_i,
\eeqa
where $\tilde\kappa_i$ is given by
\beqa
\tilde\kappa_i= d_i + \frac{3}{2} n_i + p_i - 4.
\eeqa
We are not interested here in the pion production, so the number of
external pions $E_p=0$. The number of nucleons is always conserved and
we denote it by
\beqa
N&=&2 E_n.
\eeqa
The expression for the chiral dimension is then given by
\beqa
\nu&=&4 - 3 N + \sum_i V_i \tilde\kappa_i.\label{power:counting:general}
\eeqa
As was pointed out in~\cite{Epelbaum:2007us}, Eq.~(\ref{power:counting:general}) is inconvenient since it
depends on the total number of nucleons $N$. For example, one-pion
exchange diagram in the
two-nucleon system has the chiral order $\nu=0$ since $N=2$, $\tilde\kappa_1=1$
and $V_1=2$. In the presence of a
third nucleon which acts as a spectator, it has chiral order $\nu=-3$
according to Eq.~(\ref{power:counting:general})
since $N=3$. The origin of this discrepancy lies in the different
normalization of two- and three-nucleon states:
\beqa
2N:&&\quad\langle \vec{p}_1 \,\vec{p}_2|\vec{p}_1^\prime\,
\vec{p}_2^\prime\rangle\,=\,\delta^{(3)}(\vec{p}_1^\prime - \vec{p}_1)
\delta^{(3)}(\vec{p}_2^\prime - \vec{p}_2),\nn
3N:&&\quad\langle \vec{p}_1 \,\vec{p}_2 \,\vec{p}_3|\vec{p}_1^\prime\,
\vec{p}_2^\prime \,\vec{p}_3^\prime\rangle\,=\,\nn
&&\quad\quad\delta^{(3)}(\vec{p}_1^\prime - \vec{p}_1)
\delta^{(3)}(\vec{p}_2^\prime - \vec{p}_2) \delta^{(3)}(\vec{p}_3^\prime - \vec{p}_3).\quad\quad
\eeqa
One can circumvent this if one assigns a chiral dimension to the
transition operator rather than to its matrix element in $N$-nucleon
system. In this case, we have to modify the expression for $\nu$ by
adding $3N$ to Eq.~(\ref{power:counting:general}), accounting in this
way for the normalization of the $N$-nucleon system. The expression
for $\nu$ becomes independent of $N$ but gives $\nu=6$ for one-pion-exchange. As was proposed in~\cite{Epelbaum:2007us}, we adjust the final
expression for $\nu$ by subtracting from it $6$ to get $\nu=0$ for
one-pion-exchange, which is a convention. The final expression for the
chiral dimension of the transition operator
becomes
\beqa
\nu&=&-2 + \sum_i V_i \tilde\kappa_i. \label{power:counting:final}
\eeqa
Similar to\cite{Krebs:2019aka}, we can also express the chiral dimension $\nu$
in terms of the inverse mass dimension of the coupling constant at a
vertex $"i"$
\beqa
\kappa_i&=&d_i + \frac{3}{2} n_i + p_i + s_i - 4,\label{kappa_inverse_mass_dimesion_of_coupling}
\eeqa
where $s_i$ is the number of external sources which for the current
operators can be only $0$ or $1$. Consistent with~\cite{Krebs:2019aka}, we get 
\beqa
\nu&=&-3 + \sum_i V_i \kappa_i.\label{power:counting:forces}
\eeqa
for the current operator and
\beqa
\nu&=&-2 + \sum_i V_i \kappa_i. \label{power:counting:currents}
\eeqa
for the nuclear force.
%We adopt here a two-nucleon
%power counting where we perform power counting of momenta and masses of
%light degrees of freedom in the two-nucleon system. Even if we look at
%a single-nucleon current we consider it in the presence of the second
%nucleon which plays the role of a spectator. Due to the presence of
%spectator, we have to multiply single-nucleon current operator by a
%three-dimensional delta-function which counts as $Q^{-3}$. A 
%Feynman diagram which contributes to current operators counts as $Q^\nu$ where
%\beqa
%\nu&=&-3 + \sum_i V_i\kappa_i.
%\eeqa
%Here $V_i$ denotes the number of vertices of a type $\kappa_i$ where $\kappa_i$
%denotes the inverse mass dimension of the coupling constant in the vertex
%\beqa
%\kappa&=& d + \frac{3}{2}n + p + c - 4.
%\eeqa
%Here, $d$ is the number of derivatives and/or pion masses. 
%The letters $n,p$ and $c$ denote here the number of nucleon-fields,
%pion-fields and the number of external sources, respectively.

For the counting of the nucleon mass $m$, we adopt a two-nucleon power counting where $1/m$-contributions count as two powers of Q~\cite{Weinberg:1991um}
\beqa
\frac{Q}{m}\sim\frac{Q^2}{\Lambda_\chi^2}.
\eeqa
Here $\Lambda_\chi\sim 700\,{\rm MeV}$ and $m$ are the chiral symmetry breaking
scale and the nucleon mass, respectively.
\subsection{Vector current up to order $Q$}
\label{vectorUpToOrderQ}
\subsubsection{Single-nucleon current}
\label{VectorsingleNuclCurrent}
We start our discussion with electromagnetic vector current. Leading contribution to the vector
current starts at the order $Q^{-3}$. At this order there is only a
contribution to the single-nucleon charge operator. It is well known that chiral
expansion of the single-nucleon currents does not converge well~\cite{Phillips:2016mov,Kubis:2000zd,Schindler:2005ke}. For
moderate virtualities $Q^2\sim 0.3\,{\rm GeV}^2$, an
explicit inclusion of $\rho$-meson is essential. For this reason the
usual practice is to parametrize single-nucleon vector current by
e.g. Sachs form factors and use their phenomenological form extracted
from experimental data in practical calculations~\cite{Ye:2017gyb,Ye:smallrp,Belushkin:2006qa,Lorenz:2012tm,Lorenz:2014yda}. The general form of the single-nucleon
current can be characterized by its non-relativistic one-over-nucleon-mass expansion
given symbolically by
\beqa
{ V}_{{\rm 1N}}^0 &=& { V}_{{\rm 1N:static}}^0 + {
  V}_{{\rm 1N:}1/m}^0 + { V}_{{\rm 1N:}1/m^2}^0,\\
\vec{{\fet V}}_{{\rm 1N}} &=& \vec{{\fet V}}_{{\rm 1N:static}} + \vec{{\fet
  V}}_{{\rm 1N:}1/m} + \vec{{\fet V}}_{{\rm 1N:}1/m^2} +\vec{\fet V}_{{\rm
    1N:off-shell}}.\label{singleN:general}\nonumber
\eeqa
In terms of Sachs form factors, the non-relativistic charge is
parametrized by
\beqa
{V}_{{\rm 1N:static}}^0&=&e { G}_E(Q^2),\nn
{ V}_{{\rm 1N:} 1/m}^0&=&
\frac{i\,e}{2m^2}\vec{k}\cdot(\vec{k}_1\times\vec{\sigma}) {G}_M(Q^2),\nn
{ V}_{{\rm 1N:}1/m^2}^0&=& -\frac{e}{8
  m^2}\big[Q^2+2\,i\,\vec{k}\cdot(\vec{k}_1\times\vec{\sigma}) \big]
{ G}_E(Q^2),\label{V1N:Charge:nonrel}\quad\quad
\eeqa
and the non-relativistic current is given by
\beqa
\vec{{\fet V}}_{{\rm 1N:static}}&=&- \frac{i\,e}{2
  m}\vec{k}\times\vec{\sigma}\, { G}_M(Q^2),\nn
\vec{{\fet V}}_{{\rm 1N:} 1/m}&=&\frac{e}{m}\vec{k}_1\,{ G}_E(Q^2),\label{V1N:Current:nonrel}\nn
\vec{{\fet V}}_{{\rm 1N:} 1/m^2}&=&\frac{e}{16
  m^3}\bigg[i\,\vec{k}\times\vec{\sigma}(2\vec{k}_1^{\,2}+Q^2)+2\,i\,\vec{k}\times\vec{k}_1\,\vec{k}_1\cdot\vec{\sigma}\nn
&+& 2
\vec{k}_1(i\,\vec{k}\cdot(\vec{k}_1\times\vec{\sigma})+Q^2)
-2\,\vec{k}\,\vec{k}\cdot\vec{k}_1\nn
&+& 6\,i\,\vec{k}_1\times\vec{\sigma}\,\vec{k}\cdot\vec{k}_1\bigg]{ G}_M(Q^2),
\eeqa
where $\vec{k}$ is a photon momentum, 
$\vec{k}_1=(\vec{p}^{\,\prime}+\vec{p})/2$, and
$\vec{p}^\prime( \vec{p})$ are outgoing (incoming) momenta of the single--nucleon current operator. Virtuality in our kinematics is given by
$Q^2=\vec{k}^2$. Note, that the form factors $G_E(Q^2)$ and $G_M(Q^2)$
in Eqs.~(\ref{V1N:Charge:nonrel}) and (\ref{V1N:Current:nonrel}) are
operators in isospin space. The proton and
neutron electromagnetic form factors can be extracted out of them by
projecting these to the corresponding state.

As already briefly explained in
Sec.~\ref{sec:nuclearcurrent} (see~\cite{Krebs:2019aka} for more
comprehensive discussion)  we apply unitary transformations on
the Hamilton operator which explicitly depends on an external source and thus
on time. These unitary transformations generate off-shell contributions
to the longitudinal component of the current which depend on energy transfer $k_0$ and additional
relativistic $1/m$ corrections. This contribution can also be
parametrized by Sachs form factors via
\beqa
\vec{\fet V}_{{\rm
    1N:off-shell}}&=&\vec{k}\bigg(k_0-\frac{\vec{k}\cdot\vec{k}_1}{m}\bigg)
\frac{e}{Q^2}\bigg[\big({ G}_E(Q^2)-{
  G}_E(0)\big)\nn
&+&\frac{i}{2m^2}\vec{k}\cdot(\vec{k}_1\times\vec{\sigma})\big({
  G}_M(Q^2)-{ G}_M(0)\big)\bigg].\nn
&&\label{singleN:offshell}
\eeqa
\subsubsection{Two-nucleon vector current}
\label{sec:2nvectorCurrentGeneral}
We switch now to a discussion of the two-nucleon vector current
operator. Various contributions can be characterized by the number of pion
exchanges and/or short-range interactions
\beqa
\label{vectorcurrent:rangedecomp}
V_{{\rm 2N}}^\mu&=&V_{{\rm 2N:1\pi}}^\mu+V_{{\rm 2N:2\pi}}^\mu+V_{{\rm cont}}^\mu.
\eeqa
\paragraph{One-pion-exchange vector current}
\label{sec:OPEvCurrent}

\begin{figure}
\begin{center}
\resizebox{0.27\textwidth}{!}{%
  \includegraphics{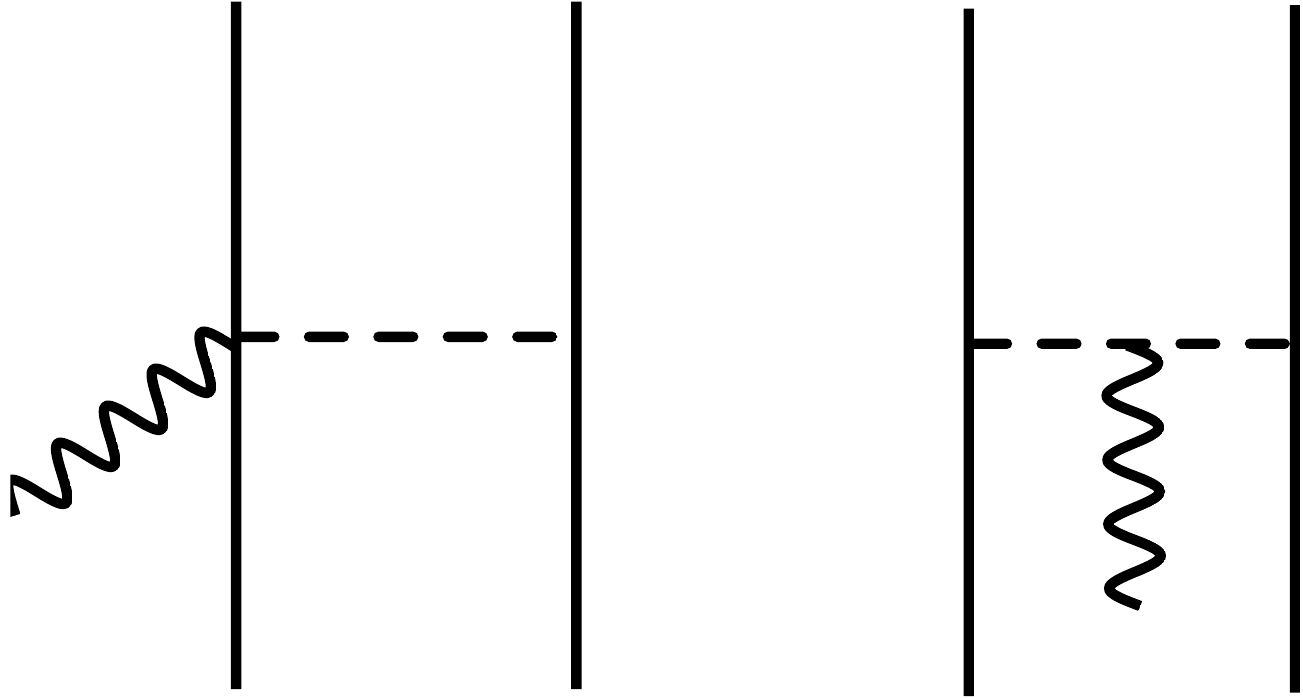}
}
\end{center}
\vspace{0.5cm}       % Give the correct figure height in cm
\caption{Feynman diagrams which contribute to the two-nucleon vector
  current operator at order $Q^{-1}$.}
\label{fig:vector_current_qToM1}       % Give a unique label
\end{figure}

Leading contribution to the one-pion-exchange (OPE) current shows up
at the order $Q^{-1}$. The corresponding Feynman diagrams are listed in
Fig.~\ref{fig:vector_current_qToM1}. 
At this
order we get a well known result for current operator
\beqa
{\fet V}_{{\rm 2N:1\pi}}^{ (Q^{-1})}&=&i \frac{e g_A^2}{4 F_\pi^2} [{\fet
  \tau}_1\times{\fet
  \tau}_2]^3\frac{\vec{q}_2\cdot\vec{\sigma}_2}{q_2^2+M_\pi^2}\bigg(\vec{q}_1\frac{\vec{q}_1\cdot\vec{\sigma}_1}{q_1^2+M_\pi^2}-\vec{\sigma}_1\bigg)\nn
&+& 1\leftrightarrow 2,\label{vector_current_qToMinumsOne}
\eeqa
and for the charge operator
\beqa
{ V}_{{\rm 2N:1\pi}}^{0, (Q^{-1})}&=&0.
\eeqa
Here $e$ is electric coupling, $g_A$ axial vector coupling to the
nucleon, $F_\pi$ pion decay constant, $M_\pi$  pion mass,
$\vec{\sigma}_i$ and $\vec{\tau}_i$ are Pauli spin and
isospin matrices with label $i=1,2$ labeling a corresponding
nucleon. Momenta $\vec{q}_{1,2}$ are defined by
\beqa
\vec{q}_i=\vec{p}_i^\prime - \vec{p}_i,
\eeqa 
where $i=1,2$ and $\vec{p}_i^\prime$ or $\vec{p}_i$ are outgoing or
incoming momenta of the $i$-th nucleon, respectively.

There is no contribution at order $Q^0$ such that the next correction
starts at the order $Q$ which are leading one-loop
contributions in the static limit and/or leading relativistic correction
to OPE current
\beqa
{V}_{{\rm 2N:1\pi}}^{\mu, (Q)}&=&{V}_{{\rm 2N:1\pi, }\,{\rm
    static}}^{\mu, (Q)}+{V}_{{\rm 2N: 1\pi, }1/m}^{\mu, (Q)}.
\eeqa
The corresponding set of diagrams can be found in~\cite{Kolling:2011mt}. The explicit form of static contributions can be given in terms of
scalar functions $f_{1\dots 6}(k)$. The vector contribution is given
by~\cite{Kolling:2011mt}
\beqa
&& \vec{V}_{{\rm 2N:1\pi,}\,{\rm static}}^{(Q)} = 
 \frac{\vec{\sigma}_2\cdot\vec{q}_2}{q_2^2+M_\pi^2}\vec{q}_1
   \times\vec{q}_2 \left[
   [{\fet \tau}_2]^3 \, f_1(k )\right.\nn
&+& \left. \vec{\tau}_1 \cdot\vec{\tau}_2\, f_2(k )
 \right] +
 \left[\vec{\tau}_1\times\vec{\tau}_2 \right]^3
  \frac{\vec{\sigma}_2\cdot\vec{q}_2}{q_2^2 + M_\pi^2} \biggl\{\nn 
&&  \vec{k}\times\left[\vec{q}_2\times\vec{\sigma}_1 
    \right] f_3(k )
+   
  \, \vec{k}\times\left[\vec{q}_1\times\vec{\sigma}_1 \right] f_4(k
  )\nn
&+&
 \vec{\sigma}_1 \cdot
    \vec{q}_1\, \left( \frac{\vec{k}}{k^2} - \frac{\vec{q}_1
      }{q_1^2+M_\pi^2}  \right) f_5(k) \nn
&+& \,\biggl[\frac{\vec{\sigma}_1\cdot
      \vec{q}_1}{q_1^2+M_\pi^2}\vec{q}_1  -  \vec{\sigma}_1
    \biggr]f_6(k )\biggl\}
+ 1\leftrightarrow 2 \,,\label{OPE:Static:Current:Parametrization}
\eeqa
where the scalar functions $f_i(k)$ are given by
\beqa
  f_1\left(k \right) & = & 2i e\frac{g_A}{F_\pi^2}\, \bar{d}_8\, ,
  \quad f_2\left(k \right)  \,=\,  2i e\frac{g_A}{F_\pi^2}\, \bar{d}_9 \, , \nn  
  f_3\left(k \right)  &=&  - ie\frac{g_A}{64 F_\pi^4\pi^2}
  \left[\,g_A^3\,\left( 2L(k)-1\right)  + 32 F_\pi^2\pi^2 \bar{d}_{21}
  \right]\, , \nn
  f_4\left(k \right) & = & - ie\frac{g_A}{4 F_\pi^2}
  \,\bar{d}_{22} \, , \nn  
  f_5\left(k \right)  &=&  - ie \frac{g_A^2}{384 F_\pi^4\pi^2}\left[2(4M_\pi^2 +
      k^2)L(k) +\left(6 \, \bar{l}_6 -\frac{5}{3}\right)k^2 \right.\nn  
&-& \left. 8M_\pi^2
    \right]\, , \nn  
  f_6\left(k \right)  &=&  - ie\frac{g_A}{ F_\pi^2}  M_\pi^2
  \,\bar{d}_{18}  \,.\label{OPE:fis:current}
\eeqa
Here $\bar{d}_i$ are low-energy constants (LEC) from the order $Q^3$ pion-nucleon
Lagrangian~\cite{Fettes:2000gb}. $\bar{l}_6$ is a LEC from $Q^4$ pion-Lagrangian~\cite{Gasser:1987rb}. Their values can be fixed from pion-nucleon scattering and
pion-photo- or electroproduction. The charge contribution is given by
\beqa
  V_{{\rm 2N:1\pi,}\,{\rm static}}^{0, (Q)} & = &
  \frac{\vec{\sigma}_2\cdot\vec{q}_2}{q_2^2 + M_\pi^2}[{\fet \tau}_2]^3\biggr[ 
  \vec{\sigma}_1\cdot\vec{k} \, \vec{q}_2 \cdot\vec{k} f_7( k ) \nn
&+&
  \vec{\sigma}_1\cdot\vec{q}_2 f_8( k )\biggl] + 1\leftrightarrow 2,\label{OPE:Static:Charge:Parametrization}
\eeqa
where 
\beqa
f_7( k ) & = & e \frac{g_A^4}{64F_\pi^4\pi} \left[A(k) +
    \frac{M_\pi-4M_\pi^2 \,A(k)}{k^2} \right]\, , \nn 
f_8( k ) &=&   e \frac{g_A^4}{64F_\pi^4\pi}  \left[(4M_\pi^2+k^2)A(k)
  - M_\pi \right] \, .
\label{OPE:fis:charge}
\eeqa
The loop function $L(k)$ and $A(k)$ are defined
\beqa
\label{loop_functions}
  L(k)  &=&  \frac{1}{2}\frac{\sqrt{k^2 + 4M_\pi^2}}{k}\log\left(\frac{\sqrt{k^2 + 4M_\pi^2}+k}{\sqrt{k^2 + 4M_\pi^2}-k} \right), \nn
  A(k) &=& \frac{1}{2k}\arctan\left(\frac{k}{2M_\pi}\right).
\eeqa
Relativistic corrections for the vector operator vanish
\beqa
{\vec V}_{{\rm 1\pi:}1/m}^{ (Q)}=0.
\eeqa
Relativistic corrections for the charge operator are
\beqa
&&{V}_{{\rm 2N: 1\pi, }1/m}^{0, (Q)}=\frac{e g_A^2}{16F_\pi^2m}
\frac{1}{q_2^2 + M_\pi^2} \biggl\{(1-2\bar\beta_9) \nn
&\times&\left([{\fet \tau}]_2^3 +
   \vec{\tau}_1\cdot\vec{\tau}_2 \right) \vec{\sigma}_1\cdot\vec{k}
 \vec{\sigma}_2\cdot\vec{q}_2 -  
  i(1+2\bar\beta_9) \left[
   \vec{\tau}_1\times\vec{\tau}_2\right]^3\nn
&\times&\biggl[  \vec{\sigma}_1\cdot\vec{k}_1
 \vec{\sigma}_2\cdot\vec{q}_2 
- \vec{\sigma}_2\cdot\vec{k}_2
 \vec{\sigma}_1\cdot\vec{q}_2 
- 2\, \frac{\vec{\sigma}_1\cdot\vec{q}_1}{q_1^2+M_\pi^2}\vec{\sigma}_2\cdot\vec{q}_2\nn
&\times&  \vec{q}_1\cdot\vec{k}_1  \biggr]\biggr\} 
+ \frac{e g_A^2\,
        }{16 F_\pi^2 m } 
\frac{ \vec{\sigma}_1\cdot\vec{q}_2  \, \vec{\sigma}_2\cdot\vec{q}_2
     }{(q_2^2+M_\pi^2)^2}  \biggl[(2\bar\beta_8-1)\nn
&\times&([{\fet \tau}_2]^3 + 
  \vec{\tau_1}\cdot\vec{\tau}_2) \vec{q}_2\cdot\vec{k} 
+   i \left[
  \vec{\tau}_1 \times \vec{\tau}_2\right]^3  
\left(\left( 2\bar\beta_8 -  1\right) \vec{q}_2 \cdot\vec{k}_1\right. \nn
&-& \left.\left(
    2\bar\beta_8 + 1\right) \vec{q}_2\cdot\vec{k}_2\,\right)\biggr]
+ 1\leftrightarrow 2.\label{vector_current_charge_one_over_m}
\eeqa
Here $\bar{\beta}_8$ and $\bar{\beta}_9$ are phases from unitary
transformations which are not fixed. The same phases show up in
nuclear forces. Usually they are fixed by requirement of minimal
non-locality of the OPE NN potential.
\paragraph{Two-pion-exchange vector current}
\label{sec:TPEvectorCurrent}
Contributions to two-pion-exchange (TPE) vector current start to show
up at order $Q$. They are parameter-free. The corresponding diagrams
can be found in~\cite{Kolling:2009iq}. Due to the coupling of the vector source to two
pions there appear loop functions which depend on three momenta
$\vec{k}, \vec{q_1}$ and $\vec{q}_2$ which are momentum transfer of
the vector source, momentum transfer of the first and second
nucleons, respectively. This leads to a somewhat lengthy expression which
have been derived in~\cite{Kolling:2009iq} and are
listed in Appendix~\ref{two_pion_excange_vector_current_app} for completeness:
\paragraph{Short-range vector current}
\label{sec:contactVectorCurrent}
The first contribution to short-range two-nucleon current shows up at
the order $Q$. The diagrams with short-range interactions at this order can be found in~\cite{Kolling:2011mt}.
There are two contributions~\cite{Kolling:2011mt}
\beqa
V_{{\rm 2N:\,cont}}^{\mu, (Q)}&=&V_{{\rm 2N:\,cont,\,tree}}^{\mu, (Q)} +
V_{{\rm 2N:\,cont,\,loop}}^{\mu, (Q)}.\label{vector_2N_cont}
\eeqa
The current contribution coming from tree-diagrams is given by
\beqa
&&{\vec V}_{{\rm 2N:\,cont,\,tree}}^{(Q)}\,=\, e\,\frac{i}{16}\left[ \vec{\tau}_1 \times \vec{\tau}_2\right]^3 \,\biggl[\left(C_2 +3C_4 + C_7\right)
    \, \vec{q}_1 \nn
&-& \left(-C_2 + C_4 + C_7 \right) \,
  \, \left(\vec{\sigma}_1 \cdot\vec{\sigma}_2 \right)\, \, \vec{q}_1 
+ C_7 \, \left(  \vec{\sigma}_2\cdot\vec{q}_1
      \,  \vec{\sigma}_1\right.\nn
&+&\left. \vec{\sigma}_1\cdot\vec{q}_1
     \,  \vec{\sigma}_2 \right)\biggr]
 - e\,\frac{C_5 \, i}{16}\, [{\fet \tau}_1]^3 \,
 \left[\left(\vec{\sigma}_1 + \vec{\sigma}_2 \right)\times\vec{q}_1
   \right] \nn
&+&  i e L_1 \, [{\fet \tau}_1]^3 \,
\left[\left(\vec{\sigma}_1 - \vec{\sigma}_2
  \right)\times\vec{k}\right] \nn
&+& i e L_2 \, \left[\left(\vec{\sigma}_1 + \vec{\sigma}_2
  \right)\times\vec{q}_1 \right] \, . \label{shortTree}
\eeqa
%\beqa
%{\vec V}_{{\rm cont:\,tree}}^{(Q)}&=& e\,\frac{i}{16}\left[ \vec{\tau}_1 \times \vec{\tau}_2\right]^3 \,\bigg%l[\left(C_2 +3C_4 + C_7\right)
%    \, \vec{q}_1 \nn
%&-& \left(-C_2 + C_4 + C_7 \right) \,
%  \, \left(\vec{\sigma}_1 \cdot\vec{\sigma}_2 \right)\, \, \vec{q}_1 \nn
%&+& C_7 \, \left(  \vec{\sigma}_2\cdot\vec{q}_1
%      \,  \vec{\sigma}_1+ \, \vec{\sigma}_1\cdot\vec{q}_1
%     \,  \vec{\sigma}_2 \right)\biggr]\nn
% &- &e\,\frac{C_5 \, i}{16}\, [{\fet \tau}_1]^3 \,
% \left[\left(\vec{\sigma}_1 + \vec{\sigma}_2 \right)\times\vec{q}_1
%   \right] \nn
%&+&  i e L_1 \, [{\fet \tau}_1]^3 \,
%\left[\left(\vec{\sigma}_1 - \vec{\sigma}_2
%  \right)\times\vec{k}\right] \nn
%&+& i e L_2 \, \left[\left(\vec{\sigma}_1 + \vec{\sigma}_2
%  \right)\times\vec{q}_1 \right] \, . \label{shortTree}
%\eeqa
As can be seen from Eq.~(\ref{shortTree}) there are $C_i$ LECs which
also contribute to the two-nucleon potential and appear here due to the
minimal coupling, and there are two additional constants $L_{1,2}$
which describe entirely electromagnetic effects. Charge short-range contribution
from tree diagrams at order $Q$ vanishes
\beqa
{V}_{{\rm 2N:\, cont,\,tree}}^{0, (Q)}&=&0.
\eeqa
There are also contributions from one-loop diagrams which include one leading-order
 two-nucleon contact interaction and  two-pion propagators. They
 only contribute to charge operator\footnote{Different conventions are being used in the literature for
  the  leading-order\,  two-nucleon contact interactions $\propto C_{S, T}$. To match the
  convention of
  Refs.~\cite{Epelbaum:2014sza,Reinert:2017usi,Epelbaum:2014efa}, the
  factors of $32 F_\pi^2$ in Eq.~(5.7) of \cite{Kolling:2011mt} should be
  replaced by $16 F_\pi^2$.}
\beqa
{V}_{{\rm 2N:\,cont,\,loop}}^{0, (Q)}&=&C_T\, [{\fet \tau}_1]^3\, \left[
      \vec{\sigma}_1\cdot\vec{k}\,\vec{\sigma}_2\cdot\vec{k}\, f_{9}(k
      ) \right.\nn
&+&\left. \vec{\sigma}_1\cdot\vec{\sigma}_2 \,
    f_{10}( k )\right]  \,,\label{em_charge_short_order_Q}
\eeqa
where 
\beqa
f_{9}( k ) &=& e \frac{g_A^2}{16F_\pi^2\pi} \left(A(k) +
    \frac{M_\pi-4M_\pi^2 \,A(k)}{k^2} \right)\, , \nn 
f_{10}( k ) &=& e\frac{g_A^2}{16 F_\pi^2\pi}\, \left(M_\pi-(4M_\pi^2 + 3k^2)\,A(k) \right)\,.
\eeqa
Corresponding contributions to the current operator vanish
\beqa
{\vec V}_{{\rm cont:\,loop}}^{ (Q)}&=&0.
\eeqa
\subsubsection{Three-nucleon vector current}
\label{sec:3NvectorCurrent}
At the order $Q$ there are first contributions to three-nucleon vector
current~\cite{Krebs:2019aka}. There are no contributions to the vector operator 
\beqa
{\vec V}_{{\rm 3N}}^{(Q)}&=&0.
\eeqa
Contributions to the charge operator can be parametrized in the form
\beqa
{V}_{{\rm 3N}}^{0,
  (Q)}&=&\bigg[\frac{(\vec{q}_1+\vec{q}_2)\cdot\vec{\sigma}_3}{(\vec{q}_1+\vec{q}_2)^2+M_\pi^2}+\frac{\vec{q}_3\cdot\vec{\sigma}_3}{q_3^2+M_\pi^2}\bigg]\nn
&\times&\big({v}_{{\rm 3N:\,long}}+{v}_{{\rm 3N:\,short}}\big) +
5\,{\rm permutations}.\label{vector_charge:3N:general}
\eeqa
where the long- and short-range
contributions are given by
\beqa
{v}_{{\rm 3N:\,long}}&=&\frac{e}{16 F_\pi^4} \frac{\vec{q}_1\cdot\vec{\sigma}_1}{q_1^2+M_\pi^2}\bigg[\big({\fet \tau}_1\cdot{\fet \tau}_3
  [{\fet \tau}_2]^3 -{\fet \tau}_2\cdot{\fet \tau}_3
  [{\fet \tau}_1]^3\big)\nn
&\times&\bigg(-2
g_A^4\frac{q_1^2+\vec{q}_1\cdot\vec{q}_2}{(\vec{q}_1+\vec{q}_2)^2+M_\pi^2}
+ g_A^2\bigg) \nn
&-& i[{\fet
  \tau}_1\times{\fet \tau}_3]^3 2
g_A^4\frac{q_1^2+\vec{q}_1\cdot\vec{q}_2}{(\vec{q}_1+\vec{q}_2)^2+M_\pi^2}\bigg]
,\label{vector_charge:3N:long}\\
{v}_{{\rm 3N:\,short}}&=&-[{\fet \tau}_1\times{\fet\tau}_3]^3
\frac{e\,g_A^2 C_T}{2 F_\pi^2}
\frac{
(\vec{q}_1+\vec{q}_2)\cdot
    (\vec{\sigma}_1\times\vec{\sigma}_2)
}{
(\vec{q}_1+\vec{q}_2)^2+M_\pi^2
}.\nn
&&\label{vector_charge:3N:short}
\eeqa
Due to the approximate spin-isospin ${\rm SU}(4)$ Wigner symmetry~\cite{Mehen:1999qs}, $C_T$ appears to be small such that we do not
expect large contributions from short-range part of the three-nucleon
vector current.
\subsection{Axial vector current up to order $Q$}
\label{axialvectorUpToOrderQ}
The weak sector of nuclear physics can be probed by a nuclear
axial-vector current. We give here its expressions up to order $Q$ in
chiral expansion.
\subsubsection{Single-nucleon axial vector current}
\label{sec:axialVSingleNucleon}
The leading-order contribution to an axial vector current shows up at
order $Q^{-3}$ where axial-vector source couples directly to a single-nucleon or to a pion, which itself propagates and couples to a single-nucleon
generating in this way a pion-pole term.  It is convenient to
parametrize the single-nucleon current by the axial and pseudoscalar form
factors. Up to the order $Q$ the parametrization is given by
\beqa
A_{{\rm 1N}}^{\mu, a}&=&A_{{\rm 1N:on-shell}}^{\mu, a}+A_{{\rm 1N:off-shell}}^{\mu, a}\;.
\eeqa
The charge operator is parametrized by
\beqa
&&{A}_{{\rm 1N:on-shell}}^{0, a}\,=\,
\frac{[{\fet \tau}]^a}{2} \bigg[-\frac{\vec{k}_1\cdot\vec{\sigma}}{m} {G}_A(t) +
\frac{\vec{k}\cdot\vec{k}_1}{m}\bigg(\frac{\vec{k}\cdot\vec{\sigma}}{4
  m^2} \nn
&&- \frac{\vec{k}\cdot\vec{\sigma}(k^2+4 k_1^2)+4
  \vec{k}\cdot\vec{k}_1 \vec{k}_1\cdot\vec{\sigma}}{32
  m^4}\bigg)\times G_P(t)\bigg],\nn
&&{A}_{{\rm 1N:off-shell}}^{0, a}\,=\,0.\label{V1N:Charge:nonrelAxial}
\eeqa
The current operator is parametrized by
\beqa
&&\vec{{A}}_{{\rm 1N:on-shell}}^{a}\,=\,\frac{[{\fet \tau}]^a}{2}\bigg[\bigg(-\vec{\sigma} +
\frac{1}{8 m^2}\Big(4\,\vec{\sigma} \,k_1^2-4\, \vec{k}_1\,
\vec{k}_1\cdot\vec{\sigma} \nn
&&+ 2\, i\,\vec{k}\times\vec{k}_1 +
\vec{k}\,\vec{k}\cdot\vec{\sigma}\Big) \bigg)
G_A(t)+\vec{k}\,\bigg(\frac{\vec{k}\cdot\vec{\sigma}}{4 m^2} \nn
&&-\frac{1}{32
  m^4}\Big(\vec{k}\cdot\vec{\sigma}\, k^2 + 4\,\vec{k}\cdot\vec{\sigma}\,
k_1^2 + 4\,
\vec{k}_1\cdot\vec{\sigma}\,\vec{k}\cdot\vec{k}_1\Big)\bigg)G_P(t)\bigg],\nn
&&\vec{A}_{{\rm
    1N:off-shell}}^a\,=\,\vec{k}\bigg(k_0-\frac{\vec{k}\cdot\vec{k}_1}{m}\bigg)
\frac{[{\fet\tau}]^a}{16 m^3}\Big[- (1+2\beta_9)\nn
&&\times\,\vec{k}_1\cdot\vec{\sigma}\,
G_P(t) + (1+2\beta_8) \vec{k}\cdot\vec{k}_1 \vec{k}\cdot\vec{\sigma}\,G_P^\prime(t)\Big],\label{singleN:axialoffshell}
\eeqa
Chiral EFT results are given by chiral expansion of the axial and
pseudoscalar form factors. Corresponding expressions are worked out
in~\cite{Krebs:2016rqz}. To make this review self-consistent we will
briefly discuss them here.

The well-known leading-order result for the axial charge and
current operators have the form
\beqa
A^{0,  a \, (Q^{-3})}_{{\rm 1N: \, static}}  &=& 0,\nn
\vec{A}^{ a \, (Q^{-3})}_{{\rm 1N: \, static}}  &=&
-\frac{g_A}{2}[{\fet\tau}_i]^a\bigg(\vec{\sigma}_i - \frac{\vec{k}\,
  \vec{k}\cdot\vec{\sigma}_i}{k^2+M_\pi^2}\bigg).\label{LOSingleNCurrentExpr}
\eeqa
There are only vanishing contributions at order $Q^{-2}$.
At order $Q^{-1}$, we encounter three kinds of
corrections. First, there are terms emerging
from the time-dependence of unitary transformations which have the
form 
\beqa
\label{SingleNChargeUTPrimeQToMinus1}
A^{0,  a \, (Q^{-1})}_{{\rm 1N: \, UT^\prime}}  &=& \frac{g_A}{2} 
\frac{k_0}{k^2+M_\pi^2}\vec{k}\cdot\vec{\sigma}_i [{\fet\tau}_i]^a,\\
\vec{A}^{ a \, (Q^{-1})}_{{\rm 1N: \, UT^\prime}}  &=&0.
\eeqa
and contribute to ${A}_{{\rm 1N:off-shell}}^{\mu, a}$. Secondly, at
order $Q^{-1}$ there are static limit contributions to $A_{{\rm
    1N:on-shell}}^{\mu, a}$ which are given by
\beqa
A^{0,  a \, (Q^{-1})}_{{\rm 1N: \, static}}  &=& 0,\nn
\vec{A}^{a \, (Q^{-1})}_{{\rm 1N: \, static}}  &=&
\frac{1}{2}\bar{d}_{22}\Big(\vec{\sigma}_i k^2-\vec{k}\,
\vec{k}\cdot\vec{\sigma}_i\Big)[{\fet\tau}_i]^a \nn
&-& \bar{d}_{18} M_\pi^2[{\fet\tau}_i]^a
\frac{\vec{k}\,\vec{k}\cdot\vec{\sigma}_i}{k^2+M_\pi^2} \,.\label{SingleNQtoMinus1StaticExpr}
\eeqa
Finally, there are leading relativistic
$1/m$-corrections which in our counting scheme start contributing at 
order $Q^{-1}$  to ${A}_{{\rm 1N:on-shell}}^{\mu, a}$ and read
\beqa
A^{0,  a \, (Q^{-1})}_{{\rm 1N:} \, 1/m}  &=& - \frac{g_A}{2 m }
[{\fet\tau}_i]^a \, \vec \sigma_i \cdot \vec k_i \,,\label{SingleNChargeOneOvermQToMinus1}\\
\vec{A}^{ a \, (Q^{-1})}_{{\rm 1N: \, 1/m}}  &=& 0\,,
\eeqa
where 
\beq
\vec{k}=\vec{p_i}^{\prime} - \vec{p}_i, \quad \vec{k}_i=\frac{\vec{p_i}^{\prime} + \vec{p}_i}{2}.
\eeq
There are no corrections to the 1N charge and current operators at the
order $Q^0$. Finally, there are various contributions at order
$Q$. The off-shell contributions coming from time-dependent unitary
transformations give different contributions. One of them is coming from
relativistic corrections which are
proportional to $k_0/m$ and are given by
\beqa
A^{0,  a \, (Q)}_{{\rm 1N:\,}1/m, {\rm UT^\prime}}  &=&0,\nn
\vec{A}^{ a \, (Q)}_{{\rm 1N:\,}1/m, {\rm UT^\prime}}  &=&
-\frac{g_A k_0}{8
  m}\frac{\vec{k}}{k^2+M_\pi^2}[{\fet\tau}_i]^a
\bigg(
2(1 + 2 \,\bar{\beta}_9)\vec \sigma_i \cdot \vec k_i \nn
&-& (1 + 2 \,\bar{\beta}_8)\vec{k}\cdot\vec{\sigma}_i\frac{p_i^{\prime\,2}-p_i^2}{k^2+M_\pi^2}\bigg).\label{k0OvermNuclcorr}
\eeqa
They explicitly depend on unitary phases $\bar{\beta}_8$ and
$\bar{\beta}_9$. Note that these are the same unitary phases that
influence a non-locality degree of
relativistic $1/m^2$ corrections of the one-pion-exchange nuclear
force. For the static part which is proportional to $k_0$, we get
nonvanishing contributions 
\beqa
\label{SingleNChargeStaticUTPrimeQTo1}
A^{0,  a \, (Q)}_{{\rm 1N:\,static,\, UT^\prime}}  &=&-k_0  \frac{[{\fet\tau}_i]^a}{2}
\vec{k}\cdot \vec{\sigma}_i \bigg[\bar{d}_{22}+ \frac{2
\bar{d}_{18} M_\pi^2}{k^2+M_\pi^2}\bigg]\,,\nn
\vec{A}^{ a \, (Q)}_{{\rm 1N:\,static,\, UT^\prime}}  &=& 0.
\eeqa
The second class of order-$Q$ contributions involves relativistic
$1/m^2$-corrections:
\beqa
A^{0,  a \, (Q)}_{{\rm 1N:} \, 1/m^2}  &=&0,\nn
\vec{A}^{ a \, (Q)}_{{\rm 1N: \, 1/m^2}}  &=&\frac{g_A}{16
  m^2}[{\fet\tau}_i]^a\bigg(\vec{k}\,\vec{k}\cdot\vec{\sigma}_i(1-2\bar{\beta}_8)\frac{(p_i^{\prime\,2}-p_i^2)^2}{(k^2+M_\pi^2)^2}\nn
&-&2 \vec{k} \frac{(p_i^{\prime\, 2} +
  p_i^2)\vec{k}\cdot\vec{\sigma}_i-2 \bar{\beta}_9 (p_i^{\prime\, 2} -
  p_i^2)\vec{k}_i\cdot\vec{\sigma}_i }{k^2+M_\pi^2}\nn
&+&2\,i\,[\vec{k}\times\vec{k}_i]+\vec{k}\,\vec{k}\cdot\vec{\sigma}_i
- 4\,\vec{k}_i\,\vec{k}_i\cdot\vec{\sigma}_i \nn
&+& \vec{\sigma}_i\Big(2(p_i^{\prime\,2}+p_i^2)-k^2\Big)\bigg)\,.\label{oneovermTo2corr}
\eeqa
These are a linear combination of on-shell and off-shell
contributions. The third kind of order-$Q$ contributions emerges from relativistic
$1/m$-corrections to the leading one-loop terms which contributes to $A_{{\rm
    1N:on-shell}}^{\mu, a}$:
\beqa
\label{SingleNChargeOneOvermNuclQTo1}
A^{0,  a \, (Q)}_{{\rm 1N:} \, 1/m}  &=&\bar{d}_{22}\vec{k}_i\cdot\vec{\sigma}_i[{\fet\tau}_i]^a\frac{k^2}{2 m},\\
\vec{A}^{ a \, (Q)}_{{\rm 1N: \, 1/m}}  &=& 0\,.
\eeqa
Finally, static two-loop contributions to the on-shell current are
given by
\beqa
A^{0,  a \, (Q)}_{{\rm 1N: \, static}}  &=& 0,\nn
\vec{A}^{a \, (Q)}_{{\rm 1N: \, static}}
&=&-\frac{1}{2}[{\fet\tau}_i]^a\vec{\sigma}_i \Big(
- f_0^A M_\pi^2 k^2 + 
f_1^A k^4 \nn
&+&
G_A^{(Q^4)}(-k^2)\Big)+\frac{1}{8}\vec{k}\,\vec{k}\cdot\vec{\sigma}_i
[{\fet\tau}_i]^a\Big(
-4 f_0^A M_\pi^2 \nn
&-& f_1^P k^2 + G_P^{(Q^2)}(-k^2)\Big)\,.\quad\quad\label{SingleNtwoLoopExpr}
\eeqa
Here we perform  the chiral expansion of the axial form factor which can be found
 e.g.~in~\cite{Bernard:1996cc,Fearing:1997dp}, 
see also~\cite{Bernard:1993bq,Schindler:2006it} for results obtained within Lorentz-invariant
formulations. Rewritten in our notation, the chiral expansion of the axial form factor is given
by 
\beqa
\label{tempGA}
G_A(t)&=&g_A+(\bar{d}_{22}+f_0^A M_\pi^2) t + f_1^A t^2 +
G_A^{(Q^4)}(t) \nn
&+& {\cal O}(Q^5),
\eeqa
where $f_i^A$ are LECs of dimension ${\rm GeV}^{-4}$ and 
\beq
G_A^{(Q^4)}(t) = \frac{t^3}{\pi}\int_{9 M_\pi^2}^\infty \frac{{\rm
    Im}\,G_A^{(Q^4)}(t^\prime\,)}{t^{\prime\,3}(t^\prime - t - i \epsilon)}d t^\prime\,,\label{TwoLoopGA}
\eeq
with the imaginary part calculated utilizing the Cutkosky rules~\cite{Bernard:1996cc}
\beqa
{\rm
    Im}\,G_A^{(Q^4)}(t)&=&\frac{g_A}{192 \pi^3 F_\pi^4}\int_{z^2 < 1} d
  \omega_1 d \omega_2\bigg[6 g_A^2 (\sqrt{t}\, \omega_1 -
  M_\pi^2)\nn
&\times&\Big(\frac{l_2}{l_1} + z\Big)\frac{\arccos(-z)}{\sqrt{1-z^2}}\nn
&+&2 g_A^2\Big(M_\pi^2 - \sqrt{t}\,\omega_1 -\omega_1^2\Big) + M_\pi^2 -
\sqrt{t}\,\omega_1\nn 
&+& 2\omega_1^2\bigg],
\eeqa
where 
\beqa
&&\omega_i=\sqrt{l_i^2 + M_\pi^2}\quad{\rm with}\quad i=1,2, \quad{\rm
  and}\nn
&&z=
  \hat{l}_1\cdot\hat{l}_2=\frac{\omega_1
    \omega_2-\sqrt{t}(\omega_1+\omega_2)+\frac{1}{2}(t+M_\pi^2)}{l_1 l_2}\,.
\eeqa
Here and in what follows, $l_i \equiv | \vec l_i |$, while $\hat l_i \equiv \vec
l_i/l_i$.  The pseudoscalar form-factor up to order $Q^4$ is given by~\cite{Kaiser:2003dr}
\beqa
G_P(t)&=&\frac{4m g_{\pi N} F_\pi}{M_\pi^2-t}-\frac{2}{3}g_A m^2
\langle r_A^2\rangle + m^2
%(f_0^P M_\pi^2 + 
f_1^P t\nn
%) 
&+& m^2 G_P^{(Q^2)}(t)+
{\cal O}(Q^3),\label{PseudoScalarFormFactorUpToQTo4}
\eeqa
where $f_i^P$ denotes the corresponding linear combinations of the  LECs of dimension ${\rm GeV}^{-4}$ from ${\cal L}_{\pi
  N}^{(5)}$ and 
\beqa
G_P^{(Q^2)}(t)&=&\frac{t^2}{\pi}\int_{9 M_\pi^2}^\infty \frac{{\rm
    Im}\,G_P^{(Q^2)}(t^\prime\,)}{t^{\prime\,2}(t^\prime - t - i \epsilon)}d t^\prime\,,\label{TwoLoopGP}
\eeqa
with the imaginary part calculated using the Cutkosky
rules~\cite{Kaiser:2003dr}
\beqa
{\rm Im}\,G_P^{(Q^2)}(t)&=&{\rm Im}\,G_P^{(1)}(t) + {\rm Im}\,G_P^{(2)}(t)
\eeqa
and
\beqa
{\rm Im}\,G_P^{(1)}(t)&=&\frac{g_A}{8\pi^3 F_\pi^4}\int_{z^2<1}
d\omega_1
d\omega_2\bigg[\frac{1}{18}-\frac{M_\pi^4}{12(t-M_\pi^2)^2}\nn
&+&\frac{4\omega_1^2-M_\pi^2}{6
  t}+\frac{\omega_1^2(3 M_\pi^2-t)}{(t-M_\pi^2)^2}\nn
&+&\frac{2 M_\pi^2
  \omega_1 \omega_2 z}{t(t-M_\pi^2)} \frac{l_2}{l_1}\bigg],\nn
{\rm Im}\,G_P^{(2)}(t)&=&\frac{g_A^3}{8\pi^3 F_\pi^4
  t}\int_{z^2<1} d\omega_1
d\omega_2\bigg[(M_\pi^2-\sqrt{t}\omega_1)\nn
&\times&\bigg(z+\frac{l_2}{l_1}\bigg)\frac{\arccos(-z)}{\sqrt{1-z^2}}
+ \frac{l_1^2}{3}+\frac{t}{9}\nn
&+&\frac{M_\pi^2}{t-M_\pi^2}\bigg(\frac{7}{8}\sqrt{t}-\omega_1-\omega_2\bigg)\bigg(2\omega_1
z
\frac{l_2}{l_1}+\sqrt{t}\nn
&+&\Big((t+M_\pi^2)(4\omega_1-\sqrt{t})-4\sqrt{t}\omega_1\omega_2\Big)\nn
&\times&\frac{\arccos(-z)}{2
  l_1 l_2 \sqrt{1-z^2}}\bigg)\bigg]\,.\quad\quad
\eeqa
It is important to note that the induced pseudoscalar form factor is
related to
the induced pseudoscalar coupling constant
\beqa
g_P&=&\frac{M_\mu}{2 m} G_P(t = -0.88 M_\mu^2),
\eeqa
which is measured in muon capture
experiment~\cite{Andreev:2012fj}. For theoretical determination of
$g_P$ by using chiral Ward identities of QCD we refer to a groundbreaking
work~\cite{Bernard:1994wn}, see also~\cite{Bernard:2001rs,Pastore:2014iga}.

In practical
calculations, alternatively to the chiral expansion of the axial and pseudoscalar form
factors, one can take their empirical parametrization~\cite{Bernard:2001rs}. This is in particular reasonable if we would like to
consider electroweak probes of nuclei without being affected by  the
convergence issue of the chiral expansion of
electroweak single-nucleon currents.
\subsubsection{Two-nucleon axial vector current}
\label{sec:2naxialvectorCurrentGeneral}
We now switch to a discussion of the two-nucleon axial vector current
operator. Various contributions can be characterized by the number of pion
exchanges and/or short-range interactions
\beqa
A_{{\rm 2N}}^{\mu,a}&=&A_{{\rm 2N:\,}1\pi}^{\mu,a}+A_{{\rm 2N:\,}2\pi}^{\mu,a}+A_{{\rm 2N:\,}{\rm cont}}^{\mu,a}.
\eeqa  
\paragraph{One-pion-exchange axial vector current}
\label{sec:OPEavCurrent}

Leading contribution to the one-pion-exchange (OPE) current shows up at the order $Q^{-1}$. At this
order we get a well known result~\cite{Kubodera:1978wr,Riska:1994gz,Ando:2001es}
\beqa
 A_{{\rm 2N:\,}1\pi }^{0, a \; ({Q^{-1}})} &=&
-\frac{i g_A 
\vec{q}_1\cdot \vec{\sigma}_1 [{\fet \tau}_1\times{\fet \tau}_2]^a}{4 
F_\pi^2 \left(q_1^2 +M_\pi^2\right)}\; + 1 \leftrightarrow
2\,,\label{AxialChargeTreeOPE}\\
\vec { A}_{{\rm 2N:\,}1\pi }^{a \; ({Q^{-1}})} &=& 0.\label{AxialVectorTreeOPE}
\eeqa

\noindent
At the order $Q^0$ there are only contributions to the vector operator
\beqa
\label{AxialCurrentTree1}
&&\vec {A}_{{\rm 2N:}\,  1\pi }^{a \; ({Q^{0}})} \,=\, \frac{g_A}{2 F_\pi^2}
\frac{\vec \sigma_1 \cdot \vec q_1}{q_1^2 + M_\pi^2} \bigg\{
[{\fet\tau}_1]^a \bigg[ - 4 c_1 M_\pi^2 \frac{\vec k}{k^2 + M_\pi^2}\nn
&&+ 2 c_3 \bigg( \vec q_1 - \frac{\vec k \, \vec k \cdot \vec q_1}{k^2 +
  M_\pi^2} \bigg) \bigg]+ c_4 [ \fet \tau_1 \times \fet \tau_2 ]^a \bigg( \vec q_1 \times
\vec \sigma_2 \nn
&&- \frac{\vec k \, \vec k \cdot \vec q_1 \times \vec
  \sigma_2}{k^2 + M_\pi^2} \bigg) 
-\frac{\kappa_v}{4 m}  [ \fet \tau_1 \times \fet \tau_2 ]^a
\vec{k}\times\vec{\sigma}_2\bigg\} \nn
&&+ 1 \leftrightarrow 2\,,
\eeqa
where $c_i$ denote the LECs from $\mathcal{L}_{\pi N}^{(2)}$
and $\kappa_v$ is the isovector anomalous magnetic moment of the nucleon.
\noindent At the order $Q$ there are leading one-loop
contributions in the static limit and/or leading relativistic correction
to OPE current 
\beqa
{A}_{{\rm 2N:1\pi}}^{\mu, a (Q)}&=&{A}_{{\rm 2N:1\pi,}\,{\rm
    static}}^{\mu,a (Q)}+{A}_{{\rm 2N:1\pi, }1/m}^{\mu,a (Q)}+{A}_{{\rm 2N:1\pi, UT}^\prime}^{\mu,a (Q)}.\quad
\eeqa
The explicit form of static contributions can be given in terms of
scalar functions $h_{1\dots 8}(q_2)$. The vector contribution is given
by
\beqa
\label{Current1pi}
&&\vec {A}_{{\rm 2N:}  1\pi,{\rm static}}^{a \, (Q)}= \frac{4 F_\pi^2}{g_A} \frac{\vec q_1
 \cdot \vec \sigma_1}{q_1^2 + M_\pi^2} \Big\{ [ \fet \tau_1 \times
\fet \tau_2 ]^a \nn
&&\Big( [\vec q_1 \times \vec \sigma_2 ] \, h_1(q_2) + [\vec
q_2 \times \vec \sigma_2 ] \, h_2(q_2)\Big) \nn
&&+ [{\fet \tau}_1]^a \big(\vec q_1 -
\vec q_2\big) h_3(q_2) \Big\}+ \frac{4 F_\pi^2}{g_A}\frac{\vec q_1
 \cdot \vec \sigma_1 \,\vec k }{(k^2 + M_\pi^2)(q_1^2 + M_\pi^2)}\nn
&&\Big\{[{\fet \tau}_1]^a h_4(q_2) +   [ \fet \tau_1 \times
\fet \tau_2 ]^a 
\vec q_1 \cdot [\vec q_2
  \times \vec \sigma_2] h_5(q_2)\Big\} \nn
&&\;+ \; 1
\leftrightarrow 2, 
\eeqa
and the charge contribution is given by 
\beqa
\label{Charge1piQTo1}
&&A_{{\rm 2N:} \, 1\pi,{\rm static}}^{0, a \, (Q)}= i \frac{4 F_\pi^2}{g_A} \frac{\vec q_1
  \cdot \vec \sigma_1}{q_1^2 + M_\pi^2} \Big\{
[ \fet \tau_1 \times
\fet \tau_2 ]^a \, \big( h_6(q_2)\nn
&&+ k^2 h_7(q_2) 
%+ \frac{1}{k^2 +
%  M_\pi^2}h_8(q_2)
\big) 
+ [{\fet\tau}_1]^a \, \vec q_1 \cdot [ \vec q_2 \times
\vec \sigma_2 ] \,h_8(q_2) 
%+ \frac{1}{k^2 + M_\pi^2}h_{10}(q_2)\big) 
\Big\} \; + \; 1
\leftrightarrow 2\,,\quad\quad 
\eeqa
where the scalar functions $h_i (q_2)$ are given by 
\beqa
h_1(q_2)&=&-\frac{g_A^6M_\pi}{128\pi F_\pi^6}, \nn
h_2(q_2)&=&\frac{g_A^4M_\pi}{256\pi F_\pi^6} + \frac{g_A^4 A(q_2)\left(4M_\pi^2+q_2^2
\right)}{256\pi F_\pi^6},\nn
h_3(q_2)&=&\frac{g_A^4\left(g_A^2+1
\right)M_\pi}{128\pi F_\pi^6} +\frac{g_A^4A(q_2)\left(2M_\pi^2+q_2^2
\right)}{128\pi F_\pi^6}, \nn
h_4(q_2)&=&\frac{g_A^4}{256\pi F_\pi^6}\Big(A(q_2)\left(2M_\pi^4+5M_\pi^2q_2^2+
2 q_2^4\right)\nn
&+&\left(4g_A^2+1\right)M_\pi^3+2\left(g_A^2+1
\right)M_\pi q_2^2\Big),\nn
h_5(q_2)&=&-\frac{g_A^4}{256\pi
  F_\pi^6}\Big(A(q_2)\left(4M_\pi^2+q_2^2\right)\nn
&+&
\left(2g_A^2+1\right)M_\pi\Big),\nn
h_6(q_2)&=&\frac{g_A^2\left(3\left(64+128 g_A^2 
\right)M_\pi^2+8\left(19 g_A^2+5\right)q_2^2\right)}{36864\pi^2F_\pi^6}\nn
&-&\frac{g_A^2}{768\pi^2F_\pi^6}
L(q_2)\big(\left(8g_A^2+4\right)M_\pi^2\nn
&+&\left(5g_A^2+1
\right)q_2^2\big)+\frac{\bar{d}_{18}g_A
  M_\pi^2}{8 F_\pi^4}-\frac{\bar{d}_{5}g_A^2M_\pi^2}{2F_\pi^4}\nn
&-&\frac{g_A^2(2\bar{d}_{2}+
\bar{d}_{6})\left(M_\pi^2+q_2^2
\right)}{16F_\pi^4},\nn
h_7(q_2)&=&\frac{g_A^2(2\bar{d}_{2}-\bar{d}_{6})}{16F_\pi^4},\nn
h_8(q_2)&=&-\frac{g_A^2(\bar{d}_{15}-2\bar{d}_{23})}{8F_\pi^4}. \label{OPEhiDefenition}
\eeqa
Here $\bar{d}_i$ are low-energy constants (LEC) from $Q^3$ pion-nucleon
Lagrangian. Their values can be fixed from pion-nucleon scattering and
axial-pion-production. 

\noindent 
The relativistic corrections for the charge operator vanish
\beq
A_{{\rm 2N:} \, 1\pi , \, 1/m}^{0,a \, (Q)}= 0,
\eeq
and for the current operator are given by
\beqa
\label{Current1piRel}
&&\vec {A}_{{\rm 2N:} \, 1\pi , \, 1/m}^{a \, (Q)}=\frac{g_A}{16 F_\pi^2 m}\bigg\{i [ \fet \tau_1 \times
\fet \tau_2 ]^a
\nn
&&\bigg[\frac{1}{(q_1^2+M_\pi^2)^2}\bigg(\vec{B}_1 - \frac{\vec k \, \vec k
  \cdot \vec{B}_1}{k^2+M_\pi^2} 
\bigg) \nn
&&+ \frac{1}{q_1^2+M_\pi^2}\bigg(
\frac{\vec{B}_2}{(k^2+M_\pi^2)^2} 
+ \frac{\vec{B}_3}{k^2+M_\pi^2} + \vec{B}_4\bigg)  \bigg]\nn
&&+ [{\fet\tau}_1]^a\bigg[ \frac{1}{(q_1^2+M_\pi^2)^2}\bigg(\vec{B}_5 -
\frac{\vec k \, \vec k \cdot \vec{B}_5}{k^2+M_\pi^2}
\bigg) \nn
&&+ \frac{1}{q_1^2+M_\pi^2}\bigg(
\frac{\vec{B}_6}{(k^2+M_\pi^2)^2} 
+ \frac{\vec{B}_7}{k^2+M_\pi^2} + \vec{B}_{8}\bigg)\bigg] \bigg\} \nn
&&+
1 \; \leftrightarrow \; 2\,,
\eeqa
where the vector-valued quantities $\vec B_i$ depend on various momenta and
the Pauli spin matrices and are given by 
\beqa
\label{BiDefinition}
\vec{B}_1&=&g_A^2\vec{q}_1\cdot\vec{\sigma}_1[
-2 (1+2\bar\beta_8)\vec{q}_1\,\vec{k}_1\cdot\vec{q}_1\nn
&-&(1-2\bar\beta_8)(2\vec{q}_1\,\vec{k}_2\cdot\vec{q}_1-i\,
\vec{q}_1\times\vec{\sigma}_2\,\vec{k}\cdot\vec{q}_1],\nn
\vec{B}_2&=&(1-2\bar{\beta}_8) g_A^2
\vec{k}\,\vec{k}\cdot\vec{q}_1\vec{q}_1\cdot\vec{\sigma}_1[2\vec{k}\cdot\vec{k}_2
-i \vec{k}\cdot\vec{q}_1\times\vec{\sigma}_2],\nonumber
\eeqa
\beqa
\vec{B}_3&=&2\vec{k}\Big[-g_A^2((1+2\bar\beta_9) \vec{k}\cdot\vec{q}_1\vec{k}_1
\cdot\vec{\sigma}_1\nn
&+&(1-2\bar\beta_9)\vec{q}_1\cdot\vec{\sigma}_1(
\vec{k}\cdot\vec{k}_2+\vec{k}_2\cdot\vec{q}_1))\nn
&+&\vec{q}_1\cdot\vec{\sigma}_1(
\vec{k}\cdot\vec{k}_2+i\,
\vec{k}\cdot\vec{q}_1\times\vec{\sigma}_2-\vec{k}_1\cdot\vec{q}_1+\vec{k}_2\cdot\vec{q}_1)
\Big],\nn
\vec{B}_4&=&g_A^2[2(1+2\bar\beta_9)\vec{q}_1\vec{k}_1\cdot\vec{\sigma}_1\nn
&+&(1-2\bar\beta_9)
\vec{q}_1\cdot\vec{\sigma}_1(2\vec{k}_2
-i\vec{k}\times\vec{\sigma}_2)]\nn
&-&2\vec{q}_1\cdot\vec{\sigma}_1(i
\vec{q}_1\times\vec{\sigma}_2 - i \vec{k}\times\vec{\sigma}_2 + 2 \vec{k}_2),\nn
\vec{B}_5&=&g_A^2\vec{q}_1\cdot\vec{\sigma}_1\Big[(1-2\bar\beta_8)(
\vec{q}_1\, \vec{k}\cdot\vec{q}_1-2i \,\vec{q}_1\times\vec{\sigma}_2\vec{k}_2\cdot\vec{q}_1
)\nn
&-&2i(1+2\bar\beta_8) \vec{q}_1\times\vec{\sigma}_2\,\vec{k}_1\cdot\vec{q}_1
\Big],\nn
\vec{B}_6&=&-(1-2\bar{\beta}_8) g_A^2
\vec{k}\,\vec{q}_1\cdot\vec{\sigma}_1[(\vec{k}\cdot\vec{q}_1)^2\nn
&-&2i\, \vec{k}\cdot\vec{k}_2
\vec{k}\cdot\vec{q}_1\times\vec{\sigma}_2],\nn
\vec{B}_7&=&g_A^2\vec{k}\Big[(1-2\bar\beta_9)\vec{q}_1\cdot\vec{\sigma}_1(
-2i(\vec{k}\cdot\vec{k}_2\times\vec{\sigma}_2\nn
&+&\vec{k}_2\cdot\vec{q}_1\times\vec{\sigma}_2)+k^2+q_1^2)\nn
&-&2i (1+2\bar\beta_9)
\vec{k}_1\cdot\vec{\sigma}_1\vec{k}\cdot\vec{q}_1\times\vec{\sigma}_2\Big],\nn
\vec{B}_{8}&=&-g_A^2[(1-2\bar\beta_9)\vec{q}_1\cdot\vec{\sigma}_1(\vec{k}
-2i\,\vec{k}_2\times\vec{\sigma}_2)\nn
&-&2i(1+2\bar\beta_9) \vec{q}_1\times\vec{\sigma}_2\,\vec{k}_1\cdot\vec{\sigma}_1].\label{OPEOneOvermNuclFFs}
\eeqa
Finally, there are also energy-transfer dependent contributions to OPE
axial vector current at order $Q$ which are given by
\beqa
&&A_{{\rm 2N:\,}1\pi, {\rm UT^\prime} }^{0, a \; ({Q})} = 0\,,\\
&&\vec {A}_{{\rm 2N:\,}1\pi, {\rm UT^\prime} }^{a \; ({Q})} =
- i \frac{g_A}{8 F_\pi^2}\frac{k_0\, \vec k \, \vec{q}_1\cdot\vec{\sigma}_1}{(k^2+M_\pi^2)(q_1^2+M_\pi^2)}\nn
&&\bigg([{\fet \tau}_1\times{\fet 
\tau}_2]^a\bigg(1-\frac{2
g_A^2\vec{k}\cdot\vec{q}_1}{k^2+M_\pi^2}\bigg)-\frac{2 g_A^2[{\fet\tau}_1]^a \vec{k}\cdot[\vec{q}_1\times\vec{\sigma}_2]}{k^2+M_\pi^2}\bigg)\nn
&&+  1 \leftrightarrow 2\,. \quad\quad
\label{NNCurrent1piUTPrime}
\eeqa
It is important to note that it is not enough to know the currents at vanishing
energy-transfer $k_0=0$. As will be demonstrated later the knowledge of
the slope in energy-transfer $k_0$ is essential for checking the
continuity equations. All expressions proportional to the energy-transfer
$k_0$ are off-shell effects which disappear in the calculation of on-shell
observables. Energy-transfer contributions 
are always accompanied by the
commutator with the effective Hamiltonian. On-shell a linear combination of
$k_0$-term and the commutator with the effective Hamiltonian 
\beqa
k_0 X - \big[H_{\rm eff}, X\big]
\eeqa
vanishes. Here $X$ stays for some operator. More on this
will be discussed in Sec.~\ref{sec:beta8beta9Dependence}.
\paragraph{Two-pion-exchange axial vector current}
\label{sec:TPEaxialvectorCurrent}
Contributions to the two-pion-exchange axial vector current start to show
up at order $Q$. These contributions are parameter-free. The final results for the two-pion exchange
operators read
\beqa
\label{Current2pi}
\vec {A}_{{\rm 2N:} \, 2\pi}^{a\, (Q)}&=& \frac{2 F_\pi^2}{g_A}\frac{\vec{k}}{k^2+M_\pi^2}
\bigg\{
[{\fet\tau}_1]^a \Big(-\vec{q}_1\cdot \vec{\sigma}_2\, \vec{q}_1\cdot\vec{k}\,
g_1(q_1) \nn
&+& \vec{q}_1\cdot \vec{\sigma}_2\, g_2(q_1) - 
\vec{k}\cdot \vec{\sigma}_2\, g_3(q_1)\Big) \nn
&+& 
[{\fet\tau}_2]^a \Big(-\vec{q}_1\cdot \vec{\sigma}_1\, \vec{q}_1\cdot\vec{k}\,
g_4(q_1)-\vec{k}\cdot \vec{\sigma}_1 \,g_5(q_1) \nn
&-&\vec{q}_1\cdot
\vec{\sigma}_2\, 
\vec{q}_1\cdot\vec{k}\,g_6(q_1)
+ \vec{q}_1\cdot \vec{\sigma}_2\,
g_7(q_1) \nn
&+& \vec{k}\cdot
\vec{\sigma}_2\, 
\vec{q}_1\cdot\vec{k}\,g_8(q_1)
-\vec{k}\cdot \vec{\sigma}_2\, g_9(q_1) \Big)\nn
&+&
 [ \fet \tau_1 \times
\fet \tau_2 ]^a \Big( - \vec q_1 \cdot [ \vec \sigma_1 \times
\vec \sigma_2 ] \, \vec{q}_1\cdot\vec{k} \,g_{10}(q_1)\nn
&+&\vec q_1 \cdot [ \vec \sigma_1 \times
\vec \sigma_2 ] \, g_{11}(q_1) \nn
&-&  \vec q_1 \cdot
\vec \sigma_2 \,
\vec{q}_1\cdot[\vec q_2 \times \vec \sigma_1 ]\,
g_{12}(q_1)\Big)\bigg\}\nn
&+&\frac{2 F_\pi^2}{g_A}\bigg\{
\vec{q}_1\Big(
[{\fet\tau}_2]^a \,\vec{q}_1\cdot \vec{\sigma}_1\,g_{13}(q_1)\nn
&+&[{\fet\tau}_1]^a \,\vec{q}_1\cdot \vec{\sigma}_2\, g_{14}(q_1)\Big)
-[{\fet\tau}_1]^a \,\vec{\sigma}_2\, g_{15}(q_1)\nn 
&-& [{\fet\tau}_2]^a \,\vec{\sigma}_2\,
g_{16}(q_1) -[{\fet\tau}_2]^a \,\vec{\sigma}_1\, g_{17}(q_1)
\bigg\}\nn [4pt]
& + & 1
\leftrightarrow 2\,, \\
A_{{\rm 2N:} \, 2\pi}^{0, a \, (Q)}&=& i \frac{2 F_\pi^2}{g_A} \bigg\{
[ \fet \tau_1 \times
\fet \tau_2 ]^a
\vec q_1 \cdot \vec \sigma_2\,  g_{18}(q_1)\nn
&+& [{\fet\tau}_2]^a \vec q_1 \cdot [ \vec \sigma_1 \times
\vec \sigma_2 ] 
g_{19}(q_1)
\bigg\} \; + \; 1
\leftrightarrow 2\,, 
\label{Charge2pi}
\eeqa
where the scalar functions $g_i(q_1)$ are defined as 
\beqa
g_1(q_1)&=&\frac{g_A^4A(q_1)\left(\left(8g_A^2-4
\right)M_\pi^2+\left(g_A^2+1\right)q_1^2\right)}{256\pi 
F_\pi^6q_1^2}\nn
&-&\frac{g_A^4M_\pi\left(\left(8g_A^2-4
\right)M_\pi^2+\left(3g_A^2-1\right)q_1^2\right)}{256\pi 
F_\pi^6q_1^2\left(4M_\pi^2+q_1^2\right)},\nn
g_2(q_1)&=&\frac{g_A^4A(q_1)\left(2M_\pi^2+q_1^2\right)}{128\pi 
F_\pi^6}+\frac{g_A^4M_\pi}{128\pi F_\pi^6} 
,\nn
g_3(q_1)&=&-\frac{g_A^4A(q_1)\left(
\left(8g_A^2-4\right)M_\pi^2+\left(3g_A^2-1\right)q_1^2
\right)}{256\pi F_\pi^6}\nn
&-&\frac{\left(3g_A^2-1\right)g_A^4M_\pi}{256\pi 
F_\pi^6},\nn
g_4(q_1)&=&-\frac{g_A^6A(q_1)}{128\pi F_\pi^6},\nn
g_5(q_1)&=&-q_1^2\,g_4(q_1),\nn [4pt]
g_6(q_1)&=& g_8(q_1)\,=\, g_{10}(q_1)\,=\, g_{12}(q_1)\,=\, 
0,\nn [4pt]
g_7(q_1)&=&\frac{g_A^4A(q_1)\left(2M_\pi^2+q_1^2\right)}{128\pi 
F_\pi^6}+\frac{\left(2g_A^2+1\right)g_A^4M_\pi}{128\pi
F_\pi^6},\nn
%g_8(q_1)&=& 0,\nn 
g_9(q_1)&=&\frac{g_A^6M_\pi}{64\pi F_\pi^6},\nonumber
\eeqa
\beqa
%g_{10}(q_1)&=&0,\nn
g_{11}(q_1)&=&-\frac{g_A^4A(q_1)\left(4M_\pi^2+q_1^2\right)}{512\pi 
F_\pi^6}-\frac{g_A^4M_\pi}{512\pi F_\pi^6},\nn
%g_{12}(q_1)&=&0,\nn
g_{13}(q_1)&=&
-\frac{g_A^6A(q_1)}{128\pi F_\pi^6},\nn
g_{14}(q_1)&=&\frac{g_A^4A(q_1)\left(\left(8g_A^2-4
\right)M_\pi^2+\left(g_A^2+1\right)q_1^2\right)}{256\pi 
F_\pi^6q_1^2}\nn
&+&\frac{g_A^4M_\pi\left(\left(4-8g_A^2
\right)M_\pi^2+\left(1-3g_A^2\right)q_1^2\right)}{256\pi 
F_\pi^6q_1^2\left(4M_\pi^2+q_1^2\right)},\nn
g_{15}(q_1)&=&\frac{g_A^4A(q_1)\left(\left(8g_A^2-4
\right)M_\pi^2+\left(3g_A^2-1\right)q_1^2\right)}{256\pi 
F_\pi^6}\nn
&+&\frac{\left(3g_A^2-1\right)g_A^4M_\pi}{256\pi F_\pi^6},\nn
g_{16}(q_1)&=&\frac{g_A^4A(q_1)\left(
2M_\pi^2+q_1^2\right)}{64\pi F_\pi^6}+\frac{g_A^4M_\pi}{64\pi F_\pi^6},\nn
g_{17}(q_1)&=&-\frac{g_A^6q_1^2A(q_1)}{128\pi F_\pi^6},\nn
g_{18}(q_1)&=&\frac{g_A^2L(q_1)\left(\left(4-8g_A^2
\right)M_\pi^2+\left(1-3g_A^2\right)q_1^2
\right)}{128\pi^2F_\pi^6\left(4M_\pi^2+q_1^2\right)}, \nn
g_{19}(q_1)&=&\frac{g_A^4L(q_1)}{32\pi^2F_\pi^6}. \label{giOfq1TPEStrF}
\eeqa
\paragraph{Short-range axial vector current}
\label{sec:contactaxialVectorCurrent}
The first contribution to short-range two-nucleon current shows up at
the order $Q^0$.
\beqa
A_{{\rm 2N:}\,  \rm cont}^{0, a \; ({Q^{0}})} &=&0,\nn
\vec {A}_{{\rm 2N:}\,  \rm cont}^{a \; ({Q^{0}})} &=& -\frac{1}{4} D\,
[{\fet\tau}_1]^a \bigg(\vec{\sigma}_1-\frac{\vec{k}\, \vec{\sigma}_1\cdot\vec{k}}{k^2+M_\pi^2}\bigg) + \; 1 \leftrightarrow 2\,,\quad\quad\label{axial_vector:2N:cont:QTo0}
\eeqa
where  $D$ denote the LEC from  $\mathcal{L}_{\pi NN}^{(1)}$.
At the order $Q$ we decompose the short-range current in three
different components
\beqa
A_{{\rm 2N:cont}}^{\mu,a, (Q)}&=&A_{{\rm 2N:cont,\,static}}^{\mu,a, (Q)} +
A_{{\rm 2N:cont\,,}1/m}^{\mu,a , (Q)} + A_{{\rm 2N:cont,\,UT}^\prime}^{\mu,a, (Q)}.\nonumber
\eeqa
Static contributions are given by
\beqa
\label{CurrentCont}
&&\vec {A}_{\rm 2N: \, cont,\,static}^{a \, (Q)}= 0,\\
&&A_{\rm 2N: \, cont,\,static}^{0, a \, (Q)} = i z_1 [\fet \tau_1 \times \fet
\tau_2]^a \, \vec \sigma_1 \cdot \vec q_2 \nn 
&&+  
 i z_2 [\fet \tau_1 \times \fet
\tau_2]^a \, \vec \sigma_1 \cdot \vec q_1\;  +\;  
 i z_3 [{\fet\tau}_1]^a \, \vec q_2 \cdot \vec \sigma_1 \times \vec \sigma_2 \nn
&& +  z_4 ([{\fet\tau}_1]^a - [{\fet\tau}_2]^a) (\vec \sigma_1 - \vec \sigma_2) \cdot \vec
k_1 \; + \; 1 \leftrightarrow 2\,,\label{ChargeContQTo1}
\eeqa
where LECs $z_i$ are unknown coefficients and have to be fitted to 
experimental data. Relativistic corrections are given by
\beqa
\label{CurrentContRel}
&&\vec {A}_{{\rm 2N: \, cont,}\, 1/m}^{a \, (Q)}=-\frac{g_A }{4m}\frac{\vec{k}}{k^2+M_\pi^2}
[{\fet\tau}_1]^a\bigg\{(1-2\bar{\beta}_9)\Big(C_S\vec{q}_2\cdot\vec{\sigma}_1\nn
&&+C_T(\vec{q}_2\cdot\vec{\sigma}_2
+2i\,\vec{k}_1\cdot\vec{\sigma}_1\times\vec{\sigma}_2)\Big)\nn
&&-\frac{1-2\bar{\beta}_8}{k^2+M_\pi^2} \Big(C_S\vec{k}\cdot\vec{q}_2\vec{k}\cdot\vec{\sigma}_1
+C_T(\vec{k}\cdot\vec{q}_2\vec{k}\cdot\vec{\sigma}_2\nn
&&+2i \,\vec{k}\cdot\vec{k}_1
\vec{k}\cdot\vec{\sigma}_1\times\vec{\sigma}_2)\Big)\bigg\} \; +\;
1\leftrightarrow 2\,.
\eeqa
Finally, the energy-transfer dependent contributions are given by
\beqa
A_{{\rm 2N:\,cont,\,UT^\prime} }^{0, a \; ({Q})}
&=&0,\nn
\vec {A}_{{\rm 2N:\,cont,\,UT^\prime}}^{a \; ({Q})} &=&-i\,k_0 \vec{k}\frac{g_A C_T
\vec{k}\cdot \vec{\sigma}_1 [{\fet 
\tau}_1\times{\fet \tau}_2]^a}{
\left(k^2+M_\pi^2\right)^2}\nn
&+& 1 \leftrightarrow 2\,.\label{NNCurrentContUTPrime}
\eeqa
\subsubsection{Three-nucleon axial vector current}
\label{sec:3NaxialvectorCurrent}
At the order $Q$ there are first contributions to three-nucleon axial vector
current. We decompose it into long and short-range contributions
\beqa
A_{{\rm 3N}\,
  }^{\mu,a \, (Q)} =A_{{\rm 3N:}\,
  \pi}^{\mu,a \, (Q)} + A_{{\rm 3N:\,cont}}^{\mu,a \, (Q)}
\eeqa
We start with the long-range part. There are no charge contributions such that
\beq
A_{{\rm 3N:}\,
  \pi}^{0,a \, (Q)} =0.
\eeq
Current contributions are given by
\beq
\label{Current3NPi}
\vec {A}_{{\rm 3N:}\,
  \pi}^{a \, (Q)} =- \frac{2 F_\pi^2}{g_A} \sum_{i=1}^{8}   \vec C_i^a
+
5\,{\rm permutations}, 
\eeq
where spin-isospin dependent vector structures, which include up to
four pion propagators are given by
\beqa
\vec C_1^a&=&\frac{g_A^6}{16 F_\pi^6} \vec{q}_2\cdot \vec{\sigma}_2 \bigg[[{\fet\tau}_2]^a \Big[
\vec{q}_3 ( (\vec{q}_2\cdot \vec{q}_3+q_2^2) 
(\fet\tau_1\cdot \fet
\tau_3\nn
&-&\vec{\sigma}_1\cdot \vec{\sigma}_3)+\vec{q}_2\cdot \vec{\sigma}_1 (\vec{q}_2\cdot 
\vec{\sigma}_3+\vec{q}_3\cdot \vec{\sigma}_3))\nn
&-&
\vec{q}_2 (\vec{q}_3\cdot \vec{\sigma}_1 (\vec{q}_2\cdot 
\vec{\sigma}_3+\vec{q}_3\cdot \vec{\sigma}_3)-(\vec{q}_2\cdot 
\vec{q}_3+q_3^2) \vec{\sigma}_1\cdot 
\vec{\sigma}_3\nn
&-&(\vec{q}_2\cdot \vec{q}_3+q_2^2) 
\fet\tau_1\cdot \fet\tau_3)
-\vec{\sigma}_3 
((\vec{q}_2\cdot \vec{q}_3+q_3^2) 
\vec{q}_2\cdot \vec{\sigma}_1\nn
&-&(\vec{q}_2\cdot 
\vec{q}_3+q_2^2) \vec{q}_3\cdot 
\vec{\sigma}_1)\Big]- \left[\fet\tau_2\times \fet\tau_3\right]^a (\vec{q}_2
\times \vec{\sigma}_1\nn
&+&\vec{q}_3\times \vec{\sigma}_1) (\vec{q}_2\cdot \vec{q}_3+q_2^2)
-(
\vec{q}_2+\vec{q}_3)\nn
&\times& ([\fet\tau_1\times 
\fet\tau_2]^a \vec{q}_2\cdot \vec{q}_3\times \vec{\sigma}_3+[{\fet\tau}_3]^a \fet
\tau_1\cdot \fet\tau_2
(\vec{q}_2\cdot \vec{q}_3+q_2^2))\bigg]\nn
&\times&\frac{1}{ [q_2^2 +M_\pi^2]
[(\vec{q}_1-\vec{k})^2 +M_\pi^2]^2},\nn
\vec C_2^a&=&\frac{g_A^4}{32 F_\pi^6} \frac{1}{ [q_2^2 +M_\pi^2]
[(\vec{q}_1-\vec{k})^2+M_\pi^2]}\nn
&\times&\vec{q}_2\cdot \vec{\sigma}_2 \big[[\fet\tau_2\times \fet\tau_3]^a 
(\vec{ k}\times \vec{\sigma}_1-\vec{q}_1
\times \vec{\sigma}_1)\nn
&+&(\vec{k}- \vec{q}_1) (
[{\fet\tau}_3]^a \fet\tau_1\cdot \fet\tau_2-[{\fet\tau}_2]^a \fet\tau_1\cdot \fet
\tau_3)\big],\nn
\vec C_3^a&=&\frac{g_A^4}{32 F_\pi^6}\frac{1}{ [q_1^2 +M_\pi^2] [
q_2^2 +M_\pi^2] [q_3^2 +M_\pi^2]}\nn
&\times& (\vec{ k}-2 
\vec{q}_3)[{\fet\tau}_3]^a \fet\tau_1\cdot \fet\tau_2\vec{q}_1\cdot \vec{\sigma}_1 \vec{q}_2\cdot 
\vec{\sigma}_2 \nn
&\times&(\vec{k}\cdot \vec{\sigma}_3-2 \vec{q}_1\cdot 
\vec{\sigma}_3),\nn
\vec C_4^a&=&\frac{g_A^4 }{32 F_\pi^6}\frac{\vec{\sigma}_1 \vec{q}_2\cdot 
\vec{\sigma}_2 \vec{q}_3\cdot \vec{\sigma}_3 ([{\fet\tau}_1]^a \fet
\tau_2\cdot \fet\tau_3-[{\fet\tau}_3]^a \fet\tau_1\cdot \fet\tau_2)}{ 
[q_2^2 + M_\pi^2] [q_3^2 +M_\pi^2]},\nonumber
\eeqa
\beqa
\vec C_5^a&=&\frac{g_A^6}{16 F_\pi^6} \frac{\vec{k} \,\vec{q}_1\cdot \vec{\sigma}_1}{ [k^2+M_\pi^2] [q_1^2 +M_\pi^2
] [(\vec{q}_1+\vec{q}_2)^2 +M_\pi^2]^2}
\nn
&\times&\bigg[-[{\fet\tau}_1]^a \vec{q}_1\cdot \vec{q}_2\times 
\vec{\sigma}_2 (\vec{k}\cdot \vec{q}_1\times \vec{\sigma}_3
+\vec{k}\cdot 
\vec{q}_2\times \vec{\sigma}_3)\nn
&-&\left[\fet
\tau_1\times \fet\tau_3\right]^a\vec{q}_1\cdot \vec{q}_2\times \vec{\sigma}_2 (\vec{k}\cdot \vec{q}_1+\vec{k}\cdot 
\vec{q}_2) \nn
&+&\left[\fet
\tau_1\times \fet\tau_2\right]^a (\vec{q}_1\cdot 
\vec{q}_2+q_1^2) (\vec{k}\cdot \vec{q}_1\times 
\vec{\sigma}_3\nn
&+&\vec{k}\cdot \vec{q}_2\times \vec{\sigma}_3)
-\left(\fet\tau_2\cdot \fet\tau_3 [{\fet\tau}_1]^a-\fet\tau_1\cdot \fet\tau_3
  [{\fet\tau}_2]^a\right) 
\nn
&\times& (\vec{q}_1\cdot 
\vec{q}_2+q_1^2) (\vec{k}\cdot \vec{q}_1+\vec{k}\cdot 
\vec{q}_2)
\bigg],\nn
\vec C_6^a &=&-\frac{g_A^4}{32 F_\pi^6} \frac{\vec{k}\, \vec{q}_1\cdot \vec{\sigma}_1}{ 
[k^2+M_\pi^2] [q_1^2 +M_\pi^2]
[(\vec{q}_1+\vec{q}_2)^2 +M_\pi^2]}\nn
&\times&\bigg[ (\vec{k}
\cdot \vec{q}_1\times \vec{\sigma}_3+\vec{k}\cdot \vec{q}_2\times 
\vec{\sigma}_3) \left[\fet\tau_1\times \fet\tau_2\right]^a\nn
&-&(\vec{k}\cdot 
\vec{q}_1+\vec{k}\cdot \vec{q}_2+\vec{q}_1\cdot 
\vec{q}_2+q_1^2) \nn
&\times&\left(\fet\tau_2\cdot \fet
\tau_3 [{\fet\tau}_1]^a-\fet\tau_1\cdot \fet\tau_3
[{\fet\tau}_2]^a\right)\nn
&-& \vec{q}_1\cdot \vec{q}_2\times \vec{\sigma}_2 
\left[\fet\tau_1\times \fet\tau_3\right]^a\bigg],\nn
\vec C_7^a&=&-\frac{g_A^4}{32 F_\pi^6}\vec{ k}\, \vec{q}_1\cdot \vec{\sigma}_1 \vec{q}_2
\cdot \vec{\sigma}_2 \vec{q}_3\cdot \vec{\sigma}_3 \fet\tau_1\cdot 
\fet\tau_2 [{\fet\tau}_3]^a\nn
&\times&\left(M_\pi^2+2 \vec{q}_1\cdot \vec{q}_2+q_1^2+q_2^2\right)\nn
&\times&\frac{1 }{ [k^2+M_\pi^2]
[q_1^2 +M_\pi^2] [q_2^2 +M_\pi^2]
[q_3^2 +M_\pi^2]},\nn
\vec C_8^a&=&-\frac{g_A^4}{64 F_\pi^6}\frac{ \vec{ k}\, \vec{q}_2\cdot \vec{\sigma}_2 
\vec{q}_3\cdot \vec{\sigma}_3}{ [k^2+M_\pi^2] 
[q_2^2 +M_\pi^2] [q_3^2 +M_\pi^2]}\nn
&\times& \left(\fet\tau_2\cdot \fet\tau_3 [{\fet\tau}_1]^a
(\vec{q}_2\cdot \vec{\sigma}_1+\vec{q}_3\cdot \vec{\sigma}_1) \right.\nn
&-&\left. 2 \fet\tau_1\cdot \fet\tau_2 [{\fet\tau}_3]^a
(\vec{q}_1\cdot \vec{\sigma}_1+\vec{q}_2\cdot \vec{\sigma}_1) 
\right)\,. \label{Ci3NDefinition}
\eeqa
The short-range contributions to  the charge operator vanish
\beqa
A_{\rm 3N:\, cont}^{0, a \, (Q)} &=&0,
\eeqa
and the short-range vector contributions are given by
\beq
\label{Current3NCont}
\vec {A}_{\rm 3N:\, cont}^{a \, (Q)} =- \frac{2 F_\pi^2}{g_A} \sum_{i=1}^{3}   \vec D_i^a
+
5\,{\rm permutations}, 
\eeq
with 
\beqa
\vec D_1^a&=&-\frac{g_A^4 C_T}{4 
F_\pi^4} \big[(\vec{k}-\vec{q}_1) 
[\fet\tau_1\times \fet\tau_3]^a (\vec{k}\cdot \vec{\sigma}_2\times 
\vec{\sigma}_3\nn
&-&\vec{q}_1\cdot \vec{\sigma}_2\times \vec{\sigma}_3)-(
[{\fet\tau}_2]^a-[{\fet\tau}_3]^a) ((
\vec{k}-\vec{q}_1)\vec{\sigma}_1\cdot \vec{\sigma}_2 \nn
&\times&(\vec{k}\cdot
\vec{\sigma}_3
-\vec{q}_1\cdot 
\vec{\sigma}_3)+\vec{\sigma}_3 ((\vec{k}\cdot 
\vec{\sigma}_1-\vec{q}_1\cdot \vec{\sigma}_1)\nn
&\times& (\vec{k}\cdot \vec{\sigma}_2-
\vec{q}_1\cdot \vec{\sigma}_2)
+(2 \vec{k}\cdot
\vec{q}_1-k^2-q_1^2) \vec{\sigma}_1\cdot
\vec{\sigma}_2))\big]\nn
&\times&\frac{1}{[(\vec{q}_1-\vec{k})^2+M_\pi^2]^2},\nonumber
\eeqa
\beqa
\vec{D}_{2}^a&=&\frac{g_A^4 C_T}{4 F_\pi^4}\frac{ \vec{ k} }{ [k^2+M_\pi^2]
[(\vec{q}_1+\vec{q}_3)^2 +M_\pi^2]^2}(\vec{q}_1\cdot
  \vec{\sigma}_1
\times \vec{\sigma}_3\nn
&+&\vec{q}_3\cdot \vec{\sigma}_1\times 
\vec{\sigma}_3) \left([{\fet\tau}_3]^a \vec{k}\cdot \vec{q}_2
\times \vec{\sigma}_2+\left[\fet\tau_2
\times \fet\tau_3\right]^a (k^2\right.\nn
&-&\left.\vec{k}\cdot \vec{q}_2) \right),\nn
\vec{D}_{3}^a&=&-\frac{g_A^2 C_T}{8 F_\pi^4 }\frac{ \vec{ k}}{
[k^2+M_\pi^2] [(\vec{q}_1+\vec{q}_3)^2 +M_\pi^2]}(\vec{q}_1\cdot
  \vec{\sigma}_1
\times \vec{\sigma}_3\nn
&+&\vec{q}_3\cdot \vec{\sigma}_1\times 
\vec{\sigma}_3) \left[\fet\tau_2\times \fet\tau_3\right]^a. \label{DiFF3NDefinition}
\eeqa
Due to the smallness of $C_T$, these contributions are expected to be small.
\subsection{Pseudoscalar current up to order $Q^0$}
\label{pseudoScalarUpToOrderQ}
Approximate chiral symmetry leads to relations between pseudoscalar current and axial-vector current. In the following, we list all
expressions for pseudoscalar current up to order $Q^0$. 
\subsubsection{Single-nucleon pseudoscalar current}
\label{sec:pseudoscalarSingleNucleon}
Single-nucleon pseudoscalar current can be parametrized by
pseudoscalar form factors and their derivatives via
\beqa
i\,m_q {P}_{{\rm 1N}}^{a}&=&i\,m_q {P}_{{\rm 1N:on-shell}}^{a}+i\,m_q {P}_{{\rm 1N:off-shell}}^{a},
\eeqa
where
\beqa
&&i\,m_q\,{P}_{{\rm 1N:on-shell}}^{a}\,=\,\frac{[{\fet\tau}_i]^a}{2}
\bigg(G_A(t)+\frac{t}{4 m^2} G_P(t)\bigg)\nn
&&\bigg(\vec{k}\cdot\vec{\sigma}\bigg(1-\frac{4\,k_1^2+k^2}{8
  m^2}\bigg)-\frac{\vec{k}\cdot\vec{k}_1\,\vec{k}_1\cdot\vec{\sigma}}{2
  m^2} + {\cal
  O}(1/m^3) \bigg),\nn
&&i\,m_q {P}_{{\rm 1N:off-shell}}^{a}\,=\,-\frac{M_\pi^2}{k^2}
\vec{k}\cdot\vec{A}_{{\rm 1N:off-shell}}^{a}.
\eeqa 
Chiral EFT results for pseudoscalar current are given by chiral expansion of the axial and
pseudoscalar form factors. Corresponding expressions are worked out
in~\cite{Krebs:2016rqz}. Here we will
briefly discuss them.

The single-nucleon pseudoscalar current starts to contribute at the order
$Q^{-4}$ and is given by 
\beqa
P^{ a \, (Q^{-4})}_{{\rm 1N: \, static}}  &=& i\,\frac{M_\pi^2 g_A}{2 m_q} [{\fet\tau}_i]^a
\frac{\vec{k}\cdot\vec{\sigma}_i}{k^2+M_\pi^2} \nn
&=& - i
\frac{1}{m_q} \, \vec k \cdot \vec {A}_{\rm 1N:\, static }^{a \; (Q^{-3})} 
,\label{LOSingleNCurrentPExpr}
\eeqa
with $\vec {A}_{\rm 1N:\, static }^{a \; (Q^{-3})} $ given in
Eq.~(\ref{LOSingleNCurrentExpr}). At the order $Q^{-3}$, there are only
vanishing contributions. At order $Q^{-2}$  there are only static limit
contributions which are given by 
\beqa
P^{  a \, (Q^{-2})}_{{\rm 1N: \, static}}  &=&  i\frac{M_\pi^2 \bar{d}_{18}}{m_q}
k^2 [{\fet\tau}_i]^a\frac{\vec{k}\cdot\vec{\sigma}_i}{k^2+M_\pi^2}\nn
&=& - i
\frac{1}{m_q} \, \vec k \cdot \vec {A}_{\rm 1N:\, static }^{a \; (Q^{-1})} 
\,,\label{SingleNPQtoMinus2StaticExpr}
\eeqa
with $\vec {A}_{\rm 1N:\, static }^{a \; (Q^{-1})} $ given in
Eq.~(\ref{SingleNQtoMinus1StaticExpr}).  There are no contributions to the single-nucleon
pseudoscalar current at order $Q^{-1}$. Finally, there are various
corrections at order $Q^0$: There are relativistic corrections that depend on
the energy transfer. Their
explicit form is given by
\beqa
P^{ a \, (Q^0)}_{{\rm 1N:\,}1/m, {\rm UT^\prime}}  &=&
i \frac{M_\pi^2}{m_q k^2} 
\, \vec k \cdot \vec {A}_{{\rm 1N:\,}1/m, {\rm UT^\prime}}^{a \; (Q)} \,,
%-i\frac{G_\pi g_A k_0}{8 F_\pi
%  m}\frac{1}{k^2+M_\pi^2}\tau_i^a
%\bigg(
%2(1 + 2 \,\bar{\beta}_9)\vec \sigma_i \cdot \vec k_i - (1 + 2
%%\,\bar{\beta}_8)\vec{k}\cdot\vec{\sigma}_i\frac{p_i^{\prime\,2}-p_i^2}{k^2+M_\pi^2}\bigg).
\label{k0OvermNuclcorrP}
\eeqa
with $\vec {A}_{{\rm 1N:\,}1/m, {\rm UT^\prime}}^{a \; (Q)}$ given in
Eq.~(\ref{k0OvermNuclcorr}). The second kind of the order-$Q^0$ contributions is given by relativistic
$1/m^2$ - corrections. 
\beqa
P^{ a \, (Q^0)}_{{\rm 1N: \, 1/m^2}}  &=& i \frac{M_\pi^2 g_A}{16 m_q
  m^2}[{\fet\tau}_i]^a\bigg(\vec{k}\cdot\vec{\sigma}_i(1-2\bar{\beta}_8)\frac{(p_i^{\prime\,2}-p_i^2)^2}{(k^2+M_\pi^2)^2}\nn
&-&2 \frac{(p_i^{\prime\, 2} +
  p_i^2)\vec{k}\cdot\vec{\sigma}_i-2 \bar{\beta}_9 (p_i^{\prime\, 2} -
  p_i^2)\vec{k}_i\cdot\vec{\sigma}_i }{k^2+M_\pi^2}\bigg)\,.\nn
&&\label{oneovermTo2corrP}
\eeqa
Notice that this expression is related to the pion-pole
terms in the corresponding axial current operator $\vec {A}_{{\rm 1N: \, 1/m^2}}^{a \; (Q)} $ whose expression is given in
Eq.~(\ref{oneovermTo2corr}),  via 
\beq
P^{ a \, (Q^0)}_{{\rm 1N: \, 1/m^2}} = i \frac{M_\pi^2}{m_q k^2} 
\, \vec k \cdot \vec {A}_{{\rm 1N: \, 1/m^2}}^{a \; (Q)} \bigg|_{\rm
  pion-pole \; terms} \,.
\eeq
The third kind of order-$Q^0$ contributions are coming from static
two-loop contributions and is given by
\beqa
P^{a\, (Q^0)}_{{\rm 1N:\, static}}&=& i\frac{1}{8
  m_q}\vec{k}\cdot\vec{\sigma}_i [{\fet\tau}_i]^a
%\bigg(f_0^P t +
\Big(4 \,G_A^{(Q^4)}(-k^2)\nn
&-&k^2 \,G_P^{(Q^2)}(-k^2)\Big)
%\bigg)
.\label{PseudoscalarTwoLoopConjecture}
\eeqa 
where $G_A^{(Q^4)}(t)$ and  $G_P^{(Q^2)}(t)$ are defined in
Eqs.~(\ref{TwoLoopGA}) and (\ref{TwoLoopGP}), respectively.

Alternatively to the chiral expansion of the axial and pseudoscalar form
factors, one can take their empirical parametrization in practical
calculations. 
\subsubsection{Two-nucleon pseudoscalar current}
\label{sec:2nPseudoscalarCurrentGeneral}
Similar to axial vector current we can characterize two-nucleon
pseudoscalar current by the number of the pion
exchanges and/or short-range interactions
\beqa
P_{{\rm 2N}}^{a}&=&P_{{\rm 2N:\,}1\pi}^{a}+P_{{\rm 2N:\,}2\pi}^{a}+P_{{\rm 2N:\,}{\rm cont}}^{a}.\label{NNPseudoscalar:General}
\eeqa  
First contributions start to show up at order $Q^{-1}$.
At this order, there are only OPE and contact contributions which can be expressed
by the longitudinal part of axial-vector current via
\beqa
\label{PseudoScalarNN:1pi:QToMinus1}
i\,m_q P_{{\rm 2N:\,}1\pi}^{a, (Q^{-1})}&=&\vec{k}\cdot \vec{A}_{{\rm
    2N:\,}1\pi}^{a, (Q^{0})},\\
i\,m_q P_{{\rm 2N:\,}{\rm cont}}^{a, (Q^{-1})}&=&\vec{k}\cdot \vec{A}_{{\rm
    2N:\, cont}}^{a, (Q^{0})},\label{PseudoScalarNN:cont:QToMinus1}
\eeqa
where $\vec{A}_{{\rm
    2N:\,}1\pi}^{a, (Q^{0})}$ and $\vec{A}_{{\rm
    2N:\, cont}}^{a, (Q^{0})}$ are defined in Eqs~(\ref{AxialCurrentTree1}) and
(\ref{axial_vector:2N:cont:QTo0}), respectively.
\paragraph{One-pion-exchange pseudoscalar current}
\label{sec:OPEpseudoscalarCurrent}
At the order $Q^0$ leading one-loop static contributions and relativistic $1/m$
corrections start to show up. 
\beqa
P_{{\rm 2N:} \, 1\pi}^{a \, (Q^0)}&=&P_{{\rm 2N:} \, 1\pi, {\rm
    static}}^{a \, (Q^0)}+P_{{\rm 2N:} \, 1\pi, 1/m}^{a \,
  (Q^0)}+P_{{\rm 2N:} \, 1\pi, {\rm UT}^\prime}^{a \, (Q^0)}.\quad\quad
\eeqa
The corresponding static expressions are 
\beqa
\label{CurrentPs1pi}
&&i\,m_q P_{{\rm 2N:} \, 1\pi, {\rm static}}^{a \, (Q^0)} =  
-\frac{4\,M_\pi^2 F_\pi^2}{g_A}\frac{\vec q_1
 \cdot \vec \sigma_1  }{q_1^2 + M_\pi^2}\bigg\{[{\fet\tau}_1]^a
\bigg[h_1^P(q_2)\nn
&&+
\frac{h_4(q_2) }{k^2 + M_\pi^2}\bigg] +   [ \fet \tau_1 \times
\fet \tau_2 ]^a \vec q_1 \cdot [\vec q_2
  \times \vec \sigma_2] \frac{h_5(q_2) }{k^2 + M_\pi^2}\bigg\} \nn
&&+  1
\leftrightarrow 2,   
\eeqa
where the scalar functions $h_4 (q_2)$ and $h_5 (q_2)$ are defined in
Eq.~(\ref{OPEhiDefenition}), while 
$h_1^P(q_2)$ is given by
\bigskip
\beqa
h_1^P(q_2)&=&\frac{g_A^4}{256\pi F_\pi^6}\Big((1-2 g_A^2) M_\pi + (2
M_\pi^2 + q_2^2)A(q_2)\Big).\nn
&&
\eeqa
The relativistic $1/m$ corrections are
\beqa
\label{CurrentPs1piRel}
&&i\,m_q P_{{\rm 2N:} \, 1\pi , \, 1/m}^{a \, (Q^0)} = - \frac{g_A M_\pi^2}{16
   F_\pi^2 \, m}\bigg\{i [ \fet \tau_1 \times
\fet \tau_2 ]^a
\nn
&&\bigg[-\frac{\vec{B}_1\cdot\vec{k}}{(q_1^2+M_\pi^2)^2(k^2+M_\pi^2)} + \frac{1}{q_1^2+M_\pi^2}\bigg(\frac{\vec{B}_2\cdot\vec{k}}{k^2(k^2+M_\pi^2)^2} \nn
&&+ \frac{\vec{B}_3\cdot\vec{k}}{k^2(k^2+M_\pi^2)}\bigg)  \bigg] +
[{\fet\tau}_1]^a\bigg[ -\frac{
  \vec{B}_5\cdot\vec{k}}{(q_1^2+M_\pi^2)^2(k^2+M_\pi^2)}\nn
&&+ \frac{1}{q_1^2+M_\pi^2}\bigg(\frac{\vec{B}_6\cdot\vec{k}}{k^2(k^2+M_\pi^2)^2} 
+ \frac{\vec{B}_7\cdot\vec{k}}{k^2(k^2+M_\pi^2)}\bigg)\bigg] \bigg\}
\nn
&&+
1 \; \leftrightarrow \; 2\,,
\eeqa
where the vector quantities $\vec{B}_i$ 
are defined in
Eq.~(\ref{OPEOneOvermNuclFFs}).

\noindent Finally, the energy-transfer dependent contributions are
\beq
i\,m_q P_{{\rm 2N:\,}1\pi, {\rm UT^\prime} }^{a \; ({Q^0})} = -
\frac{M_\pi^2}{k^2} \, \vec k \cdot\vec {A}_{{\rm 2N:\,}1\pi, {\rm UT^\prime} }^{a \; ({Q})} \,,
\label{NNCurrent1piUTPrimePs}
\eeq
where $\vec {A}_{{\rm 2N:\,}1\pi, {\rm UT^\prime} }^{a \; ({Q})}$ is given
in Eq.~(\ref{NNCurrent1piUTPrime}).

\paragraph{Two-pion-exchange pseudoscalar current}
\label{sec:TPEpseudoscalarCurrent}
Two-pion-ex- change contributions can be expressed in terms of the
longitudinal component of the axial vector current. The expression is
given by
\beq
\label{Current2piPs}
i\,m_q P_{{\rm 2N:} \, 2\pi}^{a\, (Q^0)} = -
\frac{M_\pi^2}{ k^2} \, \vec k \cdot \vec A_{{\rm 2N:} \, 2\pi}^{a\,
  (Q)} \bigg|_{g_{13} = g_{14} = g_{15} = g_{16} = g_{17} = 0} \,,
\eeq
with the scalar functions $g_1(q_1), \ldots , g_{12}(q_1)$  being
defined in Eq.~(\ref{giOfq1TPEStrF}). 
\paragraph{Short-range pseudoscalar current}
\label{sec:contactPseudoscalarCurrent}
The first contribution to short-range two-nucleon current shows up at
the order $Q^{-1}$ and is given in
Eq.~(\ref{PseudoScalarNN:cont:QToMinus1}). At order $Q^0$ we characterize
short-range contributions via
\beqa
P_{{\rm 2N:\,}{\rm cont}}^{a, (Q^0)}&=&P_{{\rm 2N:\,}{\rm cont,
    static}}^{a, (Q^0)}+P_{{\rm 2N:\,}{\rm cont,} 1/m}^{a, (Q^0)}+P_{{\rm 2N:\,}{\rm
    cont, UT}^\prime}^{a, (Q^0)}.\nn
&&
\eeqa
The static contributions at the order $Q^0$ vanish
\beqa
P_{{\rm 2N:\,}{\rm cont,
    static}}^{a, (Q^0)}&=&0.
\eeqa
The relativistic corrections are given by
\beqa
\label{NNCurrentPsContOneOverm}
&&i\, m_q\,P_{{\rm 2N:\,}{\rm cont,} 1/m}^{a, (Q^0)}= -\frac{M_\pi^2}{ k^2}
\, \vec k \cdot \vec A_{{\rm 2N: \, cont,}\, 1/m}^{a \,  (Q)} \,,
\eeqa
where $\vec A_{{\rm 2N: \, cont,}\, 1/m}^{a \,
  (Q)}$ is specified in Eq.~(\ref{CurrentContRel}).
Finally, the energy-transfer-dependent contributions are given by
\beqa
\label{NNCurrentPsContUTPrime}
i\,m_q P_{{\rm 2N:\,cont,\,UT^\prime}}^{a \; ({Q^0})} &=& -\frac{M_\pi^2}{k^2} \, \vec k \cdot \vec A_{{\rm 2N:\,cont,\,UT^\prime}}^{a \; ({Q})} \,,
\eeqa
where $\vec A_{{\rm 2N:\,cont,\,UT^\prime}}^{a \; ({Q})} $ is
specified in Eq.~(\ref{NNCurrentContUTPrime}). 

\subsubsection{Three-nucleon pseudoscalar current}
\label{sec:3NpseudoscalarCurrent}
At the order $Q^0$ there are first contributions to the three-nucleon pseudoscalar
current. Similar to the axial-vector current we decompose the current
into the long and the short-range contributions
\beqa
P_{{\rm 3N}}^{a \, (Q^0)}&=&P_{{\rm 3N:}\,
  \pi}^{a \, (Q^0)}+P_{\rm 3N:\, cont}^{a \, (Q)}.
\eeqa
The long-range contributions are given by
\beqa
\label{Current3NPiPs}
P_{{\rm 3N:}\,
  \pi}^{a \, (Q^0)} &=&- i\frac{2\, F_\pi^2 M_\pi^2}{g_A m_q} \sum_{i=5}^{8}   \frac{\vec C_i^a\cdot\vec{k}}{k^2}
+
5\,{\rm permutations}\,,\nn
&&
\eeqa
where $\vec C_i^a$ are defined in
Eq.~(\ref{Ci3NDefinition}). Short-range contributions are given by
\beqa
\label{Current3NContPs}
P_{\rm 3N:\, cont}^{a \, (Q)} &=&- i\frac{2\, F_\pi^2 M_\pi^2}{g_A m_q}\sum_{i=2}^{3}   \frac{\vec D_i^a\cdot\vec{k}}{k^2}
+
5\,{\rm permutations}\,,\nn
&&
\eeqa
with $\vec D_i^a$ defined in Eq.~(\ref{DiFF3NDefinition}).
\subsection{Scalar current up to the order $Q$}
\label{scalarUpToOrderQ}
Nuclear scalar current is important for the dark matter
searches. A scenario that dark matter is
realized by weakly interacting massive particles (WIMPs) can be tested
via nuclear recoil produced by the scattering of WIMPs off atomic
nuclei (see~\cite{Hoferichter:2019uwa} and the references therein). If
WIMP, denoted by $\chi$, is a spin-$1/2$ particle and interacts with quarks and gluons via~\cite{Goodman:2010ku,Hoferichter:2015ipa}
\beqa
{\cal L}_\chi&=&\frac{1}{\Lambda^3}\sum_q\bigg[C_q^{{\rm SS}}\bar{\chi}\chi
m_q \bar{q}q + C_q^{{\rm PS}}\bar{\chi}i\gamma_5\chi m_q \bar{q}q \nn
&+&
C_q^{{\rm SP}}\bar{\chi}\chi m_q\bar{q}i\gamma_5 q +
C_q^{{\rm PP}}\bar{\chi}i\gamma_5\chi m_q \bar{q}i\gamma_5 q\bigg]\nn
&+&\frac{1}{\Lambda^2}\sum_q\bigg[C_q^{{\rm VV}}\bar{\chi}\gamma^\mu\chi\bar{q}\gamma_\mu
q+C_q^{{\rm AV}}\bar{\chi}\gamma^\mu\gamma_5\chi \bar{q}\gamma_\mu q \nn
&+&
C_q^{{\rm VA}}\bar{\chi}\gamma^\mu\chi \bar{q}\gamma_\mu\gamma_5 q +
C_q^{{\rm AA}}\bar{\chi}\gamma^\mu\gamma_5\chi\bar{q}\gamma_\mu\gamma_5 q\bigg],\label{WIMP_coupling}
\eeqa
where $\Lambda$ is a beyond standard model scale and various Wilson
coefficients $C_q$ are dimensionless. Indices $S,P,V$ and $A$ stay for
scalar, pseudoscalar, vector and axial-vector quantum numbers,
respectively. $q$-field is a quark field and $m_q$ is a quark
mass. Eq.~(\ref{WIMP_coupling}) can be considered as a starting point
for chiral EFT analysis where one needs to study scalar-,
pseudoscalar-, vector- and axial-vector currents within chiral
EFT. While vector- and axial-vector currents have been extensively
studied within standard model, scalar-current in chiral EFT appears
first in beyond
standard model physics and is much less known.
Pioneering work towards this direction was the leading-order
calculation of the scalar current by Cirigliano et
al.~\cite{Cirigliano:2012pq} and  Hoferichter et
al.~\cite{Hoferichter:2015ipa}. Recently we derived leading one-loop corrections to these
results~\cite{Krebs:2020plh}. Although the information about
short-range physics at this order needs to be fixed these calculations
give long-range contributions in a parameter-free way. In the
following all contributions to scalar current up to order $Q^0$. 

\subsubsection{Single-nucleon scalar current}
\label{sec:scalarSingleNucleon}
Scalar current on a single-nucleon can be parametrized by a scalar
form factor which in principle can be calculated within chiral
perturbation theory. Due to the lack in convergence~\cite{Becher:1999he,Bernard:1996cc} one prefers to use
dispersion relation techniques where $\pi\pi$ scattering channel
dominates $t$-dependence of the scalar form
factor~\cite{Hoferichter:2012wf}.
\subsubsection{Two-nucleon scalar current}
\label{sec:2nscalarCurrentGeneral}
Two- and three-nucleon scalar current will be formulated here within chiral
EFT. In the following, we characterize a two-nucleon scalar current by the
range of interaction
\beqa
S_{{\rm 2N}}&=&S_{{\rm 2N:}\,1\pi}+S_{{\rm 2N:}\,2\pi}+S_{{\rm 2N:\,cont}}.
\eeqa
\paragraph{One-pion-exchange scalar current}
\label{sec:OPEscalarCurrent}
First contributions to the one-pion-exchange current show up at the order $Q^{-2}$ and are given by
\beqa
m_q S_{{\rm 2N:}\,1\pi}^{(Q^{-2})}&=&-\frac{g_A^2 M_\pi^2}{4 F_\pi^2}\frac{ \vec{q}_1\cdot \vec{\sigma}_1 
\vec{q}_{2}\cdot\vec{\sigma}_{2} {\fet \tau}_{1}\cdot{\fet 
\tau}_{2}}{ \left(M_\pi^2+q_1^2\right) 
\left(M_\pi^2+q_2^2\right)}.\label{scalar:2N:1pi:QToMinus2}
\eeqa
Next correction shows up first at the order $Q^0$. 
The result is given by
\beqa
\label{Current1piSkalar}
&&m_q S_{{\rm 2N:} \, 1\pi,}^{(Q^0)} =
\frac{\vec{q}_1\cdot\vec{\sigma}_1}{q_1^2+M_\pi^2}\bigg[\vec{q}_2\cdot\vec{\sigma}_2
\bigg(\frac{o_1(k)}{q_2^2+M_\pi^2} +
o_2(k)\bigg)\nn
&&+\vec{k}\cdot\vec{\sigma}_2\bigg(o_3(k)+q_2^2 o_4(k)\bigg)\bigg]\; + \; 1
\leftrightarrow 2\,, 
\eeqa
where the scalar functions $o_i (k)$ are given by 
\bigskip
\beqa
o_1(k)&=&\frac{g_A M_\pi^2}{128\pi^2 F_\pi^4}
 \big[64\pi^2\bar{d}_{18}
      F_\pi^2 M_\pi^2+g_A k^2\bar{l}_4\nn
&-&g_A
    L(k)\left(2k^2+M_\pi^2
\right)+g_A\left(k^2+M_\pi^2\right)\big],\nn
o_2(k)&=&\frac{g_A M_\pi^2}{64\pi^2 F_\pi^4 }
 \bigg[32\pi^2 F_\pi^2\left(2\bar{d}_{16} -\bar{d}_{18} \right)-g_A\bar{l}_4\nn
%+g_A^3
&-&\frac{4g_A^3 L(k)\left(k^2+3 M_\pi^2\right)}{k^2+4M_\pi^2}
\bigg], \nn
o_3(k)&=&- \frac{g_A M_\pi^2 }{128\pi^2 F_\pi^4 k^2 }
\bigg[128\pi^2\bar{d}_{16} F_\pi^2 k^2
+g_A^3\left(-k^2+M_\pi^2
\right)\nn
&+&2g_Ak^2
-\frac{4g_AL(k)}{k^2+4 M_\pi^2}\left(\left(2g_A^2+1
\right)k^4\right.\nn
&+&\left. \left(5g_A^2+4\right)k^2 M_\pi^2+g_A^2 M_\pi^4\right)
\bigg],\nn
o_4(k)&=&-\frac{g_A^4 M_\pi^2}{128\pi^2 F_\pi^4 k^2 }
\frac{k^2+4 M_\pi^2(1-L(k))}{k^2+4 M_\pi^2}
\,. \label{OPEhiDefenitionSkalar}
\eeqa
%\beqa
%o_1(k)&=&\frac{g_A M_\pi^2}{128\pi^2 F_\pi^4 }\Big(16\pi^2\left(4\bar{d}_{18}
%      F_\pi^2 M_\pi^2+g_A k^2\bar{l}_4\right)\nn
%&-& g_A
%    L(k)\left(2k^2+M_\pi^2
%\right)+g_A\left(k^2+M_\pi^2\right)\Big),\nn
%o_2(k)&=&\frac{g_A M_\pi^2}{64\pi^2 F_\pi^4 \left(k^2+4M_\pi^2\right)}\Big(\left(k^2+4 M_\pi^2
%\right)\left(\right.\nn
%&&\left. 16\pi^2\big(4\bar{d}_{16} F_\pi^2-2\bar{d}_{18}
%    F_\pi^2-g_A\bar{l}_4
%\big)+g_A^3\right)\nn
%&-&4g_A^3 L(k)\left(k^2+3 M_\pi^2\right)
%\Big), \nonumber
%\eeqa
%\beqa
%o_3(k)&=&\frac{g_A M_\pi^2 }{128\pi^2 F_\pi^4 k^2 \left(k^2+4
%    M_\pi^2\right)}
%\bigg(-\left(k^2+4 M_\pi^2
%\right)\nn
%&\times&\left(128\pi^2\bar{d}_{16} F_\pi^2 k^2+g_A^3\left(k^2+M_\pi^2
%\right)+2g_Ak^2\right)\nn
%&+&4g_AL(k)\left(\left(2g_A^2+1
%\right)k^4+\left(5g_A^2+4\right)k^2 M_\pi^2\right. \nn
%&+&\left. g_A^2 M_\pi^4\right)
%\bigg),\nn
%o_4(k)&=&-\frac{g_A^4 M_\pi^2\left(k^2+4 M_\pi^2(1-L(k))
%\right)}{128\pi^2 F_\pi^4 k^2 \left(k^2+4 M_\pi^2\right)}.
% \label{OPEhiDefenitionSkalar}
%\eeqa
The relativistic corrections and the energy-transfer dependent
contributions to one-pion-exchange at the order $Q^0$ vanish.
\paragraph{Two-pion-exchange scalar current}
\label{sec:TPEscalarCurrent}
Two-pion-exchange \newline
current shows up first at order $Q^0$. Due to the 
appearance of three-point functions the results are lengthy. Their
explicit form can be found in~\cite{Krebs:2020plh} and, for
completeness, is given in Appendix~\ref{app:scalar_curret:2pe}.
\paragraph{Short-range scalar current}
\label{sec:contactscalarCurrent}
Short-range two-nucleon current starts to contribute at order
$Q^{0}$. The relevant LECs come from
\beqa
\mathcal{L}_{NN}^{(0)}&=& - \frac{\overline{C}_S}{2} (N^\dagger N)^2 + 2 \overline{C}_T N^\dagger
S_\mu N N^\dagger S^\mu N,\\
\mathcal{L}_{NN}^{(2)}&=& -\frac{D_S}{8} \langle \chi_+\rangle (N^\dagger N)^2 +
\frac{D_T}{2} \langle \chi_+\rangle N^\dagger
S_\mu N N^\dagger S^\mu N\;,\nonumber
\eeqa  
where $N$ is the heavy-baryon notation for the nucleon field with
velocity $v_\mu$, $S_\mu = - \gamma_5 [\gamma_\mu, \, \gamma_\nu ]
v^\nu /4 $ is the
covariant spin-operator, $\chi_+ = 2 B \left(u^\dagger (s + i p) u^\dagger + u
(s - i p) u\right)$, $B$, $\overline{C}_{S,T}$ and $D_{S,T}$ are
LECs. $\langle \dots\rangle$ denotes the trace in the flavor space. All the non-vanishing diagrams which contribute at order $Q^0$
are listed in~\cite{Krebs:2020plh}. After the renormalization of the short-range LECs we got
\beqa
\label{CurrentContSkalar}
m_q S_{\rm 2N: \, cont}^{ (Q^0)}&=& \vec{\sigma}_1\cdot\vec{\sigma}_2 s_1(k)
+ \vec{k}\cdot\vec{\sigma}_1\vec{k}\cdot\vec{\sigma}_2 s_2(k) + s_3(k),\nn
&&\label{ChargeContQTo1Skalar}
\eeqa
with the scalar function $s_i(k)$ defined by 
\beqa
s_1(k)&=&-\frac{M_\pi^2}{8\pi^2 F_\pi^2 m_q}
  \bigg[2 g_A^2 \overline{C}_T-4\pi^2 \bar{D}_T F_\pi^2\nn
&+&\frac{g_A^2 \overline{C}_T L(k)\left(3k^2+4M_\pi^2\right)}{k^2+4M_\pi^2}
\bigg],\nn
s_2(k)&=&\frac{3 g_A^2
  \overline{C}_T M_\pi^2}{8\pi^2 F_\pi^2 k^2
m_q} \frac{k^2-4 M_\pi^2(L(k)-1)}{k^2+4M_\pi^2}, \nn
s_3(k)&=&\frac{M_\pi^2}{16\pi^2 F_\pi^2 m_q}
\bigg[g_A^2 \overline{C}_T+8\pi^2\bar{D}_S F_\pi^2\nn
&-&\frac{2g_A^2 \overline{C}_T L(k)\left(3k^2+8M_\pi^2\right)}{k^2+4M_\pi^2}
\bigg]~.
\eeqa
The renormalized, scale-independent LECs $\bar{D}_{S}$,  $\bar{D}_T$ are
related to the bare ones $D_{S}$,  $D_T$  according to
\beqa
D_i=\bar{D}_i+\frac{\beta_i^{\rm NN}}{F^4}\lambda
+\frac{\beta_i^{\rm NN}}{16\pi^2 F^4}\ln\left(\frac{M_\pi}{\mu}\right),
\eeqa
with the corresponding $\beta$-functions given by
\beqa
\beta_S^{\rm NN}&=&\frac{1}{2}\left(1+6 g_A^2 - 15 g_A^4 + 24 F^2
  g_A^2 \overline{C}_T\right), \nn
\beta_T^{\rm NN}&=&\frac{1}{4}\left(1+6 g_A^2 - 15 g_A^4 + 48 F^2
  g_A^2 \overline{C}_T\right), 
\eeqa
and the quantity $\lambda$ defined as
\beqa
\lambda&=&\frac{\mu^{d-4}}{16\pi^2}\bigg(\frac{1}{d-4}+\frac{1}{2}\big(\gamma_E-\ln
4\pi -1\big)\bigg),
\eeqa
where $\gamma_E=-\Gamma^\prime(1)\simeq 0.577$ is the Euler constant, $d$ the number of space-time
dimensions and $\mu$ is the scale of dimensional regularization. 

Notice that the LECs $\overline{C}_S$, $\overline{C}_T$, $\bar{D}_S$
and $\bar{D}_T$ also contribute to
the 2N potential. The experimental data on nucleon-nucleon
scattering, however, do not allow one to disentangle the $M_\pi$-dependence of
the contact interactions and only constrain the linear combinations of the LECs~\cite{Epelbaum:2002gb}
\beqa
C_S&=& \overline{C}_S+M_\pi^2\bar{D}_S, \quad C_T\,=\,
\overline{C}_T+M_\pi^2\bar{D}_T\;.
\eeqa
The LECs $\bar{D}_S$
and  $\bar{D}_T$ can, in principle, be determined once  reliable lattice QCD
results for two-nucleon observables such as e.g.~the $^3$S$_1$ and $^1$S$_0$ scattering lengths at
unphysical (but not too large) quark masses are available, see
Refs.~\cite{Lahde:2019yvr} and references therein for a discussion of
the current status of research along this line.   

\subsubsection{Relativistic corrections}
There are only vanishing 
$1/m$-corrections and the energy-transfer dependent contributions at
the order $Q^0$. 

\subsubsection{Three-nucleon scalar current}
\label{sec:3NscalarCurrent}
At the order $Q^0$ there are no three-nucleon current
contributions. Nonvanishing contributions start to show up first at
order $Q$. 
\paragraph{Scalar current at vanishing momentum transfer}
At the vanishing momentum transfer, one can relate the scalar current to a
quark mass derivative of the nuclear forces. On the mass-shell one gets
\beqa
\label{ScalarCurrentAtZeroMomentum}
\langle f |S(0)| i \rangle &=& \bigg\langle f \bigg| \frac{\partial H_{\rm eff}}{\partial
  m_q} \bigg| i \bigg\rangle ,
  \eeqa
where the states $|i\rangle$ and $|f\rangle$ are the eigenstates of the
Hamiltonian $H_{\rm eff}$. The on-shell condition requires that
the corresponding eigenenergies $E_i$ and $E_f$ are equal. For eigenstates $| \Psi \rangle$ corresponding to a discrete energy
$E$, 
\beq
H_{\rm
  eff} | \Psi \rangle = E  | \Psi \rangle\;,
\eeq
 the  Feynman-Hellmann
theorem allows one to interpret the scalar form factor at zero
momentum transfer in terms of the eigenenergy slope in the quark mass:
\beqa
\label{ScalarCurrentDiscreteStates}
\langle \Psi|m_q S(0)| \Psi\rangle &=& m_q \frac{\partial E(m_q)}{\partial m_q}.
\eeqa
In particular, for $|\Psi\rangle$ being a single-nucleon state
at rest, the expectation value on left-hand side of
Eq.~(\ref{ScalarCurrentDiscreteStates}) is nothing but the pion-nucleon sigma-term~\cite{RuizdeElvira:2017stg}
\beqa
\langle \Psi|m_q S(0)| \Psi\rangle \,=\, m_q \frac{\partial m(m_q)}{\partial
m_q} &\equiv & \sigma_{\pi N} \,,
\eeqa
and for an extension to resonances $|R\rangle$, see e.g.~Ref.~\cite{RuizdeElvira:2017aet}.
 As was
demonstrated in~\cite{Krebs:2020plh} we explicitly verified the
relation in Eq.~(\ref{ScalarCurrentAtZeroMomentum}) up to order $Q^0$. To get a
slope in the quark mass of the nuclear force at NLO we used the
expressions from~\cite{Epelbaum:2002gb} where the authors discussed the nuclear force
at NLO in the chiral limit. 
%\begin{figure}
%\resizebox{0.75\textwidth}{!}{%
%  \includegraphics{leer.eps}
%\vspace{5cm}       % Give the correct figure height in cm
%\end{figure}
%\begin{figure*}
%\vspace*{5cm}       % Give the correct figure height in cm
%\end{figure*}
%\begin{table}
%\begin{tabular}{lll}
%\hline\noalign{\smallskip}
%\end{tabular}
%% Or use
%\vspace*{5cm}  % with the correct table height
%\end{table}
%
% The section below may be edited at your convenience to acknowledge 
% each author's contribution to the manuscript.
% You may remove it if you are a single author.
%
\subsection{Dependence on unitary phases $\bar{\beta}_8$ and
  $\bar{\beta}_9$}
\label{sec:beta8beta9Dependence}
After renormalizability and matching constraints applied to various
nuclear currents we get in the static limit a unique
result. In the relativistic corrections of the vector and axial-vector
current, however, there remains a unitary ambiguity which is parametrized by unitary phases
$\bar{\beta}_8$ and $\bar{\beta}_9$. In
Appendix~\ref{unitary_transf_app} we give their explicit form and
briefly review all other transformations which are introduced to
renormalize nuclear currents. It is instructive to unravel how
this dependence disappears if we calculate the expectation values of the
corresponding currents. There are two different mechanisms that we
are going to discuss. In the first case, the dependence on unitary
phases is compensated by the wave function of initial and final
states. In the second case, which we call a $k_0$-dependent off-shell effect, the dependence on unitary phases is not
compensated by wave functions. However, it is proportional to
$k_0-E_\beta+E_\alpha$, where $E_\alpha$ and $E_\beta$ correspond to
the energies of the initial and final states, respectively. Since on-shell we
have
\beqa
k_0&=&E_\beta - E_\alpha,\label{OnShellCondition}
\eeqa
the unitary ambiguity disappears for observable quantities. We can explicitly disentangle contributions which
 are going to be compensated by the wave functions and which are to disappear
if the on-shell condition of Eq.~(\ref{OnShellCondition}) is
satisfied. In the case of the vector current, there are only
contributions proportional to $\bar{\beta}_{8,9}$ which are to be
compensated by the wave functions, see~\cite{Filin:2019eoe} for an explicit verification. The reason is that $k_0$-dependent
terms in Eq.~(\ref{singleN:offshell}) do not depend on
$\bar{\beta}_{8,9}$. Strictly speaking, there still remains a residual
dependence on the unitary phases $\bar{\beta}_{8,9}$ even if we calculate the expectation value of
the current operator. The reason is that the transformation associated
with phases $\bar{\beta}_{8,9}$ is only approximately unitary modulo effects of higher
order in the chiral expansion. Due to the expected higher order suppression,
the dependence on  $\bar{\beta}_{8,9}$ should be weak in practical
calculations\footnote{Provided that one works with consistent
  wave functions which incorporate the same dependence on unitary phases as
  does the vector current operator. Hybrid approach, where
  phenomenological wave functions (coming \underline{not} from chiral
  EFT, like e.g. AV18~\cite{Wiringa:1994wb} or CD Bonn~\cite{Machleidt:2000ge} potentials) are
  used, should not be used for this current operator.}.
%To see this consider
%Eq.~(\ref{singleN:offshell}) for the off-shell component of the
%single-nucleon current. 

Energy-transfer $k_0$-dependent terms generated by time-dependent unitary
transformations are, in general, cancelled on-shell by an  accompanied  commutator
with the nuclear force (see Eq.~(\ref{k0DependenceJX})) 
%So nuclear vector current considered in this
%work has a general form
\beqa
%V_\mu(k)&=& v_\mu(\vec{k}) + 
&&i\, k_0 y_\mu(\vec{k}) - i\,[H_{{\rm eff}},y_\mu(\vec{k})].  \label{k0DependenceGeneral}
\eeqa
In the case of the vector current the operator $y_\mu(\vec{k})$ can be read off from
Eq.~(\ref{singleN:offshell})
\beqa
y_0(\vec{k})&=&0,\nn
\vec{y}(\vec{k})&=&-i \,\vec{k}
\frac{e}{\vec{k}^2}\bigg[\big({ G}_E(\vec{k}^2)-{
  G}_E(0)\big)\nn
&+&\frac{i}{2m^2}\vec{k}\cdot(\vec{k}_1\times\vec{\sigma})\big({
  G}_M(\vec{k}^2)-{ G}_M(0)\big)\bigg].\quad\quad
\eeqa
Actually, Eq.~(\ref{k0DependenceGeneral}) describes only the single-nucleon
contribution to the corresponding current operator
\beqa
&&i\, k_0 y_\mu(\vec{k}) - i\,[H_0,y_\mu(\vec{k})].  \label{k0DependenceGeneralIncomplete}
\eeqa
The remaining two- and more-nucleon contributions are coming from the commutator
\beqa
[H_{\rm eff}-H_0,y_\mu(\vec{k})],\label{commHwithy}
\eeqa
and is perturbatively taken into account in $\vec{V}_{{\rm
    2N:1\pi,}\,{\rm static}}^{(Q)}$ given in
Eq.~(\ref{OPE:Static:Current:Parametrization}). In particular, the
commutator
\beqa
[H_{{\rm 1\pi}}^{(Q^0)}+H_{{\rm cont}}^{(Q^0)}, \vec{y}(\vec{k})^{[Q]}]\label{commHwithyPerturb}
\eeqa
contributes to the order $Q$ vector current which is a part of  $\vec{V}_{{\rm
    2N:1\pi,}\,{\rm static}}^{(Q)}$. Here the operator $
\vec{y}(\vec{k})^{[Q]}$ is given by
\beqa
\vec{y}(\vec{k})^{[Q]}&=&-i \,\vec{k}
\frac{e}{\vec{k}^2}\bigg[\big({ G}_E(\vec{k}^2)^{[Q^2]}-{
  G}_E(0)^{[Q^2]}\big)\bigg],\quad
\eeqa
where the index in the square brackets denotes the chiral order of the
operator. Explicit expressions for the chiral expansion of the electromagnetic form factors
can be
found e.g. in~\cite{Krebs:2019aka},
% and are given here for
%completeness:
%\beqa
%{ G}_E(\vec{k}^2)^{[Q^2]}&=&\frac{1}{6 (4\pi F_\pi)^2}\big[4(1+2
%g_A^2)M_\pi^2 \nn
%&+& (1+5 g_A^2)Q^2\big]L(| \vec k \,|) \tau_3 \nn
%&+&\frac{\tau_3}{36 (4\pi F_\pi)^2}\big[-24 (1+2 g_A^2) M_\pi^2 \nn
%&-&
%Q^2(5 + 13 g_A^2)\big] + Q^2 (2 \bar{d}_7 + \bar{d}_6
%\tau_3)
%\eeqa
and the leading order nuclear force
operator is given by 
%\beqa
%H_{{\rm eff}}^{(Q^0)}-H_0&=&H_{{\rm 1\pi}}^{(Q^0)}+H_{{\rm cont}}^{(Q^0)},
%\eeqa
%Explicit expressions for the leading order nuclear force is given by
\beqa
H_{{\rm 1\pi}}^{(Q^0)}&=&-\frac{g_A^2}{4
  F_\pi^2}\vec{\tau}_1\cdot\vec{\tau}_2\frac{\vec{\sigma}_1\cdot\vec{q}\vec{\sigma}_2\cdot\vec{q}}{q^2+M_\pi^2},\nn
H_{{\rm cont}}^{(Q^0)}&=&C_S + \vec{\sigma}_1\cdot\vec{\sigma}_2 C_T,
\eeqa
where $\vec{q}$ denotes momentum transfer between nucleons. In
practical applications one may include higher-order operators to the
two- and more-nucleon current by using the whole commutator of
Eq.~(\ref{commHwithy}), instead of the commutator of Eq.~(\ref{commHwithyPerturb}). In this case, the
expectation value of the operator of Eq.~(\ref{k0DependenceGeneral}) would not contribute
on-shell and, for this reason, does not need to be calculated
explicitly. Obviously, if one would neglect the operator of
Eq.~(\ref{k0DependenceGeneral}) one has to subtract the operator of
Eq.~(\ref{commHwithyPerturb}) from $\vec{V}_{{\rm
    2N:1\pi,}\,{\rm static}}^{(Q)}$.

Similar to the vector current, we can write an axial vector version of
Eq.~(\ref{k0DependenceGeneral}) which is given by
\beqa
%A_\mu^a(k)&=& x_\mu^a(\vec{k}) + 
&&i\, k_0 z_\mu^a(\vec{k}) - i\,[H_{{\rm eff}},z_\mu^a(\vec{k})],  \label{k0DependenceAxialGeneral}
\eeqa
where
\beqa
z_0^a(\vec{k})&=&0,\nn
\vec{z}^a(\vec{k})&=&i\,\vec{k}\frac{[{\fet\tau}]^a}{16
  m^3}\Big((1+2\bar{\beta}_9)\vec{k}_1\cdot\vec{\sigma}
G_P(-\vec{k}^2) \nn
&-& (1+2\bar{\beta}_8)
\vec{k}\cdot\vec{k}_1\vec{k}\cdot\vec{\sigma} G_P^\prime(-\vec{k}^2)\Big).
\eeqa
Replacing $H_{{\rm eff}}$ by $H_0$ in
Eq.~(\ref{k0DependenceAxialGeneral}) we reproduce the off-shell part of
Eq.~(\ref{singleN:axialoffshell}). The leading two-nucleon contribution of
Eq.~(\ref{k0DependenceAxialGeneral}) is given by the commutator
$-i\,\Big[H_{{\rm eff}}^{(Q^0)}-H_0,z_\mu(\vec{k})\Big]$ which 
depends on the unitary phases $\bar{\beta}_{8,9}$. Its explicit form is
given by
\beqa
&&\vec{A}_{{\rm 2N:\,1\pi, \,off-shell}}^{a }\,=\,-i\,\Big[H_{{\rm
    1\pi}}^{(Q^0)},\vec{z}(\vec{k})\Big]=\vec{k}\frac{g_A^2}{64 F_\pi^2
  m^3}\nn
&&\times\frac{1}{q_1^2+M_\pi^2}\bigg[[\vec{\tau}_1\times\vec{\tau}_2]^a
    \Big(-2(1+2\bar{\beta}_9)\vec{k}_2\cdot\vec{q}_1
    \vec{q}_1\cdot\vec{\sigma}_1\nn
&&\times G_P(-\vec{k}^2) +
(1+2\bar{\beta}_8)\vec{k}\cdot\vec{q}_1(2\vec{k}\cdot\vec{k}_2-i\,\vec{k}\cdot(\vec{q}_1\times\vec{\sigma}_2))\nn
&&\times\vec{q}_1\cdot\vec{\sigma}_1
G_P^\prime(-\vec{k}^2)\Big) +
[{\fet\tau}_1]^a\Big(-(1+2\bar{\beta}_9)(2\vec{k}_2\cdot(\vec{q}_1\times\vec{\sigma}_2)\nn
&&+i\,\vec{q}_1^2)\vec{q}_1\cdot\vec{\sigma}_1
G_P(-\vec{k}^2) +
(1+2\bar{\beta}_8)(i\,(\vec{k}\cdot\vec{q}_1)^2\nn
&&+2\vec{k}\cdot\vec{k}_2
\vec{k}\cdot(\vec{q}_1\times\vec{\sigma}_2)\Big)\vec{q}_1\cdot\vec{\sigma}_1
G_P^\prime(-\vec{k}^2)\bigg]+1\,\leftrightarrow \,2\,.\quad\quad
\eeqa
In addition, at order $Q$,  one also needs to take into account the
relativistic corrections to the OPE 2N current that are not associated
with the terms in Eq.~(\ref{k0DependenceAxialGeneral}) and have the form
\beqa
\label{Current1piRelOffshell}
&&\delta\vec {A}_{{\rm 2N:} \, 1\pi , \, 1/m}^{a \, (Q)}=\frac{g_A}{16
  F_\pi^2 m}\bigg\{\nn
&&i [ \fet \tau_1 \times
\fet \tau_2 ]^a
\bigg[\frac{1}{(q_1^2+M_\pi^2)^2}\bigg(\vec{B}_1 - \frac{\vec k \, \vec k
  \cdot \vec{B}_1}{k^2+M_\pi^2} 
\bigg) \nn
&&+ \frac{1}{q_1^2+M_\pi^2}\bigg(\frac{\delta\vec{B}_2}{(k^2+M_\pi^2)^2} 
+ \frac{\delta\vec{B}_3}{k^2+M_\pi^2} + \vec{B}_4\bigg)  \bigg]\nn
&&+ [{\fet\tau}_1]^a\bigg[ \frac{1}{(q_1^2+M_\pi^2)^2}\bigg(\vec{B}_5 -
\frac{\vec k \, \vec k \cdot \vec{B}_5}{k^2+M_\pi^2}
\bigg) \nn
&&+ \frac{1}{q_1^2+M_\pi^2}\bigg(
\frac{\delta\vec{B}_6}{(k^2+M_\pi^2)^2} +
\frac{\delta\vec{B}_7}{k^2+M_\pi^2} + \vec{B}_{8}\bigg)\bigg] \bigg\}
\nn
&&+\;
1 \; \leftrightarrow \; 2\,,
\eeqa
where $\vec{B}_1,\vec{B}_4,\vec{B}_5$ and $ \vec{B}_{8}$ are defined
in Eq.~(\ref{BiDefinition}), and
\beqa
\delta\vec{B}_2&=&-2 g_A^2\,\vec{k}\,
\vec{\sigma}_1\cdot\vec{q}_1\vec{k}\cdot\vec{q}_1
\big(i\,\vec{k}\cdot(\vec{q}_1\times\vec{\sigma}_2)-2\,\vec{k}\cdot\vec{k}_2\big),\nn
\delta\vec{B}_3&=&-2\,\vec{k}\,\bigg(g_A^2(1+2\bar{\beta}_9)\vec{k}\cdot\vec{q}_1\vec{k}_1\cdot\vec{\sigma}_1
+\vec{\sigma}_1\cdot\vec{q}_1\nn
&\times&\big(
-i\,\vec{k}\cdot(\vec{q}_1\times\vec{\sigma}_2)+\big(g_A^2(1-2\bar{\beta}_9)-1\big)\vec{k}\cdot\vec{k}_2\nn
&+&
\vec{k}_1\cdot\vec{q}_1+(2 g_A^2-1)\vec{k}_2\cdot\vec{q}_1\big)\bigg),\nn
\delta\vec{B}_6&=&-2 g_A^2\,\vec{k} \,\vec{\sigma}_1\cdot\vec{q}_1
\big(-2\,i\,\vec{k}\cdot(\vec{q}_1\times\vec{\sigma}_2)\vec{k}\cdot\vec{k}_2\nn
&+&(\vec{k}\cdot\vec{q}_1)^2\big),\nn
\delta\vec{B}_7&=&-g_A^2\vec{k}\bigg(2\,i\,(1+2\bar{\beta}_9)\vec{k}_1\cdot\vec{\sigma}_1\vec{k}\cdot(\vec{q}_1\times\vec{\sigma}_2)
\nn
&+&\vec{q}_1\cdot\vec{\sigma}_1\big(2\,i\,(1-2\bar{\beta}_9)\vec{k}\cdot(\vec{k}_2\times\vec{\sigma}_2)\nn
&+&4\,i\,\vec{k}_2\cdot(\vec{q}_1\times\vec{\sigma}_2)-(1-2\bar{\beta}_9)\vec{k}^2-2\vec{q}_1^2\big)\bigg).\quad\quad
\eeqa
Note that
\beqa
\vec {A}_{{\rm 2N:} \, 1\pi , \, 1/m}^{a \, (Q)}&=&\vec{A}_{{\rm
    2N:\,1\pi,\,off-shell}}^{ a } + \delta\vec {A}_{{\rm 2N:} \, 1\pi , \,
  1/m}^{a \, (Q)} \nn
&+& {\cal O}(Q^2).\quad\quad
\eeqa
In the same way, we can decompose the relativistic corrections involving the
contact
interactions:
\beqa
&&\vec{A}_{{\rm 2N:\,cont, \,off-shell}}^{a }\,=\,-i\,\Big[H_{{\rm
    cont}}^{(Q^0)},\vec{z}(\vec{k})\Big]=-\vec{k}\frac{[{\fet\tau}_1]^a}{16
  m^3}\nn
&&\bigg(-(1+2\bar{\beta}_9)\big(2\,i\,C_T
\vec{k}_1\cdot(\vec{\sigma}_1\times\vec{\sigma}_2)+C_S
\vec{q}_2\cdot\vec{\sigma}_1 \nn
&&+C_T \vec{q}_2\cdot\vec{\sigma}_2\big)G_P(-\vec{k}^2)
+(1+2\bar{\beta}_8)\big(\vec{k}\cdot\vec{q}_2(C_S
\vec{k}\cdot\vec{\sigma}_1\nn
&&+C_T\vec{k}\cdot\vec{\sigma}_2)+2\,i\,C_T\,\vec{k}\cdot\vec{k}_1
\vec{k}\cdot(\vec{\sigma}_1\times\vec{\sigma}_2)\big)G_P^\prime(-\vec{k}^2)\bigg)\nn
&&+
1 \; \leftrightarrow \; 2\,.
\eeqa
The remaining relativistic corrections to short-range 2N current at
order $Q$ are given by
\beqa
&&\delta\vec {A}_{{\rm 2N:\,cont}, \, 1/m}^{a \,
  (Q)}=-\vec{k}\frac{g_A}{2
  m}\frac{[{\fet\tau}_1]^a}{k^2+M^2}\bigg(2\,i\,C_T\vec{k}_1\cdot(\vec{\sigma}_1\times\vec{\sigma}_2)\nn
&&+C_S\vec{q}_2\cdot\vec{\sigma}_1
+C_T
\vec{q}_2\cdot\vec{\sigma}_2-\frac{1}{k^2+M^2}\big(\vec{k}\cdot\vec{q}_2(C_S\vec{k}\cdot\vec{\sigma}_1\nn
&&+ C_T
\vec{k}\cdot\vec{\sigma}_2)+2\,i\,C_T\vec{k}\cdot\vec{k}_1\vec{k}\cdot(\vec{\sigma}_1\times\vec{\sigma}_2)\big)\bigg)\nn
&&+
1 \; \leftrightarrow \; 2\,.
\eeqa
As in the case of the one-pion-exchange contributions, we have
\beqa
\vec {A}_{{\rm 2N:\,cont}, \, 1/m}^{a \, (Q)}&=&\vec{A}_{{\rm
    2N:\,cont,\,off-shell}}^{ a } + \delta\vec {A}_{{\rm 2N:\,cont} , \,
  1/m}^{a \, (Q)} \nn
&+& {\cal O}(Q^2).\quad\quad
\eeqa
The unitary ambiguity in the operator
\beqa
\vec{A}_{{\rm
    1N:off-shell}}^a + \vec{A}_{{\rm 2N:\,off-shell}}^{a }\label{axVOffShellContr}
\eeqa
is
proportional to $k_0-E_\beta+E_\alpha$ which vanishes on-shell. Therefore,
$\bar{\beta}_{8,9}$-dependence in this operators vanishes only once $k_0=E_\beta-E_\alpha$. On the
other hand, the unitary $\bar{\beta}_{8,9}$-ambiguity in $\delta\vec {A}_{{\rm 2N:} \, 1\pi , \,
  1/m}^{a \, (Q)}$ is compensated by the same unitary
ambiguity of the initial and final state wave functions. 
\subsection{Summary on Current Operators in Chiral EFT}
Let us now summarize the status of calculations of current operators
in chiral EFT.  In
tables~\ref{tab_sum_current}, \ref{tab_sum_charge},
\ref{tab_sum_ax_current}, \ref{tab_sum_ax_charge}, \ref{tab_sum_current_ps} and \ref{tab_sum_scalar}
all possible contributions up to N$^3$LO are summarized for vector,
axial vector, pseudoscalar and scalar operators. Note that the nucleon mass $m$ is counted as 
\begin{eqnarray}
\frac{p}{m}\sim \frac{p^2}{\Lambda_b^2},
\end{eqnarray}
where $p$ denotes a low momentum scale and $\Lambda_b$ the breakdown
scale of the theory~\cite{Epelbaum:2014sza}. These are complete
studies up to the order N$^3$LO in chiral expansion.
\begin{table*}[t]
\caption{Chiral expansion of the nuclear electromagnetic current operator up to
  N$^3$LO. LO, NLO, NNLO and N$^3$LO refer to chiral orders $Q^{-3}$,
  $Q^{-1}$, $Q^{0}$ and $Q$, respectively.  The single-nucleon contributions are given in
  Eqs.~(2.7) and (2.16) of~\cite{Krebs:2019aka}.
\label{tab_sum_current}}
\smallskip
\begin{tabularx}{\textwidth}{lccc}
% \begin{ruledtabular}
%\begin{tabular}{@{\extracolsep{\fill}}lrrr}
\hline
  \noalign{\smallskip}
 order &  single-nucleon  &  \hspace{1cm} two-nucleon  &
                                                            \hspace{1.6cm}three-nucleon  
\smallskip
 \\
\hline 
&&& \\
LO  & ---
             & \hspace{1cm}--- & \hspace{1.6cm}--- \\ 
&&&\\
NLO & 
\begin{minipage}[t][0.6cm][t]{3cm}
\vspace{-0.662cm}
\beqa
&&\vec{\fet V}_{{\rm 1N:  \, static}}\nn  
    &+&\vec{\fet V}_{{\rm 1N:  \, 1/m}}\nonumber
\eeqa  
\end{minipage}                                                
 & 
\hspace{1.28cm}
\begin{minipage}[t][0.6cm][t]{5cm}
\vspace{-0.662cm}
\beqa
&&\vec{\fet
                                                             V}_{{\rm
                                                             2N:  \,
                                                             1\pi}}
                                                             ,
                                                             {\rm Eq.}~(\ref{vector_current_qToMinumsOne})\nonumber
\eeqa
\end{minipage}

 & 
\hspace{1.6cm}---
 \\
&&&\\
NNLO &\hspace{0.001cm}
\begin{minipage}[t][0.6cm][t]{3cm}
\vspace{-0.662cm}
\beqa
&\,\,\,&\vec{\fet V}_{{\rm 1N:  \, static}} \nonumber
\eeqa
\end{minipage}        
                    &\hspace{1cm}--- & \hspace{1.6cm}--- \\ 
N$^3$LO & \hspace{0.28cm}
\begin{minipage}[t][1.4cm][t]{3cm}
\vspace{-0.662cm}
\beqa
&&\vec{\fet V}_{{\rm 1N:  \, static}} \nn
&+& \vec{\fet V}_{{\rm 1N:  \, 1/m}}\nn
&+&\vec{\fet V}_{{\rm 1N:  \, off-shell}}\nonumber
\eeqa
\end{minipage}
& \hspace{1cm}
\begin{minipage}[t][1.4cm][t]{5cm}
\vspace{-0.662cm}
\beqa
&&\vec{\fet V}_{{\rm
                                                           2N:  \,
                                                           1\pi}} ,
   {\rm Eq.}~(\ref{OPE:Static:Current:Parametrization})\nn
&+&
                                                                  \vec{\fet
                                                           V}_{{\rm
                                                           2N:  \,
                                                           2\pi}} ,
                                                              {\rm Eq.}~(\ref{Jmom_def})\nn
&+&
                                                                  \vec{\fet
                                                           V}_{{\rm
                                                           2N:  \,
                                                           cont}},
                                                       {\rm Eq.}~(\ref{vector_2N_cont})
    \nonumber
\eeqa
\end{minipage}
                                            &  \hspace{1.6cm}---
                                                           \\
                                                             \hline
\end{tabularx}
%\end{ruledtabular}
\end{table*}

\begin{table*}
\caption{Chiral expansion of the nuclear electromagnetic charge operator up to
  N$^3$LO. LO, NLO, NNLO and N$^3$LO refer to chiral orders $Q^{-3}$,
  $Q^{-1}$, $Q^{0}$ and $Q$, respectively. The single-nucleon contributions are given in
  Eq.~(2.6) of~\cite{Krebs:2019aka}.
\label{tab_sum_charge}}
\smallskip
%\begin{ruledtabular}
%\begin{tabular}{@{\extracolsep{\fill}}lrrr}
\begin{tabularx}{\textwidth}{lccc}
  \hline 
\noalign{\smallskip}
 order &  single-nucleon  &   \hspace{1cm} two-nucleon  &
                                                       \hspace{1.6cm}   three-nucleon  
\smallskip
 \\
\hline 
&&& \\
LO & \hspace{0.9cm}${\bf V}^0_{{\rm 1N:  \, static}} \;$ 
             & \hspace{1cm}--- 
&  \hspace{1.6cm}--- 
\\ 
&&& \\
NLO &  \hspace{0.6cm}
\begin{minipage}[t][0.6cm][t]{3cm}
\vspace{-0.662cm}
\beqa
&&{\bf V}^0_{{\rm 1N:  \, static}}, \nonumber
\eeqa
\end{minipage}
                           &\hspace{1cm}--- 
& \hspace{1.6cm} --- \\
&&& \\
NNLO & \hspace{0.5cm}
\begin{minipage}[t][0.6cm][t]{3cm}
\vspace{-0.662cm}
\beqa
&&{\bf V}^0_{{\rm 1N:  \, static}} \nonumber
\eeqa
\end{minipage} 
                 & \hspace{1cm}--- 
& \hspace{1.6cm} --- \\
&&& \\
N$^3$LO & 
\hspace{0.28cm}
\begin{minipage}[t][1.4cm][t]{3cm}
\vspace{-0.662cm}
\beqa
&&{\bf V}^0_{{\rm 1N:  \, static}} \nn
&+& {\bf V}^0_{{\rm 1N:  \, 1/m}}\nn
&+& {\bf V}^0_{{\rm 1N:  \, 1/m^2}}\nonumber
\eeqa
\end{minipage}
&\hspace{1cm}
\begin{minipage}[t][1.4cm][t]{5cm}
\vspace{-0.662cm}
\beqa
&&{\bf V}^0_{{\rm 2N:  \,1\pi}}, {\rm Eq.}~(\ref{OPE:Static:Charge:Parametrization})\nn
&+& {\bf V}^0_{{\rm 2N:  \,2\pi}}, {\rm Eq.}~(\ref{J0mom_def})\nn
&+&{\bf V}^0_{{\rm 2N:  \, cont}}, {\rm Eq.}~(\ref{vector_2N_cont})\nn
&+&{\bf V}^0_{{\rm 2N:  \, 1\pi , \, 1/m}} , {\rm Eq.}~(\ref{vector_current_charge_one_over_m})\nonumber
\eeqa
\end{minipage}
&\hspace{1cm}
\begin{minipage}[t][1.4cm][t]{5cm}
\vspace{-0.662cm}
\beqa
&&{\bf V}^0_{{\rm 3N:\, \pi}}, {\rm Eq.}~(\ref{vector_charge:3N:general}),~(\ref{vector_charge:3N:long})\nn
&+& {\bf V}^0_{{\rm 3N:\, cont}},{\rm Eq.}~(\ref{vector_charge:3N:general}),~(\ref{vector_charge:3N:short})\nonumber
\eeqa
\end{minipage}\\
&&&\\
\hline
\end{tabularx}
%\end{ruledtabular}
\end{table*}

\begin{table*}
\caption{Chiral expansion of the nuclear axial current operator up to
  N$^3$LO. LO, NLO, NNLO and N$^3$LO refer to chiral orders $Q^{-3}$,
  $Q^{-1}$, $Q^{0}$ and $Q$, respectively. 
\label{tab_sum_ax_current}}
\smallskip
%\begin{ruledtabular}
%\begin{tabular}{@{\extracolsep{\fill}}lrrr}
\begin{tabularx}{\textwidth}{lccc}
  \hline 
\noalign{\smallskip}
 order &  single-nucleon  &   \hspace{1cm} two-nucleon  &
                                                       \hspace{1.6cm}   three-nucleon  
\smallskip
 \\
\hline 
&&& \\
LO & \hspace{1.2cm}$\vec{\bf A}_{{\rm 1N:  \, static}}, \;$
                Eq.~(\ref{LOSingleNCurrentExpr})
             & \hspace{1cm}--- 
&  \hspace{1.6cm}--- 
\\ 
&&& \\
NLO &  \hspace{0.75cm}
\begin{minipage}[t][0.2cm][t]{3cm}
\vspace{-0.662cm}
\beqa
&&\vec{\bf A}_{{\rm 1N:  \, static}}, {\rm  Eq.}~(\ref{SingleNQtoMinus1StaticExpr})\nonumber
\eeqa
\end{minipage}
                           &\hspace{1cm}--- 
& \hspace{1.6cm} --- \\
&&& \\
NNLO & \hspace{0.5cm}
---
                 & \hspace{0.1cm}
\begin{minipage}[t][0.5cm][t]{3cm}
\vspace{-0.662cm}
\beqa
&&\vec{\bf A}_{{\rm 2N:  \, 1\pi}}, {\rm Eq.}~(\ref{AxialCurrentTree1})  \nn
&+& \vec{\bf A}_{{\rm 2N:  \, cont}}, {\rm Eq.}~(\ref{axial_vector:2N:cont:QTo0})\nonumber
\eeqa
\end{minipage}
& \hspace{1.6cm} --- \\
&&& \\
N$^3$LO & 
\hspace{0.28cm}
\begin{minipage}[t][1.4cm][t]{3cm}
\vspace{-0.662cm}
\beqa
&&\vec{\bf A}_{{\rm 1N:  \, static}},{\rm Eq.}~(\ref{SingleNtwoLoopExpr})\nn
&+& \vec{\bf A}_{{\rm 1N:\,}1/m, {\rm UT^\prime}}, {\rm Eq.~(\ref{k0OvermNuclcorr})}\nn
&+& \vec{\bf A}_{{\rm 1N: \, 1/m^2}}, {\rm Eq.}~(\ref{oneovermTo2corr})\nonumber
\eeqa
\end{minipage}
&\hspace{1cm}
\begin{minipage}[t][2.3cm][t]{5cm}
\vspace{-0.662cm}
\beqa
&&\vec{\bf A}_{{\rm 2N:  \, 1\pi}}, {\rm Eq.}~(\ref{Current1pi})\nn
&+& \vec {\bf A}_{{\rm 2N:\,}1\pi, {\rm UT^\prime} }, {\rm Eq.}~(\ref{NNCurrent1piUTPrime}) \nn
&+& \vec{\bf A}_{{\rm 2N:} \, 1\pi , \, 1/m }, {\rm Eq.}~(\ref{Current1piRel}) \nn
&+&\vec{\bf A}_{{\rm 2N:} \, 2\pi }, {\rm Eq.}~(\ref{Current2pi})\nn
&+& \vec {\bf A}_{{\rm 2N:\,cont,\,UT^\prime}}, {\rm Eq.}~(\ref{NNCurrentContUTPrime})\nn
&+& \vec {\bf A}_{{\rm 2N:\,cont,\, 1/m}}, {\rm Eq.}~(\ref{CurrentContRel})\nonumber
\eeqa
\end{minipage}
&\hspace{1cm}
\begin{minipage}[t][2.3cm][t]{5cm}
\vspace{-0.662cm}
\beqa
&&\vec{\bf A}_{{\rm 3N:\, \pi}}, {\rm Eq.}~(\ref{Current3NPi}) \nn
&+&\vec {\bf A}_{{\rm 3N: \, cont}}, {\rm Eq.}~(\ref{Current3NCont}) \nonumber
\eeqa
\end{minipage}\\
&&&\\
\hline
\end{tabularx}
%\end{ruledtabular}
\end{table*}

\begin{table*}
\caption{Chiral expansion of the nuclear axial charge operator up to N$^3$LO.  LO, NLO, NNLO and N$^3$LO refer to chiral orders $Q^{-3}$,
  $Q^{-1}$, $Q^{0}$ and $Q$, respectively. 
\label{tab_sum_ax_charge}}
\smallskip
\begin{tabularx}{\textwidth}{lccc}
% \begin{ruledtabular}
%\begin{tabular}{@{\extracolsep{\fill}}lrrr}
\hline
  \noalign{\smallskip}
 order &  single-nucleon  &  \hspace{1cm} two-nucleon  &
                                                            \hspace{1.6cm}three-nucleon  
\smallskip
 \\
\hline 
&&& \\
LO  & ---
             & \hspace{1cm}--- & \hspace{1.6cm}--- \\ 
&&&\\
NLO & \hspace{0.01cm}
\begin{minipage}[t][0.6cm][t]{2cm}
\vspace{-0.662cm}
\beqa
&&{\bf A}_{{\rm 1N: \,UT^\prime}}^{0}, {\rm Eq.}~(\ref{SingleNChargeUTPrimeQToMinus1})\nn  
&+&{\bf A}_{{\rm 1N: \,}1/m}^{0}, {\rm Eq.}~(\ref{SingleNChargeOneOvermQToMinus1})\nonumber
\eeqa  
\end{minipage}                                                
 & 
\hspace{1.28cm}
\begin{minipage}[t][0.6cm][t]{5cm}
\vspace{-0.662cm}
\beqa
&&{\bf A}^{0}_{{\rm 2N:}\,1\pi}, {\rm Eq.~(\ref{AxialChargeTreeOPE})}\nonumber
\eeqa
\end{minipage}

 & 
\hspace{1.6cm}---
 \\
&&&\\
NNLO &\hspace{2.6cm}
\begin{minipage}[t][0.6cm][t]{2cm}
---
\end{minipage}     
                    &\hspace{1cm}--- & \hspace{1.6cm}--- \\ 
N$^3$LO & \hspace{0.01cm}
\begin{minipage}[t][1.25cm][t]{2cm}
\vspace{-0.662cm}
\beqa
&&{\bf A}_{{\rm 1N:\,static, \,UT^\prime}}^{0}, {\rm Eq.}~(\ref{SingleNChargeStaticUTPrimeQTo1}) \nn
&+& {\bf A}_{{\rm 1N:\,} 1/m}^{0 }, {\rm Eq.}~(\ref{SingleNChargeOneOvermNuclQTo1})\nonumber
\eeqa
\end{minipage}
& \hspace{1.4cm}
\begin{minipage}[t][1.25cm][t]{5cm}
\vspace{-0.662cm}
\beqa
&&{\bf A}^{0}_{{\rm 2N:}\,1\pi}, {\rm Eq.}~(\ref{Charge1piQTo1})\nn
&+& {\bf A}^{0}_{{\rm 2N:}\, 2\pi}, {\rm Eq.}~(\ref{Charge2pi})\nn
&+&{\bf A}^{0}_{\rm 2N:\, cont}, {\rm Eq.}~(\ref{ChargeContQTo1})
    \nonumber
\eeqa
\end{minipage}
                                            &  \hspace{1.6cm}---
                                                           \\
                                                             \hline
\end{tabularx}
%\end{ruledtabular}
\end{table*}

\begin{table*}
\caption{Chiral expansion of the nuclear pseudoscalar operator up to N$^3$LO.  LO, NLO, NNLO and N$^3$LO refer to chiral orders $Q^{-4}$,
  $Q^{-2}$, $Q^{-1}$ and $Q^0$, respectively. . 
\label{tab_sum_current_ps}}
\smallskip
%\begin{ruledtabular}
%\begin{tabular}{@{\extracolsep{\fill}}lrrr}
\begin{tabularx}{\textwidth}{lccc}
  \hline 
\noalign{\smallskip}
 order &  single-nucleon  &   \hspace{1cm} two-nucleon  &
                                                       \hspace{1.6cm}   three-nucleon  
\smallskip
 \\
\hline 
&&& \\
LO & \hspace{1.2cm}$P^{a}_{{\rm 1N:  \, static}}, \;$
                Eq.~(\ref{LOSingleNCurrentPExpr}) 
             & \hspace{1cm}--- 
&  \hspace{1.6cm}--- 
\\ 
&&& \\
NLO &  \hspace{0.75cm}
\begin{minipage}[t][0.2cm][t]{3cm}
\vspace{-0.662cm}
\beqa
&&P^{a}_{{\rm 1N:  \, static}},, {\rm  Eq.}~(\ref{SingleNPQtoMinus2StaticExpr})\nonumber
\eeqa
\end{minipage}
                           &\hspace{1cm}--- 
& \hspace{1.6cm} --- \\
&&& \\
NNLO & \hspace{0.5cm}
---
                 & \hspace{0.1cm}
\begin{minipage}[t][0.5cm][t]{3cm}
\vspace{-0.662cm}
\beqa
&&P^{a}_{{\rm 2N:  \, 1\pi}}, {\rm Eq.}~(\ref{PseudoScalarNN:1pi:QToMinus1})  \nn
&+&  P^{a}_{{\rm 2N:  \, cont}}, {\rm Eq.}~(\ref{PseudoScalarNN:cont:QToMinus1})\nonumber
\eeqa
\end{minipage}
& \hspace{1.6cm} --- \\
&&& \\
N$^3$LO & 
\hspace{0.28cm}
\begin{minipage}[t][1.4cm][t]{3cm}
\vspace{-0.662cm}
\beqa
&&P^{a}_{{\rm 1N:  \, static}},{\rm Eq.}~(\ref{PseudoscalarTwoLoopConjecture})\nn
&+& P^{ a}_{{\rm 1N:\,}1/m, {\rm UT^\prime}}, {\rm Eq.}~(\ref{k0OvermNuclcorrP})\nn
&+& P^{ a }_{{\rm 1N: \, 1/m^2}}, {\rm Eq.}~(\ref{oneovermTo2corrP})\nonumber
\eeqa
\end{minipage}
&\hspace{1cm}
\begin{minipage}[t][2.3cm][t]{5cm}
\vspace{-0.662cm}
\beqa
&&P^{a}_{{\rm 2N:  \, 1\pi , {\rm static}}}, {\rm Eq.}~(\ref{CurrentPs1pi})\nn
&+&  P_{{\rm 2N:\,}1\pi, {\rm UT^\prime} }^{a}, {\rm Eq.}~(\ref{NNCurrent1piUTPrimePs}) \nn
&+& P^{a}_{{\rm 2N:} \, 1\pi , \, 1/m }, {\rm Eq.}~(\ref{CurrentPs1piRel}) \nn
&+& P^{a}_{{\rm 2N:} \, 2\pi }, {\rm Eq.}~(\ref{Current2piPs})\nn
&+&  P_{{\rm 2N:\,cont,\,UT^\prime}}^{a}, {\rm Eq.}~(\ref{NNCurrentPsContUTPrime})\nn
&+& P_{{\rm 2N:\,cont,\, 1/m}}^{a}, {\rm Eq.}~(\ref{NNCurrentPsContOneOverm})\nonumber
\eeqa
\end{minipage}
&\hspace{1cm}
\begin{minipage}[t][2.3cm][t]{5cm}
\vspace{-0.662cm}
\beqa
&&P^{a}_{{\rm 3N:\, \pi}}, {\rm Eq.}~(\ref{Current3NPiPs}) \nn
&+&P^{a}_{{\rm 3N: \, cont}}, {\rm Eq.}~(\ref{Current3NContPs}) \nonumber
\eeqa
\end{minipage}\\
&&&\\
\hline
\end{tabularx}
%\end{ruledtabular}
\end{table*}

\begin{table*}
\caption{Chiral expansion of the nuclear scalar operator up to
  N$^3$LO. LO, NLO, NNLO and N$^3$LO refer to chiral orders $Q^{-3}$,
  $Q^{-2}$, $Q^{-1}$ and $Q^0$, respectively. The single-nucleon
  contributions are from dispersion relation studies where $\pi\pi$ scattering channel
dominates $t$-dependence of the scalar form factor~\cite{Hoferichter:2012wf}.
\label{tab_sum_scalar}}
\smallskip
%\begin{ruledtabular}
%\begin{tabular}{@{\extracolsep{\fill}}lrrr}
\begin{tabularx}{\textwidth}{lccc}
% \begin{ruledtabular}
%\begin{tabular}{@{\extracolsep{\fill}}lrrr}
\hline
  \noalign{\smallskip}
 order &  single-nucleon  &  \hspace{1cm} two-nucleon  &
                                                            \hspace{1.6cm}three-nucleon  
\smallskip
 \\
\hline 
&&& \\
LO  & \hspace{0.34cm}\begin{minipage}[t][0.3cm][t]{3cm}
\vspace{-0.662cm}
\beqa
&&S_{{\rm 1N:  \, static}}\nonumber
\eeqa  
\end{minipage}   
             & \hspace{1cm}--- & \hspace{1.6cm}--- \\ 
&&&\\
NLO & \hspace{0.26cm}
\begin{minipage}[t][0.3cm][t]{3cm}
\vspace{-0.662cm}
\beqa
&&S_{{\rm 1N:  \, static}}\nonumber
\eeqa  
\end{minipage}                                                
 & 
\hspace{1.26cm}
\begin{minipage}[t][0.3cm][t]{5cm}
\vspace{-0.662cm}
\beqa
&&S_{{\rm 2N:  \,1\pi}}, {\rm Eq.}~(\ref{scalar:2N:1pi:QToMinus2})\nonumber
\eeqa
\end{minipage}

 & 
\hspace{1.6cm}---
 \\
&&&\\
NNLO &\hspace{0.001cm}
\begin{minipage}[t][1.2cm][t]{3cm}
\vspace{-0.662cm}
\beqa
&\,\,\,&S_{{\rm 1N:  \, static}} \nn  
    &+&S_{{\rm 1N:  \, 1/m}}\nonumber
\eeqa
\end{minipage}        
                    &\hspace{1cm}--- & \hspace{1.6cm}--- \\ 
N$^3$LO & \hspace{0.27cm}
\begin{minipage}[t][1.4cm][t]{3cm}
\vspace{-0.662cm}
\beqa
&&S_{{\rm 1N:  \, static}} \nonumber
\eeqa
\end{minipage}
& \hspace{1cm}
\begin{minipage}[t][1.4cm][t]{5cm}
\vspace{-0.662cm}
\beqa
&&S_{{\rm 2N:  \,1\pi}} ,
   {\rm Eq.}~(\ref{Current1piSkalar})\nn
&+& S_{{\rm 2N:  \,2\pi}} , {\rm Eq.}~(\ref{ScalarCurrent2pi})\nn
&+&S_{{\rm 2N:  \,cont}}, {\rm Eq.}~(\ref{CurrentContSkalar})
    \nonumber
\eeqa
\end{minipage}
                                            &  \hspace{1.6cm}---
                                                           \\
                                                             \hline
\end{tabularx}
%\end{ruledtabular}
\end{table*}
At this order, various LECs appear which need to be fixed. In the
following, we give a summary of contributing LECs. Single-nucleon
sector is usually phenomenologically parametrized by form factors. For
this reason, we concentrate here on two-nucleon contributions.\newline

\noindent {\bf Vector current}: At the order $Q$,
  $\bar{d}_8, \bar{d}_9, \bar{d}_{21}, \bar{d}_{22}$ from ${\cal L}_{\pi N}^{(3)}$ contribute to one
  pion-exchange current operator. These can be determined from pion
  photo- and electroproduction~\cite{Bernard:2007zu,Navarro:2019iqj}. As can be seen from Eq.~(\ref{OPE:fis:current}) there is
  also a contribution of $\bar{d}_{18}$ LEC which accounts for
  Goldberger-Treiman discrepancy. One can either directly determine
  $\bar{d}_{18}$ from the Goldberger-Treiman
  discrepancy~\cite{Bernard:1995dp} or one can express vector
  current operator in terms of effective axial coupling 
\beqa
g_A^{\rm eff}&=&\frac{F_\pi g_{\pi N}}{m},
\eeqa
where pion-nucleon coupling constant up to given order is given by
\beqa
g_{\pi N}&=&\frac{g_A m}{F_\pi}\left(1-\frac{2 M_\pi^2 \bar{d}_{18}}{g_A}\right)\;.
\eeqa
For the most recent determination of the pion-nucleon coupling
constant with included isospin-breaking effects see~\cite{Reinert:2020mcu}.
Replacement of $g_A$ by $g_A^{\rm
  eff}$ and neglect of $\bar{d}_{18}$ in
Eqs.~(\ref{vector_current_qToMinumsOne}) and (\ref{OPE:fis:current})
changes the current first at the order higher than
$Q$~\footnote{Similar procedure was adopted in pion-nucleon scattering
amplitude~\cite{Fettes:1998ud,Krebs:2012yv,Siemens:2016jwj}}. Finally in addition to
various short-range $C_i$ LECs which  appear in the nuclear forces, there are also
short-range contributions of LECs $L_1$ and $L_2$ which are to be
determined from isovector and isoscalar two- or three-nucleon
observables like magnetic moment of the deuteron, $np\to d\gamma$ radiative capture
or isovector combination of trinucleons magnetic
moments~\cite{Kolling:2012cs,Piarulli:2012bn}. \newline

\noindent {\bf Axial vector current}: Apart from the
$\bar{d}_{18}$-dependence, one-pion-exchange
contributions to the axial vector current depend on $\bar{d}_{2},
\bar{d}_5, \bar{d}_6$ and a linear combination
$\bar{d}_{15}-2\bar{d}_{23}$, see
Eq.~(\ref{OPEhiDefenition}). $\bar{d}_5$ LECs is determined from
pion-nucleon scattering~\cite{Fettes:1998ud,Krebs:2012yv}. Other LECs
can appear in the weak pion production amplitude. Their numerical
values are less known due to the lack of precise experimental data. A
recent analysis shows that an assumption of the natural size of these
LECs leads to a satisfactory description of the available
data~\cite{Yao:2018pzc,Yao:2019avf}. Short-range dependence on LECs is
given in Eq.~(\ref{ChargeContQTo1}). Apart from LECs coming from nuclear forces
there are additional contributions of $z_1, z_2, z_3, z_4$ short-range
LECs. Those can/should be fitted to weak reactions in two-nucleon
sector.\newline

\noindent{\bf Pseudoscalar current}: Due to the continuity equation, the
pseudoscalar current is directly related to the longitudinal component
of the axial-vector current. Once LECs of the axial-vector current are fixed
there are no new parameters in pseudoscalar current.\newline

\noindent{\bf Scalar current}: One-pion-exchange contribution to scalar
current is given in Eq.~(\ref{OPEhiDefenitionSkalar}). As can be directly read off from Eq.~(\ref{OPEhiDefenitionSkalar}) it depends on
$\bar{d}_{16}, \bar{d}_{18}$ LECs from ${\cal L}_{\pi N}^{(3)}$ and
$\bar{l}_4$ LEC from ${\cal L}_\pi^{(4)}$. Mesonic $l_4$ LEC
contributes to the scalar radius of the pion and can be found
in~\cite{Bijnens:2014lea}. $\bar{d}_{16}$ LEC contributes to the quark
mass dependence of the axial coupling $g_A$. For this reason, careful
chiral extrapolation of lattice QCD data is needed for its
determination. A recent lattice QCD determination of $g_A$ can be
found in~\cite{Chang:2018uxx}. Short-range part of the scalar current
depends only on short-range LECs that already appear in nuclear
forces. However, the LECs $\bar{D}_S$ and $\bar{D}_T$ describe quark mass
dependence of the LECs $C_S$ and $C_T$ which appear in nuclear forces
at LO. So careful studies of chiral extrapolations of nuclear forces
are needed to get numerical values of $\bar{D}_S$ and $\bar{D}_T$.

%Single nucleon vector current is usually
%  parametrized in terms of electromagnetic form factors. If, however,
%  one is interested in chiral expansion of the electromagnetic form factors the appearing
%  LECs up to full one-loop order $Q^3$ are $\bar{d}_6, \bar{d}_7$ from ${\cal L}_{\pi N}^{(3)}$ and
%  $\bar{e}_{54}, \bar{e}_{74}$ from ${\cal L}_{\pi N}^{(4)}$ chiral
%  Lagrangians. $\bar{d}_6$ and $\bar{d}_7$ are fixed by isoscalar and
%  isovector electric radii, respectively. $\bar{e}_{54}$ and
%  $\bar{e}_{74}$ are fixed by isoscalar and
%  isovector magnetic radii, respectively~\cite{Kubis:2000zd}. One
%  pion-exchange current operator at order $Q$ depends on $\bar{l}_6$
%  from ${\cal L}_{\pi}^{(4)}$
%  wich is related to pion charge
%  radius~\cite{Bijnens:1998fm,Djukanovic:2014rua,Bijnens:2014lea}.
\section{Compare with JLab-Pisa Currents}
\label{compare:with:pisa}
As already mentioned previously, in our developments of chiral nuclear currents we used the
technique of unitary transformation. In parallel to our activities
there is a different derivation of nuclear currents where time-ordered
perturbation theory combined with the transfer-matrix
inversion technique has been used~(see
\cite{Riska:2016cud,Piarulli:2012bn} and references therein). In this
chapter we would like to briefly discuss their method and 
compare vector and axial-vector currents from two different derivation
methods. 
\subsection{Inversion of transfer matrix method} 
The JLab-Pisa-group used the inversion-of-the-transfer-matrix technique 
to derive energy-independent potential and current operators. They
start with the T-matrix and, by using a naive dimensional analysis (NDA), introduce a power counting for operators that
appear in the T-matrix. Nucleon mass in their approach is counted as a
hard chiral symmetry breaking scale $m\sim\Lambda_\chi$. The
starting point of TOPT-method of JLab-Pisa-group is Eq.~(6) of~\cite{Pastore:2011ip}
given by 
\beqa
\langle f|T|i\rangle&=&\langle f|H_1\sum_{n=1}^\infty\bigg(\frac{1}{E_i-H_0+i\eta}\bigg)^{n-1}|i\rangle,\label{t_matrix_half_off_shell}
\eeqa
where $E_i$ and $E_f$ are eigenenergies of initial and final
states $|i\rangle$ and $|f\rangle$, respectively. The full Hamiltonian
$H$ is decomposed here in a free part $H_0$
and an interacting part $H_1$
\beqa
H&=&H_0+H_1.
\eeqa
 For $E_i\neq E_f$
this is a half-off-shell T-matrix. For $E_i=E_f$ this is an on-shell
T-matrix. The T-matrix of Eq.~(\ref{t_matrix_half_off_shell})
can be than decomposed in chiral orders
\beqa
T&=&T^{(0)} + T^{(1)} + T^{(2)} + \dots\;,
\eeqa
where the indices $n$ of $T^{(n)}$-matrix operator denote their chiral
order $Q^n$ and dots denote higher than $Q^2$ order operators. The same NDA can be used for nuclear force operators
\beqa
v&=&v^{(0)} + v^{(1)} +v^{(2)} + \dots\;.
\eeqa
Inversion of the Lippmann-Schwinger equation can be used iteratively to
calculate the effective potential.
\beqa
v^{(0)}&=&T^{(0)}\;\nn
v^{(1)}&=& T^{(1)} - v^{(0)} G_0 v^{(0)}\;,\label{inversion:Tmatrix}
\eeqa
and so on. The first iteration of leading-order potentials leads
within NDA to order $Q$ contribution: Loop integration gives order $Q^3$ and a
free Green function gives order $Q^{-2}$ contributions corresponding to the inverse sum of
kinetic energies of the nucleons. For the half-off-shell T-matrix
the potential from Eq.~(\ref{inversion:Tmatrix}) is equivalent
to the potential from the folded-diagram technique, which is manifestly
non-hermitian. Equivalence of inversion of T-matrix technique and
folded-diagram approach is demonstrated in
Appendices~\ref{folded_diagram} and~\ref{app:folded:diagr}. On top of the folded-diagram technique,
the authors of \cite{Pastore:2011ip} analyzed off-shell effects in
which they allow contributions proportional to $E_f-E_i$ in the
T-matrix~\footnote{Adding to a T-matrix contributions proportional to
  $[H_0,X]$ with $X$ some operator will not affect the on-shell T-matrix
  but will change its off-shell form. As a consequence, the effective
  Hamiltonian gets transformed by a similarity transformation. In
  Appendix~\ref{unitary_equiv_S_matrix} we demonstrate on a perturbative level that a
  unitary transformation of time-dependent Hamiltonians does
  not affect the on-shell behavior of the T-matrix. Similar arguments should
  work for a similarity transformation.}. This has an influence on the form of the effective potential and
current operators and introduces differences between strict
folded-diagram and JLab-Pisa group techniques. 

For vector- and axial-vector currents JLab-Pisa group proceeds in a
similar way. They write a first-order perturbation theory for T-matrix
in the presence of vector- or axial-vector source. For electromagnetic
current, for example, the T-matrix can be organized via
\beqa
T_\gamma&=&T_\gamma^{(-3)} + T_\gamma^{(-2)} + T_\gamma^{(-1)} + \dots,
\eeqa
where $T_\gamma^{(n)}$ is of order $e Q^n$ with $e$ the electric
charge. Introducing 
\beqa
v_\gamma &=& \vec{V}_\mu\cdot\vec{j}^\mu,
\eeqa
where $V_\mu^a$ is a nuclear vector current and $j_\mu^a$ a
corresponding vector source they perform an inversion of a T-matrix to
derive nuclear current operator order by order:
\beqa
v_\gamma^{-3}&=&T_\gamma^{(-3)}\,\nn
v_\gamma^{-2}&=&T_\gamma^{(-2)} - \bigg[v_\gamma^{(-3)}G_0 v^{(0)} +
v^{(0)} G_0 v_\gamma^{(-3)}\bigg].\label{inversion_technique_v_gamma}
\eeqa

In the following, we will discuss the differences between JLab-Pisa and
Bochum-Bonn-Currents. The differences in the two methods appear first at the order
$Q$.
\subsection{Difference in the vector current at order $Q$}
At order $Q$, we decompose the current  in the number of
pion-exchange via Eq.~(\ref{vectorcurrent:rangedecomp}).
\subsubsection{One-pion-exchange contributions to the vector current}
We start with the static OPE contributions. OPE contributions of the JLab-Pisa
group have been presented in~\cite{Pastore:2009is} where they give only
unrenormalized results. Even relaxing constraints on beta functions of
$d_i$-LECs from ${\cal L}_{\pi N}^{(3)}$ Lagrangian we
are unable to renormalize their Eqs. $(3.30)-(3.42)$. In a later publication~\cite{Piarulli:2012bn} JLab-Pisa
group decided to parametrize the OPE in terms of form
factors, see Eqs.$(2.22)-(2.25)$ of \cite{Piarulli:2012bn}. Since they use this form in all their numerical calculations,
we compare if our results are compatible with their
parametrization. Replacing the form factor by unity (which does not
affect order-$Q$ results) we consider the  difference between
our results and that of JLab-Pisa group
\beqa
\delta V_{{\rm 2N:1\pi,static}}^{(Q),\mu}&=&V_{{\rm
    2N:1\pi,static}}^{{\rm(Q)},\mu} - V_{{\rm 2N:1\pi,static}}^{{\rm JLab-Pisa (Q)},\mu}.
\eeqa
The OPE contributions can be written in the form of
Eq.~(\ref{OPE:Static:Current:Parametrization}) for current and
Eq.~(\ref{OPE:Static:Charge:Parametrization}) for charge contributions, respectively. For this reason, it is enough to
give the difference in terms of scalar functions
\beqa
\delta f_i(k)&=& f_i(k) - f_i^{{\rm JLab-Pisa}}(k), \quad i=1,\dots,8,
\eeqa
where the functions $f_i(k)$ are defined in Eq.~(\ref{OPE:fis:current})
and Eq.~(\ref{OPE:fis:charge}). $f_i^{{\rm JLab-Pisa}}(k)$ are extracted out
of Eq.~(2.22) and (2.24) of~\cite{Piarulli:2012bn}. The difference is
given by
\beqa
&&\delta f_1(k)\,=\,\delta f_2(k)\,=\,0,\nn
&&\delta f_3(k)\,=\,-i\frac{e}{64 \pi^2
  F_\pi^4}\bigg(g_A^4(2 L(k)-1)+(4\pi F_\pi)^2 g_A \bar{d}_{22}\bigg),\nn
&&\delta f_4(k)\,=\,-i\frac{e g_A}{4 F_\pi^2}\bar{d}_{22},\quad \delta
f_6(k)\,=\,-\frac{i e g_A M_\pi^2}{F_\pi^2}\bar{d}_{18},\nn
&&\delta f_5(k)\,=\,-\frac{i e g_A^2}{1152 \pi^2 F_\pi^4}\bigg(-24
M_\pi^2+k^2(-5+288\pi^2 \bar{l}_6)\nn
&&+6 (k^2+4 M_\pi^2)L(k)\bigg).
\eeqa
Note, that Piarulli et al.~\cite{Piarulli:2012bn} discuss only tree-level contributions to
OPE and make a phenomenological extension of these results. They
multiply the tree-diagram results proportional to $\bar{d}_i$'s with
$G_{\gamma N\Delta}(k)/\mu_{\gamma N\Delta}$ and with $G_{\gamma N \rho}(k)$ form factors which
are set here to unity. All other terms of these form factors
contribute to orders higher than $Q$. In contrast to the statement of
Piarulli et al.~\cite{Piarulli:2012bn}, the difference $\delta
\vec{V}_{{\rm 2N:1\pi,static}}^{(Q)}$ contributes to the magnetic moment
operator
$\vec{\mu}=-(i/2)\vec{\nabla}_k\times\vec{V}|_{k=0}$. The difference
for the magnetic moment operator is given by
\beqa
\delta\vec{\mu}&=&-(i/2)\vec{\nabla}_k\times\delta\vec{V}^{(Q)}|_{k=0}\,=\,\frac{e
  g_A^4}{64 \pi^2 F_\pi^4}\frac{\left[\vec{\tau}_1\times\vec{\tau}_2
  \right]^3}{q_1^2+M_\pi^2}\nn
&\times&\bigg([\vec{q}_1\times\vec{\sigma}_2]\vec{q}_1\cdot\vec{\sigma}_1-[\vec{q}_1\times\vec{\sigma}_1]\vec{q}_1\cdot\vec{\sigma}_2\bigg).
\eeqa
This difference, however, comes from $\delta f_3(k)$ and can be absorbed
into $\bar{d}_{21}$ if one makes a shift 
\beq
\bar{d}_{21}\rightarrow \bar{d}_{21} +\frac{g_A^3}{32\pi^2 F_\pi^2}.
\eeq
Charge operator contributions have also been discussed by JLab-Pisa
group. Since they do not mention any static contributions to OPE
charge at order $Q$ we conclude that 
\beq
f_7^{{\rm JLab-Pisa}}(k)=f_8^{{\rm JLab-Pisa}}(k)=0,
\eeq
in clear disagreement to our results of Eq.~(\ref{OPE:fis:charge}).

At the same order $Q$ there are also relativistic corrections to OPE
charge and current operators. Both groups give vanishing results for
relativistic corrections~\footnote{Piarulli et al. discuss $1/m^2$
  and $1/m^3$ corrections to OPE current operator. Although they do
  not vanish they contribute to higher orders in the power counting
  adopted by our group.} for current operators such that there is no
disagreement on the current level for relativistic
corrections. Relativistic corrections to charge
contributions agree only if one fixes unitary phases to
\beq
\beta_8\,=\,\frac{\nu}{2}, \quad \beta_9\,=\,-\frac{1}{2},\label{MatchingToPisa}
\eeq
which means that they are unitary equivalent. Parameters $\beta_8$ and
$\beta_9$ are directly related 
to unitary phase parameters $\mu$ and $\nu$ introduced by
Friar~\cite{Friar:1999sj,Friar:1977xh,Friar:1979by} to describe an
off-shell dependence of relativistic $1/m^2$ corrections to OPE NN
potential:
\beq
\mu\,=\,4 \bar{\beta}_9 +1, \quad \nu\,=\,2\bar{\beta}_8.
\eeq
 These parameters are usually set to
\beq
\bar{\beta}_8\,=\,\frac{1}{4}, \quad \bar{\beta}_9\,=\,-\frac{1}{4},
\eeq
to achieve a minimal non-locality form of OPE NN
potential~\cite{Epelbaum:2014efa}. From Eq.~(\ref{MatchingToPisa}) we
conclude that the charge operator used by JLab-Pisa group does not
correspond to the minimal-nonlocality choice even for $\nu=1/2$. For this
reason, this charge operator should not be convoluted with the available
chiral nuclear forces where minimal non-locality is used. 
\subsubsection{Two-pion-exchange contributions to vector current}
At the level of the two-pion-exchange, there is an agreement between our and
JLab-Pisa group results on the current
but not on the charge operator. This disagreement was addressed by JLab-Pisa
group in~\cite{Pastore:2011ip} where they extensively discuss the TPE
charge operator. They claim that there exists a unitary transformation
that makes two charges unitary equivalent. 
%They start their
%presentation with the discussion of relativistic one-pion-exchange
%corrections, see Eq.~(16) of~\cite{Pastore:2011ip}, given by
%\beqa
%v_\pi^{(2)}(\nu=0)&=&v_\pi^{(0)}(\vec{k})\frac{(E_1^\prime-E_1)^2+(E_2^\prime
%  - E_2)^2}{2\omega_k^2},
%\eeqa
%where $v_\pi^{(0)}$ is LO OPE potential
%\beqa
%v_\pi^{(0)}(\vec{k})&=&-\frac{g_A^2}{F_\pi^2}\vec{\tau}_1\cdot\vec{\tau}_2 \frac{\vec{\sigma}_1\cdot\vec{k}\vec{\sigma}_2\cdot\vec{k}}{\omega_k^2},
%\eeqa
%and $\vec{k}=\vec{p}_1 - \vec{p}_1^\prime=\vec{p}_2^\prime -\vec{p}_2$ and $\omega_k$ is pion energy. $\vec{p}_i$ and
%$\vec{p}_i^\prime$ are initial and final momenta of $i$-th nucleon,
%respectively. This result one gets immediately after summing two
%possible time-orderings of OPE diagram with the free Green-function
%depending on initial energy $E_1+E_2$ and performing
%$1/m$-expansion. This, however, is not the only possibility for recoil
%corrections to OPE. A unitary transformation of the form
%\beqa
%
%\eeqa
%
%However, this happens on
%expense of using different forces for both currents with different
%values for $\nu$-parameter. JLab/Pisa group reproduce our results for
%$\nu=0$ case and claim the inconsistency of our approach since we use
%minimal non-locality choice in the NN force which is $\nu=1/2$. In the
%following we want to go through their arguments and show that the
%transformation they find is non-unitary. The correct statement is that
%there exists a similarity transformation which transforms our nuclear
%forces to JLab/Pisa forces.

%Nevertheless the
It is important to note that the potentials presented in \cite{Pastore:2011ip} are manifestly non-hermitian.
One can see this directly in
Eq.~(20) of \cite{Pastore:2011ip} where one finds a non-hermitian
contribution to their effective potential for the off-shell parameter
$\nu=1$. They also give a similarity transformation (erroneously called a
``unitary transformation''), see Eq.(25) of \cite{Pastore:2011ip} given by
\beqa
H(\nu)&=& e^{-i\,U(\nu)} H(\nu=0)e^{i U(\nu)}
\eeqa
  which transforms
$\nu=1$  into $\nu=0$ potential. One can immediately see from Eq.(28)
of \cite{Pastore:2011ip} given by
\beqa
i\,U^{(1)}(\nu,\vec{p}^\prime,\vec{p})&=&-\frac{\nu}{2}\int_s
\frac{v_\pi^{(0)}(\vec{p}^\prime -
  \vec{s})v_\pi^{(0)}(\vec{s}-\vec{p})}{(\vec{p}^\prime - \vec{s})^2+m_\pi^2}
\eeqa
that $i\,U^{(1)}(\nu,\vec{p}^\prime,\vec{p})$ is {\it not}
antihermitian. So $H(\nu=1)$ and $H(\nu=0)$ potentials are not unitary
equivalent but can be transformed into each other by a similarity
transformation~\footnote{In a later publication \cite{Piarulli:2012bn}  after
Eq.(2.17) the authors remark that ``expressions for $U^{(1)}(\nu)$,
contain a typographical error: the imaginary unit on the left-hand
side should be removed''. Without imaginary unit Eq.(28) of
\cite{Pastore:2011ip} leads indeed to a unitary transformation which,
however, does not transform $H(\nu=1)$ into $H(\nu=0)$. This is
obvious since $H(\nu=1)$ is non-hermitian and $H(\nu=0)$ is
hermitian and there is no unitary transformation which transforms a
non-hermitian operator into a hermitian one.}. 

Applying the same transformations on electromagnetic current operator
Pastore et al.~\cite{Pastore:2011ip} show that the Bochum-Bonn and JLab-Pisa
charge operators can be transformed into each other. However, since
the transformation given in Eq.~(28) of ~\cite{Pastore:2011ip} is not
unitary the charges are not unitary equivalent. This is a similarity
transformation that does not change the spectrum of the nuclear
force and the on-shell T-matrix but changes, in general, the normalization of the wave function.

\subsubsection{One-pion-exchange contributions to vector current}
For axial vector current at order $Q$ the situation is less
transparent than for vector current operator. The differences for
pion-exchange contributions have been extensively discussed
in~\cite{Krebs:2016rqz,Baroni:2018fdn}. The authors of~\cite{Baroni:2015uza} considered only static
contributions so we can not compare recoil corrections to OPE
currents. In the static limit, there is a clear disagreement between
Bochum-Bonn and JLab-Pisa currents even at vanishing momentum
transfer. To clarify if the currents are unitary equivalent we first
perform a calculation of relativistic corrections to box-diagrams for effective potential, all
contributions proportional to $g_A^4/m$. Starting with folded-diagram technique which
is equivalent to the inversion of the transfer matrix technique. The effective
potential which one gets, in this case, is non-hermitian and is in
general given by~\cite{Suzuki:PTP1983,Krebs:2004st}
\beqa
H_{\rm eff}^{\rm FD}&=&(1-A)H(1+A) ,
\eeqa  
where $A$ satisfies a non-linear decoupling
Eq.~(\ref{decoupling_eq_for_A_op}). This potential is related to a
hermitian potential via a similarity transformation
\beqa
H_{\rm eff}&=&S^{1/2} H_{\rm eff}^{\rm FD} S^{-1/2},
\eeqa
with
\beqa
S&=&1+A A^\dagger + A^\dagger A.
\eeqa
The effective potential $H_{\rm eff}$ is a standard hermitian potential of
Eq.~(\ref{UT_effective_potential_okubo}) which one
gets via Okubo transformations of Eq.~(\ref{U_expressed_in_A}). In
order to get effective potential of JLab-Pisa group for the off-shell
choice $\nu=0$, see in particular Eqs.~(19) of \cite{Pastore:2011ip} where the
pion-energy denominators $\omega_1$ and $\omega_2$ factorize, we have
to perform additional unitary transformation 
\beqa
H_{\rm eff}^{\rm JLab-Pisa}&=&U_{\rm 12}^\dagger H_{\rm eff} U_{\rm 12},
\eeqa 
with
\beqa
U_{\rm 12}&=&\exp\big(\alpha_1 S_1 + \alpha_2 S_2\big),
\eeqa
where antihermitian operators $S_1$ and $S_2$ are defined in Eq. (3.25)
of~\cite{Epelbaum:2007us}. In order to reproduce Eq. (19) of
\cite{Pastore:2011ip} we have to fix the unitary phases to
\beqa
\alpha_1&=&-\frac{1}{2}, \quad \alpha_2\,=\,\frac{1}{4},
\eeqa
which is a standard choice of unitary phases for  renormalizable
nuclear forces. With this unitary convention, relativistic corrections
to box-diagrams for nuclear forces of JLab-Pisa and Bochum-Bonn groups
coincide. In this way all transformations needed to bring folded-diagram $H_{\rm eff}^{\rm FD}$ into the form of Bochum-Bonn nuclear force are fixed. JLab-Pisa group does not apply any time-dependent unitary
transformations on nuclear forces to get a current operator. For this
reason, their current operator is transformed in the same way as nuclear
forces. For JLab-Pisa group current operator we get
\beqa
\vec{j}_{\mu}^{\rm JLab-Pisa}&=&U_{12}^\dagger S^{1/2}
\vec{j}_\mu^{\rm FD} S^{-1/2} U_{12},\label{JLab_Pisa_current_formal}
\eeqa
where the folded-diagram current can be extracted via an inversion of
the T-matrix in the presence of an axial-vector source
\beqa
t_5(E_f, E_i\,)&=& \big(1-v 
G_0(E_f)\big)^{-1}v_5(E_f-E_i\,)\nn
&\times& 
\big(1-\tilde G_0(E_i) v\big)^{-1},\label{v5_FD_inversion}
\eeqa
where the T-matrix in the presence of an axial-vector source is defined by
\beqa
T(E_f,E_i\,)&=&2\pi \delta(E_f-E_i\,) t(E_i) + t_5(E_f,E_i\,)\;.
\eeqa
$E_f$ and $E_i$ denote final and initial state energies,
respectively. Eq.~(\ref{v5_FD_inversion}) is derived in Appendix~\ref{app:transfer_matrix}.
The half-off-shell transfer matrix $t(E_i)$ here does not depend on
axial source and satisfies Lippmann-Schwinger equation
\beqa
t(E_i)&=&v + v\, \tilde G_0(E_i) \,t(E_i).
\eeqa
The axial-vector source coupling is within
\beqa
v_5(E)&=&\int d^4 x \,e^{i E x_0}\vec{j}_\mu(x)\cdot \vec{a}^\mu(x), 
\eeqa 
where $\vec{a}^\mu$ denotes the axial-vector source. In
\cite{Krebs:2020rms} we explicitly performed inversion of
Eq.~(\ref{v5_FD_inversion}) and performed a similarity transformation
of Eq.~(\ref{JLab_Pisa_current_formal}). The outcome of this
calculation is supposed to be JLab-Pisa group expressions. However, we
were unable to reproduce their results concluding that either we
misinterpret the method of JLab-Pisa group or there is an error in
their calculation which needs to be clarified in the future.\footnote{Note that
the discussion here is restricted to box-diagrams such that the
transformation of Eq.~(\ref{JLab_Pisa_current_formal}) is incomplete for
other diagrams. In particular, unitary transformations proportional to
$\beta_8, \beta_9$ are not included in Eq.~(\ref{JLab_Pisa_current_formal}).}
\section{Towards Consistent Regularization of the Currents}
\label{RegularizationPath}
So far all reported calculations of current operators have been
performed by using dimensional regularization. So one could take
these operators and start to look at their expectation values 
to study observables. This is indeed what has been done by various
calculations with JLab-Pisa TOPT-currents, see e.g. \cite{Riska:2016cud} for a
review. All these calculations should be considered as a hybrid
approach where no claim on consistency between nuclear forces and
currents is made. Even if both nuclear forces and currents are
calculated from the same framework of chiral EFT the use of different
regularizations (cutoff vs dimensional regularization) leads to  chiral symmetry violation in the very first
iteration of the current with nuclear forces. Here is the
explanation: 

To
solve the Schr\"odinger equation, nuclear forces have to be
regularized. The usual way is to use cutoff regularization. Let
us for example choose a semi-local regulator discussed
in~\cite{Reinert:2017usi}. The regularized form of the long-range part of the leading-order
nuclear force, which is one-pion-exchange diagram, is given by 
\begin{eqnarray}
V_{1\pi,\Lambda}&=&-\frac{g_A^2}{4 F_\pi^2}{\bf \tau}_1\cdot{\bf \tau_2}\frac{\vec{\sigma}_1\cdot\vec{q}\, \vec{\sigma}_2\cdot\vec{q}}{q^2+M_\pi^2}e^{-\frac{q^2+M_\pi^2}{\Lambda^2}},
\end{eqnarray}
where $\vec{q}$ denotes momentum transfer between two nucleons. A
nice property of this regulator is that it does not affect long-range
part of the nuclear force at any power of $1/\Lambda$. On  the
other hand, a pion-pole contribution proportional to $g_A$ of the relativistic correction of 
the axial-vector two-nucleon current is given by
\begin{eqnarray}
\label{Current1piRelFinal}
\vec { A}_{{\rm 2N:} \, 1\pi , \, 1/m}^{ a, (Q,g_A)}&=&\frac{g_A}{8
                                                        F_\pi^2
                                                        m}i \,
                                                       [ {\vec\tau}_1 \times
{\vec \tau}_2 ]^a
\frac{\vec{q}_1\cdot\vec{\sigma}_1}{q_1^2+M_\pi^2}
 \frac{\vec{k}}{k^2+M_\pi^2}\nn
&\times&\Big[
i\,
\vec{k}\cdot\vec{q}_1\times\vec{\sigma}_2-\vec{k}_1\cdot\vec{q}_1+\vec{k}_2\cdot(\vec{q}_1
                                                        + \vec{k})
\Big]\nonumber\\
&+&
1 \; \leftrightarrow \; 2\,,
\end{eqnarray}
where $\vec{k}$ is the momentum transfer of the axial vector current,
and other momenta are defined by
\begin{eqnarray}
\vec{q}_i&=&\vec{p}_i^\prime-\vec{p}_i, \quad
             \vec{k}_i\,=\,\frac{\vec{p}_i^\prime +\vec{p}_i}{2},
             \quad i\,=\,1,2,
\end{eqnarray}
and momenta $\vec{p}_i^\prime$ and $\vec{p}_i$ correspond to the final
and initial momenta of the $i$-th nucleon, respectively. Note that this
is not the only contribution to the relativistic corrections of the
current, but the only one proportional to $g_A$. Complete
expression (including terms proportional to $g_A^3$) for the
relativistic corrections can be found in~\cite{Krebs:2016rqz}. After
we regularized the nuclear force and the axial vector current we can
perform the first iteration and take $\Lambda\to\infty$ limit:
\beqa
&&\vec { A}_{{\rm 2N:} \, 1\pi , \, 1/m}^{a, 
  (Q,g_A)}\frac{1}{E-H_0+i\epsilon}V_{1\pi,\Lambda} \nn&&+
  V_{1\pi,\Lambda}\frac{1}{E-H_0+i\epsilon}
\vec { A}_{{\rm 2N:} \, 1\pi , \, 1/m}^{a, 
  (Q,g_A)}\,=\,\nn
&&\Lambda\frac{g_A^3}{32\sqrt{2}\pi^{3/2}F_\pi^4}([{\fet
              \tau}_1]^a-[{\fet
   \tau}_2]^a)\frac{\vec{k}}{k^2+M_\pi^2}\vec{q}_1\cdot\vec{\sigma}_1\nn
&&+1
   \; \leftrightarrow \; 2\, + {\cal O}(\Lambda^0).\label{firstiteration:axialvectorcurrent}
\eeqa
Since the one-loop amplitude should be renormalizable there should
exist a counter term which absorbs the linear singularity in
$\Lambda$. From Eq.~(\ref{firstiteration:axialvectorcurrent}) we see
that this should be a contact two-nucleon interaction with one-pion
coupling to it. However, there is no counter term like this in the
chiral EFT. The counter term like this requires derivative-less
coupling of the pion which is forbidden by the chiral symmetry: There
exists only a
counter term proportional to $\vec{k}\cdot\vec{\sigma}_1$, but there
is none which is proportional to $\vec{q}_1\cdot\vec{\sigma}_1$. Here
$\vec{k}$ is the momentum of the pion coupling to the two-nucleon
interaction.\footnote{At higher orders one can construct derivative-less
  pion-four-nucleon interactions by multiplying low energy constants
  with $M_\pi^{2}$. They are coming from the explicit chiral symmetry
  breaking by finite quark mass. However, at the order $Q$ we can not
  construct a counter term like this.} If there is no counter term
that absorbs the linear cutoff singularity there should be some
cancelation in the amplitude with other terms. Indeed the same
singularity but with the opposite sign we would get for the static limit
of the axial vector current of the 
order $Q$ if we would calculate the current by using cutoff
regularization. Axial vector current at the order $Q$, however, is
calculated by using dimensional regularization and is finite. It also
remains finite if we just multiply the current with any cutoff
regulator we want. So at the level of the amplitude, the mismatch
between cutoff and dimensional regularization used in the
construction of  operators leads to a
violation of the chiral symmetry at the one-loop level which, however, is the order of
accuracy of our calculations.
So we see that it is dangerous to
multiply the current operators calculated within dimensional
regularization by some cutoff regulator and calculate the expectation
values of this. With similar arguments one
  can show that dimensionally regularized three-nucleon  forces at the
  level of N$^3$LO, which were published in~\cite{Bernard:2007sp,Bernard:2011zr}, can not be used in combination with the cutoff regularized
  two-nucleon forces at the same order~\cite{Epelbaum:2019kcf}. The mismatch between dimensional and cutoff
  regularization will lead also in this case to a violation of the
  chiral symmetry at the  one
  loop level.

To respect the chiral symmetry we need to
calculate both nuclear forces and currents with the same regulator. On
top of it the regulator which we choose should be symmetry preserving. One
possibility to construct a regulator, which manifestly respects the chiral
symmetry  was  proposed more than four decades ago by Slavnov~\cite{Slavnov:1971aw},
where he introduced a higher derivative regularization in a
study of the non-linear sigma model. Recently, the first applications of this technique
to the chiral EFT have been discussed in the
literature~\cite{Djukanovic:2004px,Long:2016vnq}. A basic idea of
Slavnov is to change the propagator of a pion field on the Lagrangian
level. Since pion fields are organized in $U$-fields which are
elements of ${\rm SU}(2)$ group and all derivatives in the Lagrangian appear
as covariant derivatives to maintain gauge and chiral
symmetry, the modified chiral Lagrangian should be
expressed in terms of covariant derivatives of $U$-fields. In this way the
gauge- and chiral symmetry are maintained by construction. For our
purpose, we change the static pion propagator by a
regularized one
\beqa
\frac{1}{q^2+M_\pi^2}\to \frac{e^{-\frac{q^2+M_\pi^2}{\Lambda^2}}}{q^2+M_\pi^2}.\label{local_reg_pion}
\eeqa
The choice of this regulator is consistent with the semilocal regulator used
in nuclear forces where pion propagator is regularized via
Eq.~(\ref{local_reg_pion}) and contact interactions via a 
non-local Gaussian regulator. The challenge is to construct the
modified chiral
Lagrangian in terms of covariant derivatives and chiral $U$-fields
which leads to modified pion propagator of Eq.~(\ref{local_reg_pion}),
modified contact interactions and to regularized forces and currents.
Construction of
consistent nuclear forces and currents within the higher derivative
approach is a work in progress. However, the first results for the deuteron
charge in this consistent approach are already available~\cite{Filin:2019eoe} and
will be briefly discussed in the next section.
\section{Two-Nucleon Charge Operator}
\label{first:application:deuteron:charge}
As already mentioned in Sec.~\ref{RegularizationPath}, at
the moment we can not take expectation values of the current operators
at the order
$Q$ without violation of the chiral symmetry at the same order of
accuracy. This happens due to the mismatch
of dimensional and cutoff regularizations of nuclear forces and
currents. Inconsistency between regularization of forces and currents
leads in some cases to strong cut-off dependence in the observables, see~\cite{Rozpedzik:2011cx} for discussion of
photodisintegration on $^2$H and $^3$He calculated with order $Q$
vector current. In many cases, however, the inconsistent (hybrid)
approach gives a satisfactory description of the data. In the last
decade, there are various application of hybrid approach with JLab-PIsa
TOPT current operators, see~\cite{Riska:2016cud} for a recent review and \cite{King:2020wmp,Lovato:2019fiw,Schiavilla:2018udt,Lovato:2017cux}
and references therein for the most recent activities in this field. We are not going to
report here on the hybrid approach activities but rather concentrate
on the  consistent calculation of the deuteron charge form factor~\cite{Filin:2019eoe}. 
\subsection{Deuteron Charge}
\label{deuteron_charge_op}
Deuteron form factors have been extensively studied within EFT
in pion-full and pion-less approaches,
see~\cite{Marcucci:2015rca,Garcon:2001sz,Gilman:2001yh} for 
reviews. To perform a consistent calculation of deuteron form
factors, we need to construct a regularized NN potential and
electromagnetic current operator with the same off-shell
properties. Up to order $Q$ the expressions for unregularized current
have been discussed in Sec.~\ref{vectorUpToOrderQ}. At the order $Q$
meson-exchange contributions to the
deuteron charge form factor are given by just relativistic corrections
to the
one-pion-exchange charge operator,
Eq.~(\ref{vector_current_charge_one_over_m}). Such a simplification for
the charge operator appears since the deuteron is an isoscalar. All
complicated one-loop terms proportional to $[\vec{\tau}_1]^3$ and
$[\vec{\tau}_1]^3$ do not contribute to expectation values after
convolution with the deuteron wave function. The only non-vanishing order $Q$
pion-exchange contribution to the deuteron charge operator is given by
\beqa
&&{V}_{{\rm 2N: 1\pi, }1/m}^{0, (Q)}=-\frac{3 e g_A^2}{16F_\pi^2m}\bigg[
\frac{1-2\bar\beta_9}{q_2^2 + M_\pi^2} \vec{\sigma}_1\cdot\vec{k}
 \vec{\sigma}_2\cdot\vec{q}_2\nn
&+& 
\frac{ 2\bar\beta_8-1
     }{(q_2^2+M_\pi^2)^2} 
  \vec{\sigma}_1\cdot\vec{q}_2  \, \vec{\sigma}_2\cdot\vec{q}_2 \vec{q}_2\cdot\vec{k}\bigg]
+ 1\leftrightarrow 2.\label{vector_current_charge_one_over_m_for_deuteron}
\eeqa
For the single-nucleon charge operator, we use
Eq.~(\ref{V1N:Charge:nonrel}) where the charge operator is expressed in terms
of electromagnetic form factors. Due to poor convergence of chiral
expansion for electromagnetic form factors we used their
phenomenological parametrization. For the calculation of the deuteron charge
form factor we used a global analysis of the experimental data
of Refs.~\cite{Ye:2017gyb,Ye:smallrp}. To estimate a systematic error we
also used dispersive analyses of
    Refs.~\cite{Belushkin:2006qa,Lorenz:2012tm,Lorenz:2014yda}. In
    derivation of
    Eq.~(\ref{vector_current_charge_one_over_m_for_deuteron}) we
    applied unitary transformations of nuclear forces on the
    leading-order single-nucleon charge operator. Since we work now
    with the form factor
    parametrization of the single-nucleon contribution we apply the same
    transformation on the electric form factor and replace Eq.~(\ref{vector_current_charge_one_over_m_for_deuteron}) by
\beqa
&&{V}_{{\rm 2N: 1\pi, }1/m}^{0, (Q)}=-\frac{3e g_A^2}{16F_\pi^2m}\bigg[
\frac{1-2\bar\beta_9}{q_2^2 + M_\pi^2} \vec{\sigma}_1\cdot\vec{k}
 \vec{\sigma}_2\cdot\vec{q}_2\nn
&+& 
\frac{ 2\bar\beta_8-1
     }{(q_2^2+M_\pi^2)^2} 
  \vec{\sigma}_1\cdot\vec{q}_2  \, \vec{\sigma}_2\cdot\vec{q}_2
  \vec{q}_2\cdot\vec{k}\bigg] G_E^S(\vec{k}^2)
+ 1\leftrightarrow 2,\nn
&&\label{vector_current_charge_one_over_m_for_deuteron_em_ff}
\eeqa 
where $G_E^S(\vec{k}^2)$ is the isoscalar part of the electric form factor
\beqa
G_E^S(\vec{k}^2)&=&G_E^p(\vec{k}^2) + G_E^n(\vec{k}^2),
\eeqa
and $G_E^p(\vec{k}^2) $ and $G_E^n(\vec{k}^2) $ are electric form factors
of proton and neutron, respectively.
To regularize
Eq.~(\ref{vector_current_charge_one_over_m_for_deuteron_em_ff}) we
have to take the same regulator which was used in the construction of
nuclear forces. This requires to make following replacements in
Eq.~(\ref{vector_current_charge_one_over_m_for_deuteron_em_ff}) for pion-exchange propagators
\beqa
\frac{1}{\vec{q}_2^2+M_\pi^2}&\to&
\frac{e^{-\frac{\vec{q}_2^2+M_\pi^2}{\Lambda^2}}}{\vec{q}_2^2+M_\pi^2},\nn
\frac{1}{[\vec{q}_2^2+M_\pi^2]^2}&\to&\bigg(1+\frac{\vec{q}_2^2+M_\pi^2}{\Lambda^2}\bigg)\frac{e^{-\frac{\vec{q}_2^2+M_\pi^2}{\Lambda^2}}}{[\vec{q}_2^2+M_\pi^2]^2}.\label{local_reg_prescription_nn}
\eeqa
Note the dependence of the deuteron charge operator in Eq.~(\ref{vector_current_charge_one_over_m_for_deuteron_em_ff}) on
unitary phases $\bar{\beta}_8$ and $\bar{\beta}_9$. The same unitary phases appear
in relativistic $1/m^2$-corrections to one-pion-exchange in nuclear
force. They are usually chosen as
\beqa
\bar{\beta}_8&=&\frac{1}{4}, \quad \bar{\beta}_9\,=\,-\frac{1}{4}, \label{minimal_nonlocality_off_shell}
\eeqa
to maintain a minimal non-locality of
the nuclear force. However, they might be chosen differently, and any
calculated observable should only weakly depend on them in a
consistent calculation\footnote{If unitary
transformations with the phases $\bar{\beta}_8$ and $\bar{\beta}_9$
would be implemented in an exact way without any approximation the observables
would be independent on them.}. To test the consistency of our
calculation we generated two versions of nuclear forces, the
non-minimal coupling choice of Eq.~(\ref{minimal_nonlocality_off_shell}) and 
\beqa
\bar{\beta}_8&=&\bar{\beta}_9\,=\,\frac{1}{2},\label{vanishing_deuteron_ope_charge_choice}
\eeqa
for which the deuteron pion-exchange charge contribution
disappears. Both versions lead to the same results for the deuteron charge
form factor. 
It is important to note that regularization prescription in
Eq.~(\ref{local_reg_prescription_nn}) works only due to the simple structure of
the charge operator in
Eq.~(\ref{vector_current_charge_one_over_m_for_deuteron_em_ff}). For the
current operator or even for the charge operator for isovector observables the regularization
of the current operator needs more effort and can be worked out within
higher derivative regularization. This is still a work in progress.

At the order $Q$ there are no short-range contribution to
deuteron charge form factor, as can be directly followed from
Eq.~(\ref{em_charge_short_order_Q}). The first short-range
contributions show up at the order $Q^2$ and can be parametrized by three parameters multiplied with
isoscalar electric form factor
\beqa
  \label{eq:contactchargedensity}
 V_\text{2N:cont}^0
  &=&2 e G_\text{E}^\text{S}({\vec{k}}^2) \bigg(
    A \, \vec{k}^2
  + B \, \vec{k}^2 (\vec{\sigma}_1 \cdot \vec{\sigma}_2)\nn
  &+& C \, \vec{k} \cdot \vec{\sigma}_1  \vec{k} \cdot \vec{\sigma}_2
   \bigg),
\eeqa
where $A$, $B$, and $C$ are low-energy constants
(LECs)~\cite{Phillips:2016mov}. Although these are $Q^2$-order
effects, they catch the short-range
off-shell dependence of N$^4$LO nuclear forces. This is a direct
consequence of the fact that contributions in Eq.~(\ref{eq:contactchargedensity})
can be generated with unitary transformations acting on the
single-nucleon charge density.  The same unitary transformations
acting on the kinetic energy term produce off-shell short-range
contributions to nuclear force at N$^4$LO. Not all three LECs contribute
independently in the deuteron charge form factor. Only a linear combination
$ A+B+C/3$ appears as a free parameter that we fitted to
the deuteron charge form factor
data. We regularize these
charge contributions in the same non-local way as was done in N$^4$LO
nuclear force~\cite{Reinert:2017usi}
\beqa
  \label{eq:contactchargedensityReg}
 V_\text{2N:cont}^{0,\text{reg}}
&=&2 e G_\text{E}^\text{S}({\vec{k}}^2) \nn
&\times&\bigg[ (A + B \,  (\vec{\sigma}_1 \cdot \vec{\sigma}_2))
    F_1 \left( \frac{\vec{p}_1-\vec{p}_2}{2}, \frac{\vec{p}'_1-\vec{p}'_2}{2}, \vec{k} \right)\nn
&+& C F_2 \left( \frac{\vec{p}_1-\vec{p}_2}{2}, \frac{\vec{p}'_1-\vec{p}'_2}{2}, \vec{k} \right)
   \bigg],
\eeqa
where the functions $F_1$ and $F_2$ are defined as
\beqa
  F_i (\vec{p}, \vec{p}', \vec{k}) &:=&
      E_i \left( \vec{p}-\frac{\vec{k}}{2}, \vec{p}' \right)
    + E_i \left( \vec{p}+\frac{\vec{k}}{2}, \vec{p}' \right)\nn
    &+& E_i \left( \vec{p}'-\frac{\vec{k}}{2}, \vec{p} \right)
    + E_i \left( \vec{p}'+\frac{\vec{k}}{2}, \vec{p} \right),\quad\quad
\\
  E_1 \left( \vec{p}, \vec{p}' \right) &:=& \left( \vec{p}^2 - \vec{p}'^2 \right)
   e^{-\frac{\vec{p}^2 + \vec{p}'^2}{ \Lambda ^2}},\nn
  E_2 \left( \vec{p}, \vec{p}' \right) &:=& \left[
    (\vec{\sigma}_1 \cdot \vec{p}) (\vec{\sigma}_2 \cdot \vec{p}) -  (\vec{\sigma}_1 \cdot \vec{p}') (\vec{\sigma}_2 \cdot \vec{p}')
   \right]\nn
   &\times&e^{-\frac{\vec{p}^2 + \vec{p}'^2}{ \Lambda ^2}}.
\eeqa

In order to give a theoretical error quantification due to truncation
of chiral expansion one can use the
algorithm proposed in~\cite{Epelbaum:2014sza}.  Namely, one can estimate the truncation error $\delta(X)^{(i)}$ of an observable $X$ at 
$i$-th order of the chiral expansion, with $i=0,2,3,\dots$. If $Q$ denotes the chiral expansion parameter,
the expressions for truncation errors are 
\small
\begin{eqnarray}
\delta(X)^{(0)} &\geq& \max \left( Q^2 \vert X^{(0)} \vert\,, \vert X^{(i \geq 0)} - X^{(j \geq 0)} \vert \right) \,,\nonumber \\
\delta(X)^{(2)} &=& \max \left( Q^{3} \vert X^{(0)} \vert \,, Q \vert \Delta X^{(2)} \vert \,, \vert X^{(i \geq 2)} - X^{(j \geq 2)} \vert \right) \,, \nonumber \\
\delta(X)^{(i)} &=& \max \left( Q^{i+1} \vert X^{(0)} \vert \,, Q^{i-1}
                    \vert \Delta X^{(2)} \vert \,, Q^{i-2} \vert
                    \Delta X^{(3)} \vert \right) \nonumber\\
&& {\rm for} \, i \geq 3\;.
\label{therrors}
\end{eqnarray}
\normalsize
In the above formulas $X^{(i)}$ is a prediction for the observable $X$ at $i$-th order, $\Delta X^{(2)} \equiv X^{(2)} - X^{(0)}$ and $\Delta X^{(i)} 
\equiv X^{(i)} - X^{(i-1)}$ for $i \geq 3$. The algorithm of
Eq.~(\ref{therrors}) is very simple. However, it does not give a statistical
interpretation of the error estimate. An interesting algorithm based
on Bayesian approach was developed
in~\cite{Furnstahl:2015rha,Melendez:2017phj,Melendez:2019izc}. Based
on these studies we employed a Bayesian model specified
in~\cite{Epelbaum:2019zqc} to give a theoretical truncation error
estimate of the deuteron charge form factor.

\begin{figure}
\begin{center}
%\hspace*{-0.3cm}\includegraphics[width=8.7cm]{figure_GC_v2.pdf}
\hspace*{-0.3cm}\includegraphics[width=8.7cm]{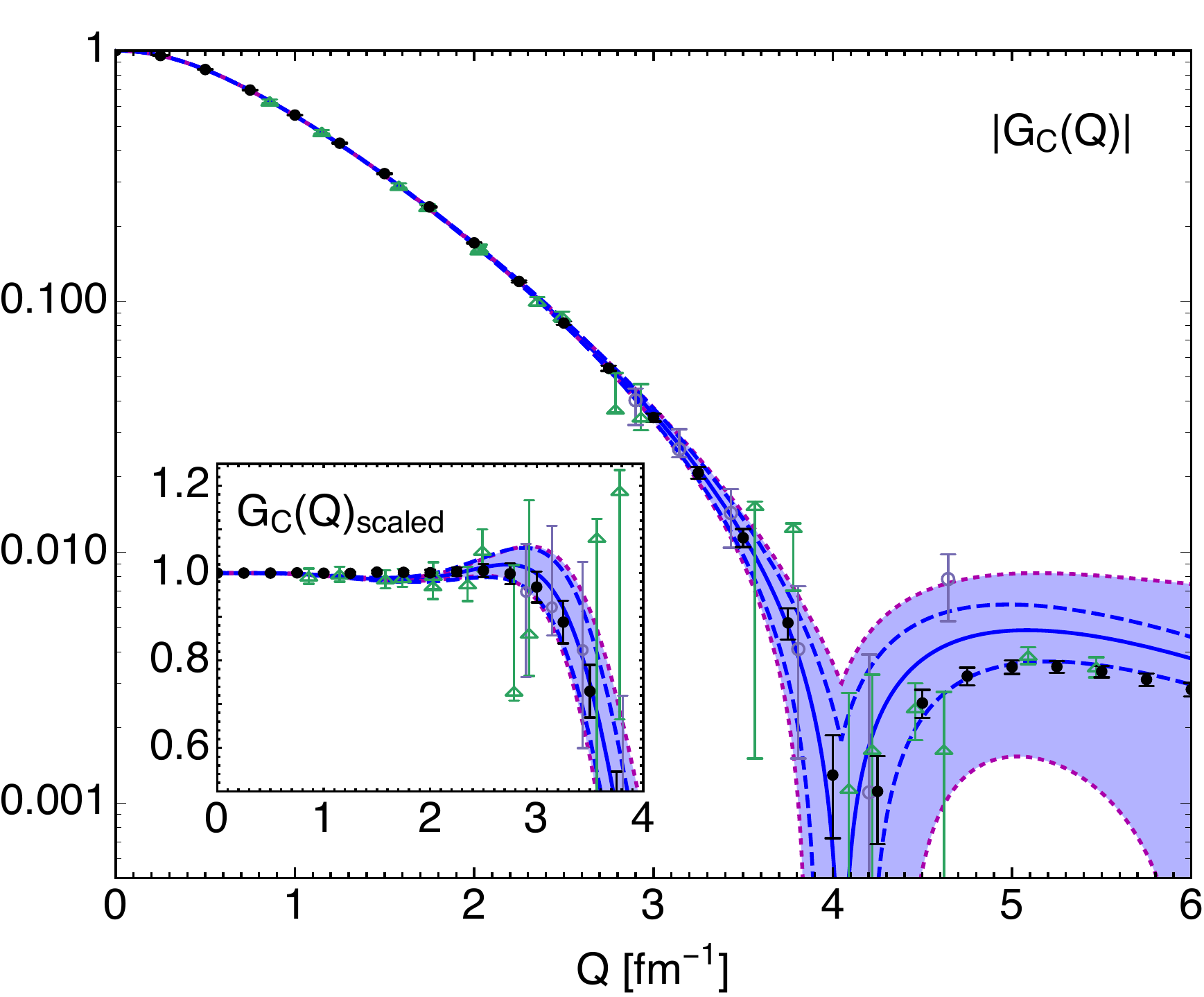}
\caption{\label{fig:GC_FFFit}
  (Color online)   Deuteron charge form factor from the best fit  to data up to $Q=4$ fm$^{-1}$ evaluated for the cutoff  $\Lambda = 500$ MeV (solid blue lines).
   Band between  dashed (blue) lines corresponds to a $1\sigma$ error in the determination of
   the short-range contribution to the charge density operator at N$^4$LO.
  Light-shaded (orange dotted) band corresponds to
  the estimated  error ($68\%$ degree of belief) from truncation of the chiral expansion at N$^4$LO.
  Open violet circles, green   triangles and blue squares are experimental data from
  Refs.~\cite{Nikolenko:2003zq},~\cite{Abbott:2000ak} and~\cite{Abbott:2000fg}, respectively.
  Black solid circles correspond to the parameterization of the
  deuteron form factors from Ref.~\cite{Marcucci:2015rca,Sick:priv}
  which  is not used in the fit and shown just for comparison. The
  rescaled charge form factor of the deuteron, $G_\text{C}{(Q)}_{\rm scaled}$, as defined in  Ref.~\cite{Marcucci:2015rca}, is shown on a linear scale.
}
\end{center}
\end{figure}

Deuteron charge form factor based on the consistently regularized charge
operator, defined in
Eqs.~(\ref{vector_current_charge_one_over_m_for_deuteron_em_ff}),
(\ref{local_reg_prescription_nn}), (\ref{eq:contactchargedensityReg})
is shown in Fig~\ref{fig:GC_FFFit} for cutoff $\Lambda=500\,{\rm
  MeV}$. A fit to the data up to $Q=4\,{\rm fm}^{-1}$
was performed to fix a linear combination of LECs $ A+B+C/3$. It was
verified that the cutoff variation in the range of $\Lambda=400\dots
500\,{MeV}$ leads to the results which are lying well within the
truncation error band. Out of the calculated deuteron charge form
factor one can extract the structure radius of the deuteron
\beqa
 r_{\rm str} = 1.9731 \substack{+0.0013\\ -0.0018}\ \text{fm},\label{structure_radius_deuteron}
\eeqa
Individual contributions to the uncertainties are given in
Table~\ref{Tab:rstr_uncert}. This structure radius was calculated with the
minimal-nonlocality choice of Eq.~(\ref{minimal_nonlocality_off_shell})
of unitary phases $\bar{\beta}_8$ and $\bar{\beta}_9$. We repeated the
calculation of the structure radius with the choice of Eq.~(\ref{vanishing_deuteron_ope_charge_choice})
for which the deuteron contributions of long-range charge operator
vanish. The result for the structure radius for this choice agreed
with the one in Eq.~(\ref{structure_radius_deuteron}) which confirms that off-shell ambiguities
do not affect the final result. With this finding, we were able to make
a prediction for neutron root-mean-square (RMS) radius. The structure radius of the deuteron
can be expressed as the RMS radius of the deuteron minus
individual nucleon contributions and  minus relativistic correction (Foldy-Darwin term):
\beqa
 r_{\rm str}^2 = r_d^2 - r_{p}^2 - r_{n}^2  - \frac{3}{4 m_p^2},
  \label{Eq:r_str}
\eeqa
where $r_d$, $r_p$ and $r_n$ are deuteron, proton and neutron
RMS charge radii, respectively. $m_p$ denotes here a proton mass.

%\begingroup
%\squeezetable
\begin{table*}[t]
\caption{\label{Tab:rstr_uncert} Deuteron structure radius squared predicted at
N$^4$LO in $\chi$EFT (1st column) and the individual contributions
to its uncertainty: from
the truncation of the chiral expansion (2nd),  the statistical error in the
short-range charge density operator extracted from $G_{\rm C} (Q^2)$
(3rd),  the errors from the
statistical uncertainty in $\pi$N LECs from the Roy-Steiner analysis of Ref.~\cite{Hoferichter:2015tha,Hoferichter:2015hva}
propagated  through the variation in the deuteron wave functions (4th),   the errors from the
statistical uncertainty in 2N LECs from the analysis of the 2N
observables of Ref.~\cite{Reinert:2017usi} (5th), the error from  the
choice of the maximal energy   in the fit (6th)
as well as the total uncertainty evaluated using the sum of these numbers in quadrature (7th).
All numbers are given in fm$^2$.
}
\small
%\smallskip
%\begin{ruledtabular} 
%@{\extracolsep{\fill}}
\begin{tabularx}{\textwidth}{@{\extracolsep{\fill}} ccccccc}
\hline
{$r_{\rm str}^2$ } & {{\rm truncation}} &  $\rho^{\rm cont}_\text{2N}$ &
                                                                    {$\pi$N LECs} & {2N LECs}                                   & $Q$-range &   {total} \\
\hline\hline
%\vspace{-1mm}
3.8933                                 & $\pm$0.0032         & $\pm$0.0037 & $\pm$0.0004             &     $ \substack{+0.0010\\ -0.0047}$       & $\pm$0.0017                                    & $ \substack{+0.0053\\ -0.0070}$               \\
\hline
\end{tabularx}
%\end{ruledtabular}
\end{table*}
%\endgroup
The  extraction of the deuteron-proton RMS
charge radii difference can be made
from the hydrogen-deuterium 1S-2S isotope shift measurements
~\cite{Pachucki:2018yxe} accompanied with an accurate QED
analysis up to three-photon exchange accuracy. According
to~\cite{Pachucki:2018yxe} the deuteron-proton RMS
charge radii difference is given by
\begin{equation}
  r_d^2 - r_{p}^2 = 3.820 70(31) \text{fm}^2.
  \label{Eq:rd-rp}
\end{equation}
See also~\cite{Jentschura:2011NOTinHep} for an earlier determination.
This relation leads immediately to a very precise prediction of the neutron
rms radius
\begin{equation}
 \label{Eq:neut_rad}
 r_n^2 = - 0.106 \substack{ +0.007\\ -0.005\\} \, \text{fm}^2.
\end{equation}
which is   $1.7\sigma$ smaller than the one given by the
PDG~\cite{Tanabashi:2018oca}. It is important to note that our results
for neutron radius rely on chiral nuclear forces fitted to the
Granada-2013 database~\cite{Perez:2013jpa}. Isospin breaking effect analysis in the nuclear force was done
along with the treatment of Nijmegens
group~\cite{Stoks:1993tb}. Isospin-breaking in one-pion-exchange due
to different pion masses $M_{\pi^0}$ and $M_{\pi^\pm}$ as well as charge
dependence of the short-range interactions in the $^1S_0$ wave was
explicitly taken into account. These are the dominant
isospin-breaking effects that are needed for a correct description of
the phase-shifts. For calculation of
scattering observables in the two-nucleon system, the isospin-breaking
due to long-range electromagnetic interaction have been
considered. Improved Coulomb potential~\cite{Austen:1983te}, the magnetic-moment
interaction~\cite{Stoks:1990us} as well as the vacuum-polarization potential~\cite{Durand:1957zz} were taken
into account~\cite{Epelbaum:2019kcf}. There are, however, further corrections that
are systematically worked out within chiral EFT. Expressions for the
leading and subleading isospin-breaking two-pion-exchange-potential
and irreducible pion-photon exchange contributions are already
available. Also charge-dependence of the pion-nucleon coupling constant 
needs to be accounted for in a systematic treatment within chiral
EFT. Chiral nuclear force which takes into account these
isospin-breaking effects was presented in~\cite{Reinert:2020mcu}.  It will be
interesting to repeat the calculation of the deuteron charge form
factor by using the new chiral nuclear force~\cite{Reinert:2020mcu}
and extract in this way the value for neutron RMS radius.

\section{Summary}

As reviewed in this article, chiral EFT provides a powerful framework
for a description of nuclear forces and 
currents in a systematically improvable way. Nuclear forces constructed up to N$^4$LO achieved already a
tremendous precision such that they describe two-nucleon data with
$\chi^2\sim 1$ and can be considered as a partial wave
analysis~\cite{Reinert:2017usi}. Nuclear vector and axial-vector
current operators have been worked out up to order $Q$ with
the leading-order starting at $Q^{-3}$. They have been
calculated by using the T-matrix inversion technique by JLab-Pisa group
and the unitary transformation technique by Bochum-Bonn
group. Pseudoscalar (scalar) current operators have been worked out up to order $Q^0$ with
the leading-order starting at $Q^{-4}$ ($Q^{-3}$). They have been
calculated within the unitary transformation technique by Bochum-Bonn group.  To get a
renormalizable current our group used time-dependent unitary
transformations which explicitly depend on external sources. This
leads to energy-transfer dependent currents with the modified continuity
equations for vector and axial-vector currents. The behavior of
four-vectors under the Lorentz transformation further constrains the
currents. Expressions for vector as
well as for axial-vector currents derived by JLab-Pisa and Bochum-Bonn group differ at order $Q$. It was shown
that for the vector current, two-pion-exchange operators can be transformed into each other
by a similarity transformation. For the axial-vector current the situation
is more confusing. We recalculated box-diagram contributions to the axial vector current, which are
proportional to $g_A^5$, by using the T-matrix
inversion technique. At the level of  Fock-space we came to the
conclusion that these currents should be unitary equivalent to
Bochum-Bonn current. However, we could not reproduce the final results of
JLab-Pisa group. This issue should be clarified in the future. 

In most available calculations of nuclear currents, dimensional
regularization has been used to regularize loop diagrams. Additional
regularization of the current operator is, however, required when it is sandwiched
between wave functions for calculation of observables. Nuclear wave functions
themselves are calculated from the solution of the Schr\"odinger equation with
an input of nuclear force where cutoff regularization was used. A naive
procedure to multiply the current operators with some cutoff
function does not work at order $Q$. In the case of axial-vector
current we have explicitly shown that this procedure leads to
violation of the chiral symmetry. This happens due to different divergent pieces in dimensional
and cutoff regularization. A completely perturbative one-loop calculation of
two-nucleon axial-vector current observable does not exist in this
case in an infinite cutoff limit. To renormalize the theory one
has to include derivative-less pion-two-nucleon vertices. Chiral
symmetry constraint, however, does not allow the existence of such a
vertex. This shows that regularization artifacts in such a procedure
are not under control. This happens just due to different
regularizations used in the calculation of currents (dimensional
regularization) and in the iteration procedure (cutoff
regularization). If one would use the same regularization in the
calculation of forces and current operators this problem would not
arise. We conclude that to achieve order-$Q$ precision, the
current operators have to be calculated with the same cutoff
regularization which was used in nuclear forces. On top of this, a chosen
cutoff regularization should preserve the chiral symmetry. A higher
derivative regularization approach, where cutoff regularization is introduced on the
Lagrangian level with all derivative operators being covariant, seems
to be a promising tool to achieve this goal. Calculation of 
nuclear current operators within this procedure is in progress. 
 
Concerning numerical studies with nuclear current operators, we restricted our discussion only to
consistently regularized current operators. It is important to mention
that deuteron form factors and many other observables for the vector and
axial-vector current have been extensively discussed in the literature
within a hybrid approach, see Sec.~\ref{intro} for references. We claim, however, that at
the order $Q$ the consistency issue becomes essential and should be carefully investigated.

As the first application of a consistently regularized electromagnetic
charge operator, we discussed the deuteron charge form factor. 
Although consistently regularized current operators are not yet
available, it is already possible to get a consistently regularized
 electromagnetic charge operator for isoscalar observables like
 deuteron charge form factor. The reason is that the meson-exchange charge operator
 at the order $Q$ has a very simple form. Only relativistic
 $1/m$ - corrections to the leading one-pion-exchange charge operator survive
 at this order. A consistent regularization of this operator is very
 simple and was discussed in Sec.~\ref{deuteron_charge_op}. On top of
 a parameter-free long-range charge operator, there is one short-range
 operator that we fitted to the deuteron charge form factor. The
 constructed charge operator inherits automatically unitary ambiguities of the
 nuclear force. Thus, a nontrivial test of the consistency is a test
 if the deuteron form factor is independent of the chosen unitary ambiguity in nuclear
 forces and currents. This consistency check was performed and we were
 able to describe the deuteron form factor with a quantified
 truncation error analysis. The Bayesian approach was used to give a
 statistical interpretation of truncation errors. It was demonstrated that working with
 consistently regularized forces and currents allows for very precise
extraction of the RMS radius of the neutron out of the deuteron
radius. At this level of precision isospin-breaking effects in nuclear forces
play a significant role. So far they have been taken into account only
partly, in the same way as was done in most phenomenological
potentials. Chiral EFT, however, allows for more precise treatment. In
particular, contributions of the
leading and subleading isospin-breaking two-pion-exchange-potential,
irreducible pion-photon exchange and charge-dependence of pion-nucleon coupling constant
needs to be accounted for in a systematic treatment within chiral
EFT. The construction of this force is by now finished. So it will be
very interesting to
repeat the calculation of the deuteron form factor with the consistently
regularized current operator and the state of the art chiral nuclear
force. This is work in progress.

\section*{Acknowledgement}
I would like to express my thanks to my collaborators Evgeny Epelbaum and
Ulf-G. Mei{\ss}ner as well as Arseniy Filin, Vadim Baru, Patrick
Reinert and Daniel M\"oller for sharing their insight on the discussed
topics. This work is supported by DFG (CRC110, ``Symmetries
and the Emergence of Structure in QCD'').
\begin{appendix}
\section{External sources in chiral EFT}
\label{ext_sources_app}
In this appendix, we briefly review the role of external sources in
chiral perturbation theory. External sources played an essential role
in the original formulation of chiral perturbation theory  by Gasser
and Leutwyler~\cite{Gasser:1983yg}. They allow for a
systematic inclusion of chiral symmetry breaking effects, they provide
a tool for generating all possible Ward-identities, and they can be
used to define nuclear current operators. Let us briefly follow the
discussion of~\cite{Gasser:1983yg} for two-flavour case. We start with QCD Lagrangian in the
chiral limit where up and down quark masses are set to zero,
\beqa
{\cal L}^{0}_{\rm QCD}&=&\sum_{\textrm{ quark flavors}}\bar{q}i
\gamma^\mu\left(\partial_\mu+i g 
\frac{\lambda_a}{2}G^{a}_{\mu}\right)q\nn
&-&\frac{1}{4}G^a_{\mu\nu}G^{a\mu\nu},
\eeqa
with the gluon-fields $G^a_\mu$ and 
\beqa
G^a_{\mu\nu}&=&\partial_\mu G^a_\nu-\partial_\nu G^a_\mu
-g f^{a b c}G^{b}_\mu G^{c}_\nu
\eeqa
the corresponding field strength tensor. 
$\lambda^a$ are $3\times 3$ 
hermitian traceless Gell-Mann matrices in color space, 
which satisfy the relations
\beqa
\textrm{ Tr}\left(\lambda_a\lambda_b\right)&=&2\delta_{a b},\\
\left\lbrack \frac{\lambda_a}{2},\frac{\lambda_b}{2}\right\rbrack&=&
i f^{a b c}\frac{\lambda_c}{2},
\eeqa
with $f^{a b c}$ the totally antisymmetric ${\rm SU}(3)$ structure
constants. It is easy to show that ${\cal L}^{0}_{QCD}$ is invariant
under the global chiral ${\rm SU}(2)_L\times {\rm SU}(2)_R$
transformation which can be decomposed in vector and axial
transformations. The invariance of ${\cal L}^{0}_{QCD}$ under vector
transformation is equivalent to isospin-symmetry. The invariance under
axial transformation is not shared by the ground state. So the axial
symmetry of QCD is spontaneously broken. As a consequence there
exist three (number of broken generators) massless Goldstone bosons which one
identifies with pions. Due to their Goldstone boson nature, the interaction
between pions vanishes if their four-momenta vanish. This gives us an
expansion parameter in low energy sector given by low momenta of pions
divided by chiral symmetry breaking scale $\Lambda_\chi$. In reality, quark masses
are not equal to zero but are much smaller than
$\Lambda_\chi$. A massive version of QCD violates chiral
symmetry. However, since the quark masses are small one can use
perturbation theory for the systematic inclusion of chiral symmetry
breaking effects. Additional to quark-gluon interaction, quarks also
interact with photons and $W^\pm, Z^0$ bosons. A convenient method for
the 
description of these interactions and incorporation of quark masses is the introduction of axial-vector,
vector, pseudoscalar, and scalar sources. One extends massless QCD to
\beqa
{\cal L}_{QCD}&=&{\cal L}^0_{QCD}+\bar{q}\gamma^\mu(v_\mu+\gamma_5 a_\mu)q
-\bar{q}(s-i\gamma_5 p)q,\quad\quad
\eeqa
where the external fields $v_\mu(x), a_\mu(x), s(x), p(x)$ are hermitian, 
color-neutral matrices in flavor space. The quark mass matrix is included in
$s(x)$.
If we require that the external fields transform under a local
chiral transformation as
\beqa
\label{CurrTransf}
v_\mu^\prime+a_\mu^\prime&=&R(v_\mu+a_\mu)R^\dagger 
+ i R\partial_\mu R^\dagger,\nonumber\\
v_\mu^\prime-a_\mu^\prime&=&L(v_\mu-a_\mu)L^\dagger 
+i L\partial_\mu L^\dagger,\nonumber\\
s^\prime+i p^\prime&=&R (s + i p)L^\dagger,
\eeqa
then ${\cal L}_{QCD}$ is invariant under a local chiral
transformation, and the generating 
functional 
\beqa
\exp(i Z\lbrack v,a,s,p\rbrack)&=&
\int \lbrack d G\rbrack\lbrack d q\rbrack\lbrack d \bar{q}\rbrack
\exp\left(i\int d^4x {\cal L}_{QCD}\right)\nonumber
\eeqa
is invariant under the
transformation given in Eq.~(\ref{CurrTransf}): 
\beqa
Z\lbrack v^\prime, a^\prime, s^\prime, p^\prime\rbrack&=&
Z\lbrack v,a,s,p\rbrack.\label{generating_qcd_functional}
\eeqa
Eq.~(\ref{generating_qcd_functional}) can be used as a master equation
for generating all possible Ward identities of QCD. In the absence of anomalies, the Ward identities are equivalent to the
statement that the generating functional is invariant under the gauge
transformation of the external fields~\cite{Leutwyler:1993iq}. In order that the
Green functions of the effective field theory
obey the Ward-Identities of QCD
it is sufficient to construct an effective generating functional 
in the presence of the same external fields such that it is invariant
under the same gauge transformation:
\begin{eqnarray}
Z_{\rm eff}\lbrack v^\prime, a^\prime, s^\prime, p^\prime\rbrack&=&
Z_{\rm eff}\lbrack v,a,s,p\rbrack.\label{generating_eff_functional}
\end{eqnarray}
The generating functional $Z_{\rm eff}$ is built out of the point-like
pion and nucleon degrees of freedom. A path-integral form of the
effective generating functional is given by
\beqa
\exp(i Z_{\rm eff}\lbrack v,a,s,p\rbrack)&=&
\int \lbrack d N\rbrack \lbrack d \bar{N}\rbrack\lbrack d U\rbrack
\exp\left(i\int d^4x {\cal L}_{\rm eff}\right),\nonumber
\eeqa
where $N$ and $U$ denote nucleon and pion fields, respectively. $U$
field is parametrized in form of a unitary matrix with $\det U=1$.
The effective Lagrangian ${\cal L}_{\rm eff}$ in
$Z_{\rm eff}$ is the most general Lagrangian compatible with Eq.~(\ref{generating_eff_functional}).
To achieve this, one replaces the usual derivative of pion fields by the covariant one:
\begin{eqnarray}
\partial_\mu U&\rightarrow& \nabla_\mu U\,=\,\partial_\mu U-i(v_\mu+a_\mu)U
+i U(v_\mu-a_\mu).\quad\quad
\end{eqnarray}
Under local chiral transformations the above term transforms as
\begin{eqnarray}
\nabla_\mu U&\rightarrow& R (\nabla_\mu U) L^\dagger.
\end{eqnarray}
The scalar and pseudoscalar external fields can be embodied in the field
\begin{eqnarray}
\chi&=&2B(s+i p),
\end{eqnarray} 
where the constant $B$ is related to the non vanishing quark condensate by
\begin{eqnarray}
\langle 0| \bar{u}u|0\rangle&=&\langle 0| \bar{d}d|0\rangle\,=\,-F_\pi^2 B +
{\cal O}(m_u,m_d).
\end{eqnarray}
The lowest order contribution to ${\cal L}_{\rm eff}$ with only pionic degrees of freedom which is invariant under a local chiral 
transformation is then given by
\begin{eqnarray}
\label{EffLag2}
{\cal L}_{\pi\pi}^{(2)}&=&\frac{F_\pi^2}{4}\textrm{ Tr}\left(
\nabla_\mu U (\nabla_\mu U)^\dagger+\chi U^\dagger+U\chi^\dagger\right).
\end{eqnarray}
The only requirement for the pion field $U$ is that it has to be a
unitary matrix with $\det U=1$. 
Their explicit parametrization affects only off-shell objects but does
not affect observables. We see that the introduction of external sources
can be used as a tool for the systematic construction of chiral
perturbation theory where explicit chiral symmetry breaking (due to
the non-vanishing quark mass) as well as
electro-weak couplings are taken into account. In the same framework,
it is natural to use external sources to derive nuclear currents where
nuclei are probed by photon or $W^\pm,Z^0$ exchange. For this purpose,
one can derive effective Hamiltonian out of the effective Lagrangian. The
effective Hamiltonian which we get in this way depends on external
sources and is time-dependent. Explicit expressions for the effective
Hamiltonian can be found in~\cite{Kolling:2009iq,Krebs:2016rqz,Baroni:2015uza}. Nuclear currents are derived out of
the quantized effective Hamiltonian by taking functional derivatives
in external sources.
%%%%%%%%%%%%%%%%%%%%%%%%%%%%%%%%%%%%%%%%%%%%%%%%%%%%%%%%%%%%%%%%%%%%%%%
\section{Continuity equation}
\label{app:continuity_equation}
Here we derive a continuity equation which is a direct consequence of
chiral or U$(1)_V$ symmetry. The effective Hamiltonian in
the presence of external sources can be written in the form
\beqa
&&H_{\rm eff}[s,p,a,v]=H_{\rm eff}+\int d^3 x\Big[
S(\vec{x})f_{s}\left(i\frac{\partial}{\partial
  t}\right)(s(x)-m_q) \nn
&&+ 
\fet{P}(\vec{x})f_{p}\left(i\frac{\partial}{\partial
  t}\right)\cdot \fet{p}(x)+ 
\fet{A}_\mu(\vec{x})f_{a}\left(i\frac{\partial}{\partial
  t}\right)\cdot \fet{a}^\mu(x)\nn 
&&+ 
\fet{V}_\mu(\vec{x})f_{v}\left(i\frac{\partial}{\partial
  t}\right)\cdot \fet{v}^\mu(x)\Big] + {\cal O}({\rm source}^2),
\eeqa  
where ``source'' denotes one of the sources $s,p,a$ or $v$. The
sources depend on a four-vector $x=(t,\fet{x})$. Note that we allow
here a dependence of the Hamiltonian not only on the sources but also
on arbitrary many time-derivatives of the sources which we denote by
functions $f_s, f_p,f_a$, and $f_v$. Under an infinitesimal local chiral
transformation, the sources transform via
\beqa
{\fet v}_\mu&\to& {\fet v}_\mu^\prime = {\fet v}_\mu + {\fet
  v}_\mu\times{\fet \epsilon}_V + {\fet a}_\mu\times{\fet \epsilon}_A
+ \partial_\mu{\fet \epsilon}_V,\nn
{\fet a}_\mu&\to&{\fet a}_\mu^\prime ={\fet a}_\mu + {\fet
  a}_\mu\times{\fet \epsilon}_V + {\fet v}_\mu\times{\fet \epsilon}_A
+\partial_\mu{\fet \epsilon}_A,\nn
s_0&\to&s_0^\prime=s_0 - {\fet p}\cdot{\fet \epsilon}_A, \nn
{\fet s}&\to&{\fet s}^\prime={\fet s} + {\fet s}\times{\fet \epsilon}_V - p_0
{\fet \epsilon}_A,\nn
i\,p_0&\to&i\,p_0^\prime = i (p_0 +{\fet s}\cdot{\fet \epsilon}_A),\nn
i\,{\fet p}&\to&i\,{\fet p}^\prime = i({\fet p}+{\fet p}\times{\fet
  \epsilon}_V+ s_0\,{\fet \epsilon}_A).
\eeqa
Due to the chiral symmetry, there exists a unitary transformation $U(t)$ such that
\beqa
H_{\rm eff}[s^\prime,p^\prime,a^\prime,v^\prime]&=&U(t)^\dagger H_{\rm
  eff}[s,p,a,v] U(t)\nn
&+& \left(i\,\frac{\partial}{\partial t}U(t)^\dagger\right)U(t).
\eeqa
We make an ansatz
\beqa
U(t)&=&\exp\Bigg(i\int d^3x\Big[{\fet R}^v(\vec{x},i\partial/\partial
    t)\cdot{\fet \epsilon}_V(\vec{x},t)\nn
&+&{\fet R}^a(\vec{x},i\partial/\partial
    t)\cdot{\fet \epsilon}_A(\vec{x},t)\Big]\Bigg) 
\eeqa
We concentrate now on the vector part and set $\vec{\epsilon}_A=0$. The
derivation of the continuity equation for the axial-vector follows the
same path. Keeping only linear terms in ${\fet\epsilon}_V$ and setting
all sources to zero, besides $s_0$ which is set to $s_0=m_q$, we get
\beqa
&&\int d^3x\Big[{\fet V}_\mu(\vec{x}) f_{v}\left(i\frac{\partial}{\partial
  t}\right)\partial^\mu+i\Big[{\fet R}^v(\vec{x},i\partial/\partial
    t),H_{\rm eff}\Big]\nn
&&-{\fet R}^v(\vec{x},i\partial/\partial
    t)\frac{\partial}{\partial t}\Big]\cdot{\fet\epsilon}_V(\vec{x},t)=0,
\label{vector_constraint_cont_eq}
\eeqa
for arbitrary ${\fet \epsilon}_V(\vec{x},t)$. We can solve this
equation perturbatively in $i\,\partial/\partial t$ writing
\beqa
{\fet R}^v(\vec{x},i\partial/\partial
    t)&=&\sum_{n=0}^\infty
    \vec{R}_n^v(\vec{x})
\left(i\frac{\partial}{\partial
          t}\right)^n
,\nn
f_v\left(i\partial/\partial
    t\right)&=&\sum_{n=0}^\infty
   f_v^n\left(i\frac{\partial}{\partial
          t}\right)^n.
\eeqa
With this ansatz we can rewrite Eq.~(\ref{vector_constraint_cont_eq})
into a series of equations
\beqa
\int d^3 x &&\Big[-i\,{\fet V}_0(\vec{x})f_v^{n-1} -\partial^j {\fet V}_j(\vec{x}) f_v^n+i\Big[{\fet
  R}_n^v(\vec{x}),H_{\rm eff}\Big]\nn
&&+i\, {\fet
  R}_{n-1}^v(\vec{x})\Big]\cdot \left(i\frac{\partial}{\partial
    t}\right)^n{\fet \epsilon}_V(\vec{x},t)=0,
\label{recursive_cont_eq}
\eeqa
with $f_v^{-1}={\fet
  R}_{-1}^v(\vec{x})=0$. Eq.~(\ref{recursive_cont_eq}) gives a recursive
definition of $ {\fet
  R}_{n-1}^v(\vec{x})$ operator. The vector current operator in
momentum space is given by
\beqa
\tilde{\fet V}_\mu(\vec{k},k_0)&=&
\int d^4x {\vec V}_\mu(\vec{x}) f_{v}\left(i\frac{\partial}{\partial
  t}\right) \exp(-i \,k\cdot x) \nn
&=&\tilde{\vec V}_\mu(\vec{k}) f_{v}\left(k_0\right).
\eeqa
Rewritten in momentum space
Eq.~(\ref{recursive_cont_eq}) is given by
\beqa
&&\frac{1}{n!}\frac{\partial^n}{\partial
  k_0^n}\Bigg|_{k_0=0}k^\mu \tilde{\fet
    V}_\mu(k)+\Big[H_{\rm eff}, \tilde{\fet R}_n^v(\vec{k})\Big]-
  \tilde{\fet R}_{n-1}^v(\vec{k})=0,  \nonumber
\eeqa
One can also write this in a following form
\beqa
&&k^\mu \tilde{\fet
    V}_\mu(k)|_{k_0=0}+\Big[H_{\rm
    eff},\tilde{\fet R}_1^v(\vec{k})\Big]=0,\nn
&&\tilde{\fet R}_1^v(\vec{k})=\frac{\partial}{\partial
  k_0}\Bigg|_{k_0=0} k^\mu \tilde{\fet
    V}_\mu(k)+\Big[H_{\rm
    eff}, \tilde{\fet R}_2^v(\vec{k})\Big],\nn
&&\tilde{\fet R}_2^v(\vec{k})=\frac{1}{2!}\frac{\partial^2}{\partial
  k_0^2}\Bigg|_{k_0=0} k^\mu \tilde{\fet
    V}_\mu(k)+\Big[H_{\rm
    eff},\tilde{\fet R}_3^v(\vec{k})\Big],\nonumber
\eeqa
and so on. So altogether we get
\beqa
\sum_{n=0}^\infty\frac{1}{n!}\Big[H_{\rm eff}, \frac{\partial^n}{\partial
  k_0^n}\Bigg|_{k_0=0}k^\mu \tilde{\fet
    V}_\mu(k)\Big]_n=0,\label{continuity_equation_written_out}
\eeqa
where we used a definition of $n$-th commutator
\beqa
\big[A,B\big]_n=\big[A,\big[A,B\big]_{n-1}\big], {\rm and}\,\,\big[A,B]_0=B.\label{nth_commutator_definition}
\eeqa
Using the Baker-Campbell-Hausdorff formula
\beqa
\exp(A)B\exp(-A)=\sum_{n=0}^\infty \frac{1}{n!}[A,B]_n,
\eeqa
we can write the continuity
equation~(\ref{continuity_equation_written_out}) in a more compact form
\beqa
&&\exp\left(H_{\rm eff}\frac{\overrightarrow{\partial}}{\partial k_0}\right)k^\mu \tilde{\fet
    V}_\mu(k)\exp\left(-H_{\rm eff}\frac{\overleftarrow{\partial}}{\partial k_0}\right)\Bigg|_{k_0=0}=0.\nn
&&\label{continuity_vector}
\eeqa
In a similar way one can derive the continuity equation for the 
axial-vector current
\beqa
&&\exp\left(H_{\rm eff}\frac{\overrightarrow{\partial}}{\partial k_0}\right)\Big[k^\mu \tilde{\fet
    A}_\mu(k)+i\,m_q \tilde{\fet P}(k)\Big]\nn
&&\times\exp\left(-H_{\rm eff}\frac{\overleftarrow{\partial}}{\partial
    k_0}\right)\Bigg|_{k_0=0}=0.
\label{continuity_axial}
\eeqa
Note that without exponential operators
Eqs.~(\ref{continuity_vector}) and (\ref{continuity_axial}) reduce to the on-shell continuity equations. So 
the exponential operators in Eqs.~(\ref{continuity_vector}) and
(\ref{continuity_axial}) seem to switch on the on-shell condition. In
order to
prove that this is indeed the case we put
Eqs.~(\ref{continuity_vector}) between 
initial and final states which are the eigenstates of the nuclear force
\beqa
&& H_{\rm eff}|i\rangle\,=\,E_i|i\rangle, \quad H_{\rm eff}|f\rangle\,=\,E_f|f\rangle.\label{eigenstate_rel}
\eeqa
Using Eq.~(\ref{eigenstate_rel}) we get
\beqa
&&\exp\left(E_f\frac{\overrightarrow{\partial}}{\partial k_0}\right) \langle f| k^\mu \tilde{\fet
    V}_\mu(k)|i\rangle\exp\left(-E_i\frac{\overleftarrow{\partial}}{\partial
      k_0}\right)\Bigg|_{k_0=0}=0\nn
&&\label{continuity_eq_on_shell}
\eeqa
Since the exponential operators in Eq.~(\ref{continuity_eq_on_shell})
are just the translation operators with the property
\beqa
&&\exp\left(E_f\frac{\overrightarrow{\partial}}{\partial k_0}\right)F(k_0)\,=\,F(k_0+E_f),
\eeqa
for any smooth function $F(k_0)$. Applying exponential operators from the
left and right hand sides we get
\beqa
&&\exp\left(E_f\frac{\overrightarrow{\partial}}{\partial k_0}\right)F(k_0)\exp\left(-E_i\frac{\overleftarrow{\partial}}{\partial
      k_0}\right)\Bigg|_{k_0=0}\nn
&&=\,F(k_0+E_f-E_i) \Bigg|_{k_0=0}\,=\, F(k_0) \Bigg|_{k_0=E_f-E_i}.
\eeqa
Applying this result to the continuity equation for the vector current we get
\beqa
&&\langle f| k^\mu \tilde{\fet
    V}_\mu(k)|i\rangle\Bigg|_{k_0=E_f-E_i}=0.\label{continuity_eq_on_shell_condition}
\eeqa
But the condition on the left-hand side of
Eq.~(\ref{continuity_eq_on_shell_condition}) is just the on-shell
condition. So we see that the exponential operators in
Eqs.~(\ref{continuity_vector}) and (\ref{continuity_axial}) do just
switch on the on-shell condition. Note that for the energy-transfer
independent currents Eqs.~(\ref{continuity_vector}) and
(\ref{continuity_axial}) reduce to the ordinary continuity equations 
\beqa
\Big[H_{\rm eff}, \tilde{\fet
    V}_0({\vec{k}})\Big]&=&\vec{k}\cdot \tilde{\fet
    V}(\vec{k}), \nn
\Big[H_{\rm eff}, \tilde{\fet
    A}_0(\vec{k})\Big]&=&\vec{k}\cdot \tilde{\fet
    A}({\vec{k}})-i\,m_q \tilde{\fet P}(\vec{k}) , 
\eeqa
For the linear dependence on the energy-transfer
Eqs.~(\ref{continuity_vector}) and (\ref{continuity_axial}) reduce to
Eq.~(\ref{continuity_va}). For the quadratic and the higher-order dependence
on the energy-transfer Eqs.~(\ref{continuity_vector}) and
(\ref{continuity_axial}) produce an increasing number of commutators
which, however, is always finite, as long as the currents are
polynomials of the energy transfer.
%%%%%%%%%%%%%%%%%%%%%%%%%%%%%%%%%%%%%%%%%%%%%%%%%%%%%%%%%%%%%%%%%%%%%%%%%%%%%%%%%%%% 
\section{Unitary transformations}
\label{unitary_transf_app}
In this appendix we review all unitary transformations which are needed
for derivation of renormalizable nuclear forces and currents. As already
explained in section~\ref{sec:nuclearcurrent}, the unitary
transformations which bring nuclear Hamiltonian into a block-diagonal
renormalizable form are time-independent. The renormalizability of the
chiral nuclear force constrains strongly the choice of the unitary
phases and leaves only two phases $\bar{\beta}_8$ and $\bar{\beta}_9$
unfixed. These unitary transformations can be
parametrized via
\beqa
U^{\eta}&=&\exp\left(\sum_{i=1}^6\alpha_i S_i\right),
\eeqa
where $S_i$ are antihermitian operators. Their explicit form as well
as the values of the phases $\alpha_i$ can be
found in~\cite{Epelbaum:2007us}. These transformations act on
$\eta$-space only, and are applied to the
effective Hamiltonian on top of Okubo transformations of
Eq.~(\ref{U_expressed_in_A}). $U^\eta$ are needed to make nuclear forces
renormalizable. 

Additionally to these transformations
there are further two time-independent transformations which
contribute to relativistic corrections:
\beqa
U^{\eta,1/m}&=&\exp\left(\bar{\beta}_8 S_8+\bar{\beta}_9 S_9\right),
\eeqa
where antihermitian operators $S_8$ and $S_9$ are given by~\cite{Kolling:2011mt,Krebs:2019aka}
\beqa
S_8&=&\eta \tilde H_{20}^{(2)}\eta
H_{21}^{(1)}\frac{\lambda^1}{E_\pi^3}H_{21}^{(1)} - h.c.,\nn
S_9&=&\eta \tilde H_{21}^{(3)}\frac{\lambda^1}{E_\pi^2}H_{21}^{(1)} - h.c..
\eeqa
The operators $H_{a b}^{(\kappa)}$ refer to the vertices in the effective
chiral Hamiltonian with $a$ nucleon and $b$ pion fields. $\kappa$
denotes the inverse mass dimension of the coupling constant in the
given vertex, see
Eq.~(\ref{kappa_inverse_mass_dimesion_of_coupling}). The operators
$\tilde H_{a b}^{(\kappa)}$ refer to the $1/m$-correction of the
corresponding vertices. The phases $\bar{\beta}_8$ and $\bar{\beta}_9$
are known to be responsible for the degree of non-locality of the
relativistic $1/m^2$-correction to the one-pion exchange. Minimal
nonlocality is achieved for the values in
Eq.~(\ref{minimal_nonlocality_off_shell}).

There are various time-dependent unitary transformations which depend
on external sources. Those are extensively discussed
in~\cite{Kolling:2011mt,Krebs:2019aka} for vector currents and
in~\cite{Krebs:2016rqz} for axial-vector and pseudoscalar
currents. Most of the phases of these unitary transformations are either
fixed by renormalizability and matching to nuclear forces
requirements or they do not affect the final
expressions of the current operators.  
%%%%%%%%%%%%%%%%%%%%%%%%%%%%%%%%%%%%%%%%%%%%%%%%%%%%%%%%%%%%%%%%%%%%%%%%%%%%%%%%%%%%
\section{Vector Current: Two-Pion-Exchange}
\label{two_pion_excange_vector_current_app}
In this appendix, we give two-pion-exchange contributions to the vector
current operator. Due to the coupling of the scalar source to two
pions there appear loop functions which depend on three momenta
$\vec{k}, \vec{q_1}$ and $\vec{q}_2$ which are momentum transfer of
the vector source, momentum transfer of the first and second
nucleons, respectively. This leads to lengthy expressions which
have been derived in \cite{Kolling:2009iq} and are
listed here for completeness:
\beqa
  \vec{J}  &=&  \sum_{i=1}^5\sum_{j=1}^{24} f_i^j\left(\vec{q}_1,
    \vec{q}_2\right) \, T_i \vec{O}_j, \label{Jmom_def}\\
J^0&=& \sum_{i=1}^5\sum_{j=1}^{8} f_i^{jS}\left(\vec{q}_1,
    \vec{q}_2\right) \, T_i O_j^S \,, \label{J0mom_def}
\eeqa
where $f_i^j\equiv f_i^j\left(\vec{q}_1 , \vec{q}_2 \right)$ are scalar
functions and the spin-momentum operators $\vec{O}_i$ and $O^S_i$ are given by  
\beqa
  \vec{O}_1    & = & \vec{q}_1+\vec{q}_2,\nn
  \vec{O}_2    & = & \vec{q}_1-\vec{q}_2,\nn
  \vec{O}_3    & = & \cp{q}{\sigma}{1}{2}+\cp{q}{\sigma}{2}{1},\nn
  \vec{O}_4    & = & \cp{q}{\sigma}{1}{2}-\cp{q}{\sigma}{2}{1},\nn
  \vec{O}_5    & = & \cp{q}{\sigma}{1}{1}+\cp{q}{\sigma}{2}{2},\nn
  \vec{O}_6    & = & \cp{q}{\sigma}{1}{1}-\cp{q}{\sigma}{2}{2},\nn
  \vec{O}_7    & = & \vec{q}_1 \spr{q}{q}{\sigma}{1}{2}{2}+\vec{q}_2 \spr{q}{q}{\sigma}{1}{2}{1},\nn
  \vec{O}_8    & = & \vec{q}_1 \spr{q}{q}{\sigma}{1}{2}{2}-\vec{q}_2
  \spr{q}{q}{\sigma}{1}{2}{1},\nn
  \vec{O}_9    & = & \vec{q}_2 \spr{q}{q}{\sigma}{1}{2}{2}+\vec{q}_1 \spr{q}{q}{\sigma}{1}{2}{1},\nn
  \vec{O}_{10} & = & \vec{q}_2 \spr{q}{q}{\sigma}{1}{2}{2}-\vec{q}_1
  \spr{q}{q}{\sigma}{1}{2}{1},\nonumber
\eeqa
\beqa
  \vec{O}_{11} & = & \left(\vec{q}_1+\vec{q}_2 \right)\vek{\sigma}{\sigma}{1}{2},\\
  \vec{O}_{12} & = & \left(\vec{q}_1-\vec{q}_2 \right)\vek{\sigma}{\sigma}{1}{2},\nn
  \vec{O}_{13} & = & \vec{q}_1\vek{q}{\sigma}{1}{1}\vek{q}{\sigma}{1}{2}+
  \vec{q}_2\vek{q}{\sigma}{2}{1}\vek{q}{\sigma}{2}{2},\nn
  \vec{O}_{14} & = & \vec{q}_1\vek{q}{\sigma}{1}{1}\vek{q}{\sigma}{1}{2}-
  \vec{q}_2\vek{q}{\sigma}{2}{1}\vek{q}{\sigma}{2}{2},\nn
  \vec{O}_{15} & = & \left(\vec{q}_1+\vec{q}_2 \right)\vek{q}{\sigma}{2}{1}\vek{q}{\sigma}{1}{2},\nn
  \vec{O}_{16} & = & \left(\vec{q}_1-\vec{q}_2 \right)\vek{q}{\sigma}{2}{1}\vek{q}{\sigma}{1}{2},\nn
  \vec{O}_{17} & = & \left(\vec{q}_1+\vec{q}_2 \right)\vek{q}{\sigma}{1}{1}\vek{q}{\sigma}{2}{2},\nn
  \vec{O}_{18} & = & \left(\vec{q}_1-\vec{q}_2 \right)\vek{q}{\sigma}{1}{1}\vek{q}{\sigma}{2}{2},\nn
  \vec{O}_{19} & = & \vec{\sigma}_1 \vek{q}{\sigma}{1}{2}+\vec{\sigma}_2\vek{q}{\sigma}{2}{1},\nn
  \vec{O}_{20} & = & \vec{\sigma}_1 \vek{q}{\sigma}{1}{2}-\vec{\sigma}_2\vek{q}{\sigma}{2}{1},\nn
  \vec{O}_{21} & = & \vec{\sigma}_1 \vek{q}{\sigma}{2}{2}+\vec{\sigma}_2\vek{q}{\sigma}{1}{1},\nn
  \vec{O}_{22} & = & \vec{\sigma}_1 \vek{q}{\sigma}{2}{2}-\vec{\sigma}_2\vek{q}{\sigma}{1}{1},\nn
  \vec{O}_{23} & = & \vec{q}_1\vek{q}{\sigma}{2}{1}\vek{q}{\sigma}{2}{2} +
  \vec{q}_2\vek{q}{\sigma}{1}{1}\vek{q}{\sigma}{1}{2},\nn
  \vec{O}_{24} & = & \vec{q}_1\vek{q}{\sigma}{2}{1}\vek{q}{\sigma}{2}{2} -
  \vec{q}_2\vek{q}{\sigma}{1}{1}\vek{q}{\sigma}{1}{2},\nonumber
\end{eqnarray}
and
\begin{eqnarray}
  O_1^S    & = & \one ,\nn
  O_2^S    & = & \vec q_1 \cdot [ \vec q_2 \times \vec \sigma_2 ] + \vec q_1
  \cdot [ \vec q_2 \times \vec \sigma_1 ]\,, \nn
    O_3^S    & = & \vec q_1 \cdot [ \vec q_2 \times \vec \sigma_2 ] - \vec q_1
  \cdot [ \vec q_2 \times \vec \sigma_1 ]\,, \nn
  O_{4}^S & = & \vec \sigma_1 \cdot \vec \sigma_2 \,, \nn
  O_{5}^S & = & \vek{q}{\sigma}{1}{2}\vek{q}{\sigma}{2}{1},\nn
  O_{6}^S & = & \vek{q}{\sigma}{1}{1}\vek{q}{\sigma}{2}{2},\nn
  O_{7}^S & = &
  \vek{q}{\sigma}{2}{1}\vek{q}{\sigma}{2}{2}+\vek{q}{\sigma}{1}{1}\vek{q}{\sigma}{1}{2},\nn
  O_{8}^S & = &
  \vek{q}{\sigma}{2}{1}\vek{q}{\sigma}{2}{2}-\vek{q}{\sigma}{1}{1}\vek{q}{\sigma}{1}{2}.
\end{eqnarray}
As a basis for the isospin operators we choose
\begin{eqnarray}
  T_1 & = & [{\fet\tau}_1]^3 + [{\fet\tau}_2]^3,\nn 
  T_2 & = & [{\fet\tau}_1]^3 - [{\fet\tau}_2]^3,\nn
  T_3 & = & \xp{\tau}{\tau}{1}{2},\nn
  T_4 & = & \vec \tau_1 \cdot \vec \tau_2 \,,\nn
  T_5 & = & \one.
\end{eqnarray}
The nonvanishing long-range contributions to the scalar functions $f_i^j\equiv
f_i^j\left(\vec{q}_1 , \vec{q}_2 \right)$ are given by 
\begin{eqnarray}
\label{fun_f}
  f_3^{1} & = & \frac{i e g_A^2 L(q_1)}{128 \pi ^2 F_\pi^4} \biggl[\frac{g_A^2 (8
      M_\pi^2 + 3 q_1^2)}{4 M_\pi^2 + q_1^2} - 1\biggr]\nn
&& + \frac{  e \pi}{F_\pi^4}
\biggl[
     g_A^4 M_\pi^4 \II{d + 2}{2}{1}{2}  + 4 \pi  g_A^4 M_\pi^2 q_1^2
     \II{d + 4}{2}{2}{2} \nn
&&- 8 \pi  g_A^4 M_\pi^2 q_1 \II{d + 4}{3}{1}{2}
(q_1 - q_2
    z) - 96 \pi ^2 g_A^4 q_1^3 q_2 z \II{d + 6}{4}{1}{2}\nn
&& + 32 \pi ^2 g_A^4 q_1^2
    q_2 \II{d + 6}{3}{2}{2} (q_1 z + q_2 z^2 + q_2)\nn
&& - 2 \pi  g_A^4 q_1
    \II{d + 4}{2}{1}{2} (q_1 + 2 q_2 z) - 2 (g_A^2 - 1) g_A^2 M_\pi^2 \nn
&&\times 
    \II{d + 2}{2}{1}{0} 
- 4 \pi  (g_A^2 - 1) g_A^2 q_1^2 \II{d + 4}{2}{2}{0} \nn
&&+ 8 \pi
    (g_A^2 - 1) g_A^2 q_1 \II{d + 4}{3}{1}{0} (q_1 - q_2 z)\nn
&&- 2 \pi
    (g_A^2 - 1)^2 \IIR{d + 4}{2}{1}{0}
  \biggr]
 -   (1 \leftrightarrow 2),
\eeqa
\beqa
  f_3^{2} & = & \frac{i e g_A^2 (g_A^2 + 1)}{256 \pi ^2
    F_\pi^4}
 + \frac{i e g_A^2 L(q_1) }{128 \pi ^2 F_\pi^4}\biggl[\frac{g_A^2 (8 M_\pi^2 + 3 q_1^2)}{4
    M_\pi^2 + q_1^2} - 1\biggr]\nn
&& + \frac{e}{8 F_\pi^4}
\biggl[ - 64 \pi ^2 g_A^4 q_1 \II{d + 4}{3}{1}{2}
(M_\pi^2 (q_1 - q_2 z) + q_1^2 q_2 z)\nn
&&
 + 16 \pi ^2 g_A^4 q_1
\II{d + 4}{2}{2}{2}  (q_1 q_2^2 (z^2 + 1) \nn
&&- 2 M_\pi^2 (q_1 - q_2 z)) - 2 \pi
g_A^4 \II{d + 2}{1}{1}{2}  (2 M_\pi^2 + q_1 q_2 z)  \nn
&&{}+ 8 \pi  g_A^4 M_\pi^2
\II{d + 2}{2}{1}{2}  (M_\pi^2 + q_1 (q_2 z - q_1)) \nn
&&- 768 \pi ^3 g_A^4 q_1^3
q_2 z \II{d + 6}{4}{1}{2}  - 256 \pi ^3 g_A^4 q_1^2 q_2 \II{d + 6}{3}{2}{2}\nn
&&\times(q_1 z - q_2 (z^2 + 1)) 
 + 16 \pi ^2 g_A^4 q_1 \II{d + 4}{2}{1}{2}  (q_1 - 2
q_2 z)\nn
&& + 2 \pi  (g_A^4 - 1) \IIR{d + 2}{1}{1}{0}  - 8 \pi  (g_A^2 - 1) g_A^2
\II{d + 2}{2}{1}{0} \nn
&&\times (2 M_\pi^2 + q_1 (q_2 z - q_1)) + 32 \pi ^2 (g_A^2 - 1)
g_A^2 q_1 \II{d + 4}{2}{2}{0} \nn
&&\times  (q_1 - q_2 z) + 64 \pi ^2 (g_A^2 - 1) g_A^2 q_1
\II{d + 4}{3}{1}{0}  (q_1 - q_2 z)\nn
&& - 16 \pi ^2 (g_A^2 - 1)^2
\IIR{d + 4}{2}{1}{0}  + g_A^4 M_\pi^4 \II{4}{1}{1}{2} \nn
&& - 2 (g_A^2 - 1) g_A^2
M_\pi^2 \II{4}{1}{1}{0} \biggr]
 +  (1 \leftrightarrow 2), 
\eeqa
\beqa
f_1^{3} & =& \frac{i e g_A^2 (g_A^2 + 1)}{256 \pi ^2 F_\pi^4}
 + \frac{i e g_A^2 L(q_1)}{128 \pi ^2 F_\pi^4} \nn
&&\times\biggl[\frac{g_A^2 (8
  M_\pi^2 + 3 q_1^2)}{4 M_\pi^2 + q_1^2} - 1\biggr] + (1
   \leftrightarrow 2),
\eeqa
\beqa
f_1^{4} & = & \frac{i e g_A^2 L(q_1)}{128 \pi ^2 F_\pi^4}
\biggl[\frac{g_A^2 (8 M_\pi^2 + 3 q_1^2)}{4 M_\pi^2 + q_1^2} - 1\biggr] -  (1
\leftrightarrow 2), \quad
\eeqa
\beqa
f_1^{5} & = &  - \frac{i e g_A^4}{128 \pi ^2 F_\pi^4}
 + \frac{i e g_A^4 L(k)}{32 \pi ^2 F_\pi^4}
 - \frac{i e g_A^4 L(q_1)}{64 \pi ^2 F_\pi^4}
 + \frac{  e g_A^2 \pi}{2 F_\pi^4}\nn
&&\times\biggl[g_A^2 ( - M_\pi^2) \II{d + 2}{1}{1}{2}
 + 4 \pi  g_A^2 q_1 \II{d + 4}{2}{1}{2}  (q_1 - q_2 z)\nn
&&{}
 + (g_A^2 - 1)
\IIR{d + 2}{1}{1}{0} \biggr]
 +  (1 \leftrightarrow 2), 
\eeqa
\beqa
f_1^{6} & = &  - \frac{i e g_A^4 L(q_1)}{64 \pi ^2 F_\pi^4}
 - \frac{2  e g_A^4 \pi ^2}{F_\pi^4} \II{d + 4}{2}{1}{2}  q_1 (q_1 +
              q_2 z)\nn
&& - 
(1 \leftrightarrow 2), 
\eeqa
\beqa
 f_1^{7} & = & \frac{  e g_A^2 \pi}{2 F_\pi^4}
\biggl[8 \pi  g_A^2 (q_1^2 - 2 M_\pi^2) \II{d + 4}{3}{1}{2}  - g_A^2
M_\pi^2 \II{d + 2}{2}{1}{2}\nn
&&  + 192 \pi ^2 g_A^2 q_1^2 \II{d + 6}{4}{1}{2}
 - 64 \pi ^2 g_A^2 q_1 q_2 z \II{d + 6}{3}{2}{2} \nn
&& + 8 \pi  g_A^2
\II{d + 4}{2}{1}{2} + (g_A^2 - 1) \II{d + 2}{2}{1}{0} \nn
&& + 16 \pi  (g_A^2 - 1)
\II{d + 4}{3}{1}{0} \biggr]
 -  (1 \leftrightarrow 2),
\eeqa
\beqa
f_1^{8} & = & \frac{i e g_A^4}{32 \pi ^2 F_\pi^4 k^2}
 - \frac{i e g_A^4 L(k)}{32 \pi ^2 F_\pi^4 k^2}
 + \frac{ e g_A^2\pi }{4 F_\pi^4}\nn
&&\times\biggl[16 \pi  g_A^2 (q_1^2 - 2 M_\pi^2) \II{d + 4}{3}{1}{2}  - 2 g_A^2
M_\pi^2 \II{d + 2}{2}{1}{2}\\
&&  + 384 \pi ^2 g_A^2 q_1^2 \II{d + 6}{4}{1}{2} 
 - 8
\pi  g_A^2 q_1 q_2 z \II{d + 4}{2}{2}{2}  \nn
&&- 128 \pi ^2 g_A^2 q_1 q_2 z
\II{d + 6}{3}{2}{2}  + g_A^2 \II{d + 2}{1}{1}{2}  + 16 \pi  g_A^2
\II{d + 4}{2}{1}{2}\nn
&&  + 2 (g_A^2 - 1) \II{d + 2}{2}{1}{0} 
 + 32 \pi
(g_A^2 - 1) \II{d + 4}{3}{1}{0} \biggr]
 +  (1 \leftrightarrow 2),\nonumber
\eeqa
\beqa
  f_1^{9} & =& \frac{ e g_A^2\pi }{2 F_\pi^4}
\biggl[g_A^2 M_\pi^2 \II{d + 2}{2}{1}{2}  - 8 \pi  g_A^2 q_1^2
\II{d + 4}{3}{1}{2} \nn
&& - 64 \pi ^2 g_A^2 q_1 \II{d + 6}{3}{2}{2}  (q_1 + q_2
z) + 4 \pi  g_A^2 \II{d + 4}{2}{1}{2}\nn
&&  + (1 - g_A^2) \II{d + 2}{2}{1}{0}
\biggr] -  (1 \leftrightarrow 2), 
\eeqa
\beqa
  f_1^{10}& = & \frac{i e g_A^4}{32 \pi ^2 F_\pi^4 k^2}
 - \frac{i e g_A^4 L(k)}{32 \pi ^2 F_\pi^4 k^2}
 - \frac{  e g_A^2\pi}{4 F_\pi^4}\nn
&&\times\biggl[ - 8 \pi  g_A^2 \II{d + 4}{2}{2}{2}  (2 M_\pi^2 + q_1 q_2 z) - 2 g_A^2
M_\pi^2 \II{d + 2}{2}{1}{2} \nn
&& + 16 \pi  g_A^2 q_1^2 \II{d + 4}{3}{1}{2} 
 + 128
\pi ^2 g_A^2 q_1 \II{d + 6}{3}{2}{2}  (q_1 - q_2 z) \nn
&&+ g_A^2
\II{d + 2}{1}{1}{2}  + 8 \pi  g_A^2 \II{d + 4}{2}{1}{2}  + 2 (g_A^2 - 1)
\II{d + 2}{2}{1}{0} \nn
&& + 16 \pi  (g_A^2 - 
1) \II{d + 4}{2}{2}{0} \biggr]
 +  (1 \leftrightarrow 2)\,.
\end{eqnarray}
These  two-pion exchange contributions have been calculated by
using dimensional regularization. 
In addition, there are nonvanishing  functions $f_2^j$ given by 
\beqa
  &&f_2^3   = f_1^4, \;\;\; f_2^4 = f_1^3, \;\;\; f_2^5 = -f_1^6, \;\;\;
  f_2^6 = - f_1^5, \nn
&&f_2^7 = f_1^8, \;\;\; f_2^8 = f_1^7, \;\;\; f_2^9 = f_1^{10}, \;\;\;
  f_2^{10} = f_1^9.
\eeqa
In the above equations, $z \equiv \hat{q}_1 \cdot\hat{q}_2$, $q_i \equiv \left|\vec{q_i}
\right|$ and the loop functions $L(q)$ and $A(q)$ are defined by
\begin{eqnarray}
\label{LundA}
  L\left( q \right) & = &
  \frac{1}{2}\frac{s}{q}\ln\left(\frac{s+q}{s-q}\right), \quad \textrm{with}
  \ s = \sqrt{q^2+4M_\pi^2} \,, \nn
A\left( q  \right) & = & \frac{1}{2q}\arctan \left( \frac{q}{2M_\pi}\right) \,.
\end{eqnarray}
Further, the functions $I$ correspond to the three-point functions via 
\beqa
  &&\II{d}{\nu_1}{\nu_2}{\nu_3}  \equiv I(d;0,1;q_1,\nu_1; -
  q_2,\nu_2; 0,\nu_3 )
\eeqa
with the notation for the momenta
\beqa
q_{i} &=& (0,\vec q_{i})
\eeqa
and a scalar one-loop three-point function in $d$ space-time dimensions
defined by
\begin{strip}
\beq
\label{def3point}
I(d;p_1,\nu_1;p_2,\nu_2;p_3,\nu_3;p_4,\nu_4)  = \mu^{4-d}\int \frac{d^d\ell}{\left(
    2\pi\right)^d} \frac{1}{[ \left(\ell+p_1\right)^2 - M_\pi^2 ]^{\nu_1} \,[ ( \ell +
    p_2)^2-M_\pi^2]^{\nu_2} [ ( \ell +
    p_3 )^2 -M_\pi^2 ]^{\nu_3} [ v \cdot \left(\ell + p_4 \right) ]^{\nu_4}} \,.
\eeq
\end{strip}
Note that in this definition higher power of covariant and
heavy-baryon propagators are allowed.
Here, all propagators are understood to have an infinitesimal  positive
imaginary part. Notice that here and in what follows, we will only need the functions $I$ for
four-momenta with vanishing zeroth component.
We further emphasize that all functions $\II{d+n}{\nu_1}{\nu_2}{\nu_3}$ which enter the above
equations except $\II{d+2}{1}{1}{0}$, $\II{d+4}{2}{1}{0}$ and
$\II{d+4}{1}{2}{0}$ are finite in dimensional regularization in the limit $d\rightarrow
4$. For these
functions, we define reduced functions by subtracting the poles in four dimensions
\begin{eqnarray}
  \IIR{d+2}{1}{1}{0} & = & \II{d+2}{1}{1}{0} - \frac{i}{4\pi}L(\mu) -
  \frac{i}{128\pi^3}\ln\left( \frac{M_\pi^2}{\mu^2}\right),\quad\\
  \IIR{d+4}{2}{1}{0} & = & \II{d+4}{2}{1}{0} + \frac{i}{48\pi^2}L(\mu) +
  \frac{i}{1536\pi^4}\ln\left( \frac{M_\pi^2}{\mu^2}\right),\nn
 \IIR{d+4}{1}{2}{0} & = & \II{d+4}{1}{2}{0} + \frac{i}{48\pi^2}L(\mu) +
  \frac{i}{1536\pi^4}\ln\left( \frac{M_\pi^2}{\mu^2}\right) \,,\nonumber
\end{eqnarray}
where 
\beq
L(\mu) =  \frac{\mu^{d-4}}{16 \pi^2}\left[ \frac{1}{d-4} +
\frac{1}{2}\left( \gamma_{\rm E} - 1 - \ln \left(4\pi \right)\right)\right].
\eeq
Here, $\mu$ is the scale introduced in dimensional regularization and
$\gamma_{\rm E} = - \Gamma^\prime (1) \simeq 0.577$.
Finally, for scalar functions contributing to the charge density we obtain the
following expressions:
\beqa
  f_3^{2S} & =& \frac{e g_A^4 A(k)}{64 \pi
    F_\pi^4}+\frac{i \pi ^2 e g_A^4}{F_\pi^4}\biggl[-2 M_\pi^2
  \II{d+4}{2}{1}{3} +16 \pi  q_1^2 \II{d+6}{3}{1}{3} \nn
&&-8 \pi  q_1 q_2 z
  \II{d+6}{2}{2}{3} +\II{d+4}{1}{1}{3} \biggr]+ (1 \leftrightarrow 2),
\eeqa
\beqa
  f_3^{3S} & = & \frac{2 i \pi ^2 e g_A^4}{F_\pi^4}\biggl[8
  \pi  q_1^2 \II{d+6}{3}{1}{3} -M_\pi^2 \II{d+4}{2}{1}{3} \biggr]- (1
  \leftrightarrow 2),\nn
&&
\eeqa
\beqa
  f_1^{1S} & = & \frac{e g_A^4 M_\pi (12 M_\pi^4+7 M_\pi^2
    q_1^2+q_1^2 q_2^2)}{64 \pi  F_\pi^4 (4 M_\pi^2+q_1^2) (4
    M_\pi^2+q_2^2)}\nn
&&-\frac{e g_A^4 A(k) (2 M_\pi^2+q_1^2)}{16 \pi  F_\pi^4}
+\frac{e g_A^4 A(q_1) (2 M_\pi^2+q_1^2)}{32 \pi  F_\pi^4}\nn
&&{}
+\frac{i   e g_A^4\pi}{F_\pi^4}
 \biggl[M_\pi^4 (-\II{d+2}{1}{1}{3} )+8 \pi  M_\pi^2 q_1
\II{d+4}{2}{1}{3}  (q_1-q_2 z)\nn
&&+64 \pi ^2 q_1^3 q_2 z \II{d+6}{3}{1}{3}
-16 \pi ^2 q_1^2 q_2^2 (z^2+1) \II{d+6}{2}{2}{3} \nn
&&+2 \pi  q_1 q_2 z
\II{d+4}{1}{1}{3} \biggr]
+ (1 \leftrightarrow 2), 
\eeqa
\beqa
  f_1^{4S} & = & -\frac{e g_A^4 q_1^2 A(q_1)}{64 \pi
    F_\pi^4}
-\frac{2 i  e g_A^4\pi ^2}{F_\pi^4}\biggl[8 \pi  q_1^2 q_2^2 (z^2-1)
\II{d+6}{2}{2}{3} \nn
&&-q_1 q_2 z \II{d+4}{1}{1}{3} \biggr]
+ (1 \leftrightarrow 2),
\eeqa
\beqa
  f_1^{5S} & = & \frac{2 i  e g_A^4\pi^2}{F_\pi^4}
\biggl[8 \pi  q_1 q_2 z \II{d+6}{2}{2}{3} -\II{d+4}{1}{1}{3} \biggr]+
(1 \leftrightarrow 2),\nn
&&
\eeqa
\beqa
  f_1^{6S}  &=&  \frac{16 i \pi ^3 e g_A^4}{F_\pi^4}
q_1 q_2 z \II{d+6}{2}{2}{3} + (1 \leftrightarrow 2),\nn
  f_1^{7S} & = & \frac{e g_A^4 A(q_1)}{128 \pi  F_\pi^4}
-\frac{16 i \pi ^3 e g_A^4}{F_\pi^4}
q_1^2 \II{d+6}{2}{2}{3} + (1 \leftrightarrow 2), \nn
  f_1^{8S} &=&  -\frac{e g_A^4 A(q_1)}{128 \pi  F_\pi^4}
-\frac{16 i \pi ^3 e g_A^4}{F_\pi^4}q_1^2 \II{d+6}{2}{2}{3} 
- (1 \leftrightarrow 2), \nn
  f_2^{1S} & = & \frac{e g_A^4 M_\pi^3 q_1^2}{64 \pi  F_\pi^4
    (4 M_\pi^2+q_1^2) (4 M_\pi^2+q_2^2)}\\
&&+\frac{e g_A^2 (g_A^2-1) A(q_1) (2 M_\pi^2+q_1^2)}{32 \pi  F_\pi^4}-
(1 \leftrightarrow 2),\nn
  f_2^{7S} & = & -\frac{e g_A^4 A(q_1)}{128 \pi  F_\pi^4}-
  (1 \leftrightarrow 2), \nn
 f_2^{8S} &=&  \frac{e g_A^4 A(q_1)}{128 \pi  F_\pi^4}+ (1
  \leftrightarrow 2).
\end{eqnarray}
Note that we use here an overcomplete basis for three-point
functions in $d+x$ space-time dimensions. One could reduce them and
express everything through just one three-point function. In this
case, however, we would produce a lot of unphysical singularities
which would cancel in the final result but would make expressions more
lengthy. 
%%%%%%%%%%%%%%%%%%%%%%%%%%%%%%%%%%%%%%%%%%%%%%%%%%%%%%%%%%%%%%%%%%%%%%%%%%%%%%%%%%%%%
\section{Scalar Current: Two-Pion-Exchange}
\label{app:scalar_curret:2pe}
In this appendix, we give two-pion-exchange contributions to the scalar
current operator. Due to the coupling of the scalar source to two
pions there appear loop functions which depend on the three momenta
$\vec{k}, \vec{q_1}$ and $\vec{q}_2$ which denote the momentum transfer of
the scalar source, the momentum transfer of the first and the second
nucleons, respectively. This leads to lengthy expressions which
have been derived in~\cite{Krebs:2020plh}
and are
listed here for completeness:
\beqa
\label{ScalarCurrent2pi}
S_{{\rm 2N:} \, 2\pi}^{(Q^0)}&=& {\fet \tau}_1\cdot{\fet \tau}_2
\big(\vec{q}_1\cdot\vec{\sigma}_1\vec{k}\cdot\vec{\sigma}_2
t_1 + t_2\big) \nn
&+&
\vec{q}_1\cdot\vec{\sigma}_1\vec{q}_2\cdot\vec{\sigma}_2 t_3 + 
\vec{q}_2\cdot\vec{\sigma}_1\vec{q}_1\cdot\vec{\sigma}_2
t_4 \nn
&+&\vec{q}_1\cdot\vec{\sigma}_1\vec{q}_1\cdot\vec{\sigma}_2 t_5
+\vec{\sigma}_1\cdot\vec{\sigma}_2 t_6+1
\leftrightarrow 2\,, 
\eeqa
The scalar functions $t_i$ are expressed in terms of the two and 
three-point functions. The notation for the three-point function is 
\beqa
&&I(d:p_1,\nu_1;p_2,\nu_2;p_3,\nu_3; 0,\nu_4)\,=\,\nn
&&\int
\frac{d^d l}{(2\pi)^d}\prod_{j=1}^3\frac{1}{[(l+p_j)^2-M_\pi^2+i \epsilon]^{\nu_j}}
\frac{1}{[v\cdot l+i \epsilon]^{\nu_4}}\,.
\quad\quad
\eeqa
For our purpose we need only 
\beqa
&&I(4:0,1;q_1,1;k,1; 0,0)\,=\,\nn
&&\frac{i}{16\pi^2} \int_0^1d t\int_0^t dy \frac{1}{C}\frac{1}{(y-y_1)(y-y_2)},
\eeqa
with
\beqa
y_1&=&\frac{D}{2C}+\sqrt{\frac{D^2+4A C}{4 C^2}},\;
y_2\,=\,\frac{D}{2C}-\sqrt{\frac{D^2+4A C}{4 C^2}},\nonumber
\eeqa
and $A\,=\,M^2+q_1^2(1-t)t$, $B\,=\,-2\vec{q}_1\cdot\vec{q}_2$,
$C\,=\,q_2^2$ and $D\,=\, 2\vec{q}_1\cdot\vec{q}_2+q_2^2 + t B$.
For $\vec{k}=0$, the three-point function reduces to a two-point
function
\beqa
I(4:0,1;q_1,1;k,1;
0,0)\big|_{\vec{k}=0}&=&-\frac{i}{8\pi^2}\frac{L(q_1)}{4 M_\pi^2+q_1^2} \;.\quad\quad\label{three_point_f_at_0}
\eeqa
Deploying Eq.~(\ref{three_point_f_at_0}) in the following expression
for $t_i$ we get the scalar current at the vanishing momentum transfer
$\vec{k}=0$.
\beqa
&&m_q S_{\rm 2N: 2\pi}^{(Q^0)}\big|_{\vec{k}=0}\,=\,
\frac{M_\pi^2}{64\pi^2F_\pi^4\left(4M_\pi^2
+q_1^2\right)}\bigg[\frac{L(q_1)}{4M_\pi^2+q_1^2}\nn
&\times&\Big(6g_A^4\left(4M_\pi^2+q_1^2\right)\left(q_1^2
\vec{\sigma}_1\cdot\vec{\sigma}_2-q_1\cdot\vec{\sigma}_1q_1\cdot\vec{
\sigma}_2\right)\nn
&+&\big(16 M_\pi^4\left(-8g_A^4+4g_A^2+1
\right)+8 M_\pi^2q_1^2\left(-10g_A^4+5g_A^2+1\right)\nn
&+&q_1^4\left(-11g_A^4
+6g_A^2+1\right)\big)\fet\tau_1 \cdot\fet\tau_2 \Big)\nn
&-&\frac{1}{2}\big(4 M_\pi^2\left(15 g_A^4-2g_A^2+1
\right)
+q_1^2\left(1-2 g_A^2+17 g_A^4\right)\big)\nn
&\times&\fet \tau_1 \cdot\fet
\tau_2\bigg]\;.
\eeqa
\begin{strip}
Using the expressions from~\cite{Epelbaum:2002gb} one can
evaluate the slope of the nuclear forces in respect to the quark
mass. Combining these with expressions for
the scalar current at $\vec{k}=0$, it is straightforward to show the
validity of Eq.~(\ref{ScalarCurrentAtZeroMomentum}), see~\cite{Krebs:2020plh} for more details.
The final expressions for $t_i$ are given by 
\beqa
m_q
t_1&=&\frac{g_A^4M_\pi^2}{128\pi^2F_\pi^4k^2}-\frac{g_A^4M_\pi^4L(k)}{32
\pi^2F_\pi^4k^2\left(k^2+4M_\pi^2\right)},\nn
m_q t_2&=&
-\frac{\left(g_A^2-1\right) M_\pi^2}{8F_\pi^4}
\left(\frac{\left(g_A^2-1\right)k^2q_1^2
q_2^2}{4\left((\vec{q}_ 1\cdot\vec{q}_ 2)^2-q_1^2q_2^2
\right)}
+\left(3g_A^2+1\right)M_\pi^2+2g_A^2q_1^2\right)i\,I(4;0,1;q_1,1;k,1;0,0)\nn
&-&\frac{M_\pi^2
L(q_1)}{256\pi^2F_\pi^4\left((\vec{q}_1\cdot\vec{q}_2)^2-q_1^2
q_2^2\right)}\left(-\frac{g_A^4}{\left(4M_\pi^2+q_1^2\right)\left(q_1^2
q_2^2\left(k^2+4M_\pi^2\right)-4M_\pi^2(\vec{q}_1\cdot\vec{q}_ 2)^2
\right)}\Big(2k^8\left(8M_\pi^4+3M_\pi^2q_1^2
\right)\right. \nn
&-&\left. k^6\left(M_\pi^4\left(52q_1^2+64q_2^2
\right)+M_\pi^2\left(19q_1^4+36q_1^2q_2^2\right)+4q_1^4q_2^2
\right)+k^4\left(M_\pi^4\left(60q_1^4+52q_1^2q_2^2+96q_2^4
\right)\right.\right.\nn
&+&\left.\left. M_\pi^2\left(21q_1^6+35q_1^4q_2^2+64q_1^2q_2^4\right)+
5q_1^4q_2^2\left(q_1^2+2q_2^2\right)\right)+k^2
Q_{-}^2\left(M_\pi^4\left(-28q_1^4+12q_1^2q_2^2+64q_2^4
\right)\right.\right.  \nn
&+&\left.\left. M_\pi^2\left(-9q_1^6+5q_1^4q_2^2+44q_1^2q_2^4\right)+
q_1^4q_2^2\left(q_1^2+8q_2^2\right)\right)+
Q_{-}^6\left(4M_\pi^4\left(q_1^2-4q_2^2\right)+M_\pi^2\left(
q_1^4-10q_1^2q_2^2\right)-2q_1^4q_2^2\right)
\Big)\right. \nn
&-&\left. 2g_A^2\left(k^4-k^2\left(q_1^2+2q_2^2\right)-q_2^2Q_{-}^2
\right)+q_1^2\left(k^2-Q_{-}^2\right)
\right)\nn
&-&\frac{M_\pi^2
L(k)}{512\pi^2 F_\pi^4 \left((\vec{q}_1\cdot\vec{q}_2)^2-q_1^2
q_2^2\right)}\bigg(-\frac{g_A^4}{\left(k^2+4M_\pi^2\right)\left(q_1^2q_2^2\left(k^2+4M_\pi^2
\right)-4M_\pi^2(\vec{q}_1\cdot\vec{q}_2)^2
\right)}\Big(5k^{10}M_\pi^2\nn
&+& k^8\left(20M_\pi^4-46
M_\pi^2q_1^2-7q_1^2q_2^2\right)+2k^6
q_1^2\left(-92M_\pi^4+M_\pi^2\left(37q_1^2+q_2^2\right)+15q_1^2
q_2^2\right)+2k^4\left(52M_\pi^4q_2^2\left(q_1^2+3q_2^2
\right)\right.  \nn
&+&\left.  M_\pi^2\left(83q_1^4q_2^2-23q_1^6\right)-8q_1^6q_2^2+8
q_1^4q_2^4\right)-4k^2M_\pi^2q_1^4Q_{-}^2\left(58M_\pi^2-
q_1^2+27q_2^2\right)+4M_\pi^2q_1^4Q_{-}^2\left(16M_\pi^2\left(
q_1^2-3q_2^2\right)\right. \nn
&+&\left. q_1^4-2q_1^2q_2^2\right)
\Big)+8g_A^2\left(2q_1^2q_2^2+q_2^2
\vec{q}_1\cdot\vec{q}_2-(\vec{q}_1\cdot\vec{q}_2)^2\right)-2k^2\vec{q}_1\cdot\vec{q}_2
\bigg)-
\frac{\left(g_A^2+1\right)^2M_\pi^2}{128\pi^2F_\pi^4}\;,\nn
m_q t_3&=&\frac{3 i\, g_A^4\,I(4;0,1;q_1,1;k,1;0,0)M_\pi^2(\vec{q}_1\cdot\vec{q}_2)
^2}{8F_\pi^4\left(q_1^2q_2^2-(\vec{q}_1\cdot\vec{q}_2)^2\right)}+\frac{3g_A^4M_\pi^2q_1^2q_2^2\left(k^2+Q_{-}^2
\right)L(q_1)\vec{q}_1\cdot\vec{q}_2}{64\pi^2F_\pi^4\left(q_1^2q_2^2-(\vec{q}_1\cdot\vec{q}_2)^2\right)\left(q_1^2q_2^2\left(k^2
+4M_\pi^2\right)-4M_\pi^2(\vec{q}_1\cdot\vec{q}_2)^2\right)}\nn
&+&\frac{3g_A^4M_\pi^2L(k)}{64\pi^2F_\pi^4}\left(\frac{1}{k^2+4M_\pi^2}-
\frac{q_1^2q_2^2\left(k^4-Q_{-}^4\right)}{4\left(q_1^2q_2^2-(\vec{q}_1\cdot\vec{q}_2)^2
\right)\left(q_1^2q_2^2\left(k^2+4M_\pi^2\right)-4M_\pi^2(\vec{q}_1\cdot\vec{q}_2)^2\right)}
\right)\;,\nn
m_q
t_4&=&m_q t_3+\frac{3i\,g_A^4\,I(4;0,1;q_1,1;k,1;0,0)M_\pi^2(q_1^2q_2^2-(\vec{q}_1\cdot\vec{q}_2)^2)}{8F_\pi^4
\left(q_1^2q_2^2-(\vec{q}_1\cdot\vec{q}_2)^2\right)}\;,\nn
m_q
t_5&=&
\frac{3g_A^4M_\pi^2q_2^2L(k)\left(-k^6 M_\pi^2+k^4
\left(3M_\pi^2 Q_+^2+2q_1^2q_2^2
\right)-3k^2M_\pi^2Q_{-}^4+M_\pi^2Q_{-}^4Q_{+}^2\right)}{64\pi^2F_\pi^4
\left(k^2+4M_\pi^2\right)\left(q_1^2q_2^2-(\vec{q}_1\cdot\vec{q}_2)^2
\right)\left(q_1^2q_2^2\left(k^2+4M_\pi^2\right)-4M_\pi^2(\vec{q}_1\cdot\vec{q}_2)^2\right)}\nn
&-&\frac{3g_A^4M_\pi^2q_1^2q_2^4\left(k^2-Q_{-}^2
\right)L(q_2)}{64\pi^2F_\pi^4\left(q_1^2q_2^2-(\vec{q}_1\cdot\vec{q}_2)^2
\right)\left(q_1^2q_2^2\left(k^2+4M_\pi^2\right)-4M_\pi^2(\vec{q}_1\cdot\vec{q}_2)^2
\right)}-\frac{3i\,g_A^4\,I(4;0,1;q_1,1;k,1;0,0)M_\pi^2q_2^2\vec{q}_1\cdot\vec{q}_2}{4F_\pi^4\left(q_1^2q_2^2-(\vec{q}_1\cdot\vec{q}_2)^2
\right)}\nn
&-&\frac{3g_A^4M_\pi^2q_1^2q_2^4\left(k^2+Q_{-}^2
\right)L(q_1)}{64\pi^2F_\pi^4\left(q_1^2q_2^2-(\vec{q}_1\cdot\vec{q}_2)^2\right)
\left(q_1^2q_2^2\left(k^2+4M_\pi^2\right)-4M_\pi^2(\vec{q}_1\cdot\vec{q}_2)^2
\right)}\;,\nn
m_q t_6&=&-\frac{3g_A^4M_\pi^2 L(k)\left(-k^6 M_\pi^2+k^4
\left(3M_\pi^2 Q_+^2+2q_1^2q_2^2
\right)-3k^2M_\pi^2Q_{-}^4+M_\pi^2Q_{-}^4Q_{+}^2\right)}{128\pi^2F_\pi^4\left(k^2+4M_\pi^2\right)\left(q_1^2q_2^2\left(k^2+4M_\pi^2\right)-4M_\pi^2(
\vec{q}_1\cdot\vec{q}_2)^2\right)}\nonumber
\eeqa
%\lipsum[1]
\end{strip}
\beqa
&+&\frac{3g_A^4M_\pi^2q_1^2q_2^2\left(k^2+Q_{-}^2
\right)L(q_1)}{64\pi^2F_\pi^4\left(q_1^2q_2^2\left(k^2+4M_\pi^2
\right)-4M_\pi^2(\vec{q}_1\cdot\vec{q}_2)^2
\right)}\;,\quad\quad
\eeqa
where $Q_\pm^2 \equiv q_1^2 \pm q_2^2$. 
\section{Folded-diagram technique I}
\label{folded_diagram}
An energy-independent
potential can be easily derived within a Releigh-Schr\"odinger perturbation theory. We will
follow the arguments of~\cite{Suzuki:PTP1983}. The starting point is a
time-independent Schr{\"o}dinger equation
\beqa
H |\Psi\rangle&=& E |\Psi\rangle.\label{Schroedinger:timeindep}
\eeqa
We project Eq.~(\ref{Schroedinger:timeindep}) to model and rest spaces
and get
\beqa
\big(E-H_0\big)\eta|\Psi\rangle&=&\eta V \eta|\Psi\rangle+\eta V
\lambda|\Psi\rangle,\nn
\big(E-H_0\big)\lambda|\Psi\rangle&=&\lambda V \eta|\Psi\rangle+\lambda V
\lambda|\Psi\rangle.\label{Schroedinger:projected:edep}
\eeqa
Eliminating $\lambda |\Psi\rangle$ from Eq.~(\ref{Schroedinger:projected:edep}) we get
\beqa
\big(E-H_0\big)\eta|\Psi\rangle&=&Q(E)\eta|\Psi\rangle,
\eeqa
with an energy-dependent potential also called a $Q$-box defined by
\beqa
Q(E)&=&\eta V \eta + \eta V \lambda \frac{1}{E-H_0-\lambda V
  \lambda}\lambda V \eta.
\eeqa
For scattering observables we get
\beqa
|\Psi^+\rangle&=&|\phi\rangle +
\frac{1}{E-H_0+i\epsilon}V|\Psi^+\rangle\nn
&=&|\phi\rangle + \frac{1}{E-H_0+i\epsilon} T(E)|\phi\rangle\label{tmatrix:general:def}
\eeqa
with
\beqa
(H-E) |\Psi^+\rangle&=& 0 \quad{\rm and}\quad (H_0-E)|\phi\rangle\,=\,0.
\eeqa
For the initial state from the model space $|\phi\rangle=\eta|\phi\rangle$ we
project Eq.~(\ref{tmatrix:general:def}) to the model space and
get
\beqa
\eta|\Psi^+\rangle&=&\bigg(1-\frac{1}{E-H_0+i\epsilon}Q(E)\bigg)^{-1}\eta|\phi\rangle\nn
&=&\bigg(1+\frac{1}{E-H_0+i\epsilon}\eta T(E)\eta\bigg)\eta|\phi\rangle.\label{tmatrix:def:general}
\eeqa
From Eq.~(\ref{tmatrix:def:general}) we get the T-matrix
\beqa
\eta T(E) \eta
&=&(E-H_0+i\epsilon)\bigg[\bigg(1-\frac{1}{E-H_0+i\epsilon}Q(E)\bigg)^{-1}\hspace{-0.2cm}-1\bigg]\nn
&=&Q(E) + Q(E)\frac{1}{E-H_0-Q(E)+i\epsilon}Q(E). \label{tmatrix:energy:dependent:pot}
\eeqa
To define an energy-independent potential we introduce a M{\o}ller
operator $\Omega$ which is defined by
\beqa
|\Psi\rangle&=&\Omega \eta |\Psi\rangle,\label{moeller:op:def}
\eeqa
with the requirement 
\beqa
\Omega &=&\Omega  \eta.\label{moeller:eta:right}
\eeqa
So this operator reproduces the original state out of projected
state\footnote{This is only true in the restricted energy range which
  should be below the pion-production threshold.}. By projecting
Eq.~(\ref{moeller:op:def}) to the model space one
immediately gets
\beqa
\eta \Omega &=& \eta.\label{moeller:eta}
\eeqa
One applies now the operator $\Omega$ to the original Schr\"odinger
equation and gets
\beqa
\big(E-H_0\big) \Omega \eta |\Psi\rangle&=& V |\Psi\rangle,\label{HOmegaSchroedinger}\\
\big(E-\Omega H_0\big) \eta |\Psi\rangle&=& \Omega \eta V |\Psi\rangle,\label{OmegaHSchroedinger}
\eeqa
where $H_0$ denotes a free Hamiltonian.
In order to get Eq.~(\ref{OmegaHSchroedinger}) one first projects the
original Schr\"odinger equation Eq.~(\ref{Schroedinger:timeindep}) to
the model space and applies on the resulting equation the operator
$\Omega$. Subtracting Eq.~(\ref{OmegaHSchroedinger}) from
Eq.~(\ref{HOmegaSchroedinger}) we get
\beqa
\big[\Omega, H_0\big]\eta |\Psi\rangle&=& \big(V - \Omega \eta
V\big)|\Psi\rangle\nn
&=&\big(V - \Omega \eta
V\big)\Omega \eta|\Psi\rangle
\eeqa
In this way we get a non-linear equation for the $\Omega$-operator
\beqa
\big[\Omega,H_0\big]-V \Omega  + \Omega V \Omega &=&0.\label{Omega:rel}
\eeqa
We can rewrite this equation into an equation for an operator $A$ defined
by
\beqa
\Omega&=&\eta + A,
\eeqa
with $A\,=\,\lambda A \eta$. The last relation follows from
Eqs.~(\ref{moeller:eta:right})
and~(\ref{moeller:eta}). From Eq.~(\ref{Omega:rel}) we get
\beqa
\lambda\big(H +\big[H,A\big] - A
V A \big) \eta&=&0.\label{decoupling:eq:A}
\eeqa
The effective energy-independent potential is defined by
\beqa
R&=&\eta V \Omega \eta.\label{R:potential:def}
\eeqa
If one projects the Schr{\"o}dinger equation to the model space one
immediately gets
\beqa
\big(H_0 + \eta V \Omega\big)\eta|\Psi\rangle&=&\big(H_0 + R\big)\eta|\Psi\rangle \,=\,E \eta |\Psi\rangle
\eeqa
For the initial state from the model space $|\phi\rangle=\eta|\phi\rangle$ we
project Eq.~(\ref{tmatrix:general:def}) to the model space and
get
\beqa
\eta|\Psi^+\rangle&=&\bigg(1-\frac{1}{E-H_0+i\epsilon}R\bigg)^{-1}\eta|\phi\rangle\nn
&=&\bigg(1+\frac{1}{E-H_0+i\epsilon}\eta T(E)\eta\bigg)\eta|\phi\rangle.
\eeqa
For the projected transfer matrix we get in this way
\beqa
\eta T(E) \eta
&=&(E-H_0+i\epsilon)\bigg[\bigg(1-\frac{1}{E-H_0+i\epsilon}R\bigg)^{-1}-1\bigg]\nn
&=&R + R\frac{1}{E-H_0-R+i\epsilon}R.\label{tmatrix:energy:independent:pot}
\eeqa
Note that Eqs.~(\ref{tmatrix:energy:independent:pot}) and
(\ref{tmatrix:energy:dependent:pot}) describe the same
T-matrix. The right-hand sides of these equations are identical even
half-off-shell. Energy independent potential can be easily described
in the
form of $Q$-boxes by using Eq.~(\ref{decoupling:eq:A}). To do this we multiply
the energy-independent potential $R$ with an initial state from model
space $|i\rangle=\eta|i\rangle$ and get
\beqa
R|i\rangle&=&\eta V \eta |i\rangle + \eta V A \eta |i\rangle
\eeqa
Rewriting the decoupling equation Eq.~(\ref{decoupling:eq:A}) into
\beqa
\lambda\big(V +H A-A H_0-A R \big)\eta&=&0
\eeqa
and applying it to the model state $|\alpha\rangle$ we get
\beqa
A|\alpha\rangle&=&\frac{1}{E_\alpha - \lambda H \lambda}\big(\lambda V
\eta-A R\big)|\alpha\rangle,\quad\quad\label{A:aplied:to:state:alpha}
\eeqa 
with $H_0|\alpha\rangle\,=\,E_\alpha |\alpha\rangle$. After
multiplication of Eq.~(\ref{A:aplied:to:state:alpha}) with $\eta V$
from the left we get
\beqa
R|\alpha\rangle&=&Q(E_\alpha) |\alpha\rangle-\eta V\frac{1}{E_\alpha-\lambda H
  \lambda} A |\beta\rangle\langle\beta|R|\alpha\rangle. \quad\quad\label{R:aplied:to:state:alpha}
\eeqa
Here we use Einstein-convention where we sum over double appearing
states (in this case $|\beta\rangle\langle\beta|$). Eqs.~(\ref{A:aplied:to:state:alpha}) and
(\ref{R:aplied:to:state:alpha}) provide an iterative solution for
energy-independent potential. Applying
e.g. Eq.~(\ref{A:aplied:to:state:alpha}) to
Eq.~(\ref{R:aplied:to:state:alpha}) we can eliminate $A$-dependence
and get iterative non-linear solution for $R$ given by
\beqa
R|\alpha\rangle&=&Q(E_\alpha) |\alpha\rangle + Q_1(E_\alpha,E_\beta)
|\beta\rangle\langle\beta|R|\alpha\rangle\nn
&+&\eta V\frac{1}{E_\alpha-\lambda H
  \lambda} \frac{1}{E_\beta-\lambda H
  \lambda} A |\gamma\rangle\langle\gamma| R
|\beta\rangle\langle\beta|R|\alpha\rangle.\nn
&=&Q(E_\alpha) |\alpha\rangle +\sum_{n=1}^\infty
Q_n(E_\alpha,E_{\beta_1},\dots,E_{\beta_{n}})\nn
&\times&\langle\beta_{n}|R|\beta_{n-1}\rangle\langle\beta_{n-1}|R|\beta_{n-2}\rangle\dots \langle\beta_{1}|R|\alpha\rangle,
\eeqa
where higher $Q$-boxes are defined by
\beqa
Q_n(E_1,\dots,E_{n+1})&=&(-1)^n \eta
V\prod_{j=1}^{n+1}\frac{1}{E_j-\lambda H \lambda} V \eta\;.\quad\quad \label{Qn:box:simplified}
\eeqa
Up to three $Q$-boxes we get
\beqa
R|\alpha\rangle&=&Q(E_\alpha) |\alpha\rangle + Q_1(E_\alpha,E_\beta)|\beta\rangle
\langle\beta|R|\alpha\rangle\nn
&+&Q_2(E_\alpha,E_{\beta_1},E_{\beta_2})|\beta_2\rangle
\langle\beta_2|R|\beta_1\rangle\langle\beta_1|R|\alpha\rangle\nn
&=&Q(E_\alpha) |\alpha\rangle + Q_1(E_\alpha,E_\beta)|\beta\rangle
\langle\beta|Q(E_\alpha)|\alpha\rangle\nn
&+& Q_1(E_\alpha,E_\beta)|\beta\rangle
\langle\beta|Q_1(E_\alpha,E_\gamma)|\gamma\rangle\langle\gamma|Q(E_\alpha)|\alpha\rangle\nn
&+&Q_2(E_\alpha,E_{\beta_1},E_{\beta_2})|\beta_2\rangle
\langle\beta_2|Q(E_{\beta_1})|\beta_1\rangle\langle\beta_1|Q(E_\alpha)|\alpha\rangle.\nn
&&
\eeqa
One can get the same result (but derived in a more cumbersome way)
with the folded-diagram technique via
inversion of half-off-shell T-matrix, see
Appendix~\ref{app:folded:diagr} for derivation. For application of
$Q$-box formalism within chiral EFT see~\cite{Krebs:2004st}.

Note that the potential $R$ is manifestly non-hermitian. Nevertheless, the
half-off-shell T-matrix is exactly reproduced by iterations of the $R$
potential. If one prefers to work with the hermitian potential, what is
usually the case, one can derive them by applying the unitary
transformation technique. Unitary transformations do not affect the
spectrum of the Hamiltonian and under some assumptions lead to the
same scattering matrix~\cite{Polyzou:FBS1990:9:97,Polyzou:2010eq}, see
also~\cite{Epelbaum:1998na} for a perturbative proof. The perturbative proof
of this statement is simple and we show this here on the level of the
T-matrix. We start with the half off-shell T-matrix element and rewrite
it into a shorter form
\beqa
&&\langle f|T(E_i)| i \rangle\,=\,\langle f|\bigg(V +
V\frac{1}{E-H+i\epsilon}V\bigg)|i\rangle\nn
&=&\langle f|\bigg(H-E_f
+\big(H-E_f\big)\frac{1}{E_i-H+i\epsilon}\big(H-E_i\big)\bigg)|i\rangle\nn
&=&\langle
f|\big(H-E_f\big)\frac{i\epsilon}{E_i-H+i\epsilon}|i\rangle\nn
&=&\big(E_i-E_f\big)\langle f|\frac{i\epsilon}{E_i-H+i\epsilon}|i\rangle.
\eeqa
The T-matrix from the transformed Hamiltonian $U^\dagger H U$, where $U$ is
a unitary transformation is given by
\beqa
&&\langle f|T^U(E_i)| i \rangle\,=\, \big(E_i-E_f\big)\langle f|
U^\dagger
\frac{i\epsilon}{E_i-H+i\epsilon} U|i\rangle\nn
&=&\big(E_i-E_f\big)\langle f|
U^\dagger
\bigg(1-\frac{1}{E_i-H_0+i\epsilon}V\bigg)^{-1}\nn
&\times&\frac{i\epsilon}{E_i - H_0 + i\epsilon}U|i\rangle\nn
&=&\big(E_i-E_f\big)\langle f|
U^\dagger \frac{i\epsilon}{E_i - H_0 + i\epsilon}U|i\rangle\nn
&+&\big(E_i-E_f\big) \langle f|
U^\dagger\frac{1}{E_i - H_0 + i\epsilon}V \nn
&\times&\bigg(1-\frac{1}{E_i-H_0+i\epsilon}V\bigg)^{-1}\frac{i\epsilon}{E_i - H_0 + i\epsilon}U|i\rangle\label{halfoffshell:TU}
\eeqa
We require now for the unitary transformation $U=1+\delta U$ to fulfill
\beqa
H_0 \delta U|i\rangle\neq E_i |i\rangle \quad H_0 \delta U|i\rangle\neq E_f |f\rangle.
\eeqa
This is a reasonable assumption if $\delta U$ is at least of the  first
order in the interaction $V$. It is easy to see that the first term of
Eq.~(\ref{halfoffshell:TU}) does not contribute to the on-shell T-matrix,
\beqa
&&\big(E_i-E_f\big)\langle f|
U^\dagger \frac{i\epsilon}{E_i - H_0 + i\epsilon}U|i\rangle\,=\,\nn
&&\big(E_i-E_f\big)\bigg(\langle f|
U^\dagger |i\rangle +\langle f|
U^\dagger \frac{i\epsilon}{E_i - H_0 + i\epsilon}\delta
U|i\rangle\bigg)\nn
&=&\big(E_i-E_f\big)\bigg[\langle f|
U^\dagger |i\rangle +\langle f|
\delta U^\dagger \frac{i\epsilon}{E_i - H_0 + i\epsilon}\delta
U|i\rangle\bigg]\nn
&+&i\epsilon \langle f|\delta U |i\rangle.\label{onshell:vanishing:1}
\eeqa
since the term in the rectangular bracket of Eq.~(\ref{onshell:vanishing:1})
is by assumption non-singular at $E_i=E_f$. To
see what remains from the second term in Eq.~(\ref{halfoffshell:TU}) on
the energy-shell we note that
\beqa
\frac{i\epsilon}{E_i - H_0 + i\epsilon}
U|i\rangle&=& |i\rangle + \frac{i\epsilon}{E_i - H_0 + i\epsilon}\delta
U|i\rangle\label{mostrightterm:Tequiv}
\eeqa
The contribution of the second term of
Eq.~(\ref{mostrightterm:Tequiv}) to the matrix element in
Eq.~(\ref{halfoffshell:TU}) is non-singular at $E_i=E_f$ since it hits
at least once the interaction $V$ from the left. So due to
$i\epsilon$ in front of it this term vanishes. On the other hand
\beqa
&&\big(E_i-E_f\big) \langle f|
U^\dagger\frac{1}{E_i - H_0 + i\epsilon}V\,=\,-\langle f|V\nn
&+&\big(E_i-E_f\big) \langle f|
\delta U^\dagger\frac{1}{E_i - H_0 + i\epsilon}V.\label{lhsterm:TonshelEquiv}
\eeqa
Only the first term in Eq.~(\ref{lhsterm:TonshelEquiv}) survives
on the mass-shell such that we get the T-matrix equivalence
\beqa
&&\langle f|T^U(E_i)| i \rangle\,=\, \langle f|T(E_i)| i \rangle +
{\cal O}(E_i - E_f).
\eeqa
A more elegant proof where the authors use the M{\o}ller operator can be found in~\cite{Epelbaum:1998na}.
\section{Folded-diagram technique II}
\label{app:folded:diagr}
In this appendix, we would like to discuss the folded-diagram
technique 
introduced by Kuo et al.  in the shell-model
calculations, see~\cite{Kuo:1990springer} for comprehensive
introduction. In particular, we show here the transfer matrix
equivalence formulation of this technique presented
in~\cite{Kuo:2015lea}. 
We start
with Lippmann-Schwinger equation for half-off-shell T-matrix given by
\beqa
T(E_i)&=& V + V \tilde G_+(E_i) T(E_i), \label{LS:eq:original}
\eeqa
where $E_i$ denotes the energy of initial state and the free Green
function and its Fourier transform are defined by
\beqa
G_+(t)&=&-i\,\theta(t) e^{-i\,(H_0-i\,\epsilon)t},\nn
\tilde G_+(E_i)&=&\int d t \, e^{i\, E_i t} G_+(t) \,=\,\frac{1}{E_i-H_0+i\,\epsilon}.\label{Free:Green:Function:Def}
\eeqa
%Projecting the T-matrix to model and rest space we get a coupled
%channel equation
%\beqa
%\eta T(E_i) \eta &=& \eta V \eta + \eta V \tilde G_+^\eta(E_i)  T(E_i)
%\eta  + \eta V \tilde G_+^\lambda(E_i)  T(E_i) \eta, \nn
%\lambda T(E_i) \eta &=& \lambda V \eta + \lambda V \tilde G_+^\eta(E_i)
%T(E_i) \eta  + \lambda V \tilde G_+^\lambda(E_i) T(E_i) \eta.\nn
%&&
%\eeqa
%We can now eliminate $\lambda T(E_i) \eta$ 
%\beqa
%\lambda T(E_i) \eta &=& \big(1-\lambda V \tilde G_+^\lambda(E_i)\big)^{-1}\big(\lambda V \eta + \lambda V \tilde G_+^\eta(E_i)
%T(E_i) \eta\big)\;.\nn
%&&
%\eeqa
We can rewrite the transfer-matrix
into
\beqa
T(E_i) = Q(E_i) + Q(E_i) \tilde G_+^\eta(E_i)   T(E_i),\label{LS:eq:energy_dep_pot}
\eeqa
where $Q$ is an energy-dependent potential which satisfies
\beqa
Q(E_i)&=& V  + \eta V \tilde G_+^\lambda(E_i)  Q(E_i),
\eeqa
and
\beqa
\tilde G_+^\eta(E)&=&\tilde G_+(E)\eta, \quad \tilde G_+^\lambda(E)\,=\,\tilde G_+(E)\lambda.
\eeqa
This result can be directly derived by rewriting
Eq.~(\ref{LS:eq:original}) via
\beqa
\big[1-V \tilde G_+^\lambda(E_i)\big] T(E_i)&=& V + V \tilde
G_+^\eta(E_i) T(E_i). \label{LS:eq:rewritten}
\eeqa
Multiplying both sides of Eq.~(\ref{LS:eq:rewritten}) with $\big[1-V
\tilde G_+^\lambda(E_i)\big]^{-1}$ we get Eq.~(\ref{LS:eq:energy_dep_pot})
with energy-dependent potential given by
\beqa
Q(E_i)&=&\big[1-V \tilde G_+^\lambda(E_i)\big]^{-1}V.
\eeqa
Using Eq.~(\ref{Free:Green:Function:Def}) we can rewrite the effective potential into
\beqa
Q(E_i)&=&\big[1-V \lambda(E_i-H_0)^{-1}\big]^{-1}V\nn
&=&(E_i - H_0 - V \lambda + V \lambda)\frac{1}{E_i-H_0- V \lambda}
V\nn
&=& V + V\frac{1}{E_i - \lambda H \lambda} V.
\eeqa
We project Eq.~(\ref{LS:eq:energy_dep_pot}) to the model space and get
\beqa
\eta T(E_i)\eta &=&  Q^\eta(E_i) + Q^\eta(E_i) \tilde G_+^\eta(E_i)\eta
T(E_i)\eta\;,
\eeqa
with
\beqa
Q^\eta(E_i) &=&\eta Q(E_i) \eta.
\eeqa
From now on we will work only with the model projected potential
$Q^\eta(E_i)$ rather than with full $Q(E_i)$. To abbreviate
the notation we make a replacement in the notation $Q^\eta\to Q$ using
from now on
\beqa
Q(E_i)&=&\eta V \eta+ \eta V\frac{1}{E_i - \lambda H \lambda} V\eta.
\eeqa
%$Q(E_i)$ is also called a $Q$-box~\cite{Kuo:1990springer}.
The effective potential $Q(E_i)$ is energy-dependent and is difficult to
deal with in practical calculation with $A>2$. For this reason, we can
define an energy-independent potential by using the folded-diagram technique. For this purpose, we define higher $Q$-boxes by
\beqa
Q_n(E_1,\dots,E_{n+1})&=&\sum_{k=1}^{n+1}C_k(E_1,\dots,E_{n+1}) Q(E_k),\label{Qn:box:definition}\quad\quad
\eeqa 
with
\beqa
C_k(E_1,\dots,E_{n+1})&=&\prod_{i=1}^{k-1}\frac{1}{E_k-E_i}\prod_{j=k+1}^{n+1}\frac{1}{E_k-E_j}.\label{Ck:definition}\quad\quad
\eeqa
The higher $Q$-boxed can be rewritten into a simpler form of
Eq.~(\ref{Qn:box:simplified}) by using the partial
fraction decomposition 
\beqa
\sum_{k=1}^{n+1} C_k(E_1,\dots,E_{n+1})\frac{1}{E_k- \lambda H
  \lambda}&=&(-1)^{n}\prod_{k=1}^{n+1}\frac{1}{E_k - \lambda H
  \lambda}\;,\nn
&&
\eeqa
as well as
%\beqa
%Q_n(E_1,\dots,E_{n+1})&=&(-1)^n \eta
%V\prod_{j=1}^{n+1}\frac{1}{E_j-\lambda H \lambda} V \eta\;,\label{Qn:box:simplified}\quad
%\eeqa
%where we used
\beqa
\sum_{k=1}^{n+1} C_k(E_1,\dots,E_{n+1}) &=& 0\;.
\eeqa

To define a folded-diagram we follow~\cite{Kuo:2015lea} and apply
the T-matrix equivalence approach. We start with the first iteration of the
energy-dependent potential and replace the energy $E_i$ in the $Q$-box
by the energy of the state on which the $Q$-box operates. To compensate
this change we need to add to the changed expression a folded-diagram
\beqa
&&\langle f|Q(E_i)\tilde G_+^{\eta}(E_i)Q(E_i)|i\rangle\,=\,\label{folded:diagram:requirement}\\
&&\sum_{\alpha}\langle f|Q(E_\alpha)|\alpha\rangle\langle\alpha|\tilde
G_+^{\eta}(E_i)Q(E_i)|i\rangle - \langle
f|Q^\prime \!\int\! Q|i\,\rangle\,.\nonumber
\eeqa
Eq.~(\ref{folded:diagram:requirement}) defines a folded-diagram by a difference between the original and the modified
iterations. Having defined a folded-diagram we can express it in terms
of the
higher $Q$-boxes:
\beqa
&&\langle
f|Q^\prime \!\int\! Q|i\,\rangle\,=\,\sum_\alpha \langle f |\big(
Q(E_\alpha)-Q(E_i)\big)|\alpha\rangle\nn
&\times&\langle \alpha|\tilde G_+^\eta(E_i) Q(E_i)|
i\,\rangle\nn
&=&\sum_\alpha
\frac{\langle f |\big(Q(E_\alpha)-Q(E_i)\big)|\alpha\rangle}{E_i-E_\alpha}\langle \alpha|Q(E_i)|
i\,\rangle\nn
&=&-\sum_\alpha\langle f|Q_1(E_i, E_\alpha)|\alpha\rangle\langle\alpha|Q(E_i)|i\rangle\;,\quad\label{folded:diagram:def}
\eeqa
where in the last step we used the definition of the higher $Q$-boxes
given in Eqs.~(\ref{Qn:box:definition}) and (\ref{Ck:definition}).
As a direct consequence of Eq.~(\ref{Qn:box:simplified}) there are no pure
nucleon cuts in a folded-diagram. 
If we would define an effective energy-independent potential by
\beqa
\langle \alpha|V_{\rm eff}|\beta\rangle &=& \langle
\alpha|Q(E_\beta)|\beta\rangle -  \langle
\alpha|Q^\prime\!\int \! Q|\beta\rangle\;, 
\eeqa
the sum of the zeroth and the first iteration of this potential would
reproduce the original result for the T-matrix (up to the two $Q$-boxes
approximation). This idea can be
generalized to any number of iteration. Let us demonstrate this for
two iterations. We define a twice-folded-diagram $\langle f|Q^\prime\!\int\! Q \int\! Q
|i\rangle $ by 
\beqa
&&\langle f|Q(E_i)\tilde G_+^\eta(E_i)
Q(E_i) \tilde G_+^\eta(E_i)
Q(E_i)|i\rangle\nn
&=&\sum_{\alpha,\beta}\langle f|Q(E_\alpha)|\alpha\rangle\langle\alpha|\tilde G_+^\eta(E_i)Q(E_\beta)|\beta\rangle
\langle\beta| \tilde G_+^\eta(E_i)Q(E_i)|i\rangle\nn
&-&\sum_{\beta}\langle f|Q^\prime\!\int\! Q |\beta\rangle
\langle\beta| \tilde G_+^\eta(E_i)Q(E_i)|i\rangle\nn
&-&\sum_{\alpha,\beta}\langle f|Q(E_\alpha) |\alpha\rangle\langle\alpha|\tilde G_+^\eta(E_i)|\beta\rangle
\langle\beta|Q^\prime\!\int\!Q|i\rangle\nn
&+&\langle f|Q^\prime\!\int\! Q \int\! Q
|i\rangle\nn
&=&\sum_{\alpha,\beta}\langle f|Q(E_\alpha)|\alpha\rangle\frac{\langle\alpha|Q(E_\beta)|\beta\rangle}{E_i-E_\alpha}
\frac{\langle\beta| Q(E_i)|i\rangle}{E_i-E_\beta}\nn
&+&\sum_{\alpha,\beta}\langle f|Q_1(E_\beta,E_\alpha)|\alpha\rangle\langle\alpha|Q(E_\beta) |\beta\rangle
\frac{\langle\beta| Q(E_i)|i\rangle}{E_i-E_\beta}\nn
&+&\sum_{\alpha,\beta}\frac{\langle f|Q(E_\alpha) |\alpha\rangle}{E_i-E_\alpha}
\langle\alpha|Q_1(E_i,E_\beta)|\beta\rangle\langle\beta|Q(E_i)|i\rangle\nn
&+&\langle f|Q^\prime\!\int\! Q \int\! Q
|i\rangle\,.\label{twice:folded:diagram:def}
\eeqa
For twice-folded-diagram we get
\beqa
&&\langle f|Q^\prime\!\int\! Q \int\! Q
|i\rangle\,=\,\sum_{\alpha,\beta}\bigg[\frac{\langle f|Q(E_i)|\alpha\rangle}{E_i-E_\alpha}\frac{\langle\alpha|
Q(E_i)|\beta\rangle}{E_i-E_\beta}\nn
&-&\frac{\langle
  f|Q(E_\alpha)|\alpha\rangle}{E_i-E_\alpha}\frac{\langle\alpha|Q(E_\beta)|\beta\rangle}{E_i-E_\beta}
-\langle f|Q_1(E_\beta,E_\alpha)|\alpha\rangle\nn
&\times&\frac{\langle\alpha|Q(E_\beta) |\beta\rangle}{E_i-E_\beta}
-\frac{\langle f|Q(E_\alpha) |\alpha\rangle}{E_i-E_\alpha}
\langle\alpha|Q_1(E_i,E_\beta)|\beta\rangle\bigg]\nn
&\times&\langle\beta| 
Q(E_i)|i\rangle\nn
&=&\sum_{\alpha,\beta}\bigg[\bigg(\frac{\langle f|Q(E_i)|\alpha\rangle}{E_i-E_\alpha}
-\frac{\langle
  f|Q(E_\alpha)|\alpha\rangle}{E_i-E_\alpha}\nn
&-&\langle f|Q_1(E_\beta,E_\alpha)|\alpha\rangle \bigg) \frac{\langle\alpha|
Q(E_\beta)|\beta\rangle}{E_i-E_\beta}
+\langle f|Q_1(E_i,E_\alpha) |\alpha\rangle\nn
&\times&\langle\alpha|Q_1(E_i,E_\beta)|\beta\rangle\bigg]\langle\beta| 
Q(E_i)|i\rangle\nn
&=&\sum_{\alpha,\beta}\bigg[\bigg(\frac{\langle f|Q(E_i)|\alpha\rangle}{(E_i-E_\alpha)(E_i-E_\beta)}
+\frac{\langle f|Q(E_\beta)|\alpha\rangle}{(E_\beta-E_i) (E_\beta -
  E_\alpha)} \nn
&+&\frac{\langle
  f|Q(E_\alpha)|\alpha\rangle}{(E_\alpha-E_i)(E_\alpha-E_\beta)}\bigg) \langle\alpha|
Q(E_\beta)|\beta\rangle\nn
&+&\langle f|Q_1(E_i,E_\alpha) |\alpha\rangle\langle\alpha|Q_1(E_i,E_\beta)|\beta\rangle\bigg]\langle\beta| 
Q(E_i)|i\rangle\nn
&=&\sum_{\alpha,\beta}\bigg[\langle f|Q_2(E_i,E_\alpha,E_\beta)|\alpha\rangle \langle\alpha|
Q(E_\beta)|\beta\rangle \nn
&+&\langle f|Q_1(E_i,E_\alpha) |\alpha\rangle\langle\alpha|Q_1(E_i,E_\beta)|\beta\rangle\bigg]\langle\beta| 
Q(E_i)|i\rangle.
\label{twice:folded:diagram:simplified}
\eeqa
For the effective potential we get
\beqa
&&\langle \alpha|V_{\rm eff}|\beta\rangle \,=\, \langle
\alpha|Q(E_\beta)|\beta\rangle -  \langle
\alpha|Q^\prime\!\int \! Q|\beta\rangle \nn
&+& \langle f|Q^\prime\!\int\! Q \int\! Q\;
|i\rangle\;.
\eeqa
To reproduce higher number of iterations one
should include more and more foldings into the effective potential. Up to the
three $Q$-boxes one gets
\beqa
&&\langle \alpha|V_{\rm eff}|\beta\rangle = \langle
\alpha|Q(E_\beta)|\beta\rangle\nn 
&+&  \langle
\alpha|Q_1(E_\beta,E_\gamma)|\gamma\rangle\langle\gamma|Q(E_\beta)|\beta\rangle \nn
&+& \langle
\alpha|Q_2(E_\beta,E_\delta,E_\gamma)|\gamma\rangle\langle\gamma|Q(E_\delta)|\delta\rangle
\langle\delta|Q(E_\beta)|\beta\rangle\nn
&+& \langle
\alpha|Q_1(E_\beta,E_\gamma)|\gamma\rangle\langle\gamma|Q_1(E_\beta, E_\delta)|\delta\rangle
\langle\delta|Q(E_\beta)|\beta\rangle\,.\quad\quad
\eeqa
From three and a higher number of iterations one can derive in a
similar way further corrections to the energy-independent
potential. 
%%%%%%%%%%%%%%%%%%%%%%%%%%%%%%%%%%%%%%%%%%%%%%%%%%%%%%%%%%%%%%%%%%%%%%%%%%%%%%%%%%%%%
\section{Unitary equivalence of S-matrix}
\label{unitary_equiv_S_matrix}
in this appendix we want to clarify under which condition  the 
scattering matrix for original and transformed time-dependent
interactions remains the same? Assume that we have a time-dependent Hamiltonian $H(t)=H_0+V(t)$ and a state
$|\Psi(t)\rangle$ which satisfies the Schr\"odinger equation
\beqa
i\frac{\partial}{\partial t} |\Psi(t)\rangle&=&H(t)  |\Psi(t)\rangle.\label{schroedinger:eq:original}
\eeqa
Here $H_0$ denotes a free Hamiltonian~\footnote{It could be also any
  time-independent Hamiltonian which e.g. can have bound states as
  eigenstates. In our case this will be kinetic energy of nucleons and
pions.}. Let $|i(t)\rangle$ and $|f(t)\rangle$ be stationary eigenstates of the free Hamiltonian
\beqa
i\frac{\partial}{\partial t} |\alpha(t)\rangle&=&H_0
|\alpha(t)\rangle,\quad \alpha=i,f,
\eeqa
which can be written in the form
\beqa
|i(t)\rangle&=&e^{-i\,E_i t}|i\rangle, \quad |f(t)\rangle\,=\,e^{-i\,E_f t}|i\rangle
\eeqa
Let $U(t)=1+\delta U(t)$ be a time-dependent unitary transformation
with existing Fourier transform for $\delta U(t)$. This means, in
particular, that $\delta U(t)$ decreases fast enough with increasing $|t|$.
Then a transformed state 
\beqa
|\Psi^U(t)\rangle &=&U^\dagger(t)|\Psi(t)\rangle
\eeqa
satisfies the
Schr\"odinger equation
\beqa
i\frac{\partial}{\partial t} |\Psi^U(t)\rangle&=&H^U(t)  |\Psi^U(t)\rangle,\label{schroedinger:eq:transf}
\eeqa
with 
\beqa
H^U(t)&=& U^\dagger(t) H(t) U(t)+i\,\dot U^\dagger(t)U(t),
\eeqa
where dot on top of the letter denotes a time-derivative 
\beqa
\dot U^\dagger&=&\frac{\partial}{\partial t} U^\dagger(t).
\eeqa
We show now that the scattering matrices for
both Hamiltonians $H(t)$ and $H^U(t)$ are the same. To give the scattering matrix in terms of 
Hamiltonians and keep notation short we change the notation for a free
Green-function (compared to Appendix~\ref{app:folded:diagr},
see Eq.~(\ref{Free:Green:Function:Def})) which we denote now by $G_0^{(+)}(t)$. The free
Green-function satisfies
\beqa
\bigg(i\frac{\partial}{\partial t} - H_0\bigg)G_0^{(+)}(t)&=&\delta(t).
\eeqa
The formal solution of this equation is given by
\beqa
G_0^{(+)}(t)&=&-i\,\theta(t) e^{-i H_0 t},\label{retarded:green}
\eeqa
where $\theta(t)$ is a step function
\beqa
\theta(t>0)&=&1,\quad \theta(t<0)\,=\,0.
\eeqa
The state $|\Psi^{(+)}(t)\rangle$ which satisfies
\beqa
|\Psi^{(+)}(t)\rangle&=&|i(t)\rangle + \int_{-\infty}^\infty d t^\prime
  G_0^{(+)}(t-t^\prime) V(t^\prime\,)
  |\Psi^{(+)}(t^\prime\,)\rangle,\nn
&&\label{psiPlus:int:relation}
\eeqa
satisfies also the original Schr\"odinger
equation~(\ref{schroedinger:eq:original}). The
$i\epsilon$-prescription chosen in Eq.~(\ref{retarded:green}) is
chosen to make the integrand of Eq.~(\ref{psiPlus:int:relation})
vanish for $t^\prime$ approaching minus infinity. The step function in
Eq.~(\ref{psiPlus:int:relation}) causes the state
$|\Psi^{(+)}(t)\rangle$ for $t\to-\infty$ to become a free incoming asymptotic
state
\beqa
|\Psi^{(+)}(-\infty)\rangle&=&|i(-\infty)\rangle. 
\eeqa
The scattering matrix for the Hamiltonian $H(t)$ is given by~\cite{BjorkenDrell:1964}
\beqa
\langle f| S | i \rangle&=& \langle f| i \rangle - i \, \langle f| T | i \rangle,
\eeqa
where the on-shell T-matrix is defined by
\beqa
\langle f| T | i \rangle&=&\lim_{t\to\infty}i\int_{-\infty}^\infty d
t^\prime\langle f(t)| G_0^{(+)}(t-t^\prime\,)V(t^\prime\,)
|\Psi^{(+)}(t^\prime\,)\rangle.\nn
&&
\eeqa
Note that in this formulation energy-conservation delta function is
not yet extracted out of the T-matrix. A full Green-function is
defined via a differential equation
\beqa
\bigg(i\frac{\partial}{\partial t}-H(t)\bigg)G^{(+)}(t,t^\prime
\,)&=&\delta(t-t^\prime\,),\nn
G^{(+)}(t,t^\prime
\,)\bigg(-i\frac{\overleftarrow{\partial}}{\partial t^\prime}-H(t^\prime\,)\bigg)&=&\delta(t-t^\prime\,).
\eeqa
which is equivalent to an iterative solution of the integral equation
\beqa
G^{(+)}(t, t^\prime\,)&=&G_0^{(+)}(t-t^\prime) \nn
&+& \int d t_1
G_0^{(+)}(t-t_1) V(t_1) G^{(+)}(t_1, t^\prime\,).\label{full:green:function:recursion:rel}\quad\quad
\eeqa
Indeed applying $i\partial/\partial t - H(t)$ on both sides of
Eq.~(\ref{full:green:function:recursion:rel}) we get
\beqa
&&\bigg(i\frac{\partial}{\partial t}-H(t)\bigg)G^{(+)}(t,t^\prime
\,)=\delta(t-t^\prime\,)-V(t) G_0^{(+)}(t-t^\prime\,)\nn
&&+\int d t_1 \bigg(i\frac{\partial}{\partial
  t}-H(t)\bigg)G_0^{(+)}(t-t_1) V(t_1) G^{(+)}(t_1,t^\prime\,)\nn
&&=\delta(t-t^\prime\,)+V(t) \bigg(
G^{(+)}(t,t^\prime\,) - G_0^{(+)}(t-t^\prime\,) \nn
&& - \int d t_1 G_0^{(+)}(t-t_1) V(t_1)
G^{(+)}(t_1,t^\prime\,)\bigg)\nn
&&=\delta(t-t^\prime\,).\label{G:diff:eq}
\eeqa
If we know the full Green function we know the solution of the
Schr\"odinger equation
\beqa
|\Psi^{(+)}(t)\rangle&=&|i(t)\rangle + \int_{-\infty}^\infty d t^\prime
  G^{(+)}(t,t^\prime) V(t^\prime\,)
  |i(t^\prime\,)\rangle,\nn
&&\label{psiPlus:int:relation:full:green}
\eeqa
Using Eq.~(\ref{psiPlus:int:relation:full:green}) and
\beqa
\langle f(t)|G_0^{(+)}(t-t^\prime\,)&=&\langle f(t^\prime)|
\eeqa
we can rewrite the on-shell T-matrix  to
\beqa
&&\langle f| T | i \rangle\,=\,i
\int_{-\infty}^\infty d
t_1\langle f(t_1)| V(t_1)
|i(t_1)\rangle\nn
&&+\,i\int_{-\infty}^\infty d
t_1 d t_2\langle f(t_1)| V(t_1) G^{(+)}(t_1,t_2) V(t_2)
|i(t_2)\rangle.\label{tmatrix:no:limit:expr}\quad\quad
\eeqa 
We write now the step function as 
\beqa
\theta(t)&=&\lim_{\epsilon\to 0+}\int\frac{d \omega}{2\pi i}\frac{e^{i \omega t}}{\omega -
  i \epsilon}.
\eeqa
$0+$ means here that $\epsilon>0$ condition is valid during the limiting procedure. We let 
from now on $\epsilon>0$ and take the limit at the end of the
calculation. This will slightly modify the differential equation for a
Green function. To see this consider the derivative of the step
function at $\epsilon>0$
\beqa
\frac{\partial}{\partial t}\theta(t)&=&\int \frac{d \omega}{2\pi
  i}i\frac{\omega - i\epsilon + i\epsilon}{\omega
  -i\epsilon}e^{i\omega t}\nn
&=&\delta(t) -\epsilon \theta(t).
\eeqa  
In the case of a free Green function, this leads to
\beqa
\bigg(i\frac{\partial}{\partial t} - H_0 + i\epsilon\bigg) G_0^{(+)}(t)&=&\delta(t).
\eeqa
For the full Green function, we get a similar relation
\beqa
\bigg(i\frac{\partial}{\partial t} - H(t) + i\epsilon\bigg) G^{(+)}(t,t^\prime)&=&\delta(t-t^\prime).
\eeqa
We can explicitly see this if we apply the operator
$i\partial/\partial t - H(t) + i\epsilon$ to
Eq.~(\ref{full:green:function:recursion:rel})
\beqa
&&\bigg(i\frac{\partial}{\partial t} - H(t) + i\epsilon\bigg)
G^{(+)}(t,t^\prime)=\delta(t-t^\prime)\nn
&&+V(t) \bigg(G^{(+)}(t,t^\prime\,) - G_0^{(+)}(t-t^\prime\,) \nn
&&- 
\int d t_1
G_0^{(+)}(t-t_1) V(t_1) G^{(+)}(t_1,t^\prime\,)\bigg) \nn
&&=\delta(t-t^\prime\,).
\eeqa
We use now $V(t)=H(t)-H_0$ and rewrite
\beqa
&&\int d t_2 G^{(+)}(t_1,t_2)V(t_2)
|i(t_2)\rangle\nn
&=&\int d t_2 G^{(+)}(t_1,t_2)\big(H(t_2)
-H_0\big)|i(t_2)\rangle\nn
&=&\int d t_2 G^{(+)}(t_1,t_2)\bigg(H(t_2)
-i\frac{\partial}{\partial t_2}\bigg)|i(t_2)\rangle\nn
&=&\int d t_2 G^{(+)}(t_1,t_2)\bigg(H(t_2)
+i\frac{\overleftarrow{\partial}}{\partial t_2}-i\epsilon + i\epsilon\bigg)|i(t_2)\rangle\nn
&=&-\int d t_2 \delta(t_1-t_2)|i(t_2)\rangle + i\epsilon \int d t_2
G^{(+)}(t_1,t_2) |i(t_2)\rangle\nn
&=&-|i(t_1)\rangle + i\epsilon \int d t_2
G^{(+)}(t_1,t_2) |i(t_2)\rangle.\label{approximate:green:function:rel}
\eeqa
Using Eq.~(\ref{approximate:green:function:rel}) we can rewrite
T-matrix of Eq.~(\ref{tmatrix:no:limit:expr}) into
\beqa
&&\langle f| T | i \rangle\,=\,i\int_{-\infty}^\infty d
t_1 d t_2\langle f(t_1)| V(t_1) i\epsilon G^{(+)}(t_1,t_2) 
|i(t_2)\rangle.\nn
&&\label{VGPlusTmatrix}
\eeqa
With the same steps as in
Eq.~(\ref{approximate:green:function:rel}) we get
\beqa
&&\int d t_1 \langle f(t_1)| V(t_1) G^{(+)}(t_1,t_2)\nn
&&=-\langle f(t_2)| +i\epsilon \int d t_1 \langle f(t_1)| G^{(+)}(t_1,t_2).\label{VGPlusIniEps}
\eeqa
Using Eq.~(\ref{VGPlusIniEps}) we can rewrite the T-matrix into
\beqa
&&\langle f| T | i \rangle\,=\, \epsilon\int_{-\infty}^\infty d
t_1 \langle f(t_1)
|i(t_1)\rangle \nn
&&-i\epsilon^2\int_{-\infty}^\infty d
t_1 d t_2\langle f(t_1)| G^{(+)}(t_1,t_2) 
|i(t_2)\rangle\nn
&&=-i\epsilon^2\int_{-\infty}^\infty d
t_1 d t_2\langle f(t_1)| G^{(+)}(t_1,t_2) 
|i(t_2)\rangle.\label{GPlusTmatrix}
\eeqa
In the last step of Eq.~(\ref{GPlusTmatrix}) we used the fact that
\beqa
\epsilon\int_{-\infty}^\infty d
t_1 \langle f(t_1)
|i(t_1)\rangle&=&\epsilon 2\pi \delta(E_f - E_i)
\langle f
|i\rangle,
\eeqa
which vanishes in the limit $\epsilon\to 0+$. Now we introduce unity
operators on the left and the right-hand side of the Green function in
Eq.~(\ref{GPlusTmatrix}) and get
\beqa
&&\langle f| T | i \rangle\,=\,
-i\epsilon^2\int_{-\infty}^\infty d
t_1 d t_2\langle f(t_1)| U(t_1) \nn
&&\times U^\dagger(t_1) G^{(+)}(t_1,t_2) U(t_2) U^\dagger(t_2)
|i(t_2)\rangle.
\eeqa
We denote the transformed Green function by
\beqa
G_U^{(+)}(t,t^\prime\,)&=&U^\dagger(t) G^{(+)}(t, t^\prime\,)U(t^\prime\,),
\eeqa
which satisfies 
\beqa
\bigg(i\frac{\partial}{\partial t}-H^U(t)\bigg) G_U^{(+)}(t,
t^\prime\,)&=&\delta(t-t^\prime\,),\nn
G_U^{(+)}(t,
t^\prime\,) \bigg(-i\frac{\overleftarrow{\partial}}{\partial t^\prime}-H^U(t^\prime\,)\bigg) &=&\delta(t-t^\prime\,).\label{GU:diff:eq}
\eeqa
Eq.~(\ref{GU:diff:eq}) follows directly from multiplying 
Eq.~(\ref{G:diff:eq}) with $U^\dagger(t)$ and $U(t^\prime)$ from left
and right, respectively. Due to Eq.~(\ref{GU:diff:eq}) we also have
\beqa
&&G_U^{(+)}(t, t^\prime\,)=\nn
&&G_0^{(+)}(t-t^\prime)
+\int d t_1
G_0^{(+)}(t-t_1) V^U(t_1) G_U^{(+)}(t_1, t^\prime\,) =\nn
&&G_0^{(+)}(t-t^\prime) 
+ \int d t_1
G_U^{(+)}(t,t_1) V^U(t_1) G_0^{(+)}(t_1-t^\prime\,),\label{full:green:function:U:recursion:rel}\quad\quad
\eeqa
where 
\beqa
V^U(t)&=& H^U(t) - H_0.\label{VU:definition}
\eeqa
We see that the difference between $G^{(+)}$
and $G_U^{(+)}$ is in the Hamiltonian. $G^{(+)}$
and $G_U^{(+)}$  are full Green-functions with Hamiltonians $H(t)$ and
$H^U(t)$, respectively. The original T-matrix expressed in terms of
$G_U^{(+)}$ is given by
\beqa
&&\langle f| T | i \rangle\,=\,
-i\epsilon^2\int_{-\infty}^\infty d
t_1 d t_2\langle f(t_1)| U(t_1) G_U^{(+)}(t_1,t_2) \nn
&&\times U^\dagger(t_2)
|i(t_2)\rangle.
\eeqa
On the other hand the T-matrix with the transformed Hamiltonian $H_U$
is given by
\beqa
&&\langle f| T^U | i \rangle\,=\,
-i\epsilon^2\int_{-\infty}^\infty d
t_1 d t_2\langle f(t_1)| G_U^{(+)}(t_1,t_2) 
|i(t_2)\rangle.\quad\quad
\eeqa
We want to show that in the limit $\epsilon\to 0+$ the two T-matrices
are equal:
\beqa
\lim_{\epsilon\to 0+}\langle f| T | i \rangle&=&
\lim_{\epsilon\to 0+}\langle f| T^U | i \rangle,
\eeqa
if the unitary transformation satisfies
\beqa
\lim_{\epsilon\to 0+} \epsilon\int d t_2 G_0^{(+)}(t_1-t_2)\delta
U^\dagger(t_2) |i(t_2)\rangle&=&0,\nn
\lim_{\epsilon\to 0+} \epsilon\int d t_1 \langle f(t_1)|\delta U(t_1) G_0^{(+)}(t_1-t_2)&=&0.\label{deltaUCondition}
\eeqa
To do this we use Eq.~(\ref{full:green:function:U:recursion:rel}) to
rewrite the T-matrix into
\beqa
&&\langle f| T | i \rangle\,=\, -i\epsilon^2\int_{-\infty}^\infty d
t_1 d t_2\langle f(t_1)| U(t_1) \bigg[G_0^{(+)}(t_1-t_2) \nn
&& + \int d t_3 G_0^{(+)}(t_1-t_3) V^U(t_3) G_0^{(+)}(t_3-t_2) \nn
&&+ \int d t_3 d t_4 G_0^{(+)}(t_1-t_3) V^U(t_3) G_U^{(+)}(t_3,t_4)
V^U(t_4)\nn
&& G_0^{(+)}(t_4-t_2)\bigg] U^\dagger(t_2)
|i(t_2)\rangle.
\eeqa
A time integration over the time $t_2$ gives
\beqa
&&\int d t_2 G_0^{(+)}(t_1,t_2) |i(t_2)\rangle=\int d t_2
\theta(t_1-t_2)\nn
&&\times e^{-i(H_0-i\epsilon)(t_1-t_2)} e^{-i E_i t_2}|i\rangle =
|i\rangle  e^{-i E_i t_1}\nn
&&\times\int_{-\infty}^{t_1} d t_2
e^{-\epsilon(t_1-t_2)}=|i(t_1)\rangle \frac{1}{\epsilon},
\eeqa
Similarly, for the integration over $t_1$ we get
\beqa
&&\int d t_1\langle f(t_1)| G_0^{(+)}(t_1-t_2) =\langle f(t_1)| \frac{1}{\epsilon}.
\eeqa
Due to condition in Eq.~(\ref{deltaUCondition}) all contributions
proportional to $\delta U(t)$ do not generate poles in small
$\epsilon$ and for this reason vanish in the limit $\epsilon\to
0+$.\footnote{More elegant proof for time-independent unitary
  transformations can be found in~\cite{Epelbaum:1998na}. Note that we did not consider here nonperturbative
  effects but concentrated only on perturbation theory. More
  rigorous proof for time-independent unitary transformations can be
  found in~\cite{Polyzou:2010eq}.} 
\section{Transfer matrix with time-dependent interaction}
\label{app:transfer_matrix}
In this appendix we derive a transfer matrix for time-dependent
interaction. We start with the Schr\"odinger equation
\beqa
\bigg(i\,\frac{\partial}{\partial t}-H_0\bigg)|\Psi(t)\rangle&=& V(t) |\Psi(t)\rangle.
\eeqa
To keep the notation short we again change the notation for a free retarded Green function which satisfies
\beqa
\bigg(i\frac{\partial}{\partial t}-H_0\bigg)
G_+(t-t^\prime\,)&=&\delta (t-t^\prime)
\eeqa
with a constraint
\beqa
G_+(t-t^\prime\,)&=&0,\quad {\rm for}\quad t<t^\prime\;.
\eeqa
The solution is given by
\beqa
G_+(t-t^\prime\,)&=&-i\,\theta(t-t^\prime\,)e^{-i\,(H_0 - i\,\epsilon)\,(t-t^\prime)}.
\eeqa
The solution of the Schr\"oding equation can be written as
\beqa
|\Psi^+(t)\rangle&=&|\phi(t)\rangle + \int_{-\infty}^{\infty} d
t^\prime \, G_+(t-t^\prime\,)
V(t^\prime\,)|\Psi^+(t^\prime\,)\rangle,\nn
&&\label{solution_schroedinger}
\eeqa
with the state $|\phi(t)\rangle$ satisfying a free Schr\"odinger
equation
\beqa
\bigg(i\,\frac{\partial}{\partial t} - H_0\bigg)|\phi(t)\rangle &=&0.
\eeqa
We take now a Fourier transform of Eq.~(\ref{solution_schroedinger})
by multiplying both sides by $e^{i E t}$ and integrating over time:
\beqa
&&|\tilde\Psi^+(E)\rangle\,=\,|\tilde\phi(E)\rangle\nn
&&+ \int_{-\infty}^{\infty} d t\, d
t^\prime \,e^{i\, E t}G_+(t - t^\prime\,) V(t^\prime) |\Psi^+(t^\prime\,)\rangle.\label{fourier_tr:schroedinger_eq}
\eeqa
For the Fourier transform we use
\beqa
|\tilde\Psi^+(E)\rangle&=&\int_{-\infty}^{\infty} d t \, e^{i E t}
|\Psi^+(t)\rangle, \nn 
|\tilde\phi(E)\rangle&=& \int_{-\infty}^{\infty} d t \, e^{i E t} |\phi(t)\rangle.
\eeqa
To simplify Eq.~(\ref{fourier_tr:schroedinger_eq}) we Fourier-transform the
free Green-function $G_+$ 
\beqa
\tilde G_+(E)&=&\int_{-\infty}^\infty d t\, e^{i E t} G_+(t)\,=\,
-i\,\int_{0}^\infty d t \,e^{i (E - H_0+ i\,\epsilon)
  t}\nn
&=&\frac{1}{E - H_0 + i\,\epsilon}.
\eeqa
The backward Fourier transformations are given by
\beqa
G_+(t)&=&\int \frac{d E}{2\pi} e^{-i\, E t}\tilde G_+(E), \nn
|\Psi^+(t)\rangle&=&\int \frac{d E}{2\pi} e^{-i\, E t}|\tilde \Psi^+(E)\rangle\;.\label{fourier_back}
\eeqa
Using Eq.~(\ref{fourier_back}) we can rewrite
Eq.~(\ref{fourier_tr:schroedinger_eq}) into
\beqa
|\tilde\Psi^+(E)\rangle&=&|\tilde\phi(E)\rangle \nn
&+& \tilde G_+(E)\int
\frac{d E^\prime}{2\pi}\tilde V(E -
E^\prime\,)|\tilde\Psi^+(E^\prime)\rangle\label{tmatrix:prepare}\\
&=&|\tilde\phi(E)\rangle + \tilde G_+(E) \int \frac{d E^\prime}{2\pi} T(E,E^\prime)|\phi(E^\prime)\rangle.\label{tmatrix:definition}\quad\quad
\eeqa
Eq.~({\ref{tmatrix:definition}}) defines the transition matrix in the
presence of an external source. It satisfies an integral equation
\beqa
&&T(E,E^\prime)\,=\,\tilde V(E-E^\prime) \nn
&+& \int \frac{d E^{\prime\prime}}{2\pi} \tilde
V(E-E^{\prime\prime}\,)\tilde G_+(E^{\prime\prime}\,) T(E^{\prime\prime}, E^\prime\,).\label{transition_int_eq}
\eeqa
which is also equivalent to
\beqa
T(E,E^\prime)&=&\tilde V(E-E^\prime) \nn
&+& \int \frac{d E^{\prime\prime}}{2\pi} T(E, E^{\prime\prime}\,) \tilde G_+(E^{\prime\prime}\,)\tilde
V(E^{\prime\prime}-E^\prime).\label{transition_int_eq:transpose}\quad\quad
\eeqa
Rewriting Eq.~(\ref{transition_int_eq:transpose}) into 
\beqa
\tilde V(E-E^\prime\,)&=&\int \frac{d E^{\prime\prime}}{2\pi} T(E, E^{\prime\prime}\,)\bigg(2\pi\delta
(E^{\prime\prime}-E^\prime\,) \nn
&-&
\tilde G_+(E^{\prime\prime})\tilde
V(E^{\prime\prime}-E^\prime\,)\bigg),
\label{inverse_tmatrix}
\eeqa
and replacing $\tilde V(E-E^\prime\,)$ in Eq.~(\ref{tmatrix:prepare})
by the left-hand side of Eq.~(\ref{inverse_tmatrix}) we get
\beqa
&&|\tilde\Psi^+(E)\rangle\,=\,|\tilde \phi(E)\rangle  + \tilde G_+(E)\int
\frac{d E^\prime}{2\pi} \int \frac{d E^{\prime\prime}}{2\pi}
T(E,E^{\prime\prime})\nn
&\times&\bigg(2\pi \delta(E^{\prime\prime}-E^\prime\,) - \tilde G_+(E^{\prime\prime})\tilde
V(E^{\prime\prime}-E^\prime\,)\bigg)|\tilde \Psi^+(E^\prime\,)\rangle.\nonumber
\eeqa
Using 
\beqa
|\tilde \phi(E^{\prime\prime}\,)\rangle&=&\int \frac{d E^{\prime}}{2\pi}\bigg(2\pi \delta
  (E^{\prime\prime}-E^\prime\,)\nn
&-& \tilde G_+(E^{\prime\prime}\,)\tilde
  V(E^{\prime\prime}-E^\prime)\bigg)|\tilde\Psi^+(E^\prime\,)\rangle, 
\eeqa
we indeed get Eq.~(\ref{tmatrix:definition}). 

\noindent We decompose now the interaction into the time-dependent and
the time-independent parts
\beqa
V(E-E^\prime\,)&=&2\pi \delta(E-E^\prime\,) v + v_5(E-E^\prime\,),
\eeqa
where $v$ denotes the time-independent nuclear force and $v_5$ an axial
vector source dependent interaction which vanishes when the
axial-vector source is switched off. Similarly, we can decompose the
transition matrix
\beqa
T(E,E^\prime\,)&=&2\pi \delta(E-E^\prime\,) t(E) + t_5(E,E^\prime\,)\;.
\eeqa
The off-shell transfer matrix $t(E)$ satisfies the Lippmann-Schwinger equation
\beqa
t(E)&=&v + v\, \tilde G_+(E) \,t(E).
\eeqa 
$t_5$ transfer matrix satisfies
\beqa
t_5(E,E^\prime\,)&=&%v_5(E-E^\prime\,) + \int \frac{d
%  E^{\prime\prime}}{2\pi}\bigg(2\pi \delta(E-E^{\prime\prime}\,) v(E)
%+ v_5(E,E^{\prime\prime}\,)\bigg)\tilde
%G_+(E^{\prime\prime})\bigg(2\pi\delta(E^{\prime\prime}-E^\prime\,)
%t(E^\prime\,) + t_5(E^{\prime\prime},E^\prime\,)\bigg)\nn
%&=&
v_5(E-E^\prime\,) + v\tilde G_+(E) t_5(E, E^\prime\,) \nn
&+& v_5
(E-E^\prime\,)\tilde G_+(E^\prime\,) t(E^\prime) \nn
&+& \int \frac{d
  E^{\prime\prime}}{2\pi} v_5(E-E^{\prime\prime}\,) \tilde
G_+(E^{\prime\prime}\,) t_5(E^{\prime\prime},E^\prime\,).\quad\quad
\eeqa
The last term contributes only to the processes with at least two external
sources, so we neglect this term here. If we only concentrate on
one external source coupling we get 
\beqa
&&\big(1-v \tilde G_+(E)\big) t_5(E, E^\prime\,) \,=\,
v_5(E-E^\prime\,)\nn
&\times&\big(1+ \tilde G_+(E^\prime)
t(E^\prime)\big)=v_5(E-E^\prime\,)\nn
&\times&\big(1+\tilde G_+(E^\prime) v\big(1-\tilde
G_+(E^\prime) v\big)^{-1}\big)\,=\,v_5(E-E^\prime\,)\nn
&\times& \bigg(1-\tilde G_+(E^\prime) v+\tilde
G_+(E^\prime) v\bigg)\big(1-\tilde G_+(E^\prime) v\big)^{-1}\nn
&&=v_5(E-E^\prime\,) \big(1-\tilde G_+(E^\prime) v\big)^{-1}
\eeqa
So we get
\beqa
t_5(E, E^\prime\,)&=& \big(1-v \tilde
G_+(E)\big)^{-1}v_5(E-E^\prime\,) 
\big(1-\tilde G_+(E^\prime) v\big)^{-1}.\nn
&&
\eeqa
We see that all energies which appear on the left-hand side of $v_5$ are
the final state energies and all energies which appear on the right-hand
side of $v_5$ are the initial state energies.

\end{appendix}
%
% BibTeX users please use
% \bibliographystyle{}
% \bibliography{}

\end{document}